\documentclass[a4paper,11pt]{article}
\usepackage{jheppub}
\usepackage[T1]{fontenc}
\usepackage[utf8]{inputenc}
\usepackage{mathtools}
\usepackage{graphicx}
\usepackage{rotating}
\usepackage{dsfont}
\usepackage{xcolor,colortbl}
\usepackage{simplewick}
\usepackage{pgfplots}
\pgfplotsset{compat=1.12}
\usepgfplotslibrary{fillbetween}
\usepackage{pgfplotstable}
\usepackage{gnuplot-lua-tikz}
\usepackage{tabularx}
\usepackage{booktabs}
\usepackage{multirow}
\usepackage{longtable}
\usepackage{caption}
\newcommand{\deriv}{\mathrm{d}}
\newcommand{\tick}{$\times$}
\DeclareMathOperator{\tr}{tr}
\DeclareMathOperator{\diag}{diag}
\DeclareMathOperator{\arccosh}{arccosh}

\pgfplotstableread{data/lo_extrap.dat}\loextrap
\pgfplotstableread{data/nlo_extrap.dat}\nloextrap
\pgfplotstableread{data/nlo_extrap_chPT.dat}\nloextrapchpt
\pgfplotstableread{data/nlo_extrap_scale.dat}\nloextrapscale
\pgfplotstableread{data/ensdata_table.dat}\ensdata
\pgfplotstableread{data/graph_plots.dat}\graphplotdata
\pgfplotstableread[col sep=comma,string replace*={ }{}]{data/decs_cmp.dat}\deccmpdata
\pgfplotstableread[col sep=comma,string replace*={ }{}]{data/gluon_cmp.dat}\gluoncmpdata

\newcommand{\graphplotwidth}{5cm}
\newcommand{\summarygraphplotwidth}{3.3cm}
\newcommand{\gluongraphplotwidth}{3.1cm}
\newcommand{\etaerrplot}{%
  \begin{tikzpicture}[trim axis left,trim axis right]
    \begin{axis}[y=-\baselineskip,
        scale only axis,
        width             = \graphplotwidth,
        enlarge y limits  = {abs=0.5},
        axis y line*      = middle,
        y axis line style = dashed,
        ytick             = \empty,
        xtick distance    = 0.01,
        extra x tick style = {},
        axis x line*      = bottom,
        xmin              = 1.115,
        xmax              = 1.19
      ]
      \addplot+[only marks, blue, thick, mark=*, solid, skip coords between index={0}{1}, skip coords between index={6}{7}][error bars/.cd,x dir=both, x explicit, error mark=none] table [x=metatimessqrt8t0physmean,y expr=\coordindex,x error minus=metatimessqrt8t0physmerr, x error plus=metatimessqrt8t0physperr]{\graphplotdata};
      \addplot+[only marks, blue, thick, mark=o, solid, skip coords between index={1}{17}][error bars/.cd,x dir=both, x explicit, error mark=none] table [x=metatimessqrt8t0physmean,y expr=\coordindex,x error minus=metatimessqrt8t0physmerr, x error plus=metatimessqrt8t0physperr]{\graphplotdata};
      \addplot+[only marks, blue, thick, mark=star, mark size=3,solid, skip coords between index={0}{6}, skip coords between index={7}{17}][error bars/.cd,x dir=both, x explicit, error mark=none] table [x=metatimessqrt8t0physmean,y expr=\coordindex,x error minus=metatimessqrt8t0physmerr, x error plus=metatimessqrt8t0physperr]{\graphplotdata};
        \addplot+[mark=none, update limits=false,thick,color = black, name path = phys] coordinates {(1.14565572658228, -1) (1.14565572658228, \numberofrows-0.5)};
        \addplot+[mark=none, update limits=false, dashed, color = black,name path=physerrmin] coordinates {(1.12901181772152, -1) (1.12901181772152, \numberofrows-0.5)};
        \addplot+[mark=none, update limits=false, dashed, color=black,name path=physerrmax] coordinates {(1.1623, -1) (1.1623, \numberofrows-0.5)};
        \addplot+[mark=none, update limits=false, fill=black!10] fill between [of=physerrmin and physerrmax];
        \addplot+[mark=none, update limits=false, dashed, color = red,name path=errmin] coordinates {(1.1518, 0.5) (1.1518, \numberofrows-0.5)};
        \addplot+[mark=none, update limits=false, dashed, color=red,name path=errmax] coordinates {(1.1783, 0.5) (1.1783, \numberofrows-0.5)};
        \addplot+[mark=none, update limits=false, solid, color=red,name path=cv] coordinates {(1.168, 0.5) (1.168, \numberofrows-0.5)};
        \addplot+[mark=none, update limits=false, fill=red, opacity = 0.15] fill between [of=errmin and errmax];
    \end{axis}
  \end{tikzpicture}%
}
\newcommand{\etaprerrplot}{%
  \begin{tikzpicture}[trim axis left,trim axis right]
    \begin{axis}[y=-\baselineskip,
        scale only axis,
        width             = \graphplotwidth,
        enlarge y limits  = {abs=0.5},
        axis y line*      = middle,
        y axis line style = dashed,
        ytick             = \empty,
        xtick distance    = 0.02,
        axis x line*      = bottom,
        xmin              = 1.9,
        xmax              = 2.05
      ]
      \addplot+[only marks, blue, thick, mark=*, solid, skip coords between index={0}{1}, skip coords between index={6}{7}][error bars/.cd,x dir=both, x explicit, error mark=none] table [x=metaprimetimessqrt8t0physmean,y expr=\coordindex,x error minus=metaprimetimessqrt8t0physmerr, x error plus=metaprimetimessqrt8t0physperr]{\graphplotdata};
      \addplot+[only marks, blue, thick, mark=o, solid, skip coords between index={1}{17}][error bars/.cd,x dir=both, x explicit, error mark=none] table [x=metaprimetimessqrt8t0physmean,y expr=\coordindex,x error minus=metaprimetimessqrt8t0physmerr, x error plus=metaprimetimessqrt8t0physperr]{\graphplotdata};
      \addplot+[only marks, blue, thick, mark=star, mark size=3,solid, skip coords between index={0}{6}, skip coords between index={7}{17}][error bars/.cd,x dir=both, x explicit, error mark=none] table [x=metaprimetimessqrt8t0physmean,y expr=\coordindex,x error minus=metaprimetimessqrt8t0physmerr, x error plus=metaprimetimessqrt8t0physperr]{\graphplotdata};

      \addplot+[mark=none, update limits=false,color = black, thick] coordinates {(2.00285134177215, -1) (2.00285134177215, \numberofrows-0.5)};
      \addplot+[mark=none, update limits=false, dashed, color = black,name path=errmin] coordinates {(1.9737542278481, -1) (1.9737542278481, \numberofrows-0.5)};
      \addplot+[mark=none, update limits=false, dashed, color=black,name path=errmax] coordinates {(2.0319484556962, -1) (2.0319484556962, \numberofrows-0.5)};
      \addplot+[mark=none, update limits=false, fill=black!10] fill between [of=errmin and errmax];
      \addplot+[mark=none, update limits=false, dashed, color = red,name path=errmin] coordinates {(1.9425, 0.5) (1.9425, \numberofrows-0.5)};
      \addplot+[mark=none, update limits=false, dashed, color=red,name path=errmax] coordinates {(2.0141, 0.5) (2.0141, \numberofrows-0.5)};
      \addplot+[mark=none, update limits=false, solid, color=red,name path=cv] coordinates {(1.958, 0.5) (1.958, \numberofrows-0.5)};
      \addplot+[mark=none, update limits=false, fill=red, opacity = 0.15] fill between [of=errmin and errmax];
    \end{axis}
  \end{tikzpicture}%
}

\newcommand{\fetaocteterrplot}{%
  \begin{tikzpicture}[trim axis left,trim axis right]
    \begin{axis}[y=-\baselineskip,
        scale only axis,
        width             = \graphplotwidth,
        enlarge y limits  = {abs=0.5},
        axis y line*      = middle,
        y axis line style = dashed,
        ytick             = \empty,
        xtick distance    = 0.005,
        xticklabel style = {/pgf/number format/precision   = 3, /pgf/number format/fixed},
        axis x line*      = bottom
      ]
      \addplot+[only marks, blue, thick, mark=*, solid, skip coords between index={0}{1}, skip coords between index={6}{7}][error bars/.cd,x dir=both, x explicit, error mark=none] table [x=FAoctet0timessqrt8t0physmean,y expr=\coordindex,x error minus=FAoctet0timessqrt8t0physmerr, x error plus=FAoctet0timessqrt8t0physperr]{\graphplotdata};
      \addplot+[only marks, blue, thick, mark=o, solid, skip coords between index={1}{17}][error bars/.cd,x dir=both, x explicit, error mark=none] table  [x=FAoctet0timessqrt8t0physmean,y expr=\coordindex,x error minus=FAoctet0timessqrt8t0physmerr, x error plus=FAoctet0timessqrt8t0physperr]{\graphplotdata};
      \addplot+[only marks, blue, thick, mark=star, mark size=3,solid, skip coords between index={0}{6}, skip coords between index={7}{17}][error bars/.cd,x dir=both, x explicit, error mark=none] table [x=FAoctet0timessqrt8t0physmean,y expr=\coordindex,x error minus=FAoctet0timessqrt8t0physmerr, x error plus=FAoctet0timessqrt8t0physperr]{\graphplotdata};
      \addplot+[mark=none, update limits=false, dashed, color = red,name path=errmin] coordinates {(0.21749, 0.5) (0.21749, \numberofrows-0.5)};
      \addplot+[mark=none, update limits=false, dashed, color=red,name path=errmax] coordinates {(0.22457, 0.5) (0.22457, \numberofrows-0.5)};
      \addplot+[mark=none, update limits=false, solid, color=red,name path=cv] coordinates {(0.2219, 0.5) (0.2219, \numberofrows-0.5)};
      \addplot+[mark=none, update limits=false, fill=red, opacity = 0.15] fill between [of=errmin and errmax];
    \end{axis}
  \end{tikzpicture}%
}
\newcommand{\fetaprocteterrplot}{%
  \begin{tikzpicture}[trim axis left,trim axis right]
    \begin{axis}[y=-\baselineskip,
        scale only axis,
        width             = \graphplotwidth,
        enlarge y limits  = {abs=0.5},
        axis y line*      = left,
        y axis line style = dashed,
        ytick             = \empty,
        xlabel = \empty,
        xtick distance    = 0.01,
        xticklabel style = {/pgf/number format/precision   = 3, /pgf/number format/fixed},
        axis x line*      = bottom,
        xmin=-0.12, xmax=-0.07
      ]
      \addplot+[only marks, blue, thick, mark=*, solid, skip coords between index={0}{1}, skip coords between index={6}{7}][error bars/.cd,x dir=both, x explicit, error mark=none] table [x=FAoctet1timessqrt8t0physmean,y expr=\coordindex,x error minus=FAoctet1timessqrt8t0physmerr, x error plus=FAoctet1timessqrt8t0physperr]{\graphplotdata};
      \addplot+[only marks, blue, thick, mark=o, solid, skip coords between index={1}{17}][error bars/.cd,x dir=both, x explicit, error mark=none] table  [x=FAoctet1timessqrt8t0physmean,y expr=\coordindex,x error minus=FAoctet1timessqrt8t0physmerr, x error plus=FAoctet1timessqrt8t0physperr]{\graphplotdata};
      \addplot+[only marks, blue, thick, mark=star, mark size=3,solid, skip coords between index={0}{6}, skip coords between index={7}{17}][error bars/.cd,x dir=both, x explicit, error mark=none] table [x=FAoctet1timessqrt8t0physmean,y expr=\coordindex,x error minus=FAoctet1timessqrt8t0physmerr, x error plus=FAoctet1timessqrt8t0physperr]{\graphplotdata};
      \addplot+[mark=none, update limits=false, dashed, color = red,name path=errmin] coordinates {(-0.1068, 0.5) (-0.1068, \numberofrows-0.5)};
      \addplot+[mark=none, update limits=false, dashed, color=red,name path=errmax] coordinates {(-0.08332, 0.5) (-0.08332, \numberofrows-0.5)};
      \addplot+[mark=none, update limits=false, solid, color=red,name path=cv] coordinates {(-0.0939, 0.5) (-0.0939, \numberofrows-0.5)};
      \addplot+[mark=none, update limits=false, fill=red, opacity = 0.15] fill between [of=errmin and errmax];

    \end{axis}
  \end{tikzpicture}%
}
\newcommand{\fetasingleterrplot}{%
  \begin{tikzpicture}[trim axis left,trim axis right]
    \begin{axis}[y=-\baselineskip,
        scale only axis,
        width             = \graphplotwidth,
        enlarge y limits  = {abs=0.5},
        axis y line*      = middle,
        y axis line style = dashed,
        ytick             = \empty,
        xtick distance    = 0.005,
        x tick label style = {%
          /pgf/number format/fixed,
          /pgf/number format/fixed zerofill
        },
        scaled x ticks=false,
        axis x line*      = bottom,
      ]
      \addplot+[only marks, blue, thick, mark=*, solid, skip coords between index={0}{1}, skip coords between index={6}{7}][error bars/.cd,x dir=both, x explicit, error mark=none] table [x=FAsinglet0timessqrt8t0physmean,y expr=\coordindex,x error minus=FAsinglet0timessqrt8t0physmerr, x error plus=FAsinglet0timessqrt8t0physperr]{\graphplotdata};
      \addplot+[only marks, blue, thick, mark=o, solid, skip coords between index={1}{17}][error bars/.cd,x dir=both, x explicit, error mark=none] table  [x=FAsinglet0timessqrt8t0physmean,y expr=\coordindex,x error minus=FAsinglet0timessqrt8t0physmerr, x error plus=FAsinglet0timessqrt8t0physperr]{\graphplotdata};
      \addplot+[only marks, blue, thick, mark=star, mark size=3,solid, skip coords between index={0}{6}, skip coords between index={7}{17}][error bars/.cd,x dir=both, x explicit, error mark=none] table [x=FAsinglet0timessqrt8t0physmean,y expr=\coordindex,x error minus=FAsinglet0timessqrt8t0physmerr, x error plus=FAsinglet0timessqrt8t0physperr]{\graphplotdata};
      \addplot+[mark=none, update limits=false, dashed, color = red,name path=errmin] coordinates {(0.01865, 0.5) (0.01865, \numberofrows-0.5)};
      \addplot+[mark=none, update limits=false, dashed, color=red,name path=errmax] coordinates {(0.02874, 0.5) (0.02874, \numberofrows-0.5)};
      \addplot+[mark=none, update limits=false, solid, color=red,name path=cv] coordinates {(0.0224, 0.5) (0.0224, \numberofrows-0.5)};
      \addplot+[mark=none, update limits=false, fill=red, opacity = 0.15] fill between [of=errmin and errmax];
    \end{axis}
  \end{tikzpicture}%
}
\newcommand{\fetaprsingleterrplot}{%
  \begin{tikzpicture}[trim axis left,trim axis right]
    \begin{axis}[y=-\baselineskip,
        scale only axis,
        width             = \graphplotwidth,
        enlarge y limits  = {abs=0.5},
        axis y line*      = middle,
        y axis line style = dashed,
        ytick             = \empty,
        xtick distance    = 0.005,
        xticklabel style = {/pgf/number format/precision   = 3, /pgf/number format/fixed},
        axis x line*      = bottom
      ]
      \addplot+[only marks, blue, thick, mark=*, solid, skip coords between index={0}{1}, skip coords between index={6}{7}][error bars/.cd,x dir=both, x explicit, error mark=none] table [x=FAsinglet1timessqrt8t0physmean,y expr=\coordindex,x error minus=FAsinglet1timessqrt8t0physmerr, x error plus=FAsinglet1timessqrt8t0physperr]{\graphplotdata};
      \addplot+[only marks, blue, thick, mark=o, solid, skip coords between index={1}{17}][error bars/.cd,x dir=both, x explicit, error mark=none] table  [x=FAsinglet1timessqrt8t0physmean,y expr=\coordindex,x error minus=FAsinglet1timessqrt8t0physmerr, x error plus=FAsinglet1timessqrt8t0physperr]{\graphplotdata};
      \addplot+[only marks, blue, thick, mark=star, mark size=3,solid, skip coords between index={0}{6}, skip coords between index={7}{17}][error bars/.cd,x dir=both, x explicit, error mark=none] table [x=FAsinglet1timessqrt8t0physmean,y expr=\coordindex,x error minus=FAsinglet1timessqrt8t0physmerr, x error plus=FAsinglet1timessqrt8t0physperr]{\graphplotdata};
      \addplot+[mark=none, update limits=false, dashed, color = red,name path=errmin] coordinates {(0.19057, 0.5) (0.19057, \numberofrows-0.5)};
      \addplot+[mark=none, update limits=false, dashed, color=red,name path=errmax] coordinates {(0.2028, 0.5) (0.2028, \numberofrows-0.5)};
      \addplot+[mark=none, update limits=false, solid, color=red,name path=cv] coordinates {(0.1974, 0.5) (0.1974, \numberofrows-0.5)};
      \addplot+[mark=none, update limits=false, fill=red, opacity = 0.15] fill between [of=errmin and errmax];
    \end{axis}
  \end{tikzpicture}%
}
\newcommand{\flcmpplot}{%
  \begin{tikzpicture}[trim axis left,trim axis right]
    \begin{axis}[y=-\baselineskip,
        scale only axis,
        width             = \summarygraphplotwidth,
        enlarge y limits  = {abs=0.5},
        axis y line*      = middle,
        y axis line style = dashed,
        ytick             = \empty,
        xtick distance    = 5,
        extra x tick style = {},
        axis x line*      = bottom,
        xmin              = 80.1,
        xmax              = 110,
      ]
      \addplot+[only marks, blue, thick, mark=*, solid,skip coords between index={0}{1},skip coords between index={16}{20}][error bars/.cd,x dir=both, x explicit, error mark=none] table [x=Fq,y expr=\coordindex,x error minus=Fqerr, x error plus=Fqerr]{\deccmpdata};
      \addplot+[only marks, blue, thick, mark=star, mark size=3, solid,skip coords between index={0}{17}][error bars/.cd,x dir=both, x explicit, error mark=none] table [x=Fq,y expr=\numexpr\coordindex-1\relax,x error minus=Fqerr, x error plus=Fqerr]{\deccmpdata};
      \addplot+[opacity=0, mark=*, skip coords between index={1}{20}][error bars/.cd,x dir=both, x explicit, error mark=none] table [x=Fq,y expr=\coordindex,x error minus=F0err, x error plus=F0err]{\deccmpdata};
    \end{axis}
  \end{tikzpicture}%
}
\newcommand{\fscmpplot}{%
  \begin{tikzpicture}[trim axis left,trim axis right]
    \begin{axis}[y=-\baselineskip,
        scale only axis,
        width             = \summarygraphplotwidth,
        enlarge y limits  = {abs=0.5},
        axis y line*      = middle,
        y axis line style = dashed,
        ytick             = \empty,
        xtick distance    = 10,
        extra x tick style = {},
        axis x line*      = bottom,
        xmin              = 110,
        xmax              = 165,
      ]
      \addplot+[only marks, blue, thick, mark=*, solid,skip coords between index={0}{1},skip coords between index={16}{20}][error bars/.cd,x dir=both, x explicit, error mark=none] table [x=Fs,y expr=\coordindex,x error minus=Fserr, x error plus=Fserr]{\deccmpdata};
      \addplot+[only marks, blue, thick, mark=star, mark size=3, solid,skip coords between index={0}{17}][error bars/.cd,x dir=both, x explicit, error mark=none] table [x=Fs,y expr=\numexpr\coordindex-1\relax,x error minus=Fserr, x error plus=Fserr]{\deccmpdata};
      \addplot+[opacity=0, mark=*, skip coords between index={1}{20}][error bars/.cd,x dir=both, x explicit, error mark=none] table [x=Fs,y expr=\coordindex,x error minus=F0err, x error plus=F0err]{\deccmpdata};
    \end{axis}
  \end{tikzpicture}%
}
\newcommand{\philcmpplot}{%
  \begin{tikzpicture}[trim axis left,trim axis right]
    \begin{axis}[y=-\baselineskip,
        scale only axis,
        width             = \summarygraphplotwidth,
        enlarge y limits  = {abs=0.5},
        axis y line*      = middle,
        y axis line style = dashed,
        ytick             = \empty,
        xtick distance    = 2,
        extra x tick style = {},
        axis x line*      = bottom,
        xmin              = 31,
        xmax              = 46,
      ]
      \addplot+[only marks, blue, thick, mark=*, solid,skip coords between index={0}{1},skip coords between index={16}{20}][error bars/.cd,x dir=both, x explicit, error mark=none] table [x=phil,y expr=\coordindex,x error minus=philerr, x error plus=philerr]{\deccmpdata};
      \addplot+[only marks, blue, thick, mark=star,mark size=3, solid,skip coords between index={0}{17}][error bars/.cd,x dir=both, x explicit, error mark=none] table [x=phil,y expr=\numexpr\coordindex-1\relax,x error minus=philerr, x error plus=philerr]{\deccmpdata};
      \addplot+[opacity=0, mark=*, skip coords between index={2}{20}][error bars/.cd,x dir=both, x explicit, error mark=none] table [x=phil,y expr=\coordindex,x error minus=F0err, x error plus=F0err]{\deccmpdata};
    \end{axis}
  \end{tikzpicture}%
}
\newcommand{\phiscmpplot}{%
  \begin{tikzpicture}[trim axis left,trim axis right]
    \begin{axis}[y=-\baselineskip,
        scale only axis,
        width             = \summarygraphplotwidth,
        enlarge y limits  = {abs=0.5},
        axis y line*      = middle,
        y axis line style = dashed,
        ytick             = \empty,
        xtick distance    = 2,
        extra x tick style = {},
        axis x line*      = bottom,
        xmin              = 31,
        xmax              = 44,
      ]
      \addplot+[only marks, blue, thick, mark=*, solid,skip coords between index={0}{1},skip coords between index={16}{20}][error bars/.cd,x dir=both, x explicit, error mark=none] table [x=phis,y expr=\coordindex,x error minus=phiserr, x error plus=phiserr]{\deccmpdata};
      \addplot+[only marks, blue, thick, mark=star,mark size=3, solid,skip coords between index={0}{17}][error bars/.cd,x dir=both, x explicit, error mark=none] table [x=phis,y expr=\numexpr\coordindex-1\relax,x error minus=phiserr, x error plus=phiserr]{\deccmpdata};
      \addplot+[opacity=0, mark=*, skip coords between index={1}{20}][error bars/.cd,x dir=both, x explicit, error mark=none] table [x=phis,y expr=\coordindex,x error minus=F0err, x error plus=F0err]{\deccmpdata};
    \end{axis}
  \end{tikzpicture}%
}
\newcommand{\foctcmpplot}{%
  \begin{tikzpicture}[trim axis left,trim axis right]
    \begin{axis}[y=-\baselineskip,
        scale only axis,
        width             = \summarygraphplotwidth,
        enlarge y limits  = {abs=0.5},
        axis y line*      = middle,
        y axis line style = dashed,
        ytick             = \empty,
        xtick distance    = 10,
        extra x tick style = {},
        axis x line*      = bottom,
        xmin              = 101.0,
        xmax              = 144.9,
      ]
      \addplot+[only marks, blue, thick, mark=*, solid,skip coords between index={17}{20}][error bars/.cd,x dir=both, x explicit, error mark=none] table [x=F8,y expr=\coordindex,x error minus=F8err, x error plus=F8err]{\deccmpdata};
      \addplot+[only marks, blue, thick, mark=star,mark size=3, solid,skip coords between index={0}{17}][error bars/.cd,x dir=both, x explicit, error mark=none] table [x=F8,y expr=\coordindex,x error minus=F8err, x error plus=F8err]{\deccmpdata};
    \end{axis}
  \end{tikzpicture}%
}
\newcommand{\fsinglcmpplot}{%
  \begin{tikzpicture}[trim axis left,trim axis right]
    \begin{axis}[y=-\baselineskip,
        scale only axis,
        width             = \summarygraphplotwidth,
        enlarge y limits  = {abs=0.5},
        axis y line*      = middle,
        y axis line style = dashed,
        ytick             = \empty,
        xtick distance    = 10,
        extra x tick style = {},
        axis x line*      = bottom,
        xmin              = 90,
        xmax              = 124,
      ]
      \addplot+[only marks, blue, thick, mark=*, solid,skip coords between index ={17}{20}][error bars/.cd,x dir=both, x explicit, error mark=none] table [x=F0,y expr=\coordindex,x error minus=F0err, x error plus=F0err]{\deccmpdata};
      \addplot+[only marks, blue, thick, mark=star, mark size=3, solid,skip coords between index={0}{17}][error bars/.cd,x dir=both, x explicit, error mark=none] table [x=F0,y expr=\coordindex,x error minus=F0err, x error plus=F0err]{\deccmpdata};
    \end{axis}
  \end{tikzpicture}%
}
\newcommand{\phioctcmpplot}{%
  \begin{tikzpicture}[trim axis left,trim axis right]
    \begin{axis}[y=-\baselineskip,
        scale only axis,
        width             = \summarygraphplotwidth,
        enlarge y limits  = {abs=0.5},
        axis y line*      = left,
        y axis line style = dashed,
        ytick             = \empty,
        xtick distance    = 10,
        extra x tick style = {},
        axis x line*      = bottom,
        xmin              = -35,
        xmax              = -10,
      ]
      \addplot+[only marks, blue, thick, mark=*, solid, skip coords between index={17}{20}][error bars/.cd,x dir=both, x explicit, error mark=none] table [x=theta8,y expr=\coordindex,x error minus=theta8err, x error plus=theta8err]{\deccmpdata};
      \addplot+[only marks, blue, thick, mark=star,mark size=3, solid,skip coords between index={0}{17}][error bars/.cd,x dir=both, x explicit, error mark=none] table [x=theta8,y expr=\coordindex,x error minus=theta8err, x error plus=theta8err]{\deccmpdata};
    \end{axis}
  \end{tikzpicture}%
}
\newcommand{\phisinglcmpplot}{%
  \begin{tikzpicture}[trim axis left,trim axis right]
    \begin{axis}[y=-\baselineskip,
        scale only axis,
        width             = \summarygraphplotwidth,
        enlarge y limits  = {abs=0.5},
        axis y line*      = left,
        y axis line style = dashed,
        ytick             = \empty,
        xtick distance    = 10,
        extra x tick style = {},
        axis x line*      = bottom,
        xmin              = -37,
        xmax              = 2,
      ]
      \addplot+[only marks, blue, thick, mark=*, solid,skip coords between index={17}{20}][error bars/.cd,x dir=both, x explicit, error mark=none] table [x=theta0,y expr=\coordindex,x error minus=theta0err, x error plus=theta0err]{\deccmpdata};
      \addplot+[only marks, blue, thick, mark=star, mark size=3, solid,skip coords between index={0}{17}][error bars/.cd,x dir=both, x explicit, error mark=none] table [x=theta0,y expr=\coordindex,x error minus=theta0err, x error plus=theta0err]{\deccmpdata};
    \end{axis}
  \end{tikzpicture}%
}
\newcommand{\fsingllamcmpplot}{%
  \begin{tikzpicture}[trim axis left,trim axis right]
    \begin{axis}[y=-\baselineskip,
        scale only axis,
        width             = \summarygraphplotwidth,
        enlarge y limits  = {abs=0.5},
        axis y line*      = middle,
        y axis line style = dashed,
        ytick             = \empty,
        xtick distance    = 10,
        extra x tick style = {},
        axis x line*      = bottom,
        xmin              = 75,
        xmax              = 124,
      ]
      \addplot+[only marks, blue, thick, mark=*, solid,skip coords between index={17}{20}][error bars/.cd,x dir=both, x explicit, error mark=none] table [x=F0Lam,y expr=\coordindex,x error minus=F0Lamerr, x error plus=F0Lamerr]{\deccmpdata};
      \addplot+[only marks, blue, thick, mark=star,mark size=3, solid,skip coords between index={0}{17}][error bars/.cd,x dir=both, x explicit, error mark=none] table [x=F0Lam,y expr=\coordindex,x error minus=F0Lamerr, x error plus=F0Lamerr]{\deccmpdata};
    \end{axis}
  \end{tikzpicture}%
  }
\newcommand{\lamcmpplot}{%
  \begin{tikzpicture}[trim axis left,trim axis right]
    \begin{axis}[y=-\baselineskip,
        scale only axis,
        width             = \summarygraphplotwidth,
        enlarge y limits  = {abs=0.5},
        axis y line*      = middle,
        y axis line style = dashed,
        ytick             = \empty,
        xtick distance    = 0.2,
        extra x tick style = {},
        axis x line*      = bottom,
        xmin              = -0.3,
        xmax              = 0.7,
      ]
      \addplot+[only marks, blue, thick, mark=*, solid,skip coords between index={17}{20}][error bars/.cd,x dir=both, x explicit, error mark=none] table [x=Lam,y expr=\coordindex,x error minus=Lamerr, x error plus=Lamerr]{\deccmpdata};
      \addplot+[only marks, blue, thick, mark=star,mark size=3, solid,skip coords between index={0}{17}][error bars/.cd,x dir=both, x explicit, error mark=none] table [x=Lam,y expr=\coordindex,x error minus=Lamerr, x error plus=Lamerr]{\deccmpdata};
      \addplot+[opacity=0, mark=*, skip coords between index={1}{18}][error bars/.cd,x dir=both, x explicit, error mark=none] table [x=Lam,y expr=\coordindex,x error minus=Lamerr, x error plus=Lamerr]{\deccmpdata};
    \end{axis}
  \end{tikzpicture}%
}
\newcommand{\aetacmpplot}{%
  \begin{tikzpicture}[trim axis left,trim axis right]
    \begin{axis}[y=-\baselineskip,
        scale only axis,
        width             = \gluongraphplotwidth,
        enlarge y limits  = {abs=0.5},
        axis y line*      = middle,
        y axis line style = dashed,
        ytick             = \empty,
        xtick distance    = 0.01,
        extra x tick style = {},
        axis x line*      = bottom,
        x tick label style = {%
          /pgf/number format/fixed,
          /pgf/number format/fixed zerofill
        },
        scaled x ticks=false,
        xmin              = 0.0,
        xmax              = 0.030,
      ]
      \addplot+[only marks, blue, thick, mark=*, solid,skip coords between index={7}{10}][error bars/.cd,x dir=both, x explicit, error mark=none] table [x=aeta,y expr=\coordindex,x error minus=aetaerr, x error plus=aetaerr]{\gluoncmpdata};
      \addplot+[only marks, blue, thick, mark=star, mark size=3, solid,skip coords between index={0}{7}][error bars/.cd,x dir=both, x explicit, error mark=none, error mark=none] table [x=aeta,y expr=\coordindex,x error minus=aetaerr, x error plus=aetaerr]{\gluoncmpdata};
    \end{axis}
  \end{tikzpicture}%
  }
\newcommand{\aetaprimecmpplot}{%
  \begin{tikzpicture}[trim axis left,trim axis right]
    \begin{axis}[y=-\baselineskip,
        scale only axis,
        width             = \gluongraphplotwidth,
        enlarge y limits  = {abs=0.5},
        axis y line*      = middle,
        y axis line style = dashed,
        ytick             = \empty,
        xtick distance    = 0.02,
        x tick label style = {%
          /pgf/number format/fixed,
          /pgf/number format/fixed zerofill
        },
        scaled x ticks=false,
        extra x tick style = {},
        axis x line*      = bottom,
        xmin              = 0.0,
        xmax              = 0.08,
      ]
      \addplot+[only marks, blue, thick, mark=*, solid,skip coords between index={7}{10}][error bars/.cd,x dir=both, x explicit, error mark=none] table [x=aetaprime,y expr=\coordindex,x error minus=aetaprimeerr, x error plus=aetaprimeerr]{\gluoncmpdata};
      \addplot+[only marks, blue, thick, mark=star, mark size=3, solid,skip coords between index={0}{7}][error bars/.cd,x dir=both, x explicit, error mark=none] table [x=aetaprime,y expr=\coordindex,x error minus=aetaprimeerr, x error plus=aetaprimeerr]{\gluoncmpdata};
    \end{axis}
  \end{tikzpicture}%
}

\begin{document}
\pgfplotstablegetrowsof{\graphplotdata}
\let\numberofrows=\pgfplotsretval
\pgfplotstablegetrowsof{\deccmpdata}
\let\numberofcmprows=\pgfplotsretval
\def\numberofflavourcmprows{\numexpr\numberofcmprows-1\relax} 
\pgfplotstablegetrowsof{\gluoncmpdata}
\let\numberofgluoncmprows=\pgfplotsretval

\title{Masses and decay constants of the $\eta$ and $\eta^\prime$ mesons from lattice QCD}
\collaborationImg{\includegraphics[]{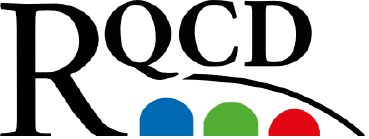}}
\author[a,b]{Gunnar~S.~Bali,}
\author[a]{Vladimir~Braun,}
\author[a]{Sara~Collins,}
\author[a]{Andreas~Sch\"afer}
\author[a]{and Jakob~Simeth}
    \emailAdd{gunnar.bali@ur.de}
    \emailAdd{sara.collins@ur.de}
    \emailAdd{jakob.simeth@ur.de}
    \affiliation[a]{Institut f\"ur Theoretische Physik, Universit\"at Regensburg, D-93040 Regensburg, Germany}
    \affiliation[b]{Tata Institute of Fundamental Research, Homi Bhabha Road, Mumbai 400005, India}

\abstract{We determine the masses, the singlet and octet decay
  constants as well as the anomalous matrix elements of the $\eta$ and
  $\eta^\prime$ mesons in $N_f=2+1$ QCD\@. The results are obtained using
  twenty-one CLS ensembles of non-perturbatively improved Wilson
  fermions that span four lattice spacings ranging from $a\approx
  0.086\,$fm down to $a\approx 0.050\,$fm.  The pion masses vary from
  $M_{\pi}=420\,$MeV to $126\,$MeV and the spatial lattice extents
  $L_s$ are such that $L_sM_\pi\gtrsim 4$, avoiding significant finite
  volume effects. The quark mass dependence of the data is tightly
  constrained by employing two trajectories in the quark mass plane,
  enabling a thorough investigation of U($3$) large-$N_c$ chiral
  perturbation theory (ChPT).  The continuum limit extrapolated data
  turn out to be reasonably well described by the next-to-leading
  order ChPT parametrization and the respective low energy constants are
  determined. The data are shown to be consistent with the singlet axial Ward identity and,
  for the first time, also the matrix elements with the topological
  charge density are computed. We also derive the corresponding
  next-to-leading order large-$N_{c}$ ChPT formulae.  We find $F^8 =
  115.0(2.8)~\text{MeV}$, $\theta_{8} = -25.8(2.3)^{\circ}$,
  $\theta_0 = -8.1(1.8)^{\circ}$ and, in the $\overline{\mathrm{MS}}$
  scheme for $N_f=3$,
  $F^{0}(\mu = 2\,\mathrm{GeV}) = 100.1(3.0)~\text{MeV}$, where
  the decay constants read $F^8_\eta=F^8\cos \theta_8$,
  $F^8_{\eta^\prime}=F^8\sin \theta_8$, $F^0_\eta=-F^0\sin \theta_0$ and
  $F^0_{\eta^\prime}=F^0\cos \theta_0$.  For
  the gluonic matrix elements, we obtain $a_{\eta}(\mu =
  2\,\mathrm{GeV}) = 0.0170(10)\,\mathrm{GeV}^{3}$ and
  $a_{\eta^{\prime}}(\mu = 2\,\mathrm{GeV}) =
  0.0381(84)\,\mathrm{GeV}^{3}$, where statistical and
  all systematic errors are added in quadrature.}
\maketitle
\section{Introduction\label{sec:intro}}
The physics of the pseudoscalar $\eta$ and $\eta^\prime$ mesons is a
fascinating area at the crossroads of many themes in hadron and
particle physics. In the exact flavour SU($3$) limit the $\eta$ meson
is part of the flavour-octet whereas the $\eta^\prime$ is a pure flavour-singlet
state whose properties are intimately related to the celebrated axial
anomaly~\cite{Witten:1978bc,Veneziano:1979ec}. However, it is known
empirically that the SU($3$) breaking effects are large and have a
non-trivial structure. These effects are usually described in terms of
a mixing scheme that considers the physical $\eta$ and $\eta^\prime$
mesons as superpositions of fundamental (e.g., flavour-singlet and
-octet) fields in a low energy effective theory. Modern
phenomenological analyses of $\eta$-$\eta^\prime$ mixing are largely based on
large-$N_c$ chiral perturbation theory
(ChPT)~\cite{DiVecchia:1980yfw,Kawarabayashi:1980dp,DiVecchia:1980vpx,Leutwyler:1997yr}
which allows for a unified treatment of the $\eta^{(\prime)}$-mesons
together with the pseudo-Goldstone octet of the lightest
pseudoscalars.  When combined with dispersion relations, this approach
provides a quantitative description of a large variety of
$\eta^{(\prime)}$ decays and low energy $\eta^{(\prime)}$ production processes,
see, e.g.,~\cite{Gan:2020aco} and references therein.

Flavour-singlet pseudoscalar mesons are a very active
area of research. Chiral dynamics has been very successful in
describing low energy pion and kaon reactions and it is natural to
attempt to generalize this to include the $\eta^{(\prime)}$
sector. Theoretical developments as well as new high-precision
experimental measurements are needed to advance this agenda.  The
study of $\eta$ and $\eta^\prime$ mesons also provides an interesting window
to beyond-the-standard-model (BSM) physics.  BSM searches in
$\eta^{(\prime)}$ decays have initially been related to flavour-conserving
tests of discrete symmetries, however, other interesting searches have
been proposed~\cite{Gan:2020aco}, e.g., for axion-like
particles. Corresponding efforts are ongoing or planned in many
experimental facilities.  A less well explored area is 
the production of $\eta$ and $\eta^\prime$ in hard processes, e.g., in $B$-meson
decays or in two-photon reactions $\gamma^\ast\gamma\to \eta^{(\prime)}$,
which constitute part of the Belle~II research
programme~\cite{Kou:2018nap}. It is not obvious whether and to what
extent the approaches based on low energy effective field theory
provide an adequate
description of such processes, that are dominated by meson wave
functions at small transverse separations, referred to as light-cone
distribution amplitudes (LCDAs). However, this is usually taken as a
working hypothesis in phenomenological applications,
see, e.g.,~\cite{Beneke:2002jn,Kroll:2002nt,Ball:2007hb,Agaev:2014wna}.  One
important issue in this context is that $\eta^{(\prime)}$ mesons, in
contrast to the pion, can contain a significant admixture of a
two-gluon component at low scales, i.e.\ a comparably large two-gluon LCDA\@.
Several different reactions were considered in an effort to extract or
at least to constrain these contributions, see,
e.g.,~\cite{Kroll:2002nt,Blechman:2004vc,Harland-Lang:2013ncy}.
However, no definite conclusion can be drawn as yet.

Lattice simulations of properties of flavour non-singlet
pseudoscalar mesons are quite advanced. Recently, continuum
limit results at physical quark masses of the first two
Gegenbauer moments of the twist-two pion and kaon LCDAs~\cite{Bali:2019dqc}
were obtained, pion transition form factors
calculated~\cite{Gerardin:2019vio} and exploratory studies of higher
twist LCDA parameters undertaken~\cite{Bali:2018spj}. However, only
a few investigations of matrix elements involving the $\eta$ and $\eta^\prime$
mesons exist to-date~\cite{Bali:2014pva,Ottnad:2017bjt}.
These are
technically demanding due to the computationally expensive
evaluation of disconnected contributions and the coupling to
the topological charge, which results in large autocorrelation times and
requires long time series to enable an adequate sampling of the topological
sectors. Moreover, the extraction of ground state
properties from correlation functions with a noise over signal
ratio that increases rapidly in Euclidean time
requires optimized methods.

Despite these challenges, steady progress has been made in computing the
masses of the $\eta$ and $\eta^\prime$ mesons, starting in the quenched
approximation~\cite{Kuramashi:1994aj,Venkataraman:1997xi,Bardeen:2000cz},
and continuing with $N_f =2$ mass-degenerate dynamical light
quarks~\cite{Venkataraman:1997xi,Struckmann:2000bt,McNeile:2000hf,Bali:2001gk,Lesk:2002gd,Hashimoto_2008,Jansen:2008wv,Sun:2017ipk,Dimopoulos:2018xkm}.
In the latter case only one $\eta$ meson exists, which
is a pure singlet state, and no flavour mixing
takes place. More realistic simulations of nature require
an additional strange
quark ($N_f
= 2+1$)~\cite{Christ:2010dd,Dudek:2011tt,Gregory:2011sg,Bali:2014pva} (see
also~\cite{Fukaya:2015ara} for a different attempt using correlators of the
topological charge density).
More recently, $N_f = 2+1+1$ results~\cite{Ottnad:2012fv,Michael:2013gka,Ottnad:2017bjt}
employing the twisted-mass fermion formulation,
using several ensembles and lattice spacings, enabled a physical point
extrapolation.
In~\cite{Kotov:2019dby} the $\eta^\prime$ mass was calculated
at non-zero temperature from topological charge density correlators.
In~\cite{Ottnad:2017bjt} also pseudoscalar matrix
elements were determined. Relating these to the four decay constants of the
$\eta/\eta^\prime$ system enabled the first lattice determination to a
precision that is on par with phenomenological studies. Another
lattice computation of these matrix elements
was carried out in $N_f=2+1$, in the context of
a calculation of the semileptonic decay form factors
$D_s\to\eta,\eta^\prime$, albeit
only on two ensembles at a single lattice spacing~\cite{Bali:2014pva}.

Here, we compute the masses, decay constants and gluonic
anomaly matrix elements of the $\eta$ and
$\eta^\prime$ mesons in $N_f=2+1$ QCD. The simulations are carried
out on twenty-one ensembles generated by the CLS (Coordinated
Lattice Simulations) initiative~\cite{Bruno:2014jqa,fixeds},
employing non-perturbatively improved Sheikholeslami-Wilson fermions.
Most of the ensembles have open boundary conditions in time, ensuring that the
topological sectors are sampled uniformly. We employ
pion masses that range from the SU($3$) symmetric point at $M_\pi
\approx 420\,\mathrm{MeV}$ down to just below the physical pion mass.
The corresponding kaon masses are tuned so that the ensembles fall
onto two distinct trajectories --- one at a constant average quark
mass and the other at an approximately constant strange quark
mass. Both lines intersect close to the physical point, aiding the
chiral interpolation. The continuum extrapolation is carried out
utilizing four lattice spacings ranging from $a\approx
0.086\,\mathrm{fm}$ down to $a\approx 0.050\,\mathrm{fm}$.

Using a combination of all-to-all propagator methods and various
interpolating operators, we obtain a correlation matrix
between pairs of interpolators as a function of
the Euclidean time separation. In addition, for each of three local currents,
axialvector, pseudoscalar and gluonic, we compute the
vector of correlators with the interpolating operators.
From these the meson masses and matrix elements are extracted
via a fit, utilizing a generalized effective mass method that
we introduce. A comparison
is made with the result of the usual generalized eigenvalue
problem (GEVP) method. The matrix elements are renormalized and
partially order $a$ improved. Remaining order $a$ terms with (as yet)
unknown coefficients as well as order $a^2$ corrections are
included in parametrizations of the lattice spacing and
quark mass dependence. Regarding the continuum limit, we are able to
simultaneously parameterize all data in terms of the six
low energy constants (LECs) of large-$N_c$ U($3$) ChPT at
next-to-leading order (NLO). Systematic errors are estimated
by carrying out a multitude of fits and also by excluding
data points at large average quark masses.

The masses of the mesons are found to be in agreement with
experiment and we determine the two decay constants
(singlet and octet) for each meson as well as the
LECs of large-$N_c$ ChPT. Due to the axial anomaly, some LECs depend
on the QCD renormalization scale, as do the singlet decay constants,
and we present our results in the $\overline{\mathrm{MS}}$ scheme
at different scales.
The gluonic matrix elements of the $\eta$ and $\eta^\prime$ mesons
and the topological susceptibility
are found to be affected by sizeable lattice cut-off effects.
In the continuum limit the topological susceptibility
is well described by the leading order (LO)
ChPT expectation. This only depends on the pion decay constant
in the chiral limit, which we obtain from our global
fits to the axial matrix elements. The continuum limit $\eta$
and $\eta^\prime$ matrix elements
satisfy the flavour-octet and flavour-singlet axial Ward
identities (AWIs) and the pseudoscalar fermionic matrix elements
are determined too.
We address implications on the phenomenology of hard processes
for the example of the $\gamma\gamma^\ast\rightarrow\eta^{(\prime)}$
transition form factors as well as radiative decays
of the $J/\psi$ to an $\eta^{(\prime)}$ meson.

The conventions and main results of this article can be found in
the following places. In sec.~\ref{sec:definitions} we detail the flavour
mixing schemes and define our normalization conventions and some
of our notations.
The basic ChPT formulae can be found in
sec.~\ref{sec:chpt}, and fig.~\ref{fig:nlodecfit} illustrates
the main results on the masses and decay constants.
Section~\ref{sec:gluonic} details the determination of the
gluonic matrix elements. The continuum limit
results are collected and discussed in sec.~\ref{sec:summary}.
Our main results on the masses, decay constants and gluonic
matrix elements are summarized
in sec.~\ref{sec:conclude}.

The remainder of the article is organized as follows.
In sec.~\ref{sec:lattice} we discuss the simulation parameters,
the lattice observables and the computational techniques used.
We then move on to introduce the generalized effective mass method,
that we employ to extract the masses in sec.~\ref{sec:extract}.
In that section we also explain the lattice evaluation of the
necessary matrix elements and our statistical analysis.
In sec.~\ref{sec:physpointextrap} we discuss the renormalization
and improvement of the lattice results and the parametrizations
of the quark mass and lattice spacing dependence.
We then determine the meson masses, decay constants and LECs,
and estimate their systematics. As mentioned above,
in sec.~\ref{sec:gluonic} we
determine the quark mass dependence of the topological
susceptibility and the gluonic matrix elements of the
$\eta$ and $\eta^\prime$ mesons. The results
are parameterized in terms of NLO large-$N_c$ U($3$) ChPT.
In sec.~\ref{sec:summary}, apart from discussing the continuum
limit results,
we also address implications on the phenomenology of hard processes
for the example of the
$\gamma\gamma^\ast\rightarrow\eta^{(\prime)}$  transition form factors,
see sec.~\ref{sec:tff}.

The article is augmented by several appendices: in app.~\ref{sec:loopfit},
we present the parametrization of the pseudoscalar loop contributions
that appear at next-to-next-to-leading order (NNLO). In app.~\ref{sec:nloglue}
we derive the dependence of the gluonic and pseudoscalar fermionic matrix elements
on the pion and kaon masses in terms
of the six NLO large-$N_c$ U($3$) LECs.
In app.~\ref{sec:lo} we show the result of a LO fit
to the $\eta$ and $\eta^\prime$ masses that we omitted from the main body of
the paper for brevity. In app.~\ref{sec:contlimitparams} we collect
the values of the parameters accompanying the lattice
spacing effects for seventeen different
fit forms. Finally, in app.~\ref{app:decayresults} we list
our results for the decay constants and mixing angles in different flavour
bases and at different scales, both in units of the gradient flow
scale $t_0$ and in physical units. A corresponding table with gluonic matrix
elements can be found in sec.~\ref{sec:gluonic}.

\section{Definitions, conventions and octet/singlet mixing\label{sec:definitions}}
The couplings between axialvector currents and
pseudoscalar states, that are also known as meson decay constants,
play a crucial role in the description of low energy physics with
$\eta$ and $\eta^\prime$ mesons.
Different normalization conventions are used throughout the literature.
Here we introduce the conventions for the local currents, decay constants
and interpolating operators that we employ in this article.
We also briefly address what is often referred to as
$\eta$-$\eta^\prime$ mixing or pseudoscalar octet/singlet mixing.

Everywhere we will assume $N_f=3$ quark flavours with the masses
$m_u$, $m_d$ and $m_s$, where we ignore the mass difference between the up
and the down quark as well as electromagnetic effects, i.e.\ we set
$m_{\ell}=m_u=m_d$. For convenience, sometimes we write out the
dependence on $N_f$ and $N_c$. Formulae without these factors
always refer to the case $N_f=N_c=3$.

We define the U($N_f$) generators $t^a$ with $t^0=\mathds{1}/\sqrt{2N_f}$,
where in the $N_f=3$
case $t^a=\lambda^a/2$ for $a\neq 0$ and
$\lambda^a$ are the eight
Gell-Mann matrices. This normalization
of the generators corresponds to $\tr\left(t^at^b\right)=\frac12\delta^{ab}$.
Using $\overline{\psi}=(\bar{u},\bar{d},\bar{s})$, we can introduce
local currents as
\begin{align}
  J^a=\overline{\psi}t^a\Gamma_J\psi,
\end{align}
where the Dirac matrix structure $\Gamma_J$ defines the current $J$,
for instance $A_{\mu}^a=\overline{\psi}t^a\gamma_{\mu}\gamma_5\psi$ and
$P^a=\overline{\psi}t^a\gamma_5\psi$.
It is also useful to define currents for individual quark flavours:
\begin{align}
  \label{eq:qcurrent}
  J^q=\bar{q}\Gamma_Jq,\quad J^{\ell}=\frac{1}{\sqrt{2}}\left(J^u+J^d\right),
\end{align}
where $q\in\{u,d,s\}$. The flavour-diagonal
singlet, triplet and octet currents $J^0$, $J^3$ and $J^8$
can be written as linear combinations
of the above flavour basis currents:
\begin{align}
  J^3&=\frac{1}{2}\left(J^u-J^d\right),\label{eq:flavour1}\\
  J^8&=\frac{1}{\sqrt{12}}\left(J^u+J^d-2J^s\right)=\frac{1}{\sqrt{6}}J^{\ell}
  -\frac{1}{\sqrt{3}}J^s,\\
  J^0&=\frac{1}{\sqrt{6}}\left(J^u+J^d+J^s\right)=\frac{1}{\sqrt{3}}J^{\ell}+\frac{1}{\sqrt{6}}J^s.\label{eq:flavour3}
\end{align}

We define the decay constants $F^a_{\mathcal{M}}$ of a pseudoscalar
meson $\mathcal{M}$:
\begin{align}
  \label{eq:defDecConst}
  \left\langle \Omega\left|A_{\mu}^a\right| \mathcal{M}(p)\right\rangle=iF^a_{\mathcal{M}}p_{\mu},
\end{align}
where $|\Omega\rangle$ denotes the vacuum and
$|\mathcal{M}(p)\rangle$ a meson state with four-momentum $p$.
Below we will
often refer to the latter as $|n\rangle$ where $n\in\mathbb{N}_0$
labels the $\eta$ meson ($n=0$), the $\eta^\prime$ meson ($n=1$)
and their excitations ($n\geq 2$).
Note that in the above normalization, at the physical point,
$F^3_{\pi^0}\approx 92\,$MeV. For $m_u=m_d=m_s$, these
conventions correspond to
$f_{\pi}=\sqrt{2}F_{\pi^0}^3=\sqrt{2}F_{\eta^8}^8=2F^1_{\pi^+}=-2i
F^2_{\pi^+}$, where $if_{\pi}p_{\mu}=\langle 0|
\bar{d}\gamma_{\mu}\gamma_5u|\pi^+(p)\rangle$.
The usual normalization for the analogous decay constants
in the light/strange flavour basis reads:
\begin{align}
  \label{eq:decayflavour}
  \left\langle \Omega\left|A_{\mu}^q\right| \mathcal{M}(p)\right\rangle=i\sqrt{2}F^q_{\mathcal{M}}p_{\mu}.
\end{align}  

Since we only consider the isospin symmetric limit,
the triplet couplings $F^3_{\mathcal{M}}$ vanish identically
for the $\eta$ and the $\eta^\prime$ mesons. Note that the singlet
couplings $F^0_{\mathcal{M}}$
in the standard $\overline{\mathrm{MS}}$ scheme acquire an anomalous
dimension due to the axial U($1$) anomaly~\cite{Kodaira:1979pa},
\begin{align}
  \mu^2 \frac{\deriv}{\deriv\mu^2} F^{0}_{\mathcal{M}}(\mu) & = -\frac{N_f}{2}
  \frac{\alpha_s^2}{\pi^2} F^{0}_{\mathcal{M}}(\mu) +\mathcal{O}(\alpha_s^3),
\end{align}
whereas the octet couplings $F^8_{\mathcal{M}}$ are scale independent.
This simplifies the scale evolution in the octet/singlet basis relative
to that in the light/strange flavour basis.

We introduce pseudoscalar interpolating operators
$\mathcal{P}_{\vec{p}}^a$ that have the same flavour structure as the local
currents $P^a$.  However, these can be non-local in space (due to
quark smearing) and are projected onto a definite spatial momentum
$\vec{p}$ in order to destroy physical states with matching quantum
numbers. For instance $\mathcal{P}_{\bf{0}}^{3\dagger}|\Omega\rangle$
gives a linear combination of the $|\pi^0\rangle$ and its excitations
(at rest).  The relation between the interpolators $\mathcal{P}^a$ and
their flavour basis counterparts $\mathcal{P}^q$ are so as to preserve
the normalization of the respective quark model wave functions,
resulting in normalizations that differ by factors $\sqrt{2}$ from the
relations between $P^a$ and $P^q$ of
eqs.~\eqref{eq:flavour1}--\eqref{eq:flavour3}.  For the pseudoscalar
singlet and octet interpolators this means that
\begin{align}
  \mathcal{P}^8&=\frac{1}{\sqrt{6}}\left(\mathcal{P}^u+\mathcal{P}^d-2\mathcal{P}^s\right)=\frac{1}{\sqrt{3}}\mathcal{P}^{\ell}
  -\sqrt{\frac{2}{3}}\mathcal{P}^s,\\
  \mathcal{P}^0&=\frac{1}{\sqrt{3}}\left(\mathcal{P}^u+\mathcal{P}^d+\mathcal{P}^s\right)=\sqrt{\frac{2}{3}}\mathcal{P}^{\ell}+\frac{1}{\sqrt{3}}\mathcal{P}^s.
\end{align}
The factor $\sqrt{2}$ within eq.~\eqref{eq:decayflavour} stems from
enforcing the above relations also between the decay constants in the octet/singlet
basis and the light/strange flavour basis:
\begin{align}
  \label{eq:convertdecay}
  \begin{pmatrix}
    F^8_{\mathcal{M}}\\F^0_{\mathcal{M}}
  \end{pmatrix}
  =\frac{1}{\sqrt{3}}
  \begin{pmatrix}
      1&-\sqrt{2}\\
      \sqrt{2}&1
  \end{pmatrix}
  \begin{pmatrix}
    F^{\ell}_{\mathcal{M}}\\F^{s}_{\mathcal{M}}
  \end{pmatrix}.
\end{align}

An interesting limit, that we also simulate here, is that of exact
SU($3$) flavour symmetry ($m_s=m_{\ell}$). In this limit
the $\eta$ meson is part of a flavour-octet, 
$\eta = \eta_8$, and $\eta^\prime$ is a flavour-singlet,
$\eta^\prime = \eta_0$. This means that the interpolator
$\mathcal{P}^0$ has no overlap with the $\eta$ meson, i.e.\
$\langle\eta(p)|\mathcal{P}_{\vec{p}}^{0\dagger}|\Omega\rangle=0$,
and the interpolator $\mathcal{P}^8$ cannot create an
$\eta^\prime$ meson. In this limit, in terms of the decay constants,
$F^8_{\eta^\prime}=F^0_{\eta}=0$. However,
SU($3$) breaking corrections are known to be large and phenomenologically
significant.

In the chiral effective theory the $\eta^\prime$ meson can be included in the framework
of the $1/N_c$ expansion~\cite{DiVecchia:1980yfw,Kawarabayashi:1980dp,DiVecchia:1980vpx,Leutwyler:1997yr}. In this approach
the axial anomaly contributes an effective mass term to
the $\eta$-$\eta^\prime$ system. This affects not only the flavour-singlet
sector but also the
flavour-diagonal octet if SU($3$) symmetry is explicitly broken by a
quark mass difference. An additional off-diagonal
contribution to the kinetic term 
$\partial_\mu \eta_8 \partial^\mu \eta_0$ shows up at the
loop-level~\cite{Leutwyler:1997yr}. As a result, the relation of the 
physical $\eta$ and $\eta^\prime$ states to the octet and singlet fields
$\eta_8$ and $\eta_0$ in the chiral Lagrangian becomes more involved,
see, e.g.,~\cite{Feldmann:1998vh,Bickert:2016fgy,Gan:2020aco}.

In general, there are four decay constants, one octet and one singlet
decay constant each for the $\eta$ and for the $\eta^\prime$.
One can always parameterize these in terms of two fundamental
decay constants $F^8$ and $F^0$ and two mixing angles
$\theta_8$ and $\theta_0$:
\begin{align}
  \label{eq:octsingletanglerep}
\begin{pmatrix}
  F_\eta^8 & F_\eta^0\\   F_{\eta^\prime}^8 & F_{\eta^\prime}^0
\end{pmatrix}
=
\begin{pmatrix}
  F^8 \cos \theta_8 & - F^0 \sin \theta_0 \\   F^8 \sin \theta_8 & F^0\cos \theta_0 
\end{pmatrix}.
\end{align}
An analogous parametrization can also be introduced in the flavour basis:
\begin{align}
  \label{eq:flavouranglerep}
\begin{pmatrix}
  F_\eta^{\ell} & F_\eta^{s}\\   F_{\eta^\prime}^{\ell} & F_{\eta^\prime}^{s}
\end{pmatrix}
=
\begin{pmatrix}
  F^{\ell} \cos \phi_{\ell} & - F^s \sin \phi_s \\   F^\ell \sin \phi_\ell & F^s\cos \phi_s 
\end{pmatrix}.
\end{align}
Obviously, in the SU($3$) limit $F^8=F_\eta^8$, $F^0=F_{\eta^\prime}^0$ and
$\theta_0=\theta_8=0$ while no such simplification exists in the
flavour basis. Note that in the standard $\overline{\mathrm{MS}}$ scheme,
within the right hand sides of the above equations, only $F^8$, $\theta_8$ and
$\theta_0$ are scale independent, whereas $F^0$,
$F^{\ell}$, $F^s$, $\phi_\ell$ and $\phi_s$ all will depend on the
QCD renormalization scale.

The above two choices of basis are essentially equivalent and the rationale for
the popularity of the flavour scheme is that the difference between $\phi_\ell$
and $\phi_s$ (which is formally a $1/N_c$ effect) 
is small and compatible with zero in phenomenological extractions from
experimental data~\cite{Schechter:1992iz,Feldmann:1998vh}.
This feature may be related to the observation that the vector mesons
$\omega$ and $\phi$ are 
to a very good approximation pure $\bar u u + \bar d d$ and $\bar s s$
states, respectively,
and the same holds for the tensor mesons $f_2(1270)$ and $f_2'(1525)$.
The smallness
of flavour mixing in these cases is a manifestation of the phenomenologically
very successful Okubo-Zweig-Iizuka (OZI) rule.
If the axial U($1$) anomaly was the only new effect
in the pseudoscalar channels,
it may be natural to assume that physical states are related to the flavour
states by an orthogonal transformation
with a single mixing angle $\phi = \phi_\ell = \phi_s$~\cite{Schechter:1992iz}.
In this approximation the relation between
the two schemes simplifies to~\cite{Feldmann:1998vh,Gan:2020aco}
\begin{align}
  & (F^8)^2 = \frac13 (F^\ell)^2 + \frac23 (F^s)^2,\quad&
    (F^0)^2 = \frac23 (F^\ell)^2 + \frac13 (F^s)^2,\\
  & \theta_8 = \phi - \arctan\left(\frac{\sqrt{2}F^s}{F^\ell}\right),\quad&
  \theta_0 = \phi - \arctan\left(\frac{\sqrt{2} F^\ell}{F^s}\right).   \label{eq:convertdecayangles}
\end{align}

Within QCD, obviously, the above relations cannot
hold at arbitrary renormalization scales.
Moreover, other matrix elements, e.g., $\langle\Omega| P^\ell|\eta^{(\prime)}\rangle$ and
$\langle\Omega| P^s|\eta^{(\prime)}\rangle$, are in general not related by the same angles.
Our lattice QCD
calculation will enable us to check the extent of the validity
of the simple mixing picture and the range of applicability
of large-$N_c$ ChPT.

\section{Lattice computation}
\label{sec:lattice}
In this section we give details of the lattice setup and gauge ensembles
and outline the construction of correlation functions, using
the local currents and interpolators defined above.
We describe the methods for the efficient evaluation of the
resulting connected and disconnected quark line diagrams.

\subsection{Gauge ensembles\label{sec:ensembles}}
\begin{figure}
  \includegraphics[width=\linewidth]{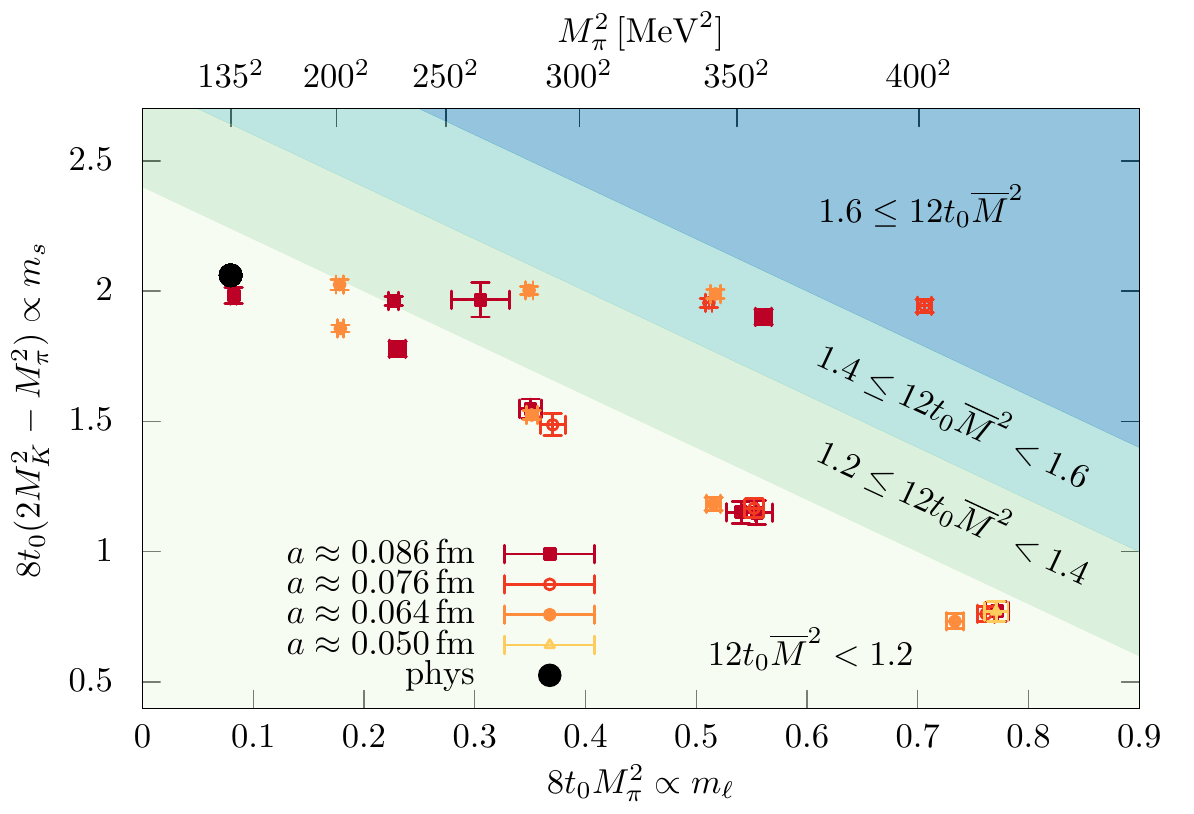}
  \caption{The positions of the analysed CLS ensembles in the quark mass plane.
    Two mass trajectories were realized that intersect approximately
    at the physical point (black circle). Along one trajectory the average
    quark mass is held fixed, along the other trajectory the strange
    quark mass is kept approximately constant.
    The symbols encode the four lattice spacings while the
    shaded areas indicate the values of the average squared pseudoscalar
    mass $\overline{M}\vphantom{M}^2$, see eq.~\eqref{eq:definembar}.
    At the physical point, $12 t_0^{\rm{ph}} \overline{M}\vphantom{M}^2 \approx 1.11$,
    where $(8t_0^{\rm{ph}})^{-1/2}\approx 0.475\,\mathrm{GeV}$~\cite{Bruno:2017lta}.\label{fig:quarkmasses}}
\end{figure}
We analyse gauge ensembles with $N_f = 2+1$ non-perturbatively improved
Wilson fermions on a L\"uscher-Weisz gauge background that were generated within
the CLS initiative~\cite{Bruno:2014jqa,fixeds}. To avoid topological
freezing at fine lattice spacings, most of the ensembles employ open boundary
conditions in time~\cite{Luscher:2011kk}. This breaks translational invariance in
that direction and introduces boundary effects, such that
measurements must be taken
in the bulk of the lattice. The fermion action ensures that hadron masses are
free of discretization effects that are linear in the lattice spacing, however, the operators also need to be
$\mathcal{O}(a)$ improved.
For the currents relevant for this study, we perform the substitutions~\cite{Bhattacharya:2005rb} ($a=1,\ldots,8$):
\begin{align}
  A^a_\mu &\mapsto A^a_\mu + ac_A\partial_{\mu}P^a,\\
  A^0_\mu &\mapsto A^0_\mu + a c_A^s\partial_{\mu}P^0,\\
  P^0    &\mapsto P^0 + a g_P \tr F_{\mu\nu}\widetilde{F}_{\mu\nu}
  =P^0 + a c_P^s\partial_\mu A_\mu^0,\label{eq:gpimp}
\end{align}
where we re-expressed the last equation using the singlet AWI
in the massless case.\footnote{Note that $c_P^s=-16\pi^2 g_P/\sqrt{6}$
  to leading order in $g^2$. This replacement
  will also affect the definitions of the mass dependent improvement coefficients
  $\bar{d}_P$ and $d_P$, relative to~\cite{Bhattacharya:2005rb}.}
In the chiral limit these replacements remove all $\mathcal{O}(a)$ effects.
The non-singlet pseudoscalar currents,
$P^a$, are already $\mathcal{O}(a)$ improved in this case. For non-vanishing
quark masses, additional mass dependent $\mathcal{O}(a)$ improvement terms
appear. These terms and the values of the associated improvement coefficients
as well as those of $c_A$ and the (unknown)
singlet coefficient $c_A^s$ and $c_P^s$ will be addressed
in secs.~\ref{sec:decaylocalmat} and~\ref{sec:contlimit} and used in the fits of sec.~\ref{sec:gluonmefermionic}.

\begin{sidewaystable}
\begin{center}
\begin{tabularx}{\textwidth}{c|c|cXXXXXXXX}
  \toprule
  $\beta$ & id & BC & trajectory & $L_t\times L_s^3/a^4$ & $M_\pi/{\rm MeV}$ & $M_K/{\rm MeV}$ & $L_s M_\pi$ & $12 t_0 \overline{M}\vphantom{M}^2$ & $8 t_0 \delta M^2$ & $t_0/a^2$\\
  \midrule
  3.4  & H101  & O & trM         & $96  \times 32^3$ & 420 & 420 & 5.85 & 1.158(16) & 0  & 2.854(8) \\
       & H102a & O & trM         & $96  \times 32^3$ & 352 & 442 & 4.90 & 1.116(16) & 0.611(14) & 2.884(9) \\
       & H102b & O & trM         & $96  \times 32^3$ & 356 & 441 & 4.96 & 1.130(18) & 0.595(11) & 2.879(9) \\
       & H105  & O & trM         & $96  \times 32^3$ & 279 & 465 & 3.89 & 1.125(16) & 1.199(12) & 2.892(6) \\
       & C101  & O & trM         & $96  \times 48^3$ & 220 & 472 & 4.60 & 1.120(12) & 1.548(14) & 2.918(4) \\
       & D150  & P & trM/ms      & $128 \times 64^3$ & 126 & 478 & 3.51 & 1.074(15) & 1.901(28) & 2.948(3) \\
       & H107  & O & ms          & $96  \times 32^3$ & 366 & 545 & 5.10 & 1.511(12) & 1.340(12) & 2.719(8) \\
       & H106  & O & ms          & $96  \times 32^3$ & 272 & 516 & 3.79 & 1.289(20) & 1.662(50) & 2.823(7) \\
       & C102  & O & ms          & $96  \times 48^3$ & 222 & 500 & 4.64 & 1.208(8)  & 1.736(11) & 2.868(5) \\\hline
  3.46 & B450  & P & trM         & $64  \times 32^3$ & 418 & 418 & 5.16 & 1.143(12) & 0 &         3.663(11)\\
       & S400  & O & trM         & $128 \times 32^3$ & 352 & 442 & 4.35 & 1.136(16) & 0.615(13) & 3.692(7) \\
       & N401  & O & trM         & $128 \times 48^3$ & 285 & 461 & 5.28 & 1.114(18) & 1.117(21) & 3.684(5) \\
       & B451  & P & ms          & $64  \times 32^3$ & 418 & 572 & 5.16 & 1.678(10) & 1.238(9)  & 3.426(7) \\
       & B452  & P & ms          & $64  \times 32^3$ & 350 & 544 & 4.32 & 1.489(8)  & 1.444(12) & 3.529(7) \\\hline
  3.55 & N202  & O & trM         & $128 \times 48^3$ & 411 & 411 & 6.43 & 1.100(12) & 0 & 5.165(14)  \\
       & N203  & O & trM         & $128 \times 48^3$ & 345 & 442 & 5.40 & 1.108(11) & 0.668(6) & 5.146(6) \\
       & N200  & O & trM         & $128 \times 48^3$ & 284 & 462 & 4.44 & 1.114(7)  & 1.174(9) & 5.160(7) \\
       & D200  & O & trM         & $128 \times 64^3$ & 201 & 480 & 4.19 & 1.107(6)  & 1.678(8) & 5.179(4) \\
       & N204  & O & ms          & $128 \times 48^3$ & 351 & 544 & 5.49 & 1.512(8)  & 1.472(6) & 4.947(8) \\
       & N201  & O & ms          & $128 \times 48^3$ & 285 & 523 & 4.46 & 1.351(7)  & 1.654(9) & 5.043(8) \\
       & D201  & O & ms          & $128 \times 64^3$ & 199 & 500 & 4.15 & 1.191(10) & 1.847(18) & 5.138(7)\\\hline
  3.7  & N300  & O & trM         & $128 \times 48^3$ & 422 & 422 & 5.10 & 1.156(16) & 0 & 8.576(21)\\
  \bottomrule
\end{tabularx}
\end{center}
\caption{Details of the CLS ensembles analysed in this study.
  Both open (``O'') and (anti-)periodic (``P'') boundary conditions
  in time are employed. The pion and kaon masses as
  well as the gradient flow scale $t_0/a^2$ are taken from~\cite{spectrum},
  while $12 t_0 \overline{M}\vphantom{M}^2$ and $8 t_0 \delta M^2$
  have been determined within this study (on a subset of the configurations
  analysed in~\cite{spectrum}). Ensembles H102a and H102b were generated with
  the same quark masses and lattice coupling but different simulation parameters
  and are therefore analysed separately.\label{tab:ensembles}
}
\end{sidewaystable}

We carry out our analysis on 21 distinct CLS ensembles that differ in terms of
the quark masses, volumes and
lattice spacings, see tab.~\ref{tab:ensembles}.
This enables us to control all sources of systematic error.
In the table we give dimensionless combinations involving
the average $\overline{M}\vphantom{M}^2$ and the difference $\delta M^2$
of the squared non-singlet pseudoscalar masses,
\begin{equation}
  \label{eq:definembar}
  12 t_0 \overline{M}\vphantom{M}^2 = 4 t_0\left(2 M_K^2 + M_\pi^2 \right),\quad
      8 t_0 \delta M^2 = 16 t_0 (M_K^2 - M_\pi^2),
\end{equation}
where $t_0$ denotes the gradient flow scale, introduced
in~\cite{Luscher:2010iy}.
The combinations $\overline{M}\vphantom{M}^2$ and $\delta M^2$  will be used in the
expressions for the
quark mass dependence of the masses and decay constants of the
$\eta$ and $\eta^\prime$ mesons in sec.~\ref{sec:chpt}.

The quark masses of the ensembles considered in this study follow two distinct
trajectories, see fig.~\ref{fig:quarkmasses}: along one trajectory the
average quark mass is kept constant~\cite{Bruno:2014jqa}, starting
from the $N_f=3$ symmetric point ($m_s=m_{\ell}$),
while along the other the renormalized strange quark mass
is held close to its physical value \cite{fixeds}.
The two trajectories intersect close to the physical point
where $12 t_0 \overline{M}\vphantom{M}^2 =
12 t_0^{\rm ph} {\overline{M}^{\rm ph}}^2 =  1.11$.
The pion masses, listed in tab.~\ref{tab:ensembles} along with the
kaon masses, range from
$422\,\rm{MeV}$ down to slightly below the physical mass.

We employ three lattice spacings, $a = 0.0859(12)\,\mathrm{fm}$, $a =
0.0761(10)\,\mathrm{fm}$ and $a =0.0643(9)\,\mathrm{fm}$, with multiple
pion and kaon masses, complemented by an
additional ensemble at a finer lattice spacing $a =
0.0497(7)\,\mathrm{fm}$~\cite{spectrum}.
Extrapolations to the physical point are performed using dimensionless
combinations with $t_0$, determined
on the same ensemble. After extrapolation, values for $t_0$ at the
physical point ($t_0^{\rm ph}$) and, for determinations of LECs, in the
SU($3$) chiral limit ($t_0^{\chi}$) are required.
Using the pion and kaon decay constants as input, \cite{Bruno:2016plf}
obtain
\begin{equation}
  \label{eq:t0ph}
  \left(8t_0^{\mathrm{ph}}\right)^{-1/2} = 475(6)\,\mathrm{MeV}.
\end{equation}
For $t_0^{\chi}=1.036(4)t_0^*$~\cite{spectrum},
using
$(8t_0^*)^{-1/2}=0.478(7)\,\mathrm{GeV}$~\cite{Bruno:2017gxd,Bruno:2016plf},
we find
\begin{equation}
  \label{eq:t0ch}
  \left(8t_0^{\chi}\right)^{-1/2}=470(7)\,\mathrm{MeV}.
\end{equation}

The line
along which we keep the sum of the quark masses constant intersects
both the physical point and a point where $m_s=m_{\ell}$. Along this
quark mass trajectory~(in the continuum limit) $\overline{M}\vphantom{M}^2$ as
well as $t_0$ are constant~\cite{Bar:2013ora} to NLO
in SU($3$) ChPT.  This motivates the definition of another
scale~\cite{Bruno:2016plf}, $t_0^*$, equating $12 t_0^*
{\overline{M}^*}^2 = 12 t_0^{\mathrm{ph}}
{\overline{M}^{\mathrm{ph}}}^2=1.11$. Naturally, this implies that
$t_0^* \approx t_0^{\mathrm{ph}}$, however, determining $t_0^*$ at
each lattice spacing does not require an extrapolation to the physical
point but just a small interpolation.  We extracted values for
$t_0^*/a^2$ (see tab.~\ref{tab:latspacings}) from a global fit to a
large number of CLS ensembles in~\cite{spectrum} and we use these
values to set the relative scale between the different lattice
spacings.

All spatial volumes are considerably larger than $(2\,\mathrm{fm})^3$ and for most
of the ensembles the dimensionless combination of the spatial lattice extent
$L_s$ and the mass of the pion $M_\pi$, $L_sM_\pi$, is larger
than four.\footnote{The only exceptions are H106 ($L_sM_\pi \approx 3.79$) and D150 ($L_sM_\pi \approx 3.51$).}
On these ensembles only very mild finite volume effects have been
observed for the non-singlet pseudoscalar masses and
decay constants~\cite{Bruno:2016plf,spectrum}. Given the
larger errors in the $\eta$-$\eta^\prime$ system compared to the pion,
we are confident that for our volumes such effects can be neglected.

\begin{table*}
  \begin{center}
    \begin{tabularx}{0.6\textwidth}{l|XXX}
      \toprule
      $\beta$ & $t_0^*/a^2$ & $a/\mathrm{fm}$ & $a^{-1}/\mathrm{GeV}$\\
      \midrule
      3.4  & 2.888(8)  & 0.0859(12) & 2.296(33) \\
      3.46 & 3.686(11) & 0.0761(11) & 2.594(38) \\
      3.55 & 5.157(15) & 0.0643(9) & 3.068(44) \\
      3.7  & 8.617(22) & 0.0497(7) & 3.966(57)\\
      \bottomrule
    \end{tabularx}
  \end{center}
  \caption{The values of $t_0^*/a^2$ and the lattice spacings employed
    in this study, as determined in~\cite{spectrum}. The lattice
    spacings were obtained using $(8t_0^{\rm{ph}})^{-1/2}\approx 0.475(6)\,\mathrm{GeV}$~\cite{Bruno:2017lta}.\label{tab:latspacings}}
\end{table*}

\subsection{Wick contractions\label{sec:wick_basis}}
The momentum projected pseudoscalar interpolators, introduced in sec.~\ref{sec:definitions},
are obtained by Fourier transforming spatially extended operators
$\mathcal{P}^a(t,\vec{x})$ that have the same flavour structure
as the local pseudoscalar currents $P^a$ and are centred about $\vec{x}$:
\begin{align}
  \label{eq:mproj}
  \mathcal{P}^a_{\vec{p}}(t)=a^3\sum_{\vec{x}}e^{-i\vec{p}\cdot\vec{x}}\mathcal{P}^a(t,\vec{x}).
\end{align}
Any linear combination of interpolators $\mathcal{P}_{\vec{p}}^{a\dagger}$ and
$\mathcal{P}_{\vec{p}}^{q\dagger}$ can be applied to the vacuum
$|\Omega\rangle$
to probe the physical eigenstates. We shall label such
linear combinations as $\mathcal{B}^{\dagger}_i(-\vec{p},t)$.
From these one can define matrices of correlation functions
\begin{align}
  \label{eq:matcor}
  C_{ij}(\vec{p},t)=\frac{1}{N_{t_{\mathrm{in}}}}\sum_{t_{\mathrm{in}}}
  \left\langle\Omega\left|\mathcal{B}_i(\vec{p},t+t_{\mathrm{in}})
  \mathcal{B}^{\dagger}_j(-\vec{p},t_{\mathrm{in}})\right|\Omega\right\rangle,
\end{align}
where $N_{t_{\mathrm{in}}}$ denotes the number of source time slices
that we average over.

Performing the Wick contractions between the quark bilinears
at the source and the sink leads to both connected and
disconnected quark line contributions $\tilde{C}$ and $\tilde{D}$,
respectively,
\begin{equation}
  \label{eq:wick}
  \contraction[1ex]{}{q}{^{f_1} \Gamma_1 \bar q^{f_2} \bar }{q}
  \contraction[1.5ex]{q^{f_1} \Gamma_1 }{\bar q}{^{f_2} \bar{q}^{f_3} \Gamma_2 }{q}
  \bcontraction[1ex]{}{q}{^{f_1} \Gamma_1 }{\bar q}
  \bcontraction[1ex]{q^{f_1} \Gamma_1 \bar q^{f_2}}{\bar{q}}{^{f_3} \Gamma_2}{q}
  q^{f_1} \Gamma_1 \bar q^{f_2} \bar{q}^{f_3} \Gamma_2 q^{f_4} = \delta_{f_1,f_3}\delta_{f_2,f_4}\tilde{C}_{f_1,f_2}^{\Gamma_1\Gamma_2} - \delta_{f_1,f_2} \delta_{f_3,f_4} \tilde{D}_{f_1,f_3}^{\Gamma_1\Gamma_2},
\end{equation}
where for brevity we only display the flavour indices
$f_i \in\{u, d, s\}$ and the Dirac structures
$\Gamma_i \in\{\gamma_5, \gamma_\mu\gamma_5\}$.
The disconnected terms only contribute
when contracting flavour-diagonal quark bilinears
and are particularly challenging to compute.

\subsection{Stochastic measurement of disconnected loops}
\label{sec:stochmeas}
The basic building blocks of disconnected correlation functions
are quark loops (one-point functions)
\begin{equation}
  L_s^{\Gamma,f}(\vec{p},t) = a^3\sum_{\vec{x},\vec{y},\vec{z}}\mathrm{tr}\left( e^{-i \vec{p}\cdot\vec{x}}\phi^{s}(x,y)\Gamma D^{-1}_f(y, z) \phi^{s}(z,x)\right)\label{eq:loopgeneral},
\end{equation}
where $f \in \{\ell, s\}$ labels the flavour of the Dirac operator $D_f$.
After performing the Wick
contractions no distinction between the mass degenerate $u$ and $d$
quarks needs to be made.
The trace is taken over the spin and colour components and
the space-time positions $x$, $y$ and $z$ share the
same time $t$, i.e.\ $x = (t, \vec{x}), y = (t, \vec{y})$
and $z = (t, \vec{z})$.
To ensure ground state dominance, we apply the Wuppertal smearing kernel~\cite{Gusken:1989ad}
\begin{equation}
  \label{eq:wupsmear}
  \phi(x,y) = \frac{1}{1+6\delta}\left( \delta(x,y) + \delta \sum_{j=\pm 1}^{\pm 3} U_j(x) \delta(x+a\hat{\jmath},y)\right)
\end{equation}
to the source and the sink of the quark propagators.
Above, $U_j(x)$ is a (spatially APE smeared~\cite{Falcioni:1984ei}) gauge
link at $x$, pointing in the
spatial direction $\hat{\jmath}$ and
$U_{-j}(x)=U^{\dagger}_j(x-a\hat\jmath)$.
The number of smearing iterations $s$ as well as the
parameter $\delta = 0.25$ determine the smearing radius. 

Since $D_f^{-1}$ is a very large matrix, the trace in eq.~\eqref{eq:loopgeneral}
cannot be computed exactly but must be estimated stochastically. To do so, we
start by constructing time-partitioned (also referred to as ``diluted'' in the
literature) sources~\cite{Bernardson:1993he}:
\begin{equation}
  \label{eq:stochsource}
  \rho_{\tau,i}(x, \alpha, a) = \begin{cases}  \frac{r_i(x, \alpha, a)}{\sqrt{2}} & \mathrm{mod}(t, \Delta_t) = \tau\,\,\text{and}\,\,b \le t < L_t - b, \\
                                  0 & \text{otherwise} \end{cases},
\end{equation}
where $L_t$ is the temporal lattice extent and $r_i(x, \alpha, a) \in
\mathbb{Z}_2\times i \mathds{Z}_2$ are random numbers drawn
independently for every site $x$, spin $\alpha$ and colour component
$a$.  $\Delta_t$ is the distance between timeslices on
which the source has support. On
lattices with open boundary conditions in time, we set $b>0$ in order
to suppress boundary effects. These random sources span a space in the
bulk of the lattice,
\begin{align}
  \sum_{\tau=0}^{\Delta_t/a-1}\sum_{i=0}^{N_{\rm{stoch}}-1}| \rho_{\tau,i} \rangle \langle \rho_{\tau,i} | = &N_{\rm{stoch}}\,\diag(\underbrace{ 0,\dots,0 }_{b/a},\underbrace{1,\dots,1}_{L_t/a-2b/a},\underbrace{ 0,\dots,0 }_{b/a}) \otimes \mathds{1}_{12 V_3/a^3}\nonumber\\
  &+ \mathcal{O}(1/\sqrt{N_{\rm{stoch}}}),\\
  a^4 \langle \rho_{\tau,i} | \rho_{\tau^\prime,j} \rangle = & (L_t - 2b)V_3 \,\delta_{i,j} \delta_{\tau,\tau^\prime},
\end{align}
where $V_3=L_s^3$ and  $N_{\rm{stoch}}$ such sources
are created for every dilution index $\tau = 0,\dots,\Delta_t/a-1$.
The lattice Dirac equation 
\begin{equation}
  D_f \sigma^{f}_{\tau,i} = \rho_{\tau,i}
\end{equation}
is solved for each fermion flavour $f$ and source $\rho_{\tau,i}$,
labelled by a stochastic index $i$ and
time partition $\tau$, to obtain 
the solution $\sigma^{f}_{\tau,i}$.

Summing over the dilution index $\tau$, we obtain the estimate of the loop
computed on the $i$-th stochastic source,
\begin{equation}
  L^{\Gamma,f}_{i,s}(\vec{p},t) = a^3\sum_{\vec{x}} \sum_{\tau=0}^{\Delta_t/a-1} \mathrm{tr}\left( e^{-i \vec{p}\cdot \vec{x}}(\rho_{\tau,i}\phi^{s})(x) \Gamma (\sigma^{f}_{\tau,i}\phi^{s})(x) \right).
  \label{eq:loopcomp}
\end{equation}
After averaging over the stochastic estimates,
we obtain an estimate of axialvector and
pseudoscalar loops of a particular flavour for a given gauge field configuration
and smearing,
\begin{equation}
  L^{\Gamma,f}_{s}(\vec{p}, t) = \frac{1}{N_{\rm{stoch}}} \sum_{i=0}^{N_{\rm stoch}-1} L_{i,s}^{\Gamma,f}(\vec{p}, t) + \mathcal{O}\left(\frac{1}{\sqrt{N_{\rm stoch}}}\right).
\end{equation}
This requires $N_{\rm stoch}\times \Delta_t/a$ inversions for each
flavour.  To extend our basis of interpolators, we compute loops with
different levels of smearing, i.e.\ $s\in\{0,s_1,s_2\}$. Unlike in the
connected case, this does not require any additional inversions: due
to its Hermiticity the smearing operator can be applied to the
stochastic sources and solutions after the inversion, as indicated in
eq.~\eqref{eq:loopcomp}.

The inverse of the Wilson Dirac operator $D_f = \frac{1}{2\kappa_f}(\mathds{1} - \kappa_f H)$
within the trace of eq.~\eqref{eq:loopgeneral}
can be expanded for small values of the hopping parameter $\kappa_f$.
This yields a geometric series in terms of the
nearest-neighbour hopping term $H$~\cite{Thron:1997iy,Bali:2005fu,Bali:2009hu},
\begin{equation}
  \label{eq:hpe}
  \mathrm{tr} \left( \Gamma D_f^{-1} \right) = 2 \kappa_f \sum_{i=0}^{\infty} \kappa_f^i \mathrm{tr} \left( \Gamma H^i \right) = 2 \kappa_f \sum_{i=0}^{n-1}\kappa_f^i \mathrm{tr}(\Gamma H^i) + \kappa_f^n \mathrm{tr}\left( \Gamma H^n D_f^{-1} \right).
\end{equation}
Above we restricted ourselves to the case without smearing.  On the
right hand side of the equation we have split the series into the
first $n$ terms for which $\tr\Gamma H^i=0$ and a remainder. Note that
the value of $n$ depends on $\Gamma$ and the fermion action employed.
In our case, in the absence of smearing, $n(\gamma_5) = 2$ for
pseudoscalar and $n(\gamma_\mu \gamma_5) = 4$ for axialvector loops.
In the stochastic estimation the first sum
only contributes to the noise.  Hence, we can obtain an improved
estimate of the trace, by applying the Dirac operator $n$ times to the
solution, replacing
\begin{equation}
  \label{eq:hpeapply}
  \Gamma \sigma^{f}_{\tau,i} \mapsto  \Gamma (1 - 2 \kappa_f D_f)^{n(\Gamma)}\sigma^{f}_{\tau,i}
\end{equation}
in eq.~\eqref{eq:loopcomp}.

In summary, we apply two noise reduction techniques, time partitioning
which eliminates noise from neighbouring time slices at the expense of
additional inversions and the hopping parameter expansion that also
reduces short-distance noise by exploiting the locality of the Dirac
operator. The latter is only applied to unsmeared loops and is most
effective for small values of the hopping parameter (corresponding to
a large quark mass).  We remark that due to the use of a highly
efficient multigrid
solver~\cite{Frommer:2013fsa,Heybrock:2014iga,Georg:2017zua}, we do
not benefit from the truncated solver method~\cite{Bali:2009hu} within
our setup. This method involves computing many (computationally cheap)
approximate solves, each of which needs to be contracted according to
eq.~\eqref{eq:loopcomp}, also applying smearing. The latter would
dominate the cost even though the implementation is highly optimized
for the hardware available to us. In our implementation, where we only
use exact solves, the computational cost for the smearing and the
contractions still accounts for roughly a third of the total computing
time.

\subsection{Measurements of connected and disconnected correlation functions\label{sec:corrmeas}}
In order to
estimate the disconnected two-point function
appearing in eq.~\eqref{eq:wick}, we
correlate and average two of the loops, defined in eq.~\eqref{eq:loopcomp}:
\begin{equation}
  \label{eq:disconcorr}
  \tilde{D}_{\substack{f_1,f_2\\s_1,s_2}}^{\Gamma_1\Gamma_2}(\vec{p}, t) = \frac{1}{N_{\rm{stoch}}(N_{\rm{stoch}}-1)}\frac{a}{L_t}\sum_{\substack{i,j=0\\i \ne j}}^{N_{\rm stoch}-1}\sum_{t_{\mathrm{in}}=0}^{L_t-a}\left\langle L^{\Gamma_1, f_1}_{i,s_1}(\vec{p},t_{\mathrm{in}}+t) L^{\Gamma_2, f_2}_{j,s_2}(-\vec{p},t_{\mathrm{in}})\right\rangle.
\end{equation}
Note that we are only allowed to sum over products of loops that
have been obtained on different random sources, hence $i\neq j$.
Equation~\eqref{eq:disconcorr} applies to periodic lattices, where $b=0$ and the
correlation functions wrap around the lattice (the periodicity of the loop is implicit,
$L_{i,s}^{\Gamma,f}(\vec{p}, t+L_t) = L_{i,s}^{\Gamma,f}(\vec{p}, t)$).
It is straightforward to adapt the above equation to lattices with open boundaries by
restricting the sum over $t_{\mathrm{in}}$ such that both $t_{\mathrm{in}}$
and $t_{\mathrm{in}}+t$ remain in the bulk of the
lattice~(defined to be a distance $b$
away from the boundaries).

We implement forward-backward averaging for the disconnected two-point functions by
simply symmetrizing with respect to the ordering of the source and sink operators:
\begin{equation}
  \label{eq:fwdbwdavg}
  \overline{D}_{\substack{f_1,f_2\\s_1,s_2}}^{\Gamma_1\Gamma_2}(\vec{p}, t) = \frac{1}{2} \left(  \tilde D_{\substack{f_1,f_2\\s_1,s_2}}^{\Gamma_1\Gamma_2}(\vec{p}, t) + \mathrm{sgn}(\Gamma_1,\Gamma_2)\tilde D_{\substack{f_2,f_1\\s_2,s_1}}^{\Gamma_2\Gamma_1}(\vec{p}, t)\right)\text{,}
\end{equation}
where $\rm{sgn}(\gamma_5,\gamma_5) =
\rm{sgn}(\gamma_\mu\gamma_5,\gamma_\mu\gamma_5) = 1$ and $\rm{sgn}(\gamma_\mu\gamma_5,\gamma_5) = -1$.
We use the same random sources for light and strange quark inversions,
preserving the correlations between light-light, strange-strange as well as
light-strange disconnected correlation functions. This is beneficial when computing
differences of disconnected correlation functions which appear after
the Wick contraction of
some of the basis states.

The expression~\eqref{eq:wick} also contains
the connected correlation function matrix $\tilde{C}$. This can be computed
by exploiting the $\gamma_5$-Hermiticity of the Dirac operator and inverting on
a smeared point source $P^s_{\alpha,a}(x) = \phi^{s}(x,y)\delta(y,
x_{\mathrm{in}})\delta(\alpha, \alpha_{\mathrm{in}}) \delta(a, a_0)$.
This yields the point-to-all propagator $M_f^s(x) =
D_f^{-1}(x,y) P^s(y)$, which is a matrix in Dirac and
colour space and can be used to construct connected
correlation functions,
starting at a fixed source position $x_{\mathrm{in}}$:
\begin{equation}
  \tilde{C}^{\Gamma_1,\Gamma_2}_{\substack{f_1,f_2\\s_1,s_2}}(\vec{p}, t) = V_3a^3\sum_{\vec{x}} e^{-i\vec{p}\cdot(\vec{x}-\vec{x}_{\mathrm{in}})}\left\langle\Gamma_1\gamma_5 \phi^{s_2}M_{f_1}^{s_1}(x) \gamma_5 \Gamma_2 \phi^{s_2}M_{f_2}^{s_1}(x)\right\rangle.
\end{equation}
In our case, $f_1 = f_2$.
Due to the reduced error compared to the disconnected correlation functions, it
is sufficient to employ only a single source position per configuration at
$x_{\mathrm{in}} = (b, \vec{0})$, leaving $(L_s-2 b)/a$ timeslices for the extraction of the
physical states. Unlike on lattices with periodic boundary conditions,
in the case of open boundary conditions, in this case we do not carry out the
forward-backward averaging of eq.~\eqref{eq:fwdbwdavg}.

Based on previous experience with the pion correlation function regarding
the boundary effects~\cite{spectrum}, we fix $b$ such that these effects are
smaller than the statistical error of the pion correlation function at
the corresponding timeslice $b/a$. This is a
conservative choice, given the comparably large errors of the disconnected
contributions. 

We implement three levels of smearing, $s \in S = \{0, s_1, s_2\}$, which
allows us to analyse local matrix elements as well as to extend our basis of interpolators.
The ratio $s_2/s_1$ is kept approximately  constant on all ensembles and the
number of smearing iterations is increased with decreasing pion mass and lattice
spacing.
In the  measurement of the disconnected loops, we choose a time separation of
$\Delta_t = 4a$ in the stochastic
dilution, except for the finest lattice spacing
where we used
$\Delta_t =6a$. These choices are listed in tab.~\ref{tab:measparams}, as well
as the number of configurations analysed and the distance between consecutive
measurements in Hybrid Monte-Carlo molecular dynamics time units.

\begin{table*}
\begin{center}
\begin{tabularx}{\textwidth}{l|cccXcccccX}
  \toprule
  id & $N_{\mathrm{conf}}$  & $\Delta \mathrm{MDU}$ & $\Delta_{\mathrm{bin}}$ & $S$ & $b/a$ & $\Delta_t / a$ & $N_{\rm stoch}$ & $ N_{\mathrm{solves}}/ 10^3$\\
  \midrule
  H101 & 963  & 8 & 4  & $\{0, 55, 92\}$   & 30 & 4 & 96 & 370\\
  H102a & 490  & 8 & 4  & $\{0, 63, 104\}$  & 30 & 4 & 96 & 376\\
  H102b & 491  & 8 & 4  & $\{0, 63, 104\}$  & 30 & 4 & 96 & 377\\
  H105 & 899  & 8 & 4  & $\{0, 75, 125\}$  & 30 & 4 & 96 & 690\\
  C101 & 504  & 8 & 4  & $\{0, 88, 146\}$  & 30 & 4 & 96 & 387\\
  D150 & 502  & 4 & 8  & $\{0, 125, 208\}$ & 0  & 4 & 96 & 386\\
  H107 & 778  & 8 & 4  & $\{0, 63, 107\}$  & 30 & 4 & 96 & 598\\
  H106 & 754  & 8 & 4  & $\{0, 63, 104\}$  & 30 & 4 & 96 & 579\\
  C102 & 729  & 8 & 4  & $\{0, 88, 146\}$  & 30 & 4 & 96 & 560\\
  B450 & 794  & 8 & 4  & $\{0, 68, 113\}$  & 0  & 4 & 96 & 305\\
  S400 & 796  & 8 & 4  & $\{0, 78, 129\}$  & 30 & 4 & 96 & 611\\
  N401 & 500  & 8 & 4  & $\{0, 94, 156\}$  & 34 & 4 & 96 & 384\\
  B451 & 1000 & 8 & 4  & $\{0, 68, 113\}$  & 0  & 4 & 96 & 768\\
  B452 & 962  & 8 & 4  & $\{0, 83, 129\}$  & 0  & 4 & 96 & 739\\
  N202 & 440  & 8 & 6  & $\{0, 98, 163\}$  & 30 & 4 & 96 & 169\\
  N203 & 563  & 8 & 6  & $\{0, 111, 185\}$ & 30 & 4 & 96 & 432\\
  N200 & 853  & 8 & 6  & $\{0, 135, 225\}$ & 30 & 4 & 96 & 655\\
  D200 & 582  & 8 & 8  & $\{0, 165, 275\}$ & 30 & 4 & 96 & 447\\
  N204 & 745  & 8 & 6  & $\{0, 111, 185\}$ & 30 & 4 & 96 & 572\\
  N201 & 757  & 8 & 6  & $\{0, 135, 225\}$ & 30 & 4 & 96 & 581\\
  D201 & 535  & 8 & 8  & $\{0, 165, 275\}$ & 34 & 4 & 96 & 411\\
  N300 & 754  & 8 & 10 & $\{0, 165, 275\}$ & 49 & 6 & 96 & 434\\
  \bottomrule
\end{tabularx}
\end{center}
\caption{Parameters related to the measurement of the correlation functions: the number of analysed configurations $N_{\rm{conf}}$, their separation in molecular dynamics units $\Delta \mathrm{MDU}$, the choice of binning $\Delta_{\mathrm{bin}}$ to account for autocorrelation effects in the statistical analysis, the numbers of smearing iterations $s\in S$, the distance from the temporal boundaries $b$ (in the case of open boundary conditions) and the time partitioning separation $\Delta_t$. In the last column we display the total number of individual Dirac vector solves carried out on each ensemble to compute the disconnected correlation functions. The number of solves needed for the connected part is much smaller (72: 2 quark masses $\times$ 3 smearing levels $\times$ source spin-colour). Ensembles H102a and H102b were generated with the same quark masses and lattice coupling but different simulation parameters and are therefore analysed separately.
  \label{tab:measparams}
}
\end{table*}

\section{Extraction of mass eigenstates and local matrix elements\label{sec:massanalysis}}
\label{sec:extract}
The masses of the $\eta$ and $\eta^\prime$ are extracted from the
matrix of correlation functions $C(t)$, eq.~\eqref{eq:matcor}.
Usually, this is done by solving a generalized eigenvalue problem
(GEVP) and fitting to the resulting
eigenvalues~\cite{Michael:1985ne,Luscher:1990ck}.  Here, we follow a
different route and directly fit to the elements of $C(t)$ or its
time-derivative, $\partial_tC(t)$. The latter reduces correlations in
the Euclidean time $t$.  This also allows us to adjust the fit ranges for
the entries $C_{ij}(t)$ individually. The matrix
analogue of effective masses is introduced and  the statistical precision
is improved by incorporating data at non-vanishing momentum. Details
of the mass determination are given and the results are compared to
those obtained by employing the GEVP. In addition, we discuss how the
decay constants of the $\eta$ and $\eta^\prime$ mesons are obtained
from combined fits including correlation functions constructed with local
currents at the sink.

\subsection{Fitting to matrices of correlation functions}
In the limit of infinite statistics, $C$ (eq.~\eqref{eq:matcor}) is a real
symmetric positive-definite $M\times M$ matrix.  The spectral decomposition 
gives
\begin{align}
  \label{eq:spectraldecomp}
  C_{ij}(t) &=  \sum_{n=0}^{\infty} \frac{1}{2 E_n V_3}
  \left\langle\Omega\left|\mathcal{B}_i(t)\vphantom{B_i^{\dagger}} \right| n \right\rangle \left\langle n \left|
  \mathcal{B}^{\dagger}_j(0)\right|\Omega\right\rangle,\\
  &=  \sum_{n=0}^{\infty} \frac{1}{2 E_n V_3} \exp(-E_n t)
  \left\langle\Omega\left|\mathcal{B}_i(0)\vphantom{B_i^{\dagger}} \right| n \right\rangle \left\langle n \left|
  \mathcal{B}^{\dagger}_j(0)\right|\Omega\right\rangle,
\end{align}
where we suppress the momentum argument and only consider a single source at $t_{\mathrm{in}}=0$.
The lowest energy states correspond to the ground states of the
$\eta/\eta^\prime$ system, $|n = 0 \rangle = | \eta \rangle$ and $| n = 1
\rangle = | \eta^\prime\rangle$.
Equation~\eqref{eq:spectraldecomp} can be written as
\begin{equation}
  \label{eq:zdz}
  C(t) = \widehat{Z} \widehat{D}(t) \widehat{Z}^{\intercal},
\end{equation}
where
$\widehat{D}(t) = \diag(\exp(-E_n t))$ for $n = 0,\ldots$ is time dependent, while
\begin{align}
  \label{eq:zoverlap}
  \widehat{Z}_{in}=\frac{1}{\sqrt{2 E_n V_3}}\langle \Omega | \mathcal{B}_i(0) | n  \rangle
  \end{align} are time
independent amplitudes (that depend on the smearing
and momentum). In practice, we truncate the infinite sum
to determine only the lowest $N$ states, hence,
\begin{equation}
  \label{eq:zdzexc}
  C(t) = Z D(t) Z^{\intercal} + \mathcal{O}(\exp(-E_{N}t)),
\end{equation}
where $D\in \mathds{R}^{N\times N}$ and we assume phase conventions
such that $Z \in \mathds{R}^{M\times N}$ with positive entries on the
diagonal. 

In nature, there are a multitude of resonances with the same
quantum numbers lying just above the $\eta^\prime(958)$: the
$\eta(1295)$, $\eta(1405)$ and $\eta(1475)$ are all close-by and will,
in general, contribute as excited states to $C(t)$.  It is therefore
important to include at least a third state, $N \ge 3$, in our
analysis to provide an effective parametrization of the contributions
of these states.  In principle, also strong decays of the $\eta$ and
$\eta^\prime$ should be taken into account. For the $\eta^{\prime}$,
the dominant decay is $\eta^\prime \to \eta \pi^+ \pi^-$ (branching
ratio 42.5\,\%~\cite{PDG}). This is kinematically only possible on
ensemble D150, for our lightest quark mass. Its decay width, however,
is about $80\,\mathrm{keV}$, which would be very difficult to
resolve considering the statistical precision we achieve. Other
channels have even smaller decay rates and many,
such as $\eta \to 3 \pi^0$, are forbidden in the isospin limit of QCD
that we simulate.

One can use eq.~\eqref{eq:zdzexc} to perform a combined fit to $C(t)$,
restricting the fit range to times large enough so that any
contributions from higher excited states $n \ge N$ fall below the
statistical precision.  The $M\times N$ amplitudes $Z_{in}$ and $N$
masses are fitted simultaneously to the $(M+1)\times M/2$ independent
components of $C_{ij}$. The bases of interpolating operators used for
each ensemble are detailed in tab.~\ref{tab:massresults}. We find fits
to be most stable for $N=M=3$. These involve 12 free parameters.  We
deviate from this choice for the $m_s=m_\ell$ ensembles, for which
there is no mixing between the singlet and octet sectors. In this case,
the matrix of correlation functions is block diagonal and we choose
$N=M=4$, such that the problem decomposes into two independent singlet and
octet $N=M=2$ fits.

On the ensembles with open boundary conditions, we take boundary
effects into account when computing the loops and connected
correlation functions, using sources and sinks that only have support
in the bulk of the lattice, see sec.~\ref{sec:corrmeas}. This allows
the simple ansatz $D(t) = \diag (\exp(- E_n t))$ for the
time dependent matrix in eq.~\eqref{eq:zdzexc}.
On lattices with \mbox{(anti-)}periodic
boundary conditions in time, states can propagate across the
boundary and we modify $D$ to take the backwards-propagating
states into account:
\begin{equation}
  \label{eq:periodicD}
  D(t) = \diag \left[ 2 \exp\left(-E_n \frac{ L_t }{2}\right) \cosh\left(-E_n\left(t-\frac{L_t}{2}\right)\right) \right].
\end{equation}

\begin{figure}[tph]
  \includegraphics[width=0.49\linewidth]{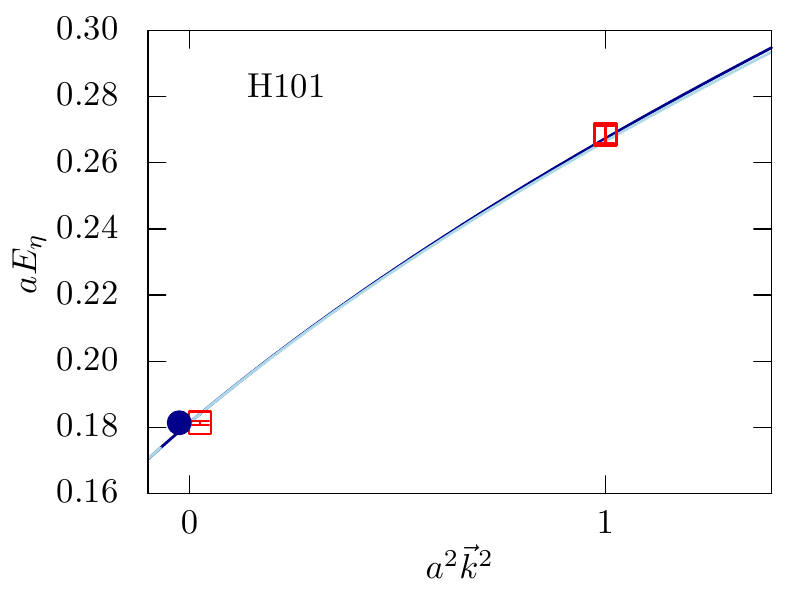}
  \includegraphics[width=0.49\linewidth]{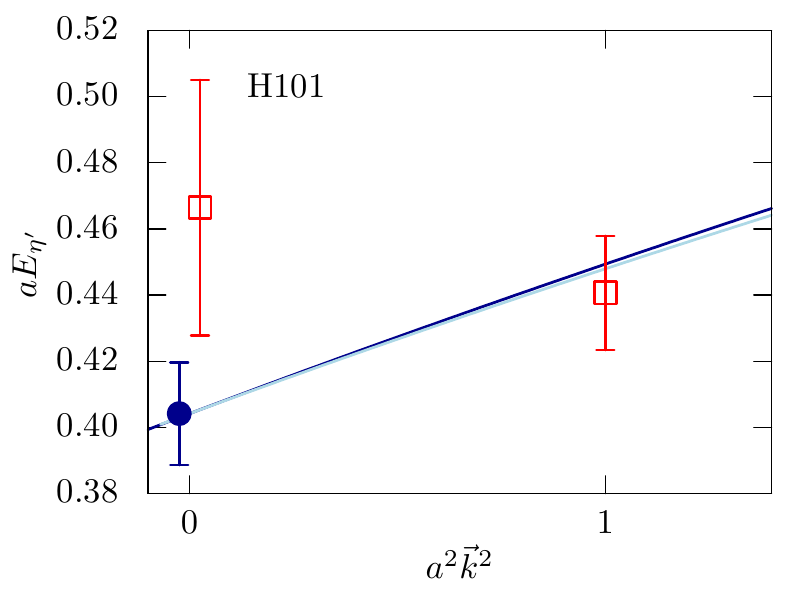}\\
  \includegraphics[width=0.49\linewidth]{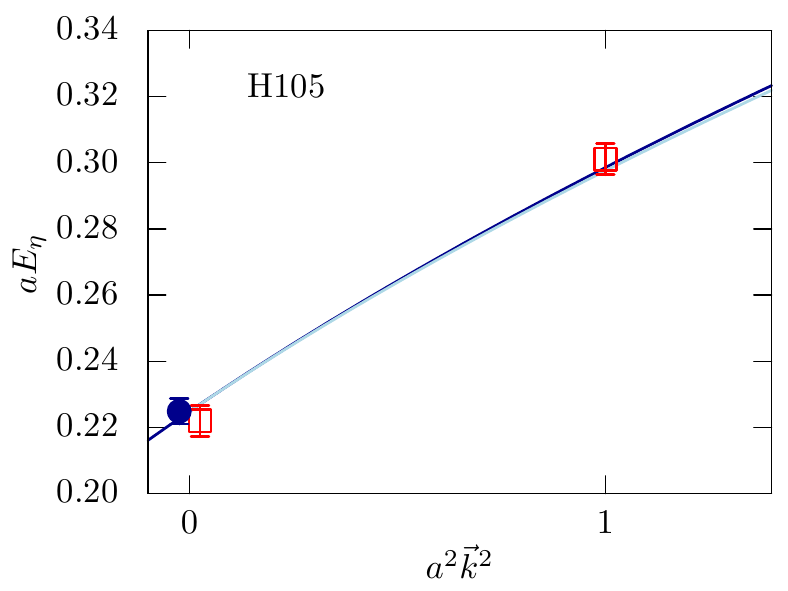}
  \includegraphics[width=0.49\linewidth]{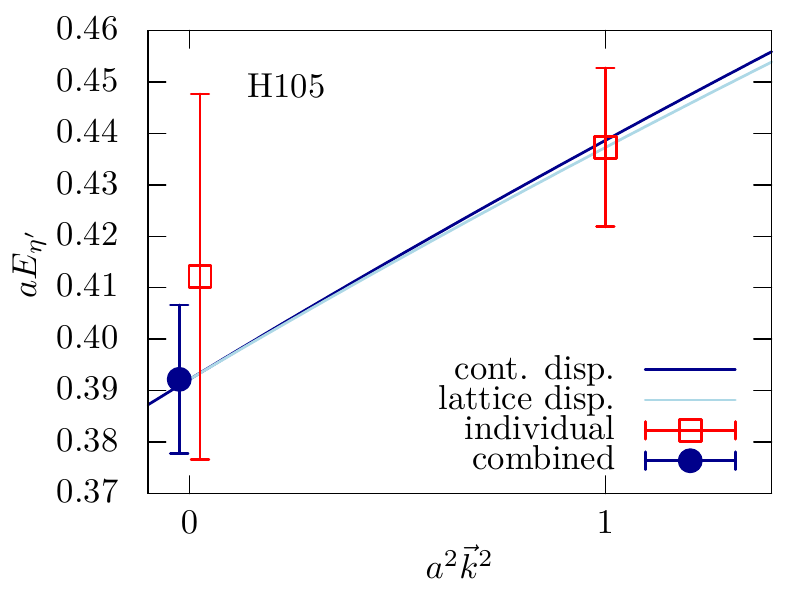}
  \caption{Energies of the $\eta$ (left panels) and $\eta^\prime$
    (right panels) mesons determined on ensembles H101 (top) and H105
    (bottom). The red squares display the energies with $a^2 \vec{k}^2 =
    0$ and $a^2 \vec{k}^2 = 1$ extracted from individual fits, while the
    blue filled symbols show the masses determined from a combined fit
    assuming the continuum dispersion relation (dark blue). The
    lattice dispersion relation~(light blue) obtained using the masses
    extracted from the combined fit is also displayed. The data points
    at $a^2\vec{k}^2 = 0$ have been shifted slightly for better
    visibility. Note that $E_\eta = E_\pi$ on the symmetric~($m_s=m_\ell$)
    ensemble H101.\label{fig:disprel}}
\end{figure}

We also include data with non-vanishing momentum in the fit, assuming
the continuum dispersion relation
\begin{equation}
  \label{eq:disprel}
  aE_n(\vec{p}) = \sqrt{a^2 M_n^2 + a^2 \vec{p}^2},
\end{equation}
where $M_n=E_n(\vec{0})$ is the mass of the $n$-th eigenstate. On the
lattice the momentum components are quantized: $p_j=2\pi ak_j/L_s$
where $k_j$ are integer multiples of $a^{-1}$.
We average over the six smallest non-trivial lattice momenta
($a^2\vec{k}^2 = 1$) and carry out a combined fit with the $\vec{k}=\vec{0}$
data, assuming the same masses $M_n$.
In fig.~\ref{fig:disprel}  we show examples of these fits
(see also fig.~3 of~\cite{Bali:2017qce}). In addition, we plot the naive lattice
dispersion relation for a free scalar particle,
\begin{equation}
  \label{eq:latdis}
aE_n(\vec{p}) = \arccosh\left(\cosh(aM_n)+\sum_j 2 \sin^2(a p_j/2)\right).
\end{equation}
Since within the relevant momentum range the differences between the
two curves~\eqref{eq:disprel} and~\eqref{eq:latdis} are much smaller
than the errors of the data, we conclude that assuming
eq.~\eqref{eq:disprel} will not bias our results. Moreover, we
find all data to be well described by this ansatz. The combined
analysis of zero and non-zero momentum data indeed reduces the
statistical error, in particular, for the $\eta^\prime$ mass. This is  in part due  to
the fact that the zero momentum data couple to the slowly fluctuating
topological charge density and exhibit longer autocorrelations,
see~\cite{Bali:2014pva}. In total we fit to $l\times M(M+1)/2$
correlation functions and the
number of fit parameters is increased to $(l \times M+1)\times N$,
where $l$ is the number of momenta. Specifically, for $M=N=3$, by setting
$l=2$, we increase the number of correlation functions from 6 to 12
and the number of fit parameters from 12 to 21.

\subsection{The generalized effective mass method}
\begin{figure}
  \includegraphics[width=0.49\linewidth]{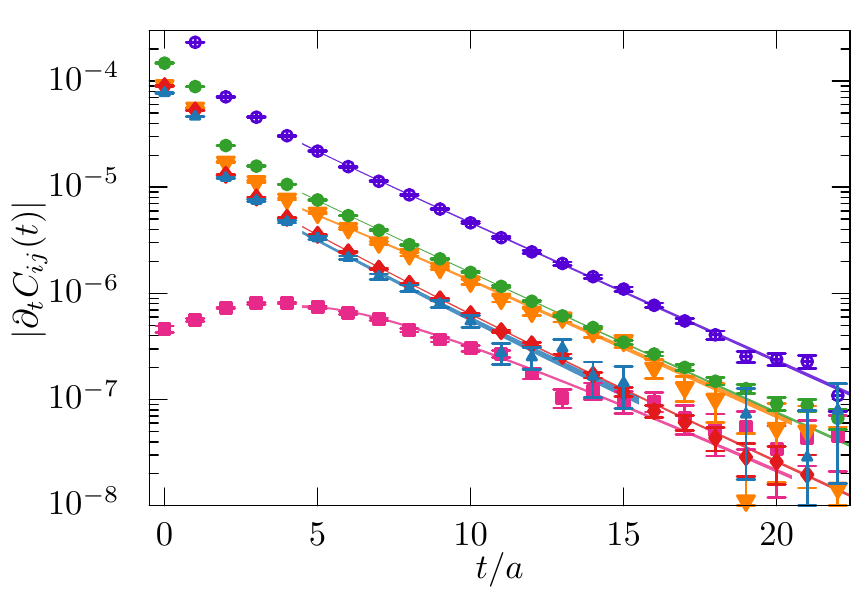}
  \includegraphics[width=0.49\linewidth]{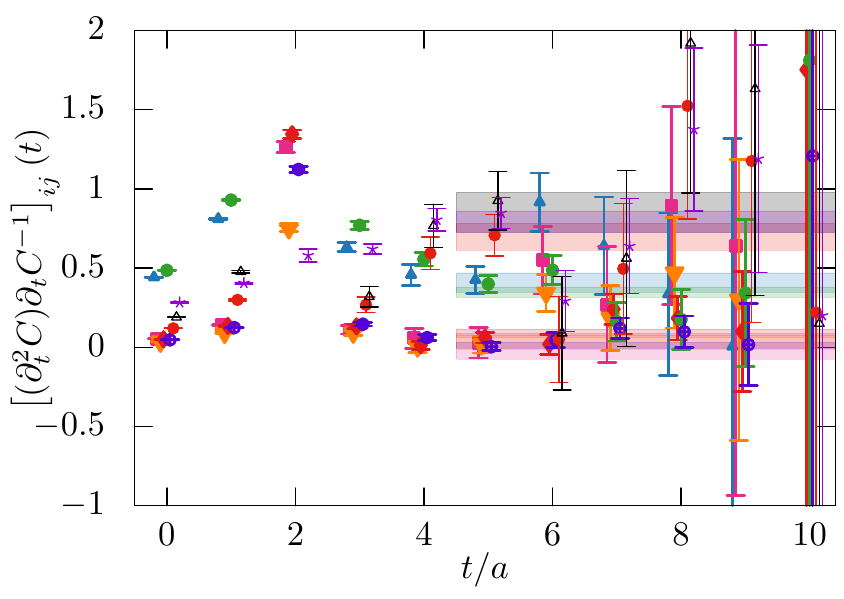}
  \caption{\label{fig:correlators} (Left) The moduli of $C_{ij}(t)$
    determined on ensemble H105 at momentum $a^2\vec{k}^2=1$. (Right)
    The elements of the generalized effective mass matrix,
    eq.~\eqref{eq:generalizedeffectivemassWDeriv}. The shaded regions
    in both panels correspond to the results of a simultaneous fit to
    eqs.~\eqref{eq:dtrick} and
    \eqref{eq:generalizedeffectivemassWDeriv}. The widths of the regions
    indicate the fit ranges.  }
\end{figure}

As suggested in~\cite{Feng:2009ij,Umeda:2007hy}, it is
advantageous to reduce the correlations between time slices by fitting to
the temporal derivative of the correlation functions. The fit form, eq.~\eqref{eq:zdz}, is modified to
\begin{equation}
  \label{eq:dtrick}
  \partial_t C(t) \sim Z\left( \partial_t D(t) \right)Z^{\intercal},
\end{equation}
where $\partial_t C(t) = \left( C(t+a) - C(t-a) \right)/(2a)$ is the
symmetric discretized derivative and
\begin{align}
  \partial_t D(t) &= -\diag\left[ E_n \exp\left(-E_n t\right) \right]\quad&\text{(open boundaries),}\\
  \partial_t D(t) &= - 2\,\diag\left[ E_n \exp\left(-E_n \frac{L_T}{2}\right)\sinh\left(-E_n \left(t-\frac{L_T}{2}\right)\right) \right]\quad&\text{(periodic boundaries).}
\end{align}
We find that this modification enables fits to discriminate between
the $\eta$ and $\eta^\prime$ contributions more easily, as long as
higher excited states are either sufficiently well parameterized (by
including them in the fit) or suppressed by the choice of the fit
window. In addition, potential constant shifts in the correlation
functions (arising from finite volume effects related to incomplete
sampling of the topological sectors, see,
e.g.,~\cite{Bali:2014pva,Ottnad:2017bjt}) are automatically
removed. Although we do
not encounter any significant shifts within our data, we observe that
utilizing eq.~\eqref{eq:dtrick} leads to decreased correlations and more
stable fit results.

The fit form involves a sum over $N$ exponentials for each of the
$M(M+1)/2$ independent entries of $C(t)$.  As the number of
states~($N$) included increases, the fits become more unstable and
sensitive to the choice of the initial guesses.  This motivates us to
define a matrix analogue of the effective mass~(for $N = M$),
\begin{align}
  \partial_t\log C(t) = & (\partial_t C(t))C^{-1}(t)\\
  = & \left( \widehat{Z} \partial_t \widehat{D}(t) \widehat{Z}^{\intercal} \right)\left( \widehat{Z} \widehat{D}(t) \widehat{Z}^{\intercal}\right)^{-1}\\
  = & - Z \, \diag_{n=0}^{N-1}(E_n)\, Z^{-1} + \mathcal{O}\left[\exp\left(-\left(E_{N}-E_{N-1}\right)t\right)\right],
  \label{eq:generalizedeffectivemass}
\end{align}
which is
constant in time~(up to excited states corrections and statistical noise).\footnote{We remark that this construction is easily generalizable to the
case $N\neq M$, employing a singular value decomposition of
$ZD(t)Z^{\intercal}$. It should be noted, however, that the leading truncation
errors then depend on the ${\rm min}(N,M)$ non-singular values.
} Since $C(t)$ is non-singular, $(\partial_t C) C^{-1}$ is an
unambiguous expression and can readily be computed. One can
easily repeat this procedure and take the second derivative, leading to
\begin{equation}
  \label{eq:generalizedeffectivemassWDeriv}
  (\partial_t^2C) (\partial_t C)^{-1} = - Z\, \diag(E_n)\, Z^{-1} + \mathcal{O}\left[\exp\left(-\left(E_{N}-E_{N-1}\right)t\right)\right].
\end{equation}
This alters the contributions from higher excited states but leaves the large-time
behaviour unaffected.

Note that $(\partial_t C) C^{-1}$ and $(\partial_t^2C)(\partial_t C)$ are
not symmetric and their $M^2$ elements converge to constant values at large
times. In order to resolve $N$ different states, $N(M+1)$ parameters
($Z_{in}$ and $E_n$) need to be determined. The $M^2$ asymptotic
values are not sufficient for this, unless $N\le M^2/(M+1)$,
which excludes the quadratic case $N=M$.\footnote{Setting $M=N+1=4$ allows
to determine all the parameters, however, this choice was found to result in larger
errors than a combined fit to eqs.~\eqref{eq:dtrick} and
\eqref{eq:generalizedeffectivemassWDeriv}.}   In this case, simultaneous fits are performed
to eqs.~\eqref{eq:dtrick} and~\eqref{eq:generalizedeffectivemassWDeriv}, where
the latter enables the fit to unambiguously resolve the spectrum of states.

To summarize our fitting strategy: we simultaneously fit the correlation
functions with two momenta $a^2\vec{k}^2 = 0$ and $a^2\vec{k}^2 = 1$
to eqs.~\eqref{eq:dtrick} and
\eqref{eq:generalizedeffectivemassWDeriv}.  Correlations between all
entries of $(\partial_t C) C^{-1}$ and $\partial_t^2C(\partial_tC)^{-1}$
at each time slice are taken into account, whereas correlations
between time slices can be neglected due to fitting to derivatives of
$C$~(we have checked that this is indeed the case).  A
typical fit is shown in fig.~\ref{fig:correlators}.  The resulting
$\eta$ and $\eta^\prime$ masses for all the ensembles are collected in
tab.~\ref{tab:massresults}, along with the $\chi^2/N_{\rm df}$ of the
fits, where in most cases we achieve $\chi^2/N_{\rm df}\approx 1$.

\begin{table*}
\begin{center}
\pgfplotstabletypeset[
columns={ensid,basis,ameta,metatimessqrt8t0,ametaprime,metaprimetimessqrt8t0,chi2pdofmass},
  every head row/.style = {before row=\toprule, after row=\midrule},
  every last row/.style = {after row=\bottomrule},
  every column/.style = {string type, column type=l},
  columns/ensid/.style = {string type, column name=id, column type=l|},
  columns/basis/.style = {string type, column name=basis, column type=l|},
  columns/ameta/.style = {string type, column name=$a\,M_\eta$},
  columns/ametaprime/.style = {string type, column name=$a\,M_{ \eta^\prime }$},
  columns/metatimessqrt8t0/.style = {string type, column name=$\sqrt{8 t_0}M_\eta$},
  columns/metaprimetimessqrt8t0/.style = {string type, column name=$\sqrt{8 t_0}M_{\eta^{\prime}}$},
  columns/chi2pdofmass/.style = {string type, column name=$\chi^2/N_{\mathrm{df}}$},
]{\ensdata}
\end{center}
\caption{Masses of the $\eta$ and $\eta^\prime$ mesons obtained from
  fits to eqs.~\eqref{eq:dtrick} and
  \eqref{eq:generalizedeffectivemassWDeriv} in lattice units and in
  units of the gradient flow scale, $\sqrt{8 t_0}$~(determined on the
  same ensemble). See tab.~\ref{tab:ensembles} for the corresponding
  pion and kaon masses and tab.~\ref{tab:latspacings} for the lattice
  spacings. We also give the smearing bases used in the construction
  of the matrix of correlation functions,
  eq.~\eqref{eq:spectraldecomp}, where $\ell, s, 8, 0$ refer to the
  light, strange, octet and singlet combinations of the pseudoscalar
  interpolating operators, respectively, and the superscript labels
  the smearing applied~(element of $S$), see tab.~\ref{tab:measparams}.
  The resulting $\chi^2/N_{\rm df}$ of the partially correlated
  fits are also given. For the ensembles with $m_s=m_\ell$, where we carry out
  two independent fits, we give both $\chi^2/N_{\rm df}$ values for
  the octet (first) and singlet (second) cases. Ensembles H102a and H102b were generated with
    the same quark masses and lattice coupling but different simulation parameters
    and are therefore analysed separately.
  \label{tab:massresults}}
\end{table*}

\subsection{Comparison to the GEVP method\label{sec:cmptogevp}}
A standard way to extract the masses of the $\eta$ and $\eta^\prime$
from the matrix of correlation functions is to solve the
GEVP~\cite{Michael:1985ne,Luscher:1990ck},
\begin{equation}
  C(t) V(t,t_0) = C(t_0) V(t,t_0) \Lambda(t,t_0),
  \label{eq:gevp}
\end{equation}
where $\Lambda = \diag(\lambda_0,\ldots ,\lambda_{M-1})$ is the
diagonal matrix of the eigenvalues and $V$ is the matrix of
eigenvectors.   One then fits to the eigenvalues
$\lambda_n\propto e^{-E_nt}$ to extract the energies.

\begin{figure}
  \includegraphics[width=0.49\linewidth]{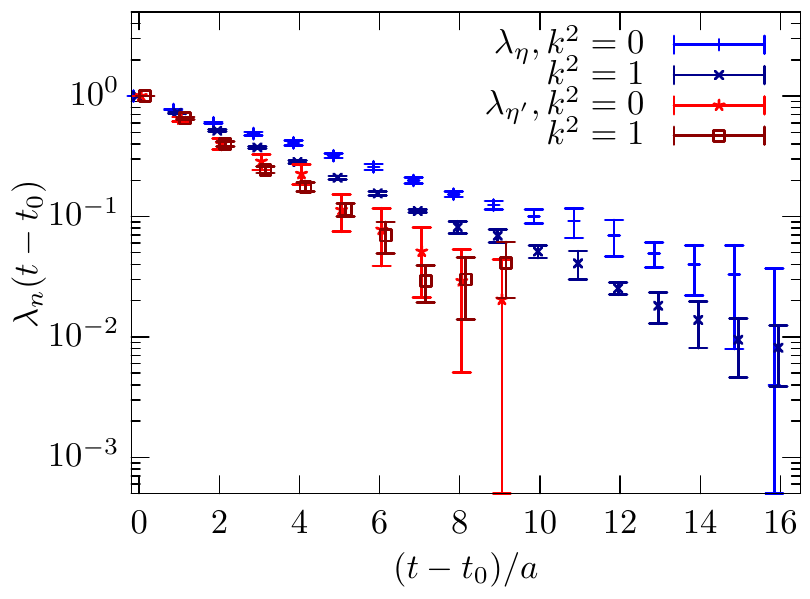}
  \includegraphics[width=0.49\linewidth]{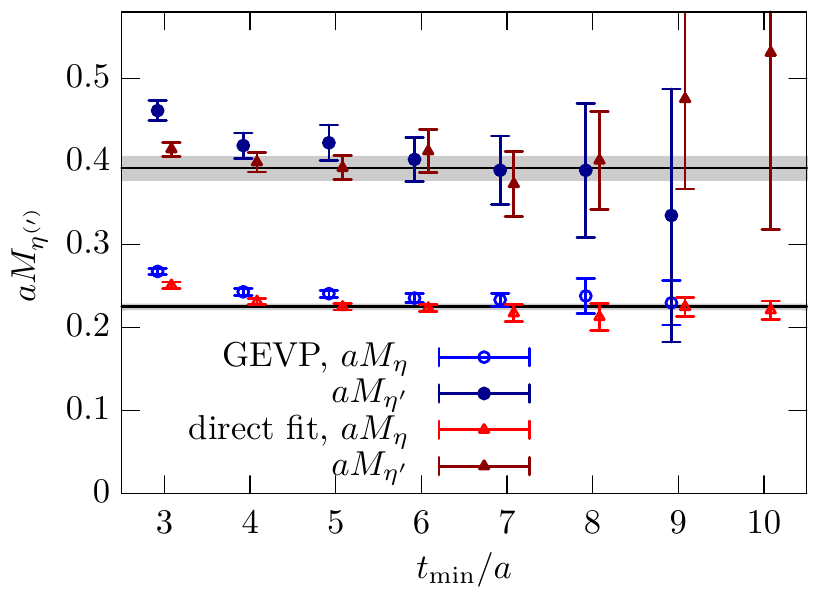}
  \caption{\label{fig:gevp} (Left) The $\eta$ and
    $\eta^\prime$ eigenvalue functions for two momenta, $a^2\vec{k}^2 = 0$ and
    $a^2\vec{k}^2 = 1$ determined on ensemble H105, obtained by solving
    the GEVP, with
    the reference time slice $t_0=5a$.  For better visibility, some points
    are shifted slightly and the data for the $\eta^\prime$ are
    omitted for $(t-t_0)/a \ge 10$ due to the large errors.
    (Right) Results for the masses determined by fitting
    to the lowest two eigenvalues of the GEVP and from direct fits
    (eqs.~\eqref{eq:dtrick} and
    \eqref{eq:generalizedeffectivemassWDeriv}) as functions of the
    starting point of the fit, $t_{\mathrm{min}}$.  The analysis is
    similar in both cases, employing the same basis of interpolators
    and incorporating data at
    two momenta. The horizontal lines and grey error bands indicate
    the final results.  These have been obtained from a slightly
    different fit, employing different $t_{\mathrm{min}}$
    for different elements of the correlation matrix. For the GEVP results, $t_0 / a = t_{\rm min}/a - 1$.}
\end{figure}

The reference timeslice $t_0$ needs to be chosen large enough~\cite{Blossier:2009kd} so that
contributions from states with $n\ge M$ are sufficiently suppressed.
In our setup, we found it hard to disentangle the excited state
contributions from the lowest two eigenvalues, having only a limited
number of timeslices $t > t_0$ available before the signal of the
heavier $\eta^\prime$ vanishes in the statistical noise.  In
particular at larger times, it also becomes increasingly difficult to
assign the correct physical states to the eigenvectors.  We compare
the GEVP with the results obtained from the fit strategy
described in the previous section
in fig.~\ref{fig:gevp}. While the two methods generally agree, the
plateau regions start earlier when using the generalized effective
mass fit method. This enables us to extract results with an increased
statistical precision compared to using the GEVP. We remark that our
fit method also allows us to extract amplitudes directly, in a
straightforward manner, as is discussed below.

\subsection{Decay constants and local matrix elements}
\label{sec:decaylocalmat}

The decay constants of pseudoscalar mesons are defined in
eq.~\eqref{eq:defDecConst}. The singlet ($a=0$) and
octet ($a=8$) decay constants
of the $\eta$ ($n=0$) and $\eta^\prime$ ($n=1$) mesons
can be obtained via
\begin{align}
  \left\langle \Omega\left|A_{\mu}^a\right| n\right\rangle=iF^a_np_{\mu},
\end{align}
where the local currents $A_{\mu}^a=\overline{\psi}t^a\gamma_\mu\gamma_5\psi$ are
introduced in eqs.~\eqref{eq:qcurrent}--\eqref{eq:flavour3}.
In addition to the singlet/octet basis, one can also define decay
constants in the flavour basis according to eq.~\eqref{eq:decayflavour}.
The two sets of decay constants are related via a
SO($2$) rotation (see eq.~\eqref{eq:convertdecay}):
\begin{align}
  \label{eq:flavourdecayconsts}
  \begin{pmatrix}
    F^{\ell}_n\\F^{s}_n
  \end{pmatrix}
  =\frac{1}{\sqrt{3}}
  \begin{pmatrix}
      1&\sqrt{2}\\
      -\sqrt{2}&1
  \end{pmatrix}
  \begin{pmatrix}
    F^{8}_n\\F^0_n
  \end{pmatrix}.
\end{align}
It is also useful to define the pseudoscalar matrix elements, 
\begin{equation}
  \label{eq:psdecs}
   \langle \Omega | P^a | n \rangle =H_n^a.
\end{equation}
We remark that for the two-point functions one can write~(in the limit of large times)
\begin{align}
  \left\langle\Omega \left|\partial_tP^a(t)\mathcal{B}_i(0)\right|\Omega\right\rangle&=
  \frac{1}{2a}\left\langle\Omega \left|\left(P(t+a)-P(t-a)\right)\mathcal{B}_i(0)\right|\Omega\right\rangle\nonumber\\
  &=-\frac{\sinh(aE_n)}{a}\left\langle\Omega \left|P^a(t)\mathcal{B}_i(0)\right|\Omega\right\rangle\nonumber\\
  &\approx-E_n\left\langle\Omega \left|P^a(t)\mathcal{B}_i(0)\right|\Omega\right\rangle,
\end{align}
where $E_n$ is the energy of the lowest state with a non-vanishing overlap
$Z_{in}=\langle\Omega|\mathcal{B}_i|n\rangle\neq 0$ (as well as
$\langle\Omega|P^a|n\rangle\neq 0$).  Using appropriate
linear combinations of the interpolators $\mathcal{B}_i$ as described
above, it is possible to project onto $n=0$ ($\eta$) as well
as onto $n=1$ ($\eta^\prime$). In this sense, we can substitute
$\langle\Omega|\partial_tP^a|n\rangle$
for $-E_nH^a_n$,  up to order $a^2$ corrections.

We define partially $\mathcal{O}(a)$ improved, unrenormalized lattice
decay constants:
\begin{equation}
  \label{eq:latdecay}
  \tilde{F}_n^aE_n = \langle\Omega| A^a_0 + a c_A^a \partial_t P^a | n \rangle
\end{equation}
with the improvement coefficients
$c_A^a=c_A$ for $a\neq 0$ and $c_A^a=c_A^s$ for $a=0$.
For the moment being, mass dependent order $a$ terms are ignored.
While the
non-singlet improvement coefficient $c_A$ has been determined
non-perturbatively~\cite{Bulava:2015bxa},
its singlet equivalent $c_A^s$ is unknown
and we parameterize it as $c_A^s =  c_A + \delta c_A$, where
$\delta c_A$ is of order $g^4$.
For the singlet case we can rewrite eq.~\eqref{eq:latdecay} as
\begin{align}
  \tilde{F}_n^0 &= \check{F}_n^0 + a\,\delta c_A \check{H}_n^0,\label{eq:singletcaimpr}
\\
  \check{F}_n^0E_n &= \langle\Omega| A^0_0 + a c_A \partial_t P^0 | n \rangle,\label{eq:singletcaimpr2}\\
  \check{H}_n^0 &= \langle\Omega| P^0 | n \rangle,\label{eq:singletcaimpr3}
\end{align}
where even in the chiral limit $\check{F}_n^0$ is only partially order
$a$-improved since we have neglected the
difference $\delta c_A$
between the singlet and the non-singlet improvement coefficients.
We also introduced the singlet pseudoscalar matrix element
$\check{H}_n^0$.
We note that $\check{H}_n^8$ is already
order $a$ improved in the chiral limit~($\check{H}_n^8=\tilde{H}_n^8$), however, this is not the case
for the singlet pseudoscalar current~\cite{Bhattacharya:2005rb},
see eq.~\eqref{eq:gpimp}, where we have to distinguish between
$\tilde{H}_n^0$ and $\check{H}_n^0$:
\begin{equation}
  \tilde H^0_n = \check H^0_n + ac_P^s\left\langle\Omega\left|\partial_{\mu}A_{\mu}^0\right|n\right\rangle\label{eq:singletpsmeunrenorm}.
\end{equation}
Note that the above $\mathcal{O}(a)$ difference between $\tilde H^0_n$
and $\check H^0_n$ does not affect eq.~\eqref{eq:singletcaimpr}, where only
the unimproved matrix element is needed.

\subsection{Determination of the decay constants}
\label{sec:detdecay}
In order to extract matrix elements with a local current $J$, we
start from a vector of $M$ correlation functions ($i=1,\ldots, M$):
\begin{equation}
  \label{eq:localcorrs}
  C_i^{ J }(t) = \left\langle \Omega\left| J(t) \mathcal{B}^{\dagger}_i(0)\right|\Omega\right\rangle,
\end{equation}
where $\mathcal{B}_i(0)$ is an interpolator with the momentum
$\vec{p}$ inserted at
the time $t_{\mathrm{in}}=0$.
For the connected contribution we utilize the translational invariance of the expectation value
to move the momentum projection from the smeared point source to the sink,
as is common in this kind of calculation.
For the
disconnected two-point function, in
order to increase the statistics, we replace
$J(t,\vec{0})\mapsto (a^3/V_3)\sum_{\vec{x}}e^{-i\vec{p}\cdot\vec{x}}J(t,\vec{x})$,
again exploiting translational invariance.
The two-point function is then constructed
in analogy to eqs.~\eqref{eq:loopcomp} and~\eqref{eq:disconcorr},
however, without smearing at the sink and with the additional
normalization factor $1/V_3$.

We carry out a spectral decomposition similar to
that of eq.~\eqref{eq:spectraldecomp}:
\begin{align}
  C_{i}^{J}(t) \approx & \sum_{n=0}^{N-1}\frac{1}{2 E_n V_3} \exp(-E_n t)
  \left\langle \Omega\left| J(0)\vphantom{B_i^{\dagger}} \right| n \right\rangle \left\langle n \left|\mathcal{B}^{\dagger}_i(0)\right|\Omega\right\rangle\nonumber\\
   =&\sum_{n=0}^{N-1}Z_{in}D_{nn}(t)j_n,
  \label{eq:localcurrents}
\end{align}
where
\begin{equation}
  \label{eq:429}
  j_n =
  \frac{1}{\sqrt{2 E_n V_3}}\langle \Omega | J | n \rangle,
\end{equation}
$Z$ is the overlap matrix $\widehat{Z}$, defined in
eq.~\eqref{eq:zoverlap}, truncated to $N$ columns and
$D(t)=\diag(\exp(-E_nt))$ is a diagonal $N\times N$ matrix.  One can
also write this in terms of matrix multiplications, $C^J(t)\approx
ZD(t)j$, where $C^J(t)$ and $j$ are $M$- and $N$-dimensional vectors,
respectively.  Using the bootstrap samples of the previously obtained
elements of $Z$ and energies $E_n$, we carry out a fit to the above
functional form, determining the matrix elements
$\langle\Omega|J|n\rangle=\sqrt{2E_nV_3}\,j_n$.

Once the axialvector
and pseudoscalar matrix elements are obtained, we can construct
the partially improved, unrenormalized decay constants $\tilde{F}^8_n$ and
$\check{F}^0_n$ for $n\in\{\eta,\eta^\prime\}$ as well as the corresponding
pseudoscalar matrix elements $\tilde{H}^8_n$ and $\check{H}^0_n$.
Below we will discuss the remaining improvement and renormalization
steps and we will add any missing improvement coefficients,
e.g., $\delta c_A$, as fit parameters in the continuum limit
extrapolation.

\section{Physical point and continuum extrapolation\label{sec:physpointextrap}}
In this section we motivate the parametrizations of the quark mass and
lattice spacing dependence of our data and present continuum limit
results on the $\eta$ and $\eta^\prime$ meson masses and their respective
decay constants. We give the physical point values as well as the
values of the NLO large-$N_c$ ChPT LECs, and provide a detailed study of
their statistical and systematic uncertainties.

First, in sec.~\ref{sec:improvement} we detail the renormalization and
$\mathcal{O}(a)$ improvement of the octet and singlet decay
constants. This affects the functional form of our continuum limit
extrapolation since not all the improvement coefficients are
known. Different possibilities exist regarding the renormalization
scheme for the singlet decay constants. This will be discussed in
sec.~\ref{sec:singletrenorm}.  In the continuum limit, large-$N_c$
U($3$) ChPT will be used to describe the mass dependence of the data.
We summarize the relevant LO and NLO ChPT expectations in
sec.~\ref{sec:chpt}. We combine these continuum limit functions with a
parametrization of the remaining $\mathcal{O}(a)$ and
$\mathcal{O}(a^2)$ lattice spacing effects in sec.~\ref{sec:contlimit},
while in sec.~\ref{sec:fitselection} we carry out several fits to our
data in order to quantify the various systematic uncertainties. A
central fit is used to predict the values for the masses and
decay constants at the physical point and the systematic
errors are estimated by varying the fit form. The results are presented in
sec.~\ref{sec:massdec}. The parameters that describe the
continuum limit behaviour correspond to the LECs of NLO large-$N_{c}$
ChPT.  Our estimates of their values are given in
sec.~\ref{sec:lecresults}.

\begin{table*}
\begin{center}
\pgfplotstabletypeset[
  columns={ensid,ZFpartimpAoctet0timessqrt8t0,ZFpartimpAoctet1timessqrt8t0,ZFpartimpAsinglet0timessqrt8t0,ZFpartimpAsinglet1timessqrt8t0},
  column type=,
  begin table={\begin{tabularx}{0.8\textwidth}{l | X X X X}},
  end table={\end{tabularx}},
  every head row/.style = {before row=\toprule, after row=\midrule},
  every last row/.style = {after row=\bottomrule},
  every column/.style = {string type},
  columns/ensid/.style = {string type, column name=id},
  columns/phi2/.style = {string type, column name=$\phi_2$},
  columns/phi4/.style = {string type, column name=$\phi_4$},
  columns/ZFpartimpAoctet0timessqrt8t0/.style   = {string type, column name=$\sqrt{8 t_0}Z_A \tilde{F}_\eta^8$},
  columns/ZFpartimpAoctet1timessqrt8t0/.style   = {string type, column
    name=$\sqrt{8 t_0}Z_A \tilde{F}_{\eta^\prime}^8$},
  columns/ZFpartimpAsinglet0timessqrt8t0/.style = {string type, column
    name=$\sqrt{8 t_0}Z_A^{s\prime} \check{F}_\eta^0$},
  columns/ZFpartimpAsinglet1timessqrt8t0/.style = {string type, column
    name=$\sqrt{8 t_0}Z_A^{s\prime} \check{F}_{\eta^\prime}^0$},
]{\ensdata}
\end{center}
\caption{Renormalized and partially improved octet and singlet decay constants
  of the $\eta$ and $\eta^\prime$ mesons obtained from fits to
  eq.~\eqref{eq:localcurrents} in units of the gradient flow scale,
  $\sqrt{8 t_0}$~(determined on the same ensemble). Ensembles H102a and H102b were generated with
  the same quark masses and lattice coupling but different simulation parameters
  and are therefore analysed separately.
\label{tab:decconstresults1}}
\end{table*}

\subsection{Renormalization and $\mathcal{O}(a)$ improvement}
\label{sec:improvement}
The $\mathcal{O}(a)$ improvement of quark bilinears has been worked
out for Wilson fermions in~\cite{Bhattacharya:2005rb}. It turns
out  (see eq.~(15) of this reference) that even
for the improvement of the octet axialvector current, singlet currents
are required. The renormalized $\mathcal{O}(a)$ improved
octet decay constant reads
\begin{align}
  F_{\eta^{(\prime)}}^{8} = Z_A \left[ (1 + 3 a \tilde{b}_A \overline{m}) \tilde{F}_{\eta^{(\prime)}}^{8} \right. + & \frac{a}{\sqrt{3}} b_A\big(m_\ell \check{F}_{\eta^{(\prime)}}^{\ell} - \sqrt{2} m_s \check{F}_{\eta^{(\prime)}}^{s} \big) \nonumber\\
  - & \left. \sqrt{2} a f_A \delta m \check{F}_{\eta^{(\prime)}}^{0} \right] + \mathcal{O}(a^2),\label{eq:octetimprovement}
\end{align}
where $Z_A$ is the renormalization factor for non-singlet currents and
$\tilde{b}_A$,\footnote{%
  We simulate at constant values of the unimproved lattice
  coupling parameter $g^2$. The difference with respect to
  keeping the $\mathcal{O}(a)$ improved coupling fixed
  amounts to replacing the
  improvement coefficients $\bar{b}_J$ and $\bar{d}_J$
  of~\cite{Bhattacharya:2005rb} by $\tilde{b}_J$ and
  $\tilde{d}_J$. The relation between these sets of parameters
  is detailed in~\cite{Korcyl:2016ugy} and the differences
  turn out to be
  tiny at our lattice spacings~\cite{improve2}.}
$b_A$ and $f_A$ are coefficients of mass dependent improvement terms.
Note that within the $\mathcal{O}(a)$
improvement terms we can replace any unimproved lattice decay constant
by either $\tilde{F}$ or $\check{F}$ since the difference will only
have an $\mathcal{O}(a^2)$ effect on the result. These
replacements are convenient for performing the continuum
extrapolation, as will be discussed in
sec.~\ref{sec:contlimit}. Subsequently, $\check{F}^{\ell}_{\eta^{(\prime)}}$
and $\check{F}^s_{\eta^{(\prime)}}$ are obtained from
$\check{F}^0_{\eta^{(\prime)}}$ and $\check{F}^8_{\eta^{(\prime)}}$
via the rotation~\eqref{eq:flavourdecayconsts}.  The bare quark masses
($f=\ell,s$) and their average and difference are given by
\begin{equation}
  am_f = \frac{1}{2}\left( \frac{1}{\kappa_f} - \frac{1}{\kappa_{\rm{cr}}} \right),\quad
  a\overline{m} = \frac{a}{3}\left( 2 m_\ell + m_s \right),\quad
  a\delta m = am_s - am_\ell.
  \label{eq:barequarkmasses}
\end{equation}
The critical hopping parameter $\kappa_{\rm{cr}}$ was determined for
our action and lattice spacings in~\cite{fixeds,spectrum}.

We use the non-perturbatively determined values of $Z_A$ that can be
found in the $Z_{A,\mathrm{sub}}^l$ column of  tab.~6
of~\cite{DallaBrida:2018tpn}.  The improvement
coefficients $b_A$ and $\tilde{b}_A$ have been determined
non-perturbatively in~\cite{Korcyl:2016ugy,improve2}.  The sea
quark coefficients, $\tilde{b}_A$ for our $\beta$ values are
\begin{align}
  \tilde{b}_A(\beta=3.4)=-0.11(13),\qquad&
  \tilde{b}_A(\beta=3.46)= 0.10(11),\nonumber\\
  \tilde{b}_A(\beta=3.55)= -0.04(12),\qquad&
  \tilde{b}_A(\beta=3.70)= -0.05(8).
\end{align}
For the valence quark coefficient $b_A$ we use the parametrization~\cite{improve2}
\begin{equation}
b_A(g^2) = 1+0.0881\, C_F g^2+b\,g^4, \quad\text{where}\quad b = 0.0113(44)
\end{equation}
and $g^2=6/\beta$ and $C_F=4/3$.
The improvement term in eq.~\eqref{eq:octetimprovement} that is
proportional to the difference of the quark masses is only present in
flavour diagonal currents. Its coefficient, $f_A$, is unknown and
formally it is of
$\mathcal{O}(g^6)$~\cite{Bhattacharya:2005rb,Gerardin:2018kpy}. This
is the only unknown parameter needed to achieve full $\mathcal{O}(a)$
improvement of the octet decay constants and we incorporate it into the
functional form of the continuum extrapolation, see sec.~\ref{sec:contlimit}.

Regarding the improvement of the singlet decay constants, utilizing
eq.~(23) of~\cite{Bhattacharya:2005rb}, we obtain
\begin{align}
  \label{eq:singletimprovement}
  F_{\eta^{(\prime)}}^{0} = Z_A^{s} \left[ \left(1 + 3 a \tilde{d}_A \overline{m} \right) \tilde{F}_{\eta^{(\prime)}}^{0} + \frac{1}{\sqrt{3}} a d_A\left(\sqrt{2} m_l \check{F}_{\eta^{(\prime)}}^{\ell} + m_s \check{F}_{\eta^{(\prime)}}^{s}\right)\right].
\end{align}
Again, we replaced the lattice decay constants within the $\mathcal{O}(a)$
terms by partially improved ones.
The renormalization factor $Z_A^{s} \neq Z_A$ is discussed in the next
subsection. Unfortunately, both improvement coefficients $d_A=b_A+\mathcal{O}(g^4)$ and
$\tilde{d}_A=\mathcal{O}(g^4)$ are only known to $\mathcal{O}(g^2)$ in perturbation theory.
In analogy to $f_A$, we will include these parameters in the continuum
extrapolation formulae~(along with $\delta c_A$, see
eqs.~\eqref{eq:singletcaimpr}--\eqref{eq:singletcaimpr3}).
The results for the renormalized but only partially improved $\eta$
and $\eta^\prime$ singlet and octet decay constants (derived from
the fits presented in sec.~\ref{sec:detdecay}) are given in
tab.~\ref{tab:decconstresults1}. 

\subsection{Renormalization of the singlet axialvector current\label{sec:singletrenorm}}
The renormalization factor of the singlet axialvector current $Z_A^s$
in the standard $\overline{\mathrm{MS}}$
scheme~\cite{Kodaira:1979pa,Larin:1993tq} depends on the
renormalization scale.  For definiteness, we detail our conventions
for the QCD $\beta$-function and the anomalous dimension of a current
$J$:
\begin{align}
  \beta(a_s)&=\mu^2\frac{\deriv a_s}{\deriv\mu^2}=
  -\sum_{n\geq 0}\beta_na_s^{n+2},\\
  \gamma_J(a_s)&=\mu^2\frac{\deriv\ln J(a_s,\mu)}{\mu^2}
  =\mu^2\frac{\deriv Z_J}{\deriv\mu^2}
  =-\sum_{n\geq 0}\gamma_{n}a_s^{n+1},
\end{align}
where $a_s=\alpha_s/\pi=g_s^2/(4\pi^2)$ and the renormalized and
bare currents $\widehat{J}$ and $J_0$, respectively, are related by
$\widehat{J}(\mu)=Z_J(\mu)J_0$. Using these normalizations, the
first two $\beta$-function
coefficients read
\begin{align}
  \beta_0&=\frac14\left(\frac{11}{3}C_A-\frac23 N_f\right),\\
  \beta_1&=\frac{1}{16}\left[\frac{34}{3}C_A^2-\left(\frac{10}{3}C_A-\frac{5}{2}C_F\right)N_f\right],
\end{align}
while the first three $\gamma$-function coefficients for the
singlet axialvector current are~\cite{Kodaira:1979pa,Larin:1993tq}\footnote{Recently, $\gamma^s_{A3}$ has been computed too~\cite{Ahmed:2021spj}.}
\begin{align}
  \gamma^s_{A0}&=0,\\
  \gamma^s_{A1}&=\frac{3}{8}C_FN_f,\\
  \gamma^s_{A2}&=\frac{1}{64}\left[\left(\frac{142}{3}C_FC_A-18C_F^2\right)N_f-
    \frac{4}{3}C_FN_f^2\right].
\end{align}
In QCD $C_A=3$ and $C_F=4/3$.
Note that $\gamma^s_A$ vanishes for $N_f=0$ since the anomalous dimension is a
sea quark effect.

From the $\beta$- and $\gamma$-functions one can easily
derive the scale evolution of local currents:
\begin{equation}
  Z_J\left(a_s(\mu_1),\mu_1\right)=Z_J\left(a_s(\mu_0),\mu_0\right)
  \exp\left(\int_{a_s(\mu_0)}^{a_s(\mu_1)}\!\!
    \deriv a\,\frac{\gamma_J(a)}{\beta(a)}\right).
\end{equation}
Normally, to leading non-trivial order, the evolution factor is
given by $(a_s(\mu_1)/a_s(\mu_0))^{\gamma_{J0}/\beta_0}$, which diverges
if one of the scales is sent to infinity. In our case, however,
$\gamma^s_{A0}=0$, leading to a finite renormalization group evolution
\begin{align}
  \frac{Z_A^s\left(a_s(\mu_1),\mu_1\right)}{Z_A^s\left(a_s(\mu_0),\mu_0\right)}
  &=
  \exp\left\{\frac{\gamma_{A1}^s}{\beta_0}\left[\left(a_s(\mu_1)-a_s(\mu_0)
    \right)\right.\right.\nonumber\\
    &\left.\left.\qquad\qquad+\frac12\left(\frac{\gamma_{A2}^s}{\gamma_{A1}^s}-
    \frac{\beta_1}{\beta_0}\right)\left(a_s^2(\mu_1)-a_s^2(\mu_0)
    \right)+\ldots\right]
    \right\}.\label{eq:rungam}
\end{align}
This suggests a modified scheme (see, e.g.,~\cite{Zoller:2013ixa}),
where the renormalization group
running is absorbed into the renormalization constant:
\begin{equation}
  \widehat{A}'_{\mu}=\left[Z_A^s(\infty)/Z_A^s(\mu)\right]\widehat{A}_{\mu}=
  Z_A^{s\prime}A_{0\mu},
\end{equation}
and
\begin{equation}
  Z_A^{s\prime}=Z_A^s(\mu = \infty) = \left[1-\frac{\gamma_{A1}^s}{\beta_0}a_s(\mu)
    +\frac{\gamma_{A1}^s}{2\beta_0}\left(\frac{\gamma_{A1}^s}{\beta_0}
    +\frac{\beta_1}{\beta_0}-\frac{\gamma_{A2}^s}{\gamma_{A1}^s}\right)
    a^2_s(\mu)+\cdots\right]Z_A^s(\mu).
  \label{eq:singletrenormatinfscale}
\end{equation}
Similar to the renormalization group invariant (RGI) scheme,
in the above $\overline{\mathrm{MS}}'$ scheme the
current is scale independent and the corresponding $\gamma$-function is trivial:
$\gamma^{s\prime}_A=0$.
However, there are two differences: it remains a 
variant of the $\overline{\mathrm{MS}}$ scheme and there is no
multiplicative ambiguity in the definition of $Z_A^{s\prime}$.
Renormalizing the singlet axialvector current in this way corresponds to the usual
convention, setting $\mu=\infty$.

At present, the difference $Z_A^{s\prime}-Z_A$ has only been computed in perturbation
theory. Setting $c_{\mathrm{SW}}$ to its leading order value $c_{\mathrm{SW}}=1$
within eq.~(32) of~\cite{Constantinou:2016ieh}, 
we obtain for our action
\begin{equation}
  Z_A^{s}(\mu)=Z_A-a_s^2(a^{-1})\left[\gamma_{A1}\ln(\mu^2a^2)+2.834(11)\right]\label{eq:ZAsinglet},
\end{equation}
where again we use the non-perturbative $Z_A$ values
of~\cite{DallaBrida:2018tpn}. Note that we
have replaced $g^2\mapsto 4\pi^2 a_s(a^{-1})$, which is
valid to this order in perturbation theory. Within the above conversion
to the $\overline{\mathrm{MS}}$ scheme we vary the scale
$\mu\in[\frac12 a^{-1},2a^{-1}]$ in order to estimate the systematics of
omitting higher perturbative orders and 
take $\mu=a^{-1}$ as our central value. The results are then run
via eq.~\eqref{eq:rungam} (not eq.~\eqref{eq:singletrenormatinfscale})
to $\mu=\infty$ to obtain the scale independent
$\overline{\mathrm{MS}}'$ result.
This is carried out
using the  three-loop $\gamma_A^s$-function and, for the running of
$a_s(\mu)$, starting from the value determined in~\cite{Bruno:2017gxd}, the
five-loop $\beta$-function~\cite{Baikov:2014qja}~(as implemented
in version~3 of the RunDec package
for Mathematica~\cite{Herren:2017osy,Chetyrkin:2000yt}).
For convenience we also quote results in the more commonly
used scale dependent prescription
at the scales $\mu = 10\,\mathrm{GeV}$, $\mu=2\,\mathrm{GeV}$ and
$\mu=1\,\mathrm{GeV}$
in QCD with $N_f=3$ active quark flavours. The corresponding
conversion factors are listed in tab.~\ref{tab:zaconvfactors}.
\begin{table*}
\begin{center}
\begin{tabularx}{\textwidth}{l|XXXX}
  \toprule
  $\mu$            & $Z_A^s(\mu) / Z_A^{s\prime} $ & & & \\
  \midrule
  RG running       & 2-loop & 2-loop & 3-loop & 3-loop \\
  $\beta$-function & 2-loop & 5-loop & 3-loop & 5-loop \\
  \midrule
  1 GeV  & $1.0881(\substack{ 28\\27})$& $1.0913(\substack{ 29\\28})$& $1.1387(\substack{ 68\\63})$  &$1.1383(\substack{ 70\\64})$ \\
  2 GeV  & $1.0565(\substack{ 10\\10})$& $1.0590(\substack{ 10\\10})$& $1.0754(\substack{ 16\\16})$  &$1.0752(\substack{ 16\\16})$ \\
  10 GeV & $1.0329(\substack{ 3 \\3 })$& $1.0341(\substack{  3\\ 3})$& $1.0390(\substack{  4\\4 })$  &$1.0389(\substack{ 4 \\ 4})$ \\
  \bottomrule
\end{tabularx}
\end{center}
  \caption{Conversion factors
    $Z_A^s(\mu) / Z_A^{s\prime} =Z_A^s(\mu) / Z_A^{s}(\infty)$, computed
    according to
    eq.~\eqref{eq:rungam} for $N_f=3$, combining different orders of
    the renormalization group running with different
    orders of the running of the coupling. The errors
    reflect the uncertainty in the $\Lambda$-parameter of $N_f=3$
    QCD~\cite{Bruno:2017lta}.\label{tab:zaconvfactors}
  }
\end{table*}

\subsection{Fit form for the chiral extrapolation\label{sec:chpt}}
We summarize the results of large-$N_c$ ChPT that will be used to
parameterize the quark mass dependence of the $\eta$ and
$\eta^\prime$ masses and decay constants when performing the
extrapolation to the physical point. We first present the general
framework, before giving explicit expressions to LO and NLO
in the following subsections.

Conventional SU($3$) ChPT entails expansions in the masses of the
octet mesons~(the pions, the kaons and the octet $\eta$) --- the Goldstone
bosons of the spontaneous breaking of SU$(3)_A$ in the QCD vacuum. To
include the singlet $\eta$, one
extends the symmetry group to U($3$) and expands 
simultaneously around the  limit $N_c \to \infty$, in
which the axial anomaly vanishes. At finite $N_c$, the singlet state acquires its
anomalous mass. Therefore, in U($3$) large-$N_c$ ChPT, the expansion
is organized in powers of
$\delta$~\cite{Gasser:1984gg,Bickert:2016fgy}, where the power counting is as follows:
\begin{equation}
  p = \mathcal{O}(\sqrt\delta),\qquad
  m = \mathcal{O}(\delta),\qquad
  1/N_c = \mathcal{O}(\delta)
\end{equation}
with $p$ being the momentum and $m$ a quark mass.

The chiral
Lagrangian at $\mathcal{O}(\delta^0)$ corresponds to massless QCD with an
infinite number of colours. At LO,
i.e.\ $\mathcal{O}(\delta^1)$, without the anomaly, the
squared pseudoscalar masses $\mu_a^2$ are related to the
quark mass matrix in the adjoint representation via the
Gell-Mann-Oakes-Renner (GMOR) relations,
$\mu^2_{ab}=2B_0\tr[t^a \diag(m_{\ell},m_{\ell},m_s) t^b]=\delta^{ab}\mu^2_a$,
where $B_0 = - \langle \overline{u}u \rangle / F^2$ is the ratio of
the chiral condensate $\langle \overline{u}u \rangle <0$ over the (squared) pion decay
constant in the SU($3$) chiral limit, $F$.
However, at this order in $\delta$ one also has to add
the $\mathcal{O}(1/N_c)$
Witten-Veneziano contribution~\cite{Witten:1978bc,Veneziano:1979ec}
to the singlet mass,  $M_0^2 = 2N_f\tau_0 / F^2$,
where $\tau_0$ denotes the quenched topological susceptibility.
Moreover, for non-degenerate quark masses, the singlet ($a=0$) and
the octet ($a=8$) pseudoscalar mesons will mix and the
corresponding non-diagonal part of the pseudoscalar mass
matrix reads, see, e.g.,~\cite{Gasser:1984gg,Bickert:2016fgy}:
\begin{equation}
  \mu^2 = \begin{pmatrix} \mu^2_8 & \mu^2_{80} \\ \mu^2_{80} & \mu^2_0\end{pmatrix}\label{eq:chptmassmatrix}.
\end{equation}
Its eigenvalues correspond to the (squared) $\eta$ and $\eta^\prime$ masses:
\begin{equation}
  \label{eq:ort}
  R \mu^2 R^{\intercal} = \begin{pmatrix} M_\eta^2 & 0 \\ 0 & M_{\eta^\prime}^2\end{pmatrix}.
\end{equation}
$R$ is an orthogonal transformation
\begin{equation}
  R  = \begin{pmatrix} \cos \theta & -\sin \theta \\ \sin\theta & \cos \theta \end{pmatrix},\label{eq:ort2}
\end{equation}
which defines the so-called mass mixing angle $\theta$. One can easily read off the relations
\begin{align}
  \mu_8^2 = & M_\eta^2 \cos^2\theta + M_{\eta^\prime}^2 \sin^2\theta,\\
  \mu_0^2 = &  M_\eta^2 \sin^2\theta + M_{\eta^\prime}^2 \cos^2\theta,\\
  \mu_{80}^2 = & (M_{\eta^\prime}^2 - M_\eta^2) \sin\theta \cos\theta,\\
  \theta = &\frac{1}{2} \arcsin\left( \frac{2 \mu_{80}^2}{\sqrt{(\mu_8^2 - \mu_0^2)^2 + 4 \mu_{80}^4}} \right) = \frac{1}{2}\arcsin\left( \frac{2 \mu_{80}^2}{M_-^2} \right),\label{eq:massangle} 
\end{align}
where
\begin{align}
  M_-^2 = & M_{\eta^\prime}^2 - M_\eta^2 = \sqrt{(\mu^2_8 - \mu^2_0)^2 + 4 \mu_{80}^4},\label{eq:mminus}\\
  M_+^2 = & M_{\eta^\prime}^2 + M_\eta^2 = \mu_8^2 + \mu_0^2.\label{eq:mplus}
\end{align}
The above relations apply to all orders in ChPT, however, the
dependencies of the mass matrix parameters $\mu_8$, $\mu_0$ and
$\mu_{80}$ on the masses of the $\eta$ and $\eta^\prime$ mesons and the
chiral anomaly vary with the order of the expansion. Also
the GMOR relations between these parameters and the
quark masses are subject to NLO corrections.
We will utilize the combinations
\begin{equation}
  M_\eta = \sqrt{\frac{1}{2}\left( M_+^2 - M_-^2\right)}
  \quad\text{and}\quad
  M_{\eta^\prime} = \sqrt{\frac{1}{2}\left( M_+^2 + M_-^2\right)}\label{eq:etaetaprmasses}
  \end{equation}
when performing the extrapolation of the $\eta$ and $\eta^\prime$ masses to the physical point.

The functions $\mu_8, \mu_0$ and $\mu_{80}$ depend on low energy
parameters and quark masses, with the latter typically being replaced
by combinations of the pion and kaon masses via the GMOR
relations. We simulate with quark masses chosen to follow two
distinct quark mass trajectories to the physical point, along one of
which the average quark mass is held constant. A more convenient
parametrization is in terms of the average and difference of the
squared pion and kaon masses,
\begin{align}
  \overline{M}\vphantom{M}^2 = & \frac{1}{3}\left(2 M_K^2 + M_\pi^2 \right)\approx 2 B_0 \overline{m},\label{eq:mbarsq}\\
  \delta M^2 = & 2 (M_K^2 - M_\pi^2) \approx 2 B_0 \delta m\label{eq:deltamsq}.
\end{align}

The computation of the decay constants is more involved. A common parametrization is
that of the two-angle mixing scheme, where the four physical decay constants are expressed in
terms of two angles $\theta_0$ and $\theta_8$ and two constants $F^0$ and $F^8$~\cite{Feldmann:1998vh,Bickert:2016fgy}~(see eq.~\eqref{eq:octsingletanglerep}),
\begin{equation}
  \begin{pmatrix}
    F^8_\eta & F^0_\eta \\
    F^8_{\eta^\prime} & F^0_{\eta^\prime}
  \end{pmatrix} = 
  \begin{pmatrix}
    F^8 \cos\theta_8 & - F^0 \sin\theta_0 \\
    F^8\sin\theta_8 & F^0 \cos\theta_0
  \end{pmatrix},
\end{equation}
leading to
\begin{align}
  F^8 = & \sqrt{(F^8_\eta)^2 + (F^8_{\eta^\prime})^2},  &
  F^0 = & \sqrt{(F^0_\eta)^2 + (F^0_{\eta^\prime})^2},\\
  \tan\theta_8 = & \frac{F^8_{\eta^\prime}}{F^8_\eta},  &
  \tan\theta_0 = & -\frac{F^0_\eta}{F^0_{ \eta^\prime }}.
\end{align}
The decay constants in the flavour basis can be expressed in the same way, 
\begin{align}
  F^s = & \sqrt{(F^s_\eta)^2 + (F^s_{\eta^\prime})^2},  &
                                                              F^\ell = & \sqrt{(F^\ell_\eta)^2 + (F^\ell_{\eta^\prime})^2},\\
  \tan\phi_s = & -\frac{F^s_{\eta^\prime}}{F^s_\eta},  &
                                                              \tan\phi_\ell = & \frac{F^\ell_\eta}{F^\ell_{ \eta^\prime }}.
\end{align}
The latter is a popular choice in phenomenological studies due to the fact that
$\phi_\ell \approx \phi_s$ at the physical point, which allows
one to express all four decay constants in terms of only three
parameters~\cite{Feldmann:1998vh}.
\subsubsection{LO large-$N_c$ ChPT\label{sec:lochpt}}
As explained above, at leading order
the elements of the pseudoscalar mass matrix are linear in the quark masses
and can be related to combinations of the non-singlet pseudoscalar meson masses
via the LO GMOR relations
\begin{align}
  (\mu_{8}^{\mathrm{LO}})^{2} = & \frac{2}{3}B_0\left( m_\ell + 2 m_s\right) = \overline{M}\vphantom{M}^2 + \frac{1}{3}\delta M^2, \label{eq:lomasses}\\
  (\mu_{0}^{\mathrm{LO}})^{2} = & \frac{2}{3}B_0\left( 2 m_\ell + m_s \right) + M_0^2 = \overline{M}\vphantom{M}^2 + M_0^2,\\
  (\mu_{80}^{\mathrm{LO}})^2 = & -\frac{2\sqrt{2}}{3} B_0 \left( m_s - m_\ell \right) = -\frac{\sqrt{2}}{3} \delta M^2,\label{eq:lomasses3}
\end{align}
where the anomalous contribution $M_0^2 = 6\tau_0 / F^2$ is
proportional to the quenched topological susceptibility $\tau_0$~\cite{Witten:1978bc,Veneziano:1979ec} and contributes at $\mathcal{O}(N_c^{-1})$ to
$\mu_0^2$, while $\overline{M}\vphantom{M}^2$ is the
${\mathcal O}(m)$ value of the squared singlet mass.

To this order, all singlet and octet decay constants can be expressed
in terms of the pion decay constant $F$ (in the chiral limit) and the
angle $\theta$, defined in eq.~\eqref{eq:massangle}:
\begin{equation}
  F^8_\eta =F^0_{\eta^\prime} = F \cos\theta,\qquad  - F^0_{\eta} = F^8_{\eta^\prime} = F \sin\
\theta,
\end{equation}
i.e.\ $F^8=F^0=F$ and $\theta_8=\theta_0=\theta$.
Note that to this order $\theta$ only depends on $M_0^2$ and $\delta M^2$.
A single mixing angle in the octet/singlet basis is not consistent with
phenomenological investigations~\cite{Feldmann:1998vh,Bickert:2016fgy} and also the results
of the present study clearly show $F^8_\eta \neq F^0_{\eta^\prime}$ and $F^0_{\eta} \neq - F^0_{\eta^\prime}$.

\subsubsection{NLO large-$N_c$ ChPT\label{sec:nlochpt}}
The large-$N_c$ ChPT expansion for the masses and decay constants has
 been worked out to NNLO in~\cite{Guo:2015xva}. Here, we use the results
of~\cite{Bickert:2016fgy} and truncate these at NLO.  To this order, only four additional LECs, $L_5$, $L_8$, $\Lambda_1(\mu)$ and $\Lambda_2(\mu)$
appear. The elements of the squared mass matrix
are given by
\begin{align}
  (\mu_8^{\mathrm{NLO}})^2 = & (\mu_8^{\mathrm{LO}})^2
  + \frac{8}{3F^2}\left( 2 L_8 - L_5 \right) \delta M^4,\label{eq:nlomu8}\\
  (\mu_0^{\mathrm{NLO}})^2 = &  (\mu_0^{\mathrm{LO}})^2
  + \frac{4}{3 F^2}\left( 2 L_8 - L_5 \right) \delta M^4 - \frac{8}{F^2} L_5 \overline{M}\vphantom{M}^2 M_0^2 - \tilde{\Lambda} \overline{M}\vphantom{M}^2 - \Lambda_1 M_0^2,\label{eq:nlomu0}\\
  (\mu_{80}^{\mathrm{NLO}})^2 = & (\mu_{80}^{\mathrm{LO}})^2
                                  - \frac{4\sqrt{2}}{3F^2}\left( 2 L_8 - L_5 \right) \delta M^4  + \frac{4 \sqrt{2}}{3F^2} L_5 M_0^2 \delta M^2 + \frac{\sqrt{2}}{6} \tilde{\Lambda} \delta M^2,\label{eq:nlomu80}
\end{align}
where we substituted $\tilde{\Lambda} = \Lambda_1 - 2 \Lambda_2$. To NLO of the
chiral expansion the latter combination 
does not depend on the QCD renormalization scale $\mu$~\cite{Bickert:2016fgy}.
In general the LECs can depend both on the QCD scale, due to the anomalous
dimension of the singlet decay constants, and the ChPT renormalization scale,
due to loop corrections.
However, in large-$N_c$ ChPT loop corrections are suppressed
by a factor of $\delta^2$ and, hence, the LECs are independent
of the ChPT scale at NLO.

The decay constants are given by
\begin{align}
  F^8_\eta & = \phantom{-}F \left[ \cos\theta + \frac{4L_5}{3F^2}\left(3 \cos\theta \overline{M}\vphantom{M}^2 + (\sqrt{2}\sin\theta + \cos\theta)\delta M^2\right) \right],\label{eq:nlodecf08}\\
  F^8_{\eta^\prime} & = \phantom{-}F \left[ \sin\theta + \frac{4L_5}{3F^2}\left( 3 \sin\theta \overline{M}\vphantom{M}^2  + (\sin\theta - \sqrt{2}\cos\theta)\delta M^2 \right)\right],\label{eq:nlodecf18}\\
  F^0_{\eta} & = - F \left[ \sin\theta \left(1 + \frac{\Lambda_1}{2}\right) + \frac{4L_5}{3F^2}\left( 3 \sin\theta \overline{M}\vphantom{M}^2  + \sqrt{2}\cos\theta \delta M^2 \right)\right],\label{eq:nlodecf00}\\
  F^0_{\eta^\prime} & = \phantom{-}F \left[ \cos\theta \left(1 + \frac{ \Lambda_1 }{2}\right)+ \frac{4L_5}{3F^2}\left(  3 \cos\theta \overline{M}\vphantom{M}^2 - \sqrt{2} \sin\theta \delta M^2 \right) \right]\label{eq:nlodecf10},
\end{align}
where $\theta$ is the mass mixing angle defined in eq.~\eqref{eq:massangle},
evaluated with the entries of the NLO mass matrix,
eqs.~\eqref{eq:nlomu8}--\eqref{eq:nlomu80}. Note that in the standard $\overline{\mathrm{MS}}$
scheme $\Lambda_1$ as well as $F^0_\eta$ and $F^0_{\eta^\prime}$ depend on $\mu$.
In general, $\theta_8\neq \theta_0\neq \theta$ to this order.

\subsubsection{Impact of the mass dependence of $t_0$ on the NLO parametrization}
In order to eliminate some of the lattice spacing effects, in
sec.~\ref{sec:fitselection} we carry out
our fits after forming dimensionless combinations $\sqrt{8t_0}M_{\mathcal{M}}$
and $\sqrt{8t_0}F^a_{\mathcal{M}}$. However, $t_0$ depends
on the pseudoscalar masses too. In the continuum limit, to leading order, this
can be parameterized as~\cite{Bar:2013ora}
\begin{equation}
  t_0(\overline{M},\delta M)=t_0^{\chi}\left(1+k\,8t_0
  \overline{M}\vphantom{M}^2\right),
\end{equation}
where $k=-0.0466(62)$~\cite{spectrum}.
This dependence somewhat alters the functional form of the
ChPT expectations for the rescaled quantities.
Within the NLO parametrization
of the squared mass matrix eqs.~\eqref{eq:nlomu8}--\eqref{eq:nlomu80},
only the element $\mu_0^2$ is affected. In this case one has to add a term
$-kM_0^2\overline{M}\vphantom{M}^28t_0$ to the parametrization,
which we do.
This is due to the fact that $M_0^2$, which is defined in the chiral
limit, appears at leading order and
$8t_0^{\chi}M_0^2=8t_0M_0^2(1-k\, 8t_0\overline{M}\vphantom{M}^2)$.
Regarding the decay constants,
rewriting $\sqrt{8t_0^{\chi}}F=\sqrt{8t_0}F(1-\frac{k}{2}8t_0
\overline{M}\vphantom{M}^2)$ means that
the term $-(k/2)F\cos\theta \overline{M}\vphantom{M}^28t_0$
needs to be added to eqs.~\eqref{eq:nlodecf08} and~\eqref{eq:nlodecf10}
while the term
$-(k/2)F\sin\theta \overline{M}\vphantom{M}^28t_0$ has to be
added to 
eq.~\eqref{eq:nlodecf18} and subtracted from
eq.~\eqref{eq:nlodecf00}.

The impact of $k\neq 0$ on the physical point masses and
decay constants turns out to be marginal. Regarding the LECs, the
biggest effect is on $F$,
$L_5$ and $L_8$, which decrease by $2.9(4)\,\mathrm{MeV}$,
by $7.7(1.0)\cdot 10^{-5}$
and by $6.2(8)\cdot 10^{-5}$, respectively, which is well
below the total errors that we find for these parameters:
$4.8\,\mathrm{MeV}$, $2.1\cdot 10^{-4}$ and $1.4\cdot 10^{-4}$.

Note that the parametrizations of the anomalous matrix elements
eqs.~\eqref{eq:aetanlo} and~\eqref{eq:aetaprimenlo}
contain leading order terms $\propto FM_0^2$,
$\propto F\overline{M}^2$ and $\propto F\delta M^2$.
This amounts to adding terms
$\propto -(3k/2)FM_0^2\overline{M}\vphantom{M}^28t_0$,
$\propto -(k/2)F\overline{M}\vphantom{M}^48t_0$
and $\propto -(k/2)F\delta M^2\overline{M}\vphantom{M}^28t_0$
in our analysis of sec~\ref{sec:gluonmefermionic}.

\subsection{Parameterizing lattice spacing effects\label{sec:contlimit}}

The lattice data do not only depend on the quark masses but also on the lattice spacing.
Here we outline our continuum limit extrapolation procedure.
We shall label the ChPT functional forms given above
as $f_O^{\mathrm{cont}}(\overline{M}\vphantom{M}^2, \delta M^2 | \ldots)$ where
the ellipses represent the fit parameters (i.e.\ the LECs)
and $O$ can be either of the two masses or four decay constants. We remind
the reader that for the decay
constants not all the $\mathcal{O}(a)$ improvement coefficients are
known. Therefore, we start from the following ansatz
\begin{align}
  \label{eq:fitansatz}
  f_O(a, \overline{M}\vphantom{M}^2, \delta M^2) & = f^{\mathrm{cont}}_O(\overline{M}\vphantom{M}^2, \delta M^2 | \ldots )
   h^{(1)}_O(a, am_\ell, am_s | \ldots) h^{(2)}_O(a^2, a^2\overline{M}\vphantom{M}^2, a^2 \delta M^2 | \ldots),
\end{align}
where $h^{(1)}_O$ contains the linear lattice spacing effects with
known or unknown coefficients and $h^{(2)}_O$ is a quadratic function
of $a$. The input data for the fits to the decay constants are the
partially improved combinations $\tilde{F}^8_n$ and
$\check{F}^0_n$~(see eqs.~\eqref{eq:latdecay}
and~\eqref{eq:singletcaimpr2}, respectively) with $n=\eta,\eta^\prime$.  In
terms of the linear lattice spacing effects, for $O=M_\eta, M_{\eta^\prime}$,
$h^{(1)}_O=1$, while for the octet decay constants these functions
contain the known parameters $b_A$, $\tilde{b}_A$ and the free parameter
$f_A$.  In the case of the singlet decay constants, within $h^{(1)}_O$
the unknown parameters $d_A$, $\tilde{d}_A$ and $\delta c_A$ appear, the
latter multiplying the term $E_n\check{H}^0_n$~(see
eqs.~\eqref{eq:singletcaimpr} and~\eqref{eq:singletcaimpr3}).  For
$h^{(2)}_O$ we make a generic quadratic ansatz.  Explicit formulae
will be given below.

Our input data are transformed into dimensionless units:
$M_n\mapsto\sqrt{8t_0}M_n$, $\check{F}^{0}_n\mapsto \sqrt{8t_0}\check{F}^{0}_n$
and $\tilde{F}^{8}_n\mapsto \sqrt{8t_0}\tilde{F}^{8}_n$,
where the scale $t_0$ is obtained on the same ensemble.  Moreover, the
parametrizations for the unrenormalized decay constants need to be
divided by $Z_A$ and $Z_A^s$, respectively.  The lattice spacing is
given in units of $t_0^*$: $a\mapsto a/\sqrt{8t_0^*}$~(see
sec.~\ref{sec:ensembles} and tab.~\ref{tab:latspacings}).  The six
parametrizations share the LECs and some of the
improvement coefficients.  Hence, we carry out simultaneous fits to
all these data. Results at the physical point can be obtained by
evaluating the continuum limit functions at
\begin{equation}
  12 t^{\rm ph}_0 {\overline{M}^{\rm ph}}^2 = 1.110\quad\text{and}\quad 8t^{\rm ph}_0\delta {M^{\rm ph}}^2 = 1.902,
\end{equation}
see sec.~\ref{sec:ensembles}.
For the linear lattice effects on the octet and singlet decay constants, we
combine the results of eqs.~\eqref{eq:octetimprovement}, \eqref{eq:singletimprovement}
and \eqref{eq:singletcaimpr}--\eqref{eq:singletcaimpr3}, to obtain the functions
\begin{align}
  h^{(1)}_{F^8_n}(a | f_A^l) &= 1 - 3 a \tilde{b}_A \overline{m}
  - ab_A\frac{Z_A}{\sqrt{3}}
  \frac{m_\ell \check{F}^\ell_n - \sqrt{2} m_s \check{F}^s_n}
       {f_{F^8_n}^{\mathrm{cont}}(\overline{M}\vphantom{M}^2, \delta M^2)}
       + \sqrt{2} af_AZ_A\frac{\delta m \check{F}^0_n}
       {f_{F^8_n}^{\mathrm{cont}}(\overline{M}\vphantom{M}^2, \delta M^2)},
          \label{eq:linearoctetaeffects}
\\
       h^{(1)}_{F^0_n}(a | d_A^l, \tilde{d}_A^l, \delta c_A^l) &= 1 - 3 a \tilde{d}_A \overline{m}
       - ad_A\frac{Z_A^s}{\sqrt{3}}\frac{\sqrt{2} m_\ell \check{F}^\ell_n + m_s \check{F}^s_n}
       {f_{F^0_n}^{\mathrm{cont}}(\overline{M}\vphantom{M}^2, \delta M^2)}
       - a \delta c_A Z_A^s\frac{\check{H}^0_n}
       {f_{F^0_n}^{\mathrm{cont}}(\overline{M}\vphantom{M}^2, \delta M^2)},
   \label{eq:linearsingletaeffects}
\end{align}
where $n=\eta,\eta^\prime$.
We have substituted the data on the decay constants
$\tilde{F}^{8}_n$ and
$\check{F}^{0}_n$ by the fitted continuum limit parametrizations
$f^{\mathrm{cont}}_{F^{a}_n}(\overline{M}\vphantom{M}^2, \delta M^2)$, which
enables us to include data points where the denominator
is small and hence carries a large relative error.
This replacement is admissible since the difference is of $\mathcal{O}(a^2)$.
Note that in ansatz~\eqref{eq:fitansatz} $h^{(1)}_O$ is multiplied
by $f^{\mathrm{cont}}_{F^{a}_n}(\overline{M}\vphantom{M}^2, \delta M^2)$.
Above, we suppressed the dependence of the improvement coefficients on $g^2$.
The only unknown functions are
$f_A(g^2)$, $d_A(g^2)$, $\tilde{d}_A(g^2)$ and $\delta c_A(g^2)$ and
we parameterize these as follows
\begin{equation}
  f_A(g^2) = f_A^l g^6, \quad
  d_A(g^2) = b_A(g^2) + d_A^l g^4, \quad
  \tilde{d}_A(g^2) = \tilde{d}_A^l g^4,\quad
  \delta c_A(g^2) = \delta c_A^l g^4,
  \label{eq:imprcoefsinglparam}
\end{equation}
such that only $f_A^l$, $d_A^l$, $\tilde{d}_A^l$ and $c_A^l$ appear
as free parameters on the left hand sides of eqs.~\eqref{eq:linearoctetaeffects}
and~\eqref{eq:linearsingletaeffects}. The above powers of $g^2$ correspond to
the first non-trivial orders of the perturbative expansions.

Turning to the quadratic lattice effects and the functions
$h^{(2)}_O$, we allow for three more fit parameters per observable $O$:
\begin{equation}
  h^{(2)}_O(a^2, t_0\overline{M}\vphantom{M}^2, t_0\delta M^2 | l_O, m_O, n_O) = 1 + a^2 \left(  l_O + m_O \overline{M}\vphantom{M}^2 + n_O\delta M^2\right).\label{eq:quadraticafx}
\end{equation}
The terms multiplied by $l_O$, $m_O$ and $n_O$ correspond to lattice spacing
effects proportional to $a^2 \Lambda^2$, $a^2 \Lambda (2 m_\ell + m_s)$ and $a^2
\Lambda (m_s - m_\ell)$, respectively, 
where $\Lambda \gg m_s \ge m_\ell$ is the QCD scale. Due to this hierarchy of
scales, other quadratic lattice spacing effects depending solely on the quark
masses like, for example, $a^2 m_\ell^2 \approx a^2 M_\pi^4/(4\,B_0^2)$,
are neglected. We remark that for the non-singlet pseudoscalar decay
constants significant $\mathcal{O}(a^2)$ effects have been
reported in lattice results determined using our action~\cite{Bruno:2016plf}.

In summary, in the simultaneous fits of the two masses and four
decay constants a total of four parameters are needed to account for the linear
cutoff effects and $6\times 3=18$ more coefficients to parameterize the
$a^2$-effects. These are in addition to the LECs $M_0$, $F_0$,
$L_5$, $L_8$, $\Lambda_1$ and $\tilde{\Lambda}$ that
appear in the continuum expressions. As will be discussed in the next
subsection, most of the lattice spacing terms cannot be resolved in our data
and the corresponding coefficients will be
set to zero in the fits that we use to determine the final results.

\subsection{Continuum and chiral extrapolation: fits and error estimates\label{sec:fitselection}}
We describe how we determine which fit parameters are
most relevant and how we estimate the systematic uncertainty
associated with the chosen set of fit forms.  Each fit is performed
simultaneously to the six observables determined on ensembles which
lie on two trajectories in the quark mass plane and span four lattice
spacings.  Correlations between the $\eta$ and $\eta^\prime$ masses and the
decay constants as well as the arguments of the fit function ($8
t_0\overline{M}\vphantom{M}^2$, $8 t_0 \delta M^2$) on each ensemble
are taken into account, the latter by employing Orear's
method~\cite{Orear:1981qt}. The fits are performed on the ensemble
averages of the data and the statistical uncertainties in the fit
parameters are obtained by repeating the fit on 500 bootstrap
samples. The statistical uncertainty is taken to be the interval that
contains the central 68.3\,\% of the 500 bootstrap values of each parameter.

The systematics associated with the continuum and quark mass
extrapolations need to be quantified. Since lattice spatial extents of
$L_sM_\pi\gtrsim 4$ are realized, finite volume effects can safely be
neglected.  In terms of the lattice spacing effects, in a first step
we establish which terms in the fit forms presented in the previous
subsection can be resolved. We start with fits to all data employing
the NLO large-$N_{c}$ ChPT continuum limit parametrization and only
include $O(a)$ terms with non-perturbatively determined coefficients,
i.e.\ those involving $b_A$, $\tilde{b}_A$ and $c_A$.  All $O(a^2)$
coefficients are omitted. For this reference fit we obtain $\chi^2 /
N_{\rm df} \approx 220 / 126 \approx 1.75$. Additional discretization terms are
subsequently included and those fits for which the coefficients can be
resolved with reasonable precision are given in tab.~\ref{tab:fitids}.
The reference fit has the id~``1'' in the table.  The LECs
extracted from these fits are collected in
tab.~\ref{tab:contlecs} and the results for the masses and decay
constants at the physical point are detailed in
tab.~\ref{tab:contphysres}. The coefficients of the discretization
terms are provided in app.~\ref{sec:contlimitparams}.

\begin{table*}
\begin{center}
  \small
  \begin{tabularx}{\textwidth}{l|XXXX|XXXXXXXXXXXX}
    \toprule
      id & $f_A^{l}$ & $d^{l}_A$ & $\tilde{d}_A$ & $\delta^{l} c_A$ & $l_{F_\eta^8}$ & $n_{F_\eta^8}$ & $l_{F_{\eta^\prime}^8}$ & $m_{F_{\eta^\prime}^8}$ & $n_{F_{\eta^\prime}^8}$ & $l_{F_\eta^0}$ & $m_{F_\eta^0}$ & $n_{F_\eta^0}$ & $l_{F_{\eta^\prime}^0}$ & $m_{F_{\eta^\prime}^0}$ & $n_{F_{\eta^\prime}^0}$\\\midrule
      1  & ---    & ---    & ---    & ---    & ---    & ---    & ---    & ---    & ---    & ---    & ---    & ---    & ---    & ---    & ---    \\
      2  & \tick & \tick & \tick & \tick & ---    & ---    & ---    & ---    & ---    & ---    & ---    & ---    & ---    & ---    & ---    \\
      3  & \tick & \tick & \tick & ---    & ---    & ---    & ---    & ---    & ---    & ---    & ---    & ---    & ---    & ---    & ---    \\
      4  & \tick & \tick & ---    & ---    & ---    & ---    & ---    & ---    & ---    & ---    & ---    & ---    & ---    & ---    & ---    \\
      5  & \tick & \tick & ---    & ---    &\tick & ---    & \tick & ---    & ---    & ---    & ---    & ---    & ---    & ---    & ---    \\
      6  & \tick & \tick & ---    & ---    & ---    & ---    & ---    & \tick & ---    & ---    & ---    & ---    & ---    & ---    & ---    \\
      7  & \tick & \tick & ---    & ---    & ---    & \tick & ---    & ---    & \tick & ---    & ---    & ---    & ---    & ---    & ---    \\
      8  & \tick & \tick & ---    & ---    &\tick & \tick & ---    & ---    & ---    & ---    & ---    & ---    & ---    & ---    & ---    \\
      9  & \tick & \tick & ---    & ---    & ---    & ---    & \tick & ---    & \tick & ---    & ---    & ---    & ---    & ---    & ---    \\
      10 & \tick & \tick & ---    & ---    & ---    & ---    & ---    & ---    & ---    & \tick & \tick & ---    & ---    & ---    & ---    \\
      11 & \tick & \tick & ---    & ---    & ---    & ---    & ---    & ---    & ---    & ---    & ---    & \tick & ---    & \tick & \tick \\
      12 & \tick & \tick & ---    & ---    & ---    & ---    & ---    & ---    & ---    & ---    & ---    & ---    & \tick & ---    & ---    \\
      13 & \tick & \tick & ---    & ---    &\tick & \tick & ---    & ---    & \tick & ---    & ---    & ---    & ---    & ---    & ---    \\
      14 & \tick & \tick & ---    & ---    &\tick & \tick & ---    & ---    & \tick & ---    & ---    & ---    & ---    & ---    & \tick \\
      15 & \tick & \tick & ---    & ---    &\tick & ---    & ---    & ---    & \tick & ---    & ---    & ---    & ---    & ---    & \tick \\
      16 & \tick & \tick & ---    & ---    & ---    & \tick & ---    & ---    & \tick & ---    & ---    & ---    & ---    & ---    & \tick \\
      17 & \tick & \tick & ---    & ---    &\tick & ---    & ---    & ---    & \tick & ---    & ---    & ---    & ---    & ---    & ---    \\
    \bottomrule
  \end{tabularx}
\end{center}

\caption{Fit forms employed to estimate the systematic uncertainty
  associated with performing the continuum limit extrapolation. A
  cross indicates the corresponding term is included in the fit
  form~(see sec.~\ref{sec:contlimit}) and the coefficient is
  reasonably well determined.
  $f_A$, $d_A$, $\tilde{d}_A$ and $\delta c_A$ are the unknown
  $\mathcal{O}$ improvement coefficients and the coefficients of
  $\mathcal{O}(a^2)$ corrections $l_O$, $m_O$ and $n_O$ are defined in
  eq.~\eqref{eq:quadraticafx}.
  The values of these coefficients can be found in 
  app.~\ref{sec:contlimitparams}.
  In all the cases, the NLO large-$N_c$
  expressions are used for the continuum part of the fit function.\label{tab:fitids} }
\end{table*}

\begin{table*}
  \centering
\pgfplotstabletypeset[
  columns={fitid,EightT0F,EightT0M0Sq,L5,L8,Lambda1,LambdaTilde},
  column type=,
  begin table={\begin{tabularx}{\textwidth}{c | XXXXXX}},
  end table={\end{tabularx}},
  every head row/.style = {before row=\toprule, after row=\midrule},
  every last row/.style = {after row=[1ex]\bottomrule},
  font=\small,
  every column/.style = {string type},
  columns/fitid/.style = {column name=id},
  columns/EightT0F/.style = {column name = $\sqrt{8 t_0^{\chi}}F$, string type},
  columns/EightT0M0Sq/.style = {column name = $8 t^{\chi}_0 M_0^2$, string type},
  columns/L5/.style = {column name = $L_5\cdot 10^{3}$, string type},
  columns/L8/.style = {column name = $L_8\cdot 10^{3}$, string type},
  columns/Lambda1/.style = {column name = $\phantom{-}\Lambda_1$, string type},
  columns/LambdaTilde/.style = {column name = $\phantom{-}\tilde{\Lambda}$, string type},
  ]{\nloextrap}
  \caption{Results for the LECs obtained when employing the fit forms
    detailed in tab.~\ref{tab:fitids}. The dimensionful LECs are given in
    units of the gradient flow scale in the chiral limit.\label{tab:contlecs}}
\end{table*}

\begin{table*}
  \centering
\pgfplotstabletypeset[
  columns={fitid,chi2perdof,metatimessqrt8t0phys,metaprimetimessqrt8t0phys,FAoctet0timessqrt8t0phys,FAoctet1timessqrt8t0phys,FAsinglet0timessqrt8t0phys,FAsinglet1timessqrt8t0phys},
  column type=,
  begin table={\begin{tabularx}{\textwidth}{c | c | XXXXXX}},
  end table={\end{tabularx}},
  every head row/.style = {before row=\toprule, after row=\midrule},
  every last row/.style = {after row=[1ex]\bottomrule},
  font=\small,
  every column/.style = {string type},
  columns/fitid/.style = {column name=id},
  columns/chi2perdof/.style = {string type, column name=$\chi^2/N_{\mathrm{df}}$},
  columns/metatimessqrt8t0phys/.style = {column name = $\sqrt{8 t_0^{\rm ph}}M_\eta$, string type},
  columns/metaprimetimessqrt8t0phys/.style = {column name = $\sqrt{8 t_0^{\rm ph}}M_{\eta^\prime}$, string type},
  columns/FAoctet0timessqrt8t0phys/.style = {column name = $\sqrt{8 t_0^{\rm ph}}F_{\eta}^{8}$, string type},
  columns/FAoctet1timessqrt8t0phys/.style = {column name = $\sqrt{8 t_0^{\rm ph}}F_{\eta^\prime}^{8}$, string type},
  columns/FAsinglet0timessqrt8t0phys/.style = {column name = $\sqrt{8 t_0^{\rm ph}}F_{\eta}^{0}$, string type},
  columns/FAsinglet1timessqrt8t0phys/.style = {column name = $\sqrt{8 t_0^{\rm ph}}F_{\eta^\prime}^{0}$, string type},
  ]{\nloextrap}
  \caption{Results for the masses and decay constants of the $\eta$ and
    $\eta^\prime$ at the physical point in units of the gradient flow
    scale obtained when employing the fit forms detailed in
    tab.~\ref{tab:fitids}.  \label{tab:contphysres}}
\end{table*}

Among the linear improvement terms (see
eqs.~\eqref{eq:linearoctetaeffects}
and~\eqref{eq:linearsingletaeffects}) those involving $f_{A}^{l}$ and
$d_{A}^{l}$ have the largest effect, shifting both the singlet and
octet decay constants considerably when they are included.  We find
the difference between the octet and the singlet quark
mass independent improvement coefficients, $\delta c_A^{l}$, is zero
within errors and $\tilde{d}_A^{l}$ is very small and only weakly
constrained by the data. We were unable to resolve discretization
effects on the masses. All fits require
the $f_{A}^{l}$ and the $d_{A}^{l}$ terms. The fits with the ids~7,
9 and 13--17 in tab.~\ref{tab:fitids}
have the lowest and very similar $\chi^2$ values,
see tab.~\ref{tab:contphysres}.
All these fits have in common that an $\mathcal{O}(a^2)$ effect
proportional to $\delta M^2$ was added
to the octet decay constant of the $\eta'$ meson ($n_{F^8_{\eta^\prime}}\neq 0$
in eq.~\eqref{eq:quadraticafx}).
In what follows we take fit~7 with $\chi^2/N_{\rm df} \approx 179/122 \approx
1.47$ as our main fit. This was selected from the fits with
$1.46\leq\chi^2/N_{\rm df}\leq 1.49$ since the resulting parameter values
are in the centre of the scatter between the different fit forms,
see tabs.~\ref{fig:fitresultslatspacingMasses}--\ref{fig:fitresultslatspacingsingldecs}. We remark again that all
correlations between observables determined on the same ensemble are
taken into account in the fits. Performing an uncorrelated fit with fit
form~7 leads to $\chi^2/N_{\rm df} \approx  155/122 \approx 1.27$.
The systematic uncertainty associated with the continuum extrapolation
is assigned to be the 68.3\,\% interval of the scatter of the central
values of the continuum limit fits performed with fit forms~2 to~17.
Fit~1 is excluded as important $\mathcal{O}(a)$ terms in the
parametrization of the octet and singlet decay constants were omitted
in this case.

\begin{table*}
  \footnotesize
  \centering
  \pgfplotstabletypeset[columns={fitid,metatimessqrt8t0physmean,metatimessqrt8t0physstr,metaprimetimessqrt8t0physmean,metaprimetimessqrt8t0physstr},
column type=,
begin table={\begin{tabularx}{\textwidth}{c p{\graphplotwidth} X p{\graphplotwidth} X}},
  end table={\end{tabularx}},
  every head row/.style = {before row=\toprule, after row=\midrule},
  every last row/.style = {after row=[3ex]\bottomrule},
  columns/fitid/.style = {column name = id},
  columns/fitname/.style = {string type, column name=Name},
  columns/metatimessqrt8t0physmean/.style = {%
    column name = {},
    assign cell content/.code = {%
    \ifnum\pgfplotstablerow=0
    \pgfkeyssetvalue{/pgfplots/table/@cell content}
    {\multirow{\numberofrows}{\graphplotwidth}{\etaerrplot}}%
    \else
    \pgfkeyssetvalue{/pgfplots/table/@cell content}{}%
    \fi
    }
  },
  columns/metatimessqrt8t0physstr/.style = {string type, column name = $\sqrt{8 t_0^{\rm{ph}}} M_\eta$},
  columns/metaprimetimessqrt8t0physmean/.style = {%
    column name = {},
    assign cell content/.code = {%
    \ifnum\pgfplotstablerow=0
    \pgfkeyssetvalue{/pgfplots/table/@cell content}
    {\multirow{\numberofrows}{\graphplotwidth}{\etaprerrplot}}%
    \else
    \pgfkeyssetvalue{/pgfplots/table/@cell content}{}%
    \fi
    }
  },
  columns/metaprimetimessqrt8t0physstr/.style = {string type, column name = $\sqrt{8 t_0^{\rm{ph}}} M_{\eta^\prime}$},
]{\graphplotdata}
\caption{Results for the masses of the $\eta$ and $\eta^\prime$ at the
  physical point in units of the gradient flow scale. The black line
  marks the experimental result converted using $(8t_0^{\rm{ph}})^{-1/2}=
  475(6)\,$MeV~\cite{Bruno:2016plf} and the grey shaded region marks
  the uncertainty due to the error on $t_0^{\rm ph}$. The red line
  indicates the central values predicted by fit~7~(see
  tab.~\ref{tab:fitids}).  The red shaded region represents the total
  uncertainty of our final results where all errors are added in quadrature~(see
  sec.~\ref{sec:massdec}). The first fit does not sufficiently parameterize the lattice
  spacing effects and is not included in the determination of the
  associated systematic error.
  \label{fig:fitresultslatspacingMasses}}
\end{table*}
\begin{table*}
  \footnotesize
  \centering
\pgfplotstabletypeset[columns={fitid,FAoctet0timessqrt8t0physmean,FAoctet0timessqrt8t0physstr,FAoctet1timessqrt8t0physmean,FAoctet1timessqrt8t0physstr},
column type=,
begin table={\begin{tabularx}{\textwidth}{c p{\graphplotwidth} X p{\graphplotwidth} X}},
  end table={\end{tabularx}},
  every head row/.style = {before row=\toprule, after row=\midrule},
  every last row/.style = {after row=[3ex]\bottomrule},
  columns/fitid/.style = {column name = id},
  columns/fitname/.style = {string type, column name=Name},
  columns/FAoctet0timessqrt8t0physmean/.style = {%
    column name = {},
    assign cell content/.code = {%
    \ifnum\pgfplotstablerow=0
    \pgfkeyssetvalue{/pgfplots/table/@cell content}
    {\multirow{\numberofrows}{\graphplotwidth}{\fetaocteterrplot}}%
    \else
    \pgfkeyssetvalue{/pgfplots/table/@cell content}{}%
    \fi
    }
  },
  columns/FAoctet0timessqrt8t0physstr/.style = {string type, column name = $\sqrt{8 t_0^{\rm{ph}}} F_{\eta}^8$},
  columns/FAoctet1timessqrt8t0physmean/.style = {%
    column name = {},
    assign cell content/.code = {%
    \ifnum\pgfplotstablerow=0
    \pgfkeyssetvalue{/pgfplots/table/@cell content}
    {\multirow{\numberofrows}{\graphplotwidth}{\fetaprocteterrplot}}%
    \else
    \pgfkeyssetvalue{/pgfplots/table/@cell content}{}%
    \fi
    }
  },
  columns/FAoctet1timessqrt8t0physstr/.style = {string type, column name =
    $\sqrt{8 t_0^{\rm{ph}}}F_{\eta^\prime}^8$},
]{\graphplotdata}
\caption{Octet decay constants of the $\eta$ and $\eta^\prime$ at the
  physical point in units of the gradient flow scale, displayed as in
  tab.~\ref{fig:fitresultslatspacingMasses}.\label{fig:fitresultslatspacingoctetdecs}}
\end{table*}
\begin{table*}
  \footnotesize
  \centering
\pgfplotstabletypeset[columns={fitid,FAsinglet0timessqrt8t0physmean,FAsinglet0timessqrt8t0physstr,FAsinglet1timessqrt8t0physmean,FAsinglet1timessqrt8t0physstr},
column type=,
begin table={\begin{tabularx}{\textwidth}{c p{\graphplotwidth} X p{\graphplotwidth} X}},
  end table={\end{tabularx}},
  every head row/.style = {before row=\toprule, after row=\midrule},
  every last row/.style = {after row=[3ex]\bottomrule},
  columns/fitid/.style = {column name = id},
  columns/fitname/.style = {string type, column name=Name},
  columns/FAsinglet0timessqrt8t0physmean/.style = {%
    column name = {},
    assign cell content/.code = {%
    \ifnum\pgfplotstablerow=0
    \pgfkeyssetvalue{/pgfplots/table/@cell content}
    {\multirow{\numberofrows}{\graphplotwidth}{\fetasingleterrplot}}%
    \else
    \pgfkeyssetvalue{/pgfplots/table/@cell content}{}%
    \fi
    }
  },
  columns/FAsinglet0timessqrt8t0physstr/.style = {string type, column name = $\sqrt{8 t_0^{\rm{ph}}} F_{\eta}^0$},
  columns/FAsinglet1timessqrt8t0physmean/.style = {%
    column name = {},
    assign cell content/.code = {%
    \ifnum\pgfplotstablerow=0
    \pgfkeyssetvalue{/pgfplots/table/@cell content}
    {\multirow{\numberofrows}{\graphplotwidth}{\fetaprsingleterrplot}}%
    \else
    \pgfkeyssetvalue{/pgfplots/table/@cell content}{}%
    \fi
    }
  },
  columns/FAsinglet1timessqrt8t0physstr/.style = {string type, column name = $\sqrt{8 t_0^{\rm{ph}}}F_{\eta^\prime}^0$},
]{\graphplotdata}
\caption{Singlet decay constants of the $\eta$ and $\eta^\prime$ at the
  physical point in units of the gradient flow scale, displayed as in
  tab.~\ref{fig:fitresultslatspacingMasses}.\label{fig:fitresultslatspacingsingldecs}}
\end{table*}

The $\chi^2/N_{\rm df}$ for our best fits are somewhat larger than one:
either have we underestimated the
errors of our masses and decay constants by about 20\,\% on average or
the functional forms employed do not describe the data
sufficiently well.
Since the lattice spacing effects seem to be relatively mild,
this suggests that NLO large-$N_c$ ChPT does not perfectly describe the data
over the range of quark masses available and higher order
contributions in the chiral expansion have to be taken into
account. The main parameter that determines the convergence of the
chiral expansion is the average pseudoscalar meson mass.  To
investigate the systematics of the chiral extrapolation, we restrict
the mass ranges of the data entering the fit, introducing the cutoffs
$12 t_0 \overline{M}\vphantom{M}^2 < c$, where $c = 1.6, 1.4,
1.2$. These values correspond to $\overline{M}\approx
  493\,\mathrm{MeV}$, $462\,\mathrm{MeV}$ and $427\,\mathrm{MeV}$,
  respectively. Note that $12 t_0 \overline{M}\vphantom{M}^2=1.11$
corresponds to the physical point and our data cover the range
$1.07\lesssim 12 t_0 \overline{M}\vphantom{M}^2 \lesssim 1.68$, see
tab.~\ref{tab:ensembles} and fig.~\ref{fig:quarkmasses}.  Applying the
cuts (successively decreasing $c$) leads to data points being removed
along the trajectory where the strange quark mass is kept constant.
For $c=1.2$ only one ensemble (D201) remains on this trajectory.  We
perform a fit for each cut using fit form~7. The results are listed in
tabs.~\ref{tab:chptphysres} and~\ref{tab:lecschptcutoffs}.  The
$\chi^2 / N_{\rm df}$ of these fits decrease down to a value of $1.25$ as the
data are restricted to smaller values of $\overline{M}\vphantom{M}^2$. This trend
suggests that higher order effects should be considered. However, the
results are all fairly independent of the cut-off. Only the central
value of $M_{\eta^\prime}$ moves upwards and $\tilde{\Lambda}$ downwards by two statistical standard deviations.

\begin{table*}
  \centering
\pgfplotstabletypeset[
  columns={masscut,chi2perdof,metatimessqrt8t0phys,metaprimetimessqrt8t0phys,FAoctet0timessqrt8t0phys,FAoctet1timessqrt8t0phys,FAsinglet0timessqrt8t0phys,FAsinglet1timessqrt8t0phys},
  column type=,
  begin table={\begin{tabularx}{\textwidth}{c | c | XXXXXX}},
  end table={\end{tabularx}},
  every head row/.style = {before row=\toprule, after row=\midrule},
  every last row/.style = {after row=[1ex]\bottomrule},
  font=\small,
  every column/.style = {string type},
  columns/masscut/.style = {column name=$c$},
  columns/chi2perdof/.style = {string type, column name=$\chi^2/N_{\mathrm{df}}$},
  columns/metatimessqrt8t0phys/.style = {column name = $\sqrt{8 t_0^{\rm ph}}M_\eta$, string type},
  columns/metaprimetimessqrt8t0phys/.style = {column name = $\sqrt{8 t_0^{\rm ph}}M_{\eta^\prime}$, string type},
  columns/FAoctet0timessqrt8t0phys/.style = {column name = $\sqrt{8 t_0^{\rm ph}}F_{\eta}^{8}$, string type},
  columns/FAoctet1timessqrt8t0phys/.style = {column name = $\sqrt{8 t_0^{\rm ph}}F_{\eta^\prime}^{8}$, string type},
  columns/FAsinglet0timessqrt8t0phys/.style = {column name = $\sqrt{8 t_0^{\rm ph}}F_{\eta}^{0}$, string type},
  columns/FAsinglet1timessqrt8t0phys/.style = {column name = $\sqrt{8 t_0^{\rm ph}}F_{\eta^\prime}^{0}$, string type},
  ]{\nloextrapchpt}
\caption{Results for masses and decay constants of the $\eta$ and
  $\eta^\prime$ at the physical point in units of the gradient flow
  scale obtained when employing fit~7 of tab.~\ref{tab:fitids} and
  imposing cut-offs $12 t_0 \overline{M}\vphantom{M}^2 <
  c$, as well as including all the data~(first row).\label{tab:chptphysres}}
\end{table*}
\begin{table*}
  \centering
\pgfplotstabletypeset[
  columns={masscut,EightT0F,EightT0M0Sq,L5,L8,Lambda1,LambdaTilde},
  column type=,
  begin table={\begin{tabularx}{\textwidth}{c | XXXXXX}},
  end table={\end{tabularx}},
  every head row/.style = {before row=\toprule, after row=\midrule},
  every last row/.style = {after row=[1ex]\bottomrule},
  font=\small,
  every column/.style = {string type},
  columns/masscut/.style = {column name=$c$},
  columns/EightT0F/.style = {column name = $\sqrt{8 t_0^{\chi}}F$, string type},
  columns/EightT0M0Sq/.style = {column name = $8 t_0^{\chi} M_0^2$, string type},
  columns/L5/.style = {column name = $L_5\cdot 10^{3}$, string type},
  columns/L8/.style = {column name = $L_8\cdot 10^{3}$, string type},
  columns/Lambda1/.style = {column name = $\phantom{-}\Lambda_1$, string type},
  columns/LambdaTilde/.style = {column name = $\phantom{-}\tilde{\Lambda}$, string type},
  ]{\nloextrapchpt}
\caption{LECs\label{tab:lecschptcutoffs} obtained for
  the fits detailed in tab~\ref{tab:chptphysres}. The dimensionful LECs are
    given in units of the gradient flow scale.}
\end{table*}

In principle, large-$N_c$ ChPT expressions for the masses and decay
constants to NNLO are available~\cite{Guo:2015xva,Bickert:2016fgy},
however, the large number of additional LECs cannot be resolved when fitting our data.
Instead, we perform a partial NNLO fit, only including the loop terms
which appear at this order. These do not involve any additional
LECs, see app.~\ref{sec:loopfit} for details on the parametrization
and the resulting LECs. However, fits to this
functional form did not improve the description of the data and our
best fit gives a $\chi^{2}/N_{\rm df} = 2.56$, indicating that
a consistent full NNLO parametrization is required.

Utilizing the available data we cannot resolve additional NNLO LECs.
The impact on our results from imposing different cut-offs on
$\overline{M}\vphantom{M}^2$ was marginal and hence, we take as our central
values the results of fit~7 to all our ensembles, where
$\chi^2/N_{\rm df} \approx 179/122 \approx 1.47$.  To account for the somewhat inferior
quality of this fit, we inflate our statistical errors by the factor
$\sqrt{\chi^2/N_{\rm df}} = 1.21$.  We also add the NLO truncation
error of large-$N_c$ ChPT as a further systematic error (with
subscript $\chi$).  This corresponds to the range of central values
resulting from the fits with different cut-offs.

\subsection{Fit results for the masses and decay constants}
\label{sec:massdec}
\paragraph{LO}
For completeness, we perform a fit to the $\eta$ and $\eta^\prime$
masses employing the LO large-$N_c$ ChPT expressions (see
sec.~\ref{sec:lochpt}). The decay constants are not included in the
analysis as our data clearly contradict the LO ChPT expectation that,
e.g., $F_{\eta}^{0} = - F_{\eta^{\prime}}^{8}$. The parametrization
of the lattice spacing effects was explored in a similar way to the
procedure described in the previous subsection. Our best fit gives
$\chi^2/N_{\rm df} \approx 91/41 \approx 2.35$ and includes the two
quark mass dependent discretization terms
$n_{M_\eta}$ and $n_{M_{\eta^\prime}}$.
This fit is displayed in fig.~\ref{fig:lomassfit} in
app.~\ref{sec:lofitres}.  The masses extracted at the physical point
read: $M_\eta = 1.024(8 t_0^{\rm ph})^{-1/2} = 487\,\mathrm{MeV}$ and
$M_{\eta^\prime} = 1.970(8 t_0^{\rm ph})^{-1/2} = 936\,\mathrm{MeV}$, where we do not
quote any errors since the fit does not describe the data sufficiently
well. The above numbers, in particular the one for the $\eta$ meson,
are significantly lower than the
corresponding experimental masses, $M_\eta\approx 548\,\mathrm{MeV}$ and
$M_{\eta^\prime}\approx 958\,\mathrm{MeV}$.
To this order, the continuum parametrization depends only on
one LEC, the (squared) anomalous mass contribution in the chiral
limit: we find $ M_0^2 = 2.787 (8 t_0^{\chi})^{-1} =
(785\,\mathrm{MeV})^2$. Since the LO fit does not describe our
data well, this value of $M_0$ should also be treated with caution.

\begin{figure}
  \centering
  \includegraphics[width=0.95\linewidth]{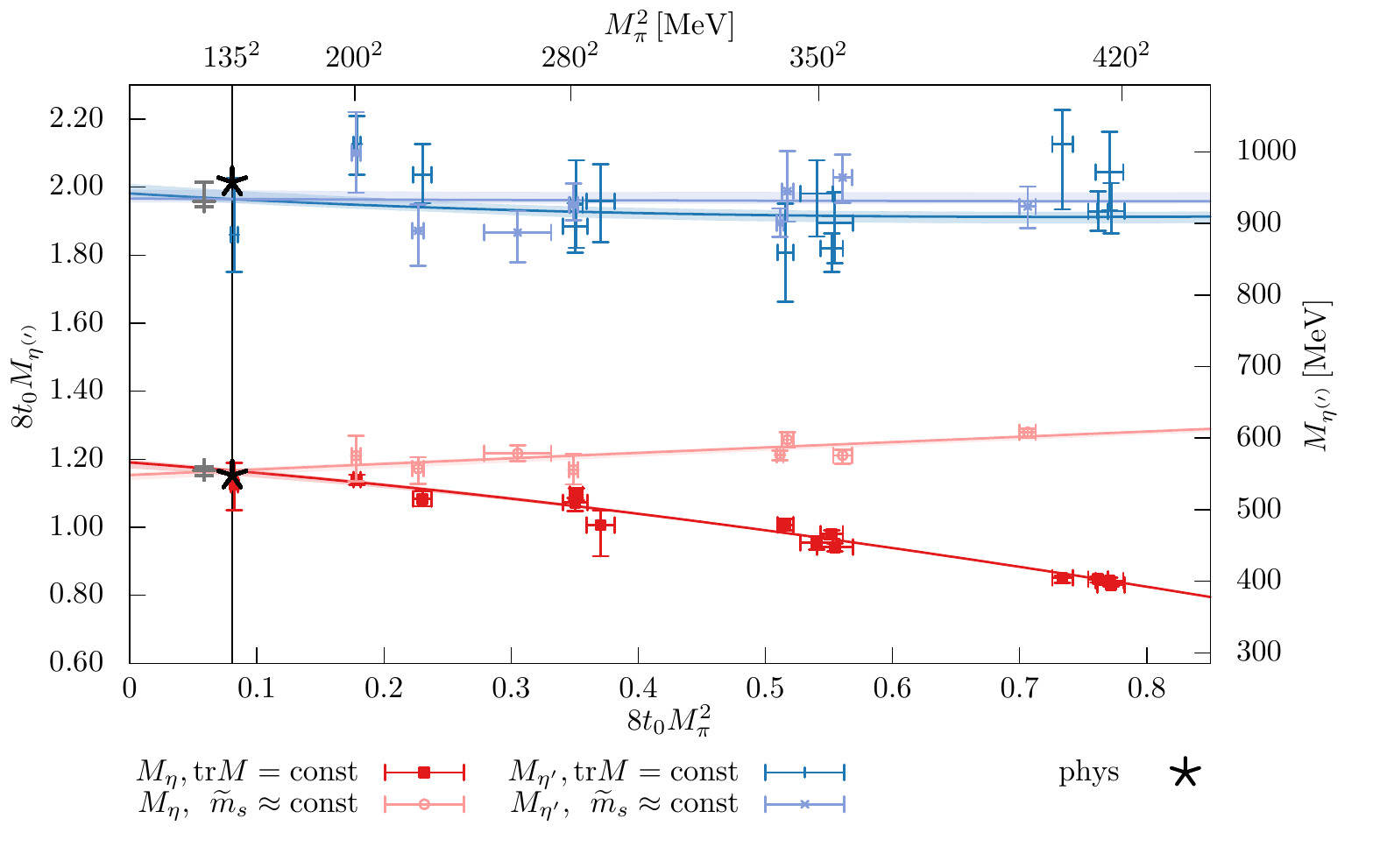}\\[-12pt]
  \includegraphics[width=0.95\linewidth]{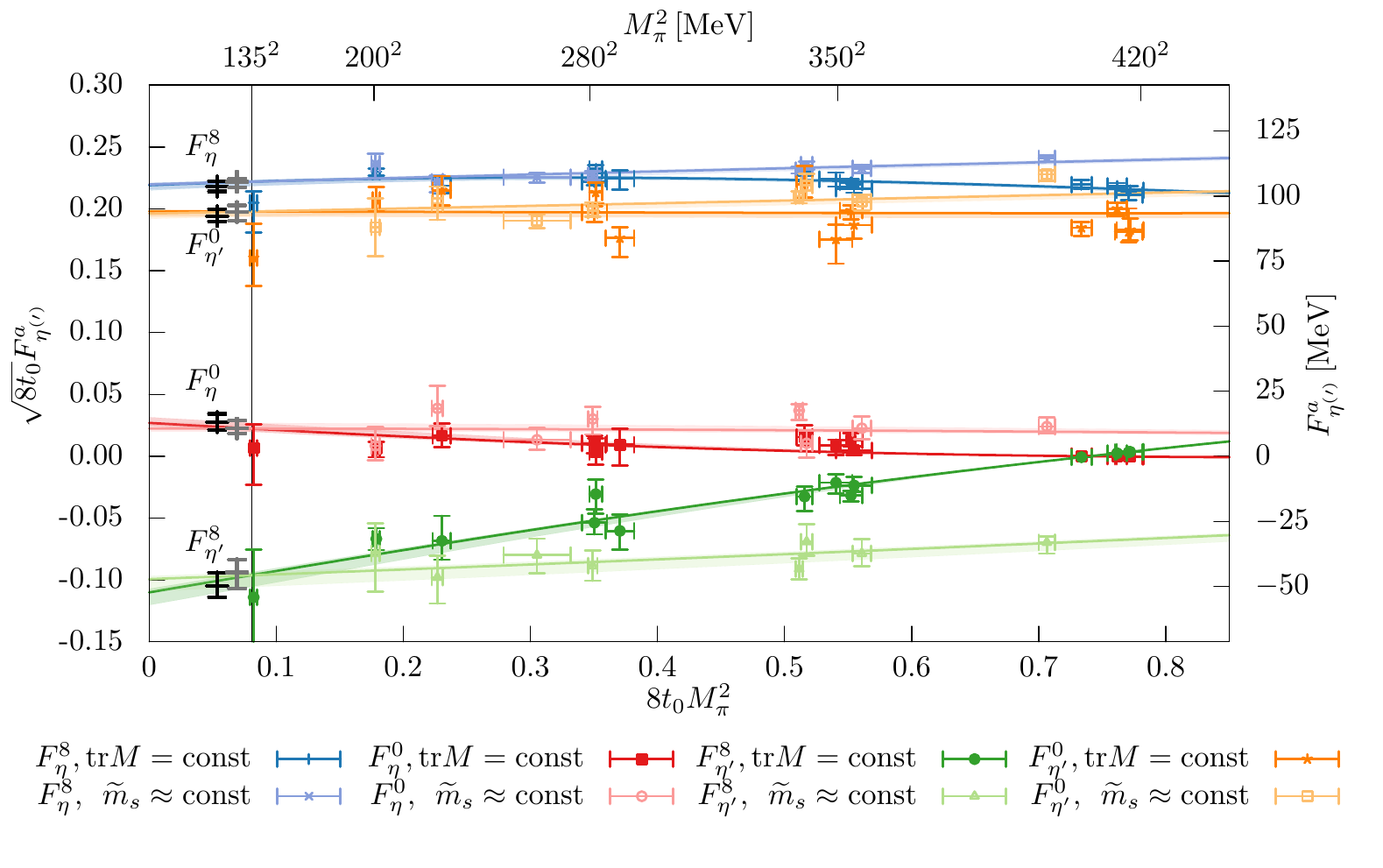}\\[-23pt]
  \caption{\label{fig:nlodecfit}Simultaneous fit to the masses (top) and
    four decay constants (bottom) of the $\eta$ and $\eta^\prime$. The
    fit form incorporates the NLO large-$N_c$ ChPT expressions and the
    discretization terms corresponding to fit~7 in
    tab.~\ref{tab:fitids}. The points have been shifted to compensate
    for lattice spacing effects and lie along two trajectories leading to the
    physical point. The continuum fit functions are indicated by the
    lines and shaded regions (statistical errors only), where the darker
    and lighter colours correspond to
    the trajectories where the flavour average quark mass is
    held constant and the strange quark mass is kept constant, respectively.
    The grey and black error bars (shifted to the left of the physical pion mass
    for better visibility) are our final results, without priors
    (grey) and
    including the experimental $\eta$ and $\eta^\prime$ masses (black stars)
    as priors (black, see sec.~\ref{sec:lecresults}). All errors
    are added in quadrature.
    }
\end{figure}

\paragraph{NLO} Our final results are
obtained employing the NLO continuum limit parametrization within
simultaneous fits to all data on the two masses and four decay
constants. This involves a total of six LECs. Lattice spacing effects
are also accounted for as discussed in
sec.~\ref{sec:fitselection}. The central values are taken from the
results of fit~7, which gave $\chi^2/N_{\rm df} \approx 179/122 \approx 1.47$.
This fit is displayed in fig.~\ref{fig:nlodecfit}.  We obtain for the masses at the physical
point, in the continuum limit
\begin{align}
  \sqrt{8t_0^{\rm ph}}M_\eta =
  1.168\left(\substack{8\\14}\right)_\mathrm{stat} \left(\substack{1\\0}\right)_a\left(\substack{5\\6}\right)_\chi
  \quad \text{and} \quad
  \sqrt{8t_0^{\rm ph}}M_{\eta^\prime} =
  1.958\left(\substack{27\\13}\right)_\mathrm{stat} \left(\substack{0\\6}\right)_a\left(\substack{48\\3}\right)_\chi,
\label{eq:massfitresults}
\end{align}
where the first error is statistical (inflated by $\sqrt{1.47}$), and
the rest are systematic errors: the second error is taken from the
spread of results when varying the parametrization of lattice spacing
effects and the third represents the uncertainty due to the
(continuum) quark mass dependence, see sec.~\ref{sec:fitselection} for
details.  The results are converted to physical units in
sec.~\ref{sec:summary}. In this section we keep all the results in units of
$8 t_0^{\rm ph}$.

\begin{table*}
  \centering
\pgfplotstabletypeset[
  columns={matching,chi2perdof,EightT0M0Sq,Lambda1,LambdaTilde,FAsinglet0timessqrt8t0phys,FAsinglet1timessqrt8t0phys},
  column type=,
  begin table={\begin{tabularx}{\textwidth}{c | c | XXXXX}},
  end table={\end{tabularx}},
  every head row/.style = {before row=\toprule, after row=\midrule},
  every last row/.style = {after row=[1ex]\bottomrule},
  font=\small,
  every column/.style = {string type},
  columns/matching/.style = {column name=$\mu_0$},
  columns/chi2perdof/.style = {string type, column name=$\chi^2/N_{\mathrm{df}}$},
  columns/EightT0M0Sq/.style = {column name = $8 t_0^{\chi} M_0^2$, string type},
  columns/Lambda1/.style = {column name = $\phantom{-}\Lambda_1$, string type},
  columns/LambdaTilde/.style = {column name = $\phantom{-}\tilde{\Lambda}$, string type},
  columns/FAsinglet0timessqrt8t0phys/.style = {column name = $\sqrt{8 t_0^{\rm ph}}F_\eta^0$},
  columns/FAsinglet1timessqrt8t0phys/.style = {column name = $\sqrt{8 t_0^{\rm ph}}F_{\eta^\prime}^0$}
  ]{\nloextrapscale}
  \caption{Results that depend on $Z_A^s$, varying the scale at which we match to perturbation theory.\label{tab:scalecutoffs}}
\end{table*}

The final results for the octet and singlet decay constants read
\begin{align}
  \sqrt{8 t_0^{\rm ph}}F^8_{\eta} &=
\phantom{-}0.2219\left(\substack{18\\37}\right)_\mathrm{stat} \left(\substack{17\\24}\right)_a\left(\substack{10\\2}\right)_\chi,\nonumber\\
  \sqrt{8 t_0^{\rm ph}}F^8_{\eta^\prime} &=
  -0.0939\left(\substack{28\\100}\right)_\mathrm{stat} \left(\substack{84\\0}\right)_a\left(\substack{58\\82}\right)_\chi,\nonumber\\
  \sqrt{8 t_0^{\rm ph}}F^0_{\eta}(\mu=\infty) &= \phantom{-}0.0224\left(\substack{53\\30}\right)_\mathrm{stat} \left(\substack{28\\0}\right)_a\left(\substack{5\\21}\right)_\chi\left(\substack{20\\8}\right)_{\mathrm{renorm}},\nonumber\\
  \sqrt{8 t_0^{\rm ph}}F^0_{\eta^\prime}(\mu=\infty) &= \phantom{-}0.1974\left(\substack{14\\48}\right)_\mathrm{stat} \left(\substack{0\\31}\right)_a\left(\substack{4\\27}\right)_\chi\left(\substack{52\\26}\right)_{\mathrm{renorm}}.
\end{align}
The singlet decay constants depend on the QCD scale. As detailed in
sec.~\ref{sec:singletrenorm}, prior to the fits we run our results
from a scale $\mu_0=a^{-1}$ up to $\mu=\infty$.  To quantify the
systematic error from the matching to the $\overline{\mathrm{MS}}$
scheme, we repeat the fits, setting $\mu_0=a^{-1}/2$ and $\mu_0=2
a^{-1}$, and add the range of results
(see~tab.~\ref{tab:scalecutoffs}) as an additional systematic
error.\footnote{ We remark that slightly different
results are also obtained for the scale independent quantities. However, the differences
are well below any other systematic error, with the exception of those for
$\tilde{\Lambda}$.}  As one may expect, this error is
dominated by the fit where we set $\mu=a^{-1}/2$, see
tab.~\ref{tab:scalecutoffs}.  The results for the decay constants can
also be converted to the strange/light flavour basis
(eq.~\eqref{eq:convertdecay}) and/or given in terms of two angles and
two dimensionful decay constants, see
eqs.~\eqref{eq:octsingletanglerep} and~\eqref{eq:flavouranglerep}.  
All the results in the different conventions and for the
additional QCD scales $\mu \in \{1\,\mathrm {GeV}, 2\,\mathrm{GeV},
10\,\mathrm{GeV}\}$ are collected in tab.~\ref{tab:alldecresults} in
app.~\ref{app:decayresults}.  We discuss the results and their scale
dependence in detail in sec.~\ref{sec:decsummary}.
\subsection{Results for the large-$N_c$ low energy constants\label{sec:lecresults}}
Our results from the fits detailed above for the large-$N_c$ LECs read
\begin{align}
  L_5 &= \phantom{-}1.58\left(\substack{17\\7}\right)_\mathrm{stat} \left(\substack{0\\22}\right)_a\left(\substack{20\\9}\right)_\chi\cdot 10^{-3},\nonumber\\
  L_8 &= \phantom{-}0.96\left(\substack{15\\6}\right)_\mathrm{stat} \left(\substack{0\\14}\right)_a\left(\substack{16\\7}\right)_\chi\cdot 10^{-3},\nonumber\\
  M_0(\mu=\infty) &= \phantom{-}1.67\left(\substack{2\\6}\right)_\mathrm{stat} \left(\substack{1\\2}\right)_a\left(\substack{0\\1}\right)_\chi\left(\substack{3\\2}\right)_{\mathrm{renorm}} \left(8t_0^{\chi}\right)^{-1/2},\nonumber\\
  F &= \phantom{-}0.1890\left(\substack{27\\37}\right)_\mathrm{stat} \left(\substack{36\\0}\right)_a\left(\substack{21\\38}\right)_\chi\left(8t_0^{\chi}\right)^{-1/2},\nonumber\\
  \Lambda_1(\mu=\infty) &= -0.22\left(\substack{1\\5}\right)_\mathrm{stat} \left(\substack{0\\3}\right)_a\left(\substack{0\\3}\right)_\chi\left(\substack{6\\3}\right)_{\mathrm{renorm}},\nonumber\\
  \tilde{\Lambda} &= -0.20\left(\substack{5\\16}\right)_\mathrm{stat} \left(\substack{19\\0}\right)_a\left(\substack{12\\29}\right)_\chi\left(\substack{3\\5}\right)_{\mathrm{renorm}}.\label{eq:lecs1}
\end{align}
The combination $\tilde{\Lambda}=\Lambda_1 - 2 \Lambda_2$ is scale invariant to
this order in ChPT~\cite{Kaiser:2000gs,Bickert:2016fgy}, however,
since its central value varies when changing $\mu_0$, see
tab.~\ref{tab:scalecutoffs}, we also assign a renormalization
error in this case. The above results give
\begin{align}
  \Lambda_2(\mu=\infty) &= -0.1\left(\substack{8\\4}\right)_\mathrm{stat} \left(\substack{0\\10}\right)_a\left(\substack{14\\8}\right)_\chi\left(\substack{5\\3}\right)_{\mathrm{renorm}}.\label{eq:lecs2}
\end{align}

The fits on which these results are based give
$\eta$ and $\eta^\prime$ masses that are
compatible, within errors,  with experiment, see above
and~sec.~\ref{sec:masssummary}.
Nevertheless, incorporating prior knowledge of the experimental masses
helps to constrain the fit and reduces the errors on the LECs.
To this end, we modify our $\chi^2$ function to penalize
fits that give masses, that are incompatible with experiment:
\begin{align}
  \chi^2_{\mathrm{priors}} = \chi^2 + &
  \left[ \frac{\sqrt{8 t^{\mathrm{ph}}_0}M^{\mathrm{ph}}_\eta - f_{M_\eta}(a=0,12 t_0\overline{M}\vphantom{M}^2, 8 t_0 \delta M^2)}{\sigma\left(\sqrt{8 t^{\mathrm{ph}}_0}M^{\mathrm{ph}}_\eta\right)} \right]^2\nonumber\\
  + & \left[ \frac{\sqrt{8 t^{\mathrm{ph}}_0}M^{\mathrm{ph}}_{\eta^\prime} - f_{M_{\eta^\prime}}(a=0,12 t_0\overline{M}\vphantom{M}^2, 8 t_0 \delta M^2)}{\sigma\left(\sqrt{8 t^{\mathrm{ph}}_0}M^{\mathrm{ph}}_{\eta^\prime}\right)} \right]^2,\label{eq:prior}
\end{align}
where we use the physical values from the Particle Data Group (PDG)~\cite{PDG} for $M_\eta$ and $M_{\eta^\prime}$, see eq.~\eqref{eq:physetaetaprimemasses}.
These are converted to dimensionless numbers,
using $(8 t_0^{\rm ph})^{-1/2} = 475(6)\,\mathrm{MeV}$~\cite{Bruno:2016plf}.
Note that the errors are dominated by the scale and are thus highly correlated. 
This is taken into account by sampling Gaussian distributed
values for $(8 t_0^{\rm ph})^{1/2}$, rather than independently
sampling the two dimensionless combinations
$(8 t_0^{\rm ph})^{1/2} M^{\rm ph}_{\eta}$ and
$(8 t_0^{\rm ph})^{1/2} M^{\rm ph}_{\eta^{\prime}}$.
If a more precise determination of $t_0^{\rm ph}$ became available, the
priors could be further constrained and the uncertainties reduced.

Repeating the whole fitting analysis, now including the priors, we obtain
results that are very similar to those of eqs.~\eqref{eq:lecs1}
and~\eqref{eq:lecs2}:
\begin{align}
  L_5 &= \phantom{-}1.66\left(\substack{12\\9}\right)_\mathrm{stat} \left(\substack{0\\26}\right)_a\left(\substack{13\\8}\right)_\chi\cdot 10^{-3},\nonumber\\
  L_8 &= \phantom{-}1.08\left(\substack{11\\6}\right)_\mathrm{stat} \left(\substack{0\\12}\right)_a\left(\substack{3\\10}\right)_\chi\cdot 10^{-3},\nonumber\\
  M_0(\mu = \infty) &= \phantom{-}1.62\left(\substack{2\\4}\right)_\mathrm{stat} \left(\substack{3\\1}\right)_a\left(\substack{3\\0}\right)_\chi\left(\substack{2\\1}\right)_{\mathrm{renorm}}\left(8t_0^{\chi}\right)^{-1/2}, \nonumber\\
  F &= \phantom{-}0.1866 \left(\substack{26\\29}\right)_\mathrm{stat} \left(\substack{54\\0}\right)_a\left(\substack{19\\16}\right)_\chi\left(8t_0^{\chi}\right)^{-1/2},\nonumber\\
  \Lambda_1(\mu = \infty) &= -0.25\left(\substack{1\\4}\right)_\mathrm{stat} \left(\substack{3\\1}\right)_a\left(\substack{1\\1}\right)_\chi\left(\substack{5\\2}\right)_{\mathrm{renorm}},\nonumber\\
  \tilde{\Lambda} &=
-0.46\left(\substack{8\\10}\right)_\mathrm{stat} \left(\substack{21\\0}\right)_a\left(\substack{9\\10}\right)_\chi\left(\substack{1\\2}\right)_{\mathrm{renorm}},\nonumber\\
  \Lambda_2(\mu = \infty) &= \phantom{-}0.11\left(\substack{5\\5}\right)_\mathrm{stat} \left(\substack{0\\9}\right)_a\left(\substack{6\\5}\right)_\chi\left(\substack{3\\2}\right)_{\mathrm{renorm}}.
  \label{eq:lecsmassprior}
\end{align}
In sec.~\ref{sec:leccomparison} we will convert
the above results into physical units
and discuss them.

In general, the large-$N_c$ LECs will differ from their SU($3$) ChPT
equivalents, see also the discussion
in~\cite{Bickert:2016fgy} and in sec.~\ref{sec:leccomparison}.
In particular, the above LECs do not depend on the ChPT scale
since chiral logarithms only appear starting at NNLO in
large-$N_c$ ChPT. As mentioned above, we checked whether such
contributions improved
the description of the data by adding the NNLO loop terms to the NLO
parametrization. However, this decreased the
quality of the fits, with the best fit giving $\chi^2/N_{\rm df} \approx 312 / 122 \approx 2.56$.
The functional form and the resulting LECs are detailed in
app.~\ref{sec:loopfit}.

\section{Gluonic matrix elements and axial Ward identities}
\label{sec:gluonic}
The AWIs are discussed and the octet
AWI is tested against our data.
We then proceed to construct the gluonic matrix elements of the
$\eta$ and $\eta^{\prime}$, using fermionic currents via the singlet AWI.
After addressing the renormalization of pseudoscalar gluonic
matrix elements, we compare the results obtained via the
singlet AWI with a direct determination. The quark mass dependence
of the topological susceptibility is also determined.
\subsection{The axial Ward identities}
\label{sec:awi2}
The AWIs between renormalized operators (indicated by a hat) read
\begin{equation}
  \partial_{\mu}\widehat{A}^{a}_{\mu}=\widehat{\left(\overline{\psi}\gamma_5\{M,t^a\}\psi\right)}
  +\sqrt{2N_f}\delta^{a0}\widehat{\omega},
  \label{eq:awi}
\end{equation}
where $M=\diag(m_\ell,m_\ell,m_s)$
is the quark mass matrix, $a\in\{0,1,\ldots,8\}$, and the topological charge
density is defined as
\begin{equation}
  \omega(x)=-\frac{1}{16\pi^2}\tr\left[F_{\mu\nu}(x)\widetilde{F}_{\mu\nu}(x)\right]=-\frac{1}{32\pi^2}F_{\mu\nu}^a(x)\widetilde{F}_{\mu\nu}^a(x)
  =-\frac{1}{64\pi^2}\epsilon_{\mu\nu\rho\sigma}
  F_{\mu\nu}^a(x)F_{\rho\sigma}^a(x).
\end{equation}
Since different conventions are used in the literature,
for clarity we have written the right hand side in three different
ways. Regarding the octet and singlet AWIs,
eq.~\eqref{eq:awi} corresponds to
\begin{equation}
  \label{eq:8awi0}
  \partial_{\mu}\widehat{A}^8_{\mu}
  =\frac{2}{3}\left(\widehat{m}_\ell + 2 \widehat{m}_s\right)\widehat{P}^8-\frac{2\sqrt{2}}{3}\delta{\widehat{m}}
  \widehat{P}^0,
\end{equation}
and
\begin{align}
  \label{eq:0awi}
  \partial_{\mu}\widehat{A}^0_{\mu}=\frac{2}{3}\left(2\widehat{m}_{\ell}
  +\widehat{m}_s\right)\widehat{P}^0-\frac{2\sqrt{2}}{3}
  \delta\widehat{m}\widehat{P}^8+\sqrt{6}\,\widehat{\omega},
\end{align}
respectively. In the octet/singlet basis only the singlet AWI receives
a contribution from the anomaly. The corresponding AWIs in the flavour
basis read
\begin{align}
  \label{eq:awiflavour}
  \partial_{\mu}\widehat{A}^s_{\mu}
  =2\widehat{m}_s\widehat{P}^s + 2\,\widehat{\omega},\quad
  \partial_{\mu}\widehat{A}^{\ell}_{\mu}
  =2\widehat{m}_{\ell}\widehat{P}^{\ell} + 2\sqrt{2}\,\widehat{\omega}.
\end{align}
These are somewhat simpler because the quark flavours decouple, up to
the anomaly contribution which now enters both AWIs.

We determine our quark masses,
using the AWIs for $a=1$ and $a=4$ in the lattice scheme:\footnote{%
The same results can be obtained in the first case for $a=2$ and the $\pi^-$
or for $a=3$ and the $\pi^0$ (where the disconnected quark contractions
cancel due to isospin symmetry), while the combination
$\widetilde{m}_{\ell}+\widetilde{m}_s$ can also be extracted
using $a=5, 6, 7$ with the appropriate kaon states.}
\begin{align}
  \label{eq:piawi}
  \partial_{\mu}\langle \Omega|\bar{d}\gamma_\mu\gamma_5u|\pi^+\rangle
  &=2\widetilde{m}_{\ell}
    \langle \Omega|\bar{d}\gamma_5u|\pi^+\rangle,\\
  \partial_{\mu}\langle \Omega|\bar{s}\gamma_\mu\gamma_5u| K^+\rangle
  &=(\widetilde{m}_{\ell}+\widetilde{m}_s)
  \langle \Omega|\bar{s}\gamma_5u| K^+\rangle.\label{eq:kawi}
\end{align}
We carry out the complete
$\mathcal{O}(a)$ improvement of the currents, so that the above relations
hold up to $\mathcal{O}(a^2)$ corrections. For this
the (combinations of) improvement coefficients
$c_A$, $b_A-b_P$ and $\tilde{b}_A-\tilde{b}_P$ are required,
all of which are known non-perturbatively.
The lattice AWI quark masses are related to the continuum
masses via
\begin{align}
  \widehat{m}_q(\mu)=\frac{Z_A}{Z_P(\mu)}\widetilde{m}_q.
\end{align}
Again $Z_A/Z_P$ is known non-perturbatively in the RGI
scheme~\cite{Campos:2018ahf}
and can be related to the $\overline{\mathrm{MS}}$ scheme at a scale
$\mu$ perturbatively at the five-loop level~\cite{Baikov:2014qja}
whenever this is needed.

The octet AWI between lattice matrix elements reads
\begin{align}
  \partial_{\mu}\left\langle\Omega\left|A^8_{\mu}\right|\mathcal{M}\right\rangle
  =\frac{2}{3}\left( \widetilde{m}_\ell + 2 \widetilde{m}_s\right)\left\langle\Omega\left|P^8\right|\mathcal{M}\right\rangle-\frac{2\sqrt{2}}{3}\delta{\widetilde{m}}\,r_P
  \left\langle\Omega\left|P^0\right|\mathcal{M}\right\rangle,
  \label{eq:8awi}
\end{align}
where $\delta\widetilde{m}=\widetilde{m}_s-\widetilde{m}_{\ell}$ denotes
the difference between the lattice AWI quark masses.
This expression is only non-trivial for $\eta$ and $\eta^\prime$ states.
The (scale independent) ratio
$r_P=Z_P^s/Z_P$ appears since the renormalization
of the singlet relative to that of the non-singlet pseudoscalar current
can differ at $\mathcal{O}(g^6)$ for Wilson fermions.
In addition to the known improvement
coefficients, also $c_P^s$ (which is equivalent to
$g_P$, see eq.~\eqref{eq:gpimp}), $d_P$ and $\tilde{d}_P$ contribute,
while for $m_s\neq m_{\ell}$, $f_A$ and $f_P$ appear too.
For $m_s=m_{\ell}$, the Ward identity is trivial if applied to
the $\eta^\prime=\eta_0$ state. Note that the left hand side of eq.~\eqref{eq:8awi}
can be replaced with
\begin{equation}
  \partial_{\mu}\left\langle\Omega\left|A^8_{\mu}\right|\mathcal{M}\right\rangle
  =Z_A^{-1}\partial_{\mu}\left\langle\Omega\left|\widehat{A}^8_{\mu}\right|\mathcal{M}\right\rangle=Z_A^{-1}M_{\mathcal{M}}^2F^8_{\mathcal{M}}.
  \label{eq:subst}
\end{equation}
Hence, the combination on the right hand side
  of eq.~\eqref{eq:8awi} does not depend on the momentum of the
  meson $\mathcal{M}$.

\begin{figure}
  \includegraphics[width=0.49\linewidth]{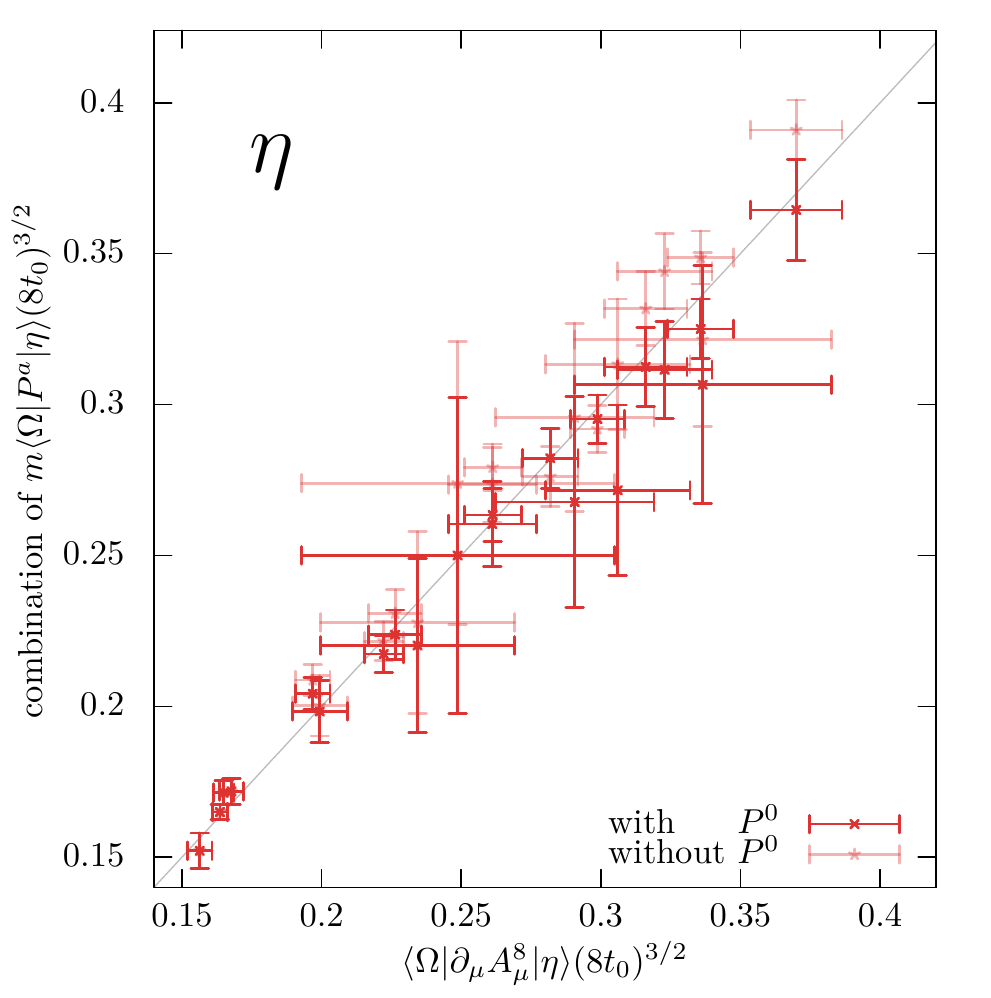}
  \includegraphics[width=0.49\linewidth]{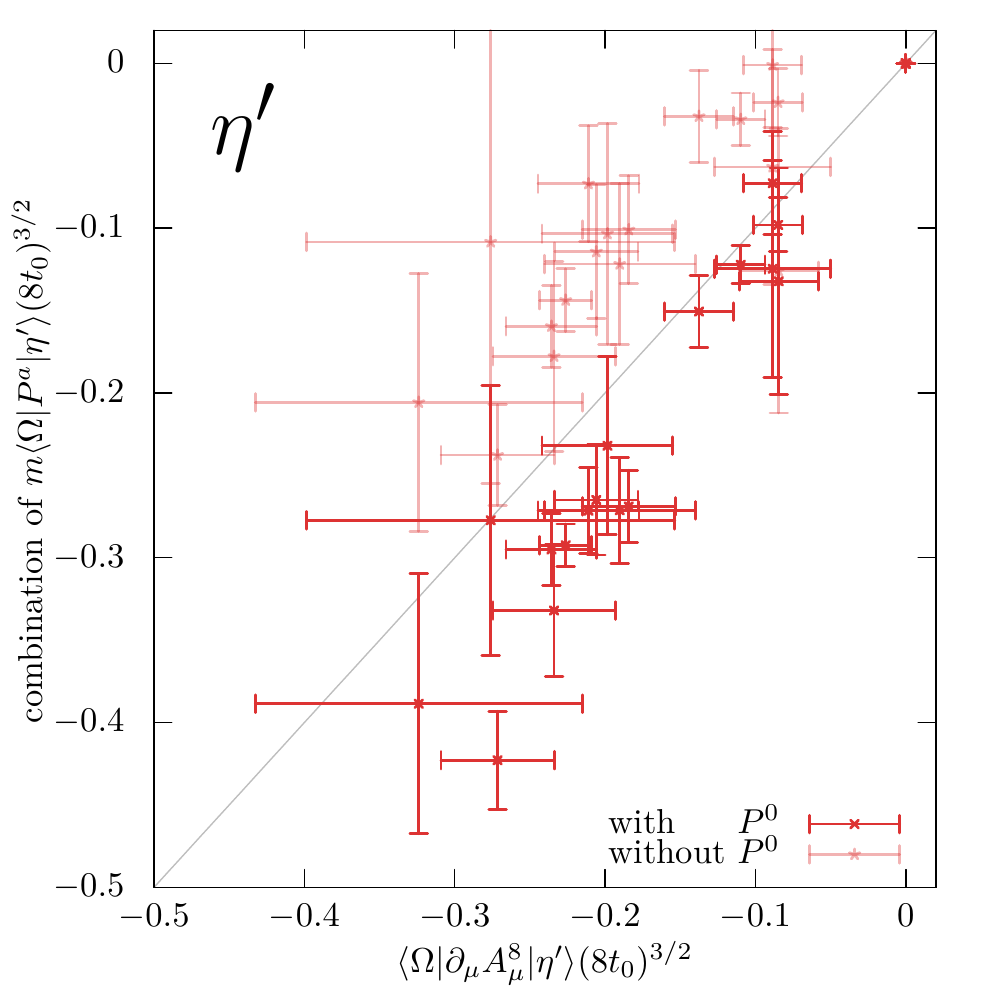}
  \caption{\label{fig:octetawi}Check of the octet AWI~\eqref{eq:8awi} for
    the $\eta$ and $\eta^\prime$ states. Light points correspond to the case when
    the singlet contribution to the octet AWI is neglected.}
\end{figure}

In fig.~\ref{fig:octetawi} we check the octet AWI~\eqref{eq:8awi} at
zero momentum directly against our data, utilizing the quark masses
computed according to eqs.~\eqref{eq:piawi} and~\eqref{eq:kawi}.  Note
that the equality should hold without any renormalization, up to the
ratio $r_P$.  For the comparison, we set $r_P=1$,
$c_P^s=d_P=\tilde{d}_P=f_P=0$ and $f_A=-0.689\,g^6$. The value of the
latter coefficient
is taken from the central fit of sec.~\ref{sec:fitselection}
(fit~7).
Throughout, we find reasonable agreement between the left and right
hand sides of the Ward identity, as shown in fig.~\ref{fig:octetawi}.  Only
the pseudoscalar combination for the $\eta^\prime$ tends to result in
slightly smaller values than those of the derivative of the
axialvector current. Within our precision, we conclude that indeed
$r_P=1$ to a good approximation and that the effect of the three (for
$m_s=m_{\ell}$) or four (for $m_s\neq m_{\ell}$) unknown improvement
terms is moderate, even for our coarsest lattice spacing.

\begin{figure}
  \includegraphics[width=0.49\linewidth]{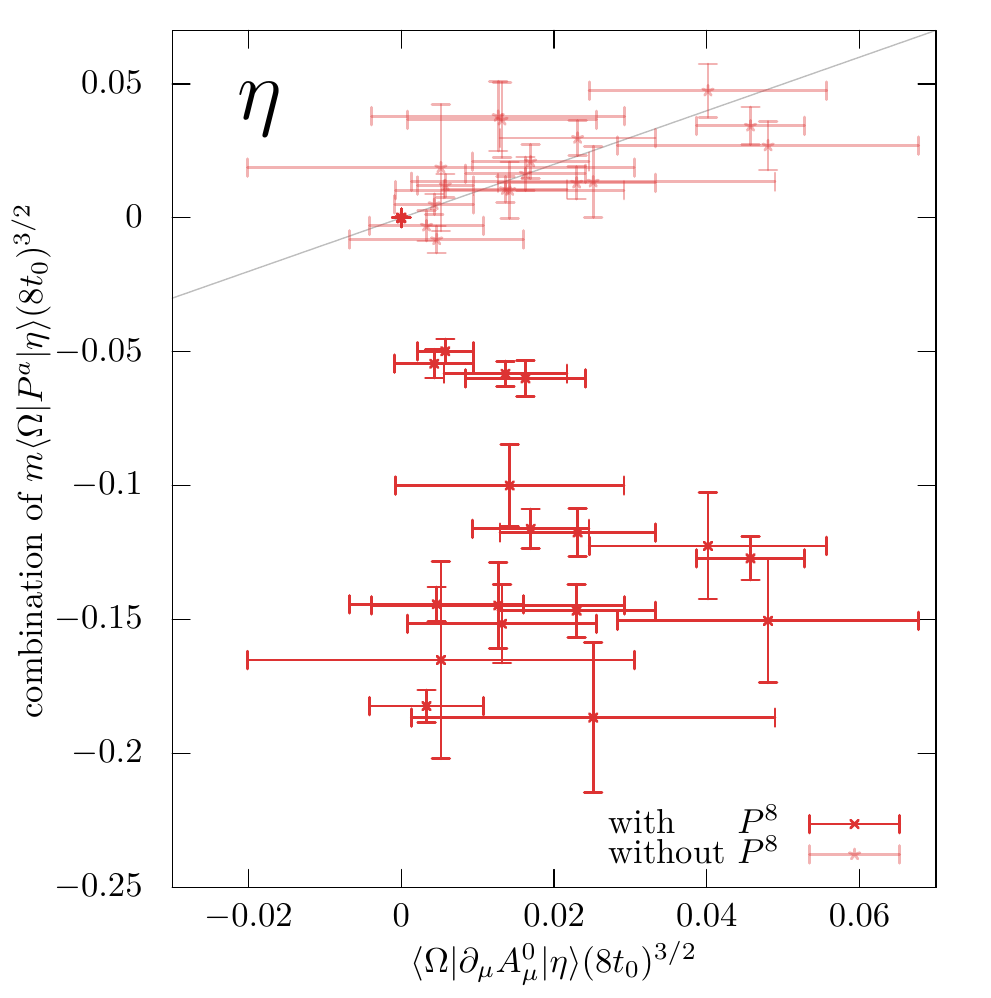}
  \includegraphics[width=0.49\linewidth]{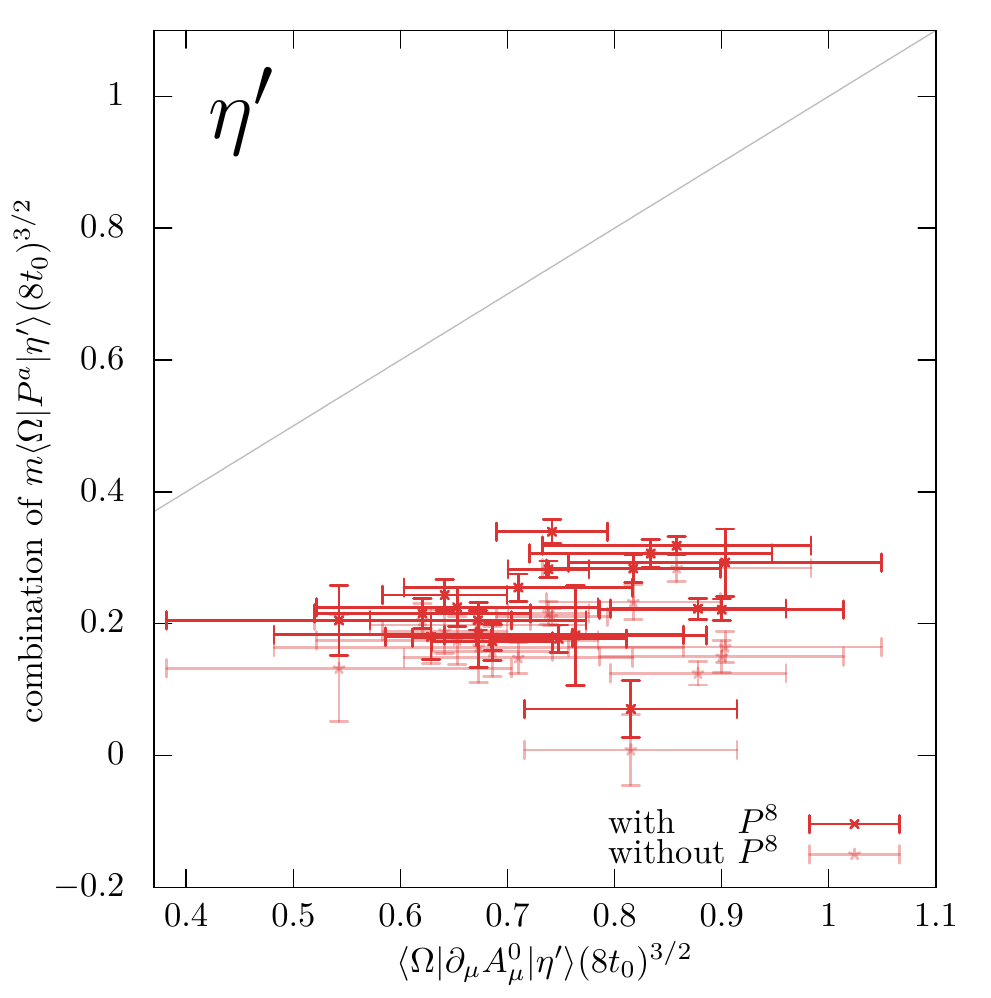}
  \caption{\label{fig:singletawi}Fermionic contributions to the singlet
    AWI for the $\eta$ and $\eta^\prime$ states. Light points correspond
    to only the singlet contribution whereas the darker points correspond to
    all pseudoscalar terms in eq.~\eqref{eq:0awi}, but the anomalous
    contribution is neglected.}
\end{figure}

In the singlet case, due to the anomaly contribution to
eq.~\eqref{eq:0awi}, we would expect that
\begin{align}
  \partial_{\mu}\left\langle\Omega\left|A^0_{\mu}\right|\mathcal{M}\right\rangle
  \neq 2\overline{\widetilde{m}}\,r_P\left\langle\Omega\left|P^0\right|\mathcal{M}\right\rangle-\frac{2\sqrt{2}}{3}\delta{\widetilde{m}}
  \left\langle\Omega\left|P^8\right|\mathcal{M}\right\rangle,
\end{align}
where $\overline{\widetilde{m}}$ denotes
the average lattice AWI quark mass.
Again we set $r_P=1$ and ignore any unknown improvement terms.
The comparison at zero momentum
is shown in fig.~\ref{fig:singletawi}.
The difference is large for both states and does not significantly depend on the
lattice spacing but mostly on the quark masses. This rules out the incomplete
singlet $\mathcal{O}(a)$ improvement as a major cause for the disagreement.
Interestingly, in the case of the $\eta$, the singlet
pseudoscalar contribution coincides with the left hand
side of eq.~\eqref{eq:0awi}: the (in this case) large octet
pseudoscalar matrix element approximately cancels
against the anomaly term. For the $\eta^\prime$ the octet
contribution is much
smaller and no such effect can be seen.
In both cases, contributions from the anomalous matrix elements
$\langle \Omega | \omega | \eta^{(\prime)}\rangle$ are large in comparison to the
terms involving pseudoscalar matrix elements and
it is clear that the anomalous term must be included.
The gluonic matrix element can be determined simply from the difference
observed in these plots, a procedure that does not involve any additional
renormalization. We follow this strategy in the next subsection.

\subsection{Fermionic determination of $\langle \Omega|2\omega|\eta\rangle$
and $\langle \Omega|2\omega|\eta^\prime\rangle$\label{sec:gluonmefermionic}}
Rather than determining the renormalized matrix elements
\begin{equation}
  a_{\mathcal{M}}=2\langle\Omega|\widehat{\omega}|\mathcal{M}\rangle
\end{equation}
directly using gluonic correlators, we first compute them via the
renormalized singlet AWI~\eqref{eq:0awi}:
\begin{align}
  a_{\mathcal{M}}(\mu)&=
    \sqrt{\frac{2}{3}}Z_A^{s}(\mu)\partial_{\mu}\left\langle
    \Omega\left|A_{\mu}^0\right|\mathcal{M}\right\rangle
    +\frac{2\sqrt{2}}{\sqrt{3}}Z_A\left[\frac{\sqrt{2}}{3}\delta\widetilde{m}
      \left\langle
    \Omega\left|P^8\right|\mathcal{M}\right\rangle
    -r_P\overline{\widetilde{m}}\left\langle
    \Omega\left|P^0\right|\mathcal{M}\right\rangle
    \right]\nonumber\\
    &=\sqrt{\frac{2}{3}}M_\mathcal{M}^2
    F^0_{\mathcal{M}}(\mu)+\frac{4}{3\sqrt{3}}\frac{Z_A}{Z_{P}}\delta\widetilde{m}H_{\mathcal{M}}^8
    -\frac{2\sqrt{2}}{\sqrt{3}}r_P\frac{Z_A}{Z_{P}}\overline{\widetilde{m}}H_{\mathcal{M}}^0,\label{eq:gluonmefermionic}
\end{align}
where $H_{\mathcal{M}}^{a}$ are the renormalized and $\mathcal{O}(a)$ improved
pseudoscalar matrix elements, in analogy
to eqs.~\eqref{eq:octetimprovement} and~\eqref{eq:singletimprovement}.
This fermionic definition has the advantage that no knowledge of the
renormalization factors $Z_{\omega A}$ and $Z_\omega$ is needed
(see sec.~\ref{sec:ren0awi} below).
Note that $a_{\mathcal{M}}$ depends on the renormalization scale
$\mu$ and only in the modified scheme, discussed in sec.~\ref{sec:singletrenorm}
(which corresponds to $\mu\rightarrow\infty$), do the
gluonic matrix elements become scale independent.

The improvement coefficients that enter the computation of the singlet decay
constants are taken from sec.~\ref{sec:fitselection} and we set
$d_A = b_A + 1.84\, g^4$, $f_{A} = -0.689\,g^{6}$ and
$\tilde{d}_A = \delta c_A = 0$ (as determined using fit~7).
Six improvement coefficients are needed for the pseudoscalar currents:
$b_P$, $\tilde{b}_P$,
$d_{P}$, $\tilde{d}_{P}$, $f_{P}$ and $c_{P}^{s}$.
The latter was defined in eq.~\eqref{eq:gpimp} and persists in the chiral limit.
We remark that by replacing 
$ag_{P}\mathrm{tr}F_{\mu \nu} \widetilde{F}_{\mu\nu} \mapsto a c_{P}^{s}\partial_{\mu}A_{\mu}^0$, the definition of the
coefficients $d_P$ and $\bar{d}_P$ (and therefore of $\tilde{d}_P$) with
respect to~\cite{Bhattacharya:2005rb} is somewhat altered.
Some of the above coefficients will be free parameters within a combined fit,
incorporating the NLO large-$N_{c}$ ChPT continuum prediction
eqs.~\eqref{eq:aetanlo} and~\eqref{eq:aetaprimenlo}, that we derive
in app.~\ref{sec:nlogluonme}.

We start by defining partially improved matrix elements at $\mu=\infty$:
\begin{align}
  \check{a}_{\eta^{(\prime)}}
    &=\sqrt{\frac{2}{3}}M_{\eta^{(\prime)}}^2
    F^0_{\eta^{(\prime)}}(\mu)+\frac{4}{3\sqrt{3}}Z_A\delta\widetilde{m}\check{H}_{\eta^{(\prime)}}^8
    -\frac{2\sqrt{2}}{\sqrt{3}}r_PZ_A\overline{\widetilde{m}}\check{H}_{\eta^{(\prime)}}^0,\label{eq:gluonmefermionicpartimp}
\end{align}
where $\check{H}^a_{\eta^{(\prime)}}=\langle\Omega| P^a|\eta^{(\prime)}\rangle$
are unimproved pseudoscalar lattice matrix elements
and we assume $r_P=1$.
We then carry out a fit according to
\begin{align}
  &\check{a}_{\eta^{(\prime)}}(a, \overline{M}\vphantom{M}^{2}, \delta M^{2}) =  a_{\eta^{(\prime)}}( \overline{M}\vphantom{M}^{2}, \delta M^{2}|\cdots) \\
  &\qquad - \frac{2\sqrt{2}Z_A}{\sqrt{3}} \overline{\widetilde{m}} \bigg[ 3 a \tilde{d}_{P} \overline{m} \check{H}_{\eta^{(\prime)}}^{0} +  a d_{P} \frac{1}{\sqrt{3}}\left(\sqrt{2}m_{\ell} \check{H}^{\ell}_{\eta^{(\prime)}} + m_{s} \check{H}^{s}_{\eta^{(\prime)}}\right) + a c_{P}^{s} M^{2}_{\eta^{(\prime)}}F^{0}_{\eta^{(\prime)}}\bigg]\nonumber\\
    &\qquad +  \frac{4Z_A}{3\sqrt{3}} \delta \widetilde{m} \left[3 a  \tilde{b}_{P} \overline{m}\check{H}_{\eta^{(\prime)}}^{8} + a b_{P}\frac{1}{\sqrt{3}} \left(m_{\ell} \check{H}^{\ell}_{\eta^{(\prime)}} - \sqrt{2}m_{s} \check{H}^{s}_{\eta^{(\prime)}}\right)+ \sqrt{2} a f_{P} \delta m \check{H}^{0}_{\eta^{(\prime)}}\right],\nonumber
\end{align}
where the continuum parametrizations
$a_{\eta^{(\prime)}}(\overline{M}\vphantom{M}^{2}, \delta M^{2}|\cdots)$ correspond to
eqs.~\eqref{eq:aetanlo}--\eqref{eq:aetaprimenlo} and the ellipses
represent the six NLO LECs. In keeping with the rest of our analysis,
all dimensionful quantities appearing within this fit are multiplied
by the appropriate powers of $\sqrt{8t_0}$.  We parameterize the
coefficients $d_{P}$, $\tilde{d}_{P}$, $f_{P}$ and $c_{P}^{s}$ (that are
functions of $g^2$) similarly to eq.~\eqref{eq:imprcoefsinglparam}
with one parameter each, while $b_P$ and $\tilde{b}_P$ are known
non-perturbatively~\cite{improve2}:
\begin{align}
  b_{P}(\beta=3.4) &=  1.622(74),
  & b_{P}(\beta=3.46) &= 1.592(213),\nonumber\\
  b_{P}(\beta=3.55) &=  1.560(165),
  & b_{P}(\beta=3.7) &= 1.696(78),\\
  \tilde{b}_{P}(\beta=3.4) &=  0.39(27),
  & \tilde{b}_{P}(\beta=3.46) &= 0.32(20),\nonumber\\
  \tilde{b}_{P}(\beta=3.55) &=  0.40(23),
  & \tilde{b}_{P}(\beta=3.7) &= 0.16(13).
\end{align}

\begin{figure}
  \includegraphics[width=0.49\linewidth]{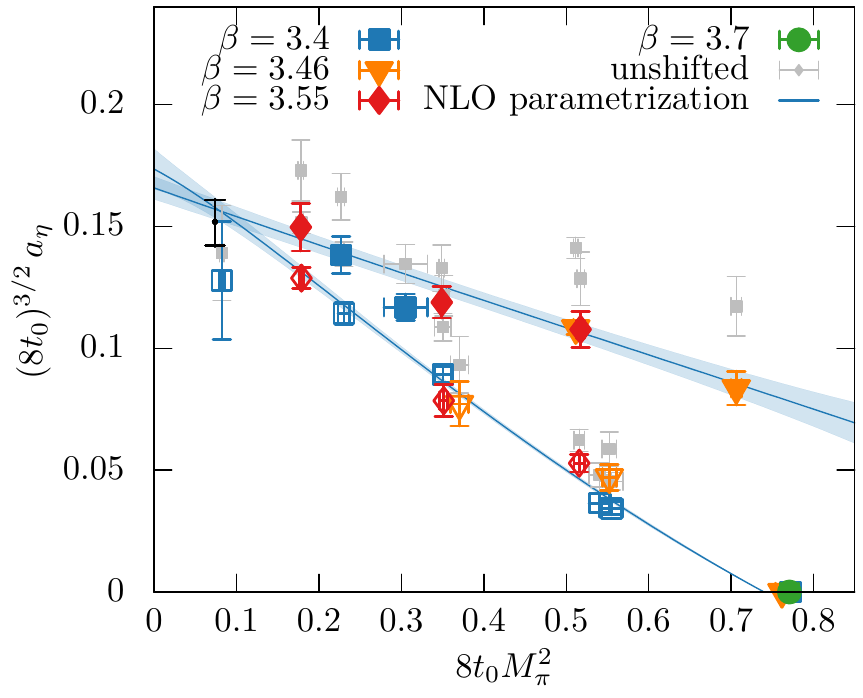}
  \includegraphics[width=0.49\linewidth]{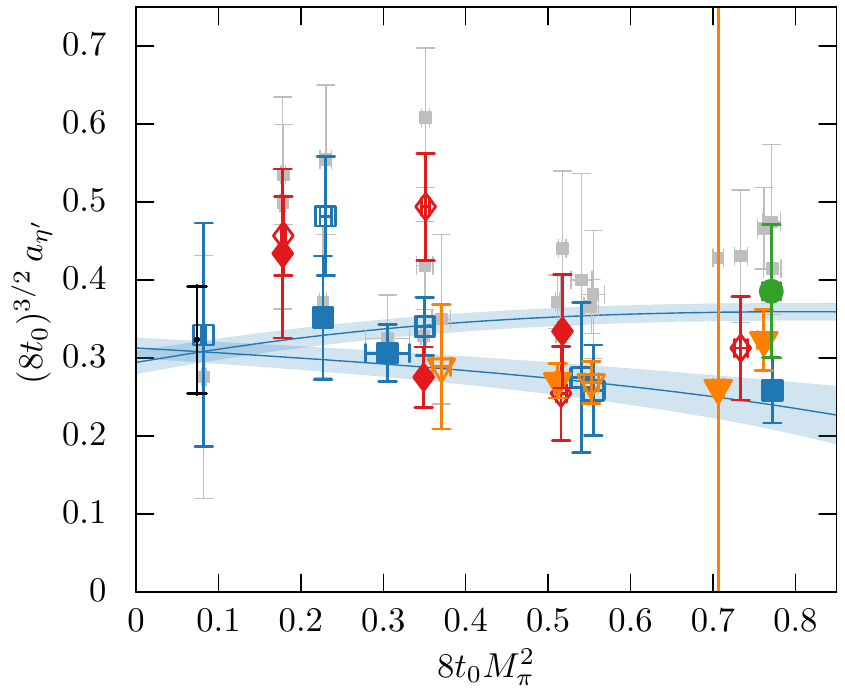}
  \caption{\label{fig:gluonme}The anomalous gluonic matrix element
    $a_\eta$ (left) and $a_{\eta^\prime}$ (right)
    determined via the singlet AWI from fermionic matrix elements,
    eq.~\eqref{eq:gluonmefermionicpartimp}. The coloured points have been
    adjusted for lattice spacing effects, while the grey points
    indicate the unshifted data. The two curves correspond to the NLO
    large-$N_c$ ChPT parametrization eqs.~\eqref{eq:aetanlo}
    and~\eqref{eq:aetaprimenlo} for trajectories with a constant average quark mass and a constant
    strange quark mass.  The black error bars indicate the final results at the physical
  point including statistical and systematic errors.}
\end{figure}

The resulting 10-parameter fit is only weakly constrained, however,
at NLO in large-$N_c$ ChPT the LECs should be identical to those
that we already determined in sec.~\ref{sec:lecresults}.
Therefore, in analogy to eq.~\eqref{eq:prior}, we add these results, given in eq.~\eqref{eq:lecsmassprior}, as priors to the $\chi^2$ function. The widths $\sigma$ are set to the statistical and systematic errors, added in quadrature.
It turns out that we are still
unable to resolve $\tilde{d}_{P}$ and fix $\tilde{d}_{P} = \tilde{b}_{P}$ instead.

The fit, shown in fig.~\ref{fig:gluonme},
gives a valid description of the data,
with a fully correlated $\chi^{2}/N_{\rm df} \approx 34 / 31 \approx 1.09$.
We obtain
\begin{equation}
  d_{P}(g^{2}) = b_{P}(g^2) + 6.6(6) g^{4}, \quad
  c_{P}^{s}(g^{2}) = -2.4(3) g^{4}\quad\text{and}\quad
  f_{P}(g^{2}) = -26(6) g^{6}
\end{equation}
for the additional improvement coefficients, setting $\tilde{d}_P=\tilde{b}_P$.
The corresponding LECs read
\begin{align}
  L_5 &= \phantom{-}1.95\left(\substack{7\\3}\right)_\mathrm{stat}\cdot 10^{-3}, &
  L_8 &= \phantom{-}0.97\left(\substack{4\\6}\right)_\mathrm{stat}\cdot 10^{-3}, &
  M_0 &= \phantom{-}1.59\left(\substack{1\\6}\right)_\mathrm{stat}\left(8 t_{0}^{\chi}\right)^{-1/2} ,\nonumber\\
F &= \phantom{-}0.1881\left(\substack{9\\24}\right)_\mathrm{stat}\left(8 t_{0}^{\chi}\right)^{-1/2} , &
  \Lambda_1 &= -0.10\left(\substack{1\\1}\right)_\mathrm{stat}, &
  \tilde{\Lambda} &= -0.21\left(\substack{3\\4}\right)_\mathrm{stat},
\end{align}
where the errors given are purely statistical and
generally small, due to the priors. In particular, $\Lambda_{1}$ is by
2.8 standard deviations larger than its input
value~\eqref{eq:lecsmassprior}, obtained from the fit to the masses
and decay constants, and $\tilde{\Lambda}$ moves up
accordingly. Also $L_5$ is larger by about $1.3\,\sigma$.
This indicates some tension between the data and the
NLO expressions.

At the physical point and $\mu = \infty$, the fit gives
\begin{equation}
  (8 t_{0}^{\rm ph})^{3/2}a_{\eta} = 0.1564\left(\substack{37\\63}\right)\quad\text{and}\quad(8 t_{0}^{\rm ph})^{3/2}a_{\eta^{\prime}} = 0.308\left(\substack{16\\17}\right).\label{eq:gluonmefitresults}
\end{equation}
The NLO prediction eqs.~\eqref{eq:aetanlo} and~\eqref{eq:aetaprimenlo},
using the LECs of eq.~\eqref{eq:lecsmassprior}, reads
\begin{equation}
  (8 t_{0}^{\rm ph})^{3/2}a_{\eta} = 0.1609\left(\substack{17\\27}\right)\quad\text{and}\quad(8 t_{0}^{\rm ph})^{3/2}a_{\eta^{\prime}} = 0.383\left(\substack{11\\17}\right).\label{eq:gluonmechptresults}
\end{equation}
Note that the latter values are based exclusively on the meson masses
and their decay constants, with no input from the data on $a_{\eta^{(\prime)}}$.
The predictions and fit results are close to each other.
However, within the relatively small
errors stated, the two results on $a_{\eta'}$ differ
by several standard deviations, which indicates
the limitations of the NLO continuum parametrization within our range of
quark masses.
Therefore, we assign the difference between eqs.~\eqref{eq:gluonmefitresults} and~\eqref{eq:gluonmechptresults} as the systematic error associated with taking the physical limit. We discuss the results and quote values in physical units in sec.~\ref{sec:gluonmesummary}.

\subsection{Renormalization of the anomaly term and the topological susceptibility}
\label{sec:ren0awi}
The singlet AWI has received a lot of attention, also in different settings, e.g.,
regarding the spin structure of the nucleon~\cite{Liang:2018pis}. It would be
desirable to validate the singlet AWI in our lattice study, also in view of confirming
a consistent continuum limit extrapolation of the lattice data. Therefore,
we will attempt to compute $a_{\eta^{(\prime)}}$ directly, destroying
$\eta^{(\prime)}$ states by the topological charge density operator.
This however requires an analysis of the renormalization of the anomaly term
and its mixing with the derivative of the axialvector current. We start from the
singlet AWI in the massless case 
\begin{equation}
  \label{eq:massl}
  \partial_{\mu}\widehat{A}^{0}_{\mu}=\sqrt{2N_f}\,\widehat{\omega}.
\end{equation}
Since $\omega(x)$ can be written
as the divergence of a topological current, the associated Pontryagin
index
\begin{equation}
  Q=\int\!\deriv^4\!x\,\omega(x)
\end{equation}
is integer-valued on
$\mathbb{R}^4$ in the continuum limit and scale independent such that
$\omega(x)$ itself will not acquire an anomalous dimension.\footnote{In our conventions the kinetic term of the Lagrangian
reads $\tfrac{1}{4g_s^2}F_{\mu\nu}^aF_{\mu\nu}^a$. In perturbative QCD the coupling
is usually not absorbed into the field and then this term amounts to
$-\tfrac{1}{4}G_{\mu\nu}^aG^{a\mu\nu}$ instead. This would have introduced
a factor of $g_s^2$ in front of $\omega(x)$ within eq.~\eqref{eq:massl} and
an additional scale dependence of $\widehat{\omega}$, governed by the
QCD $\beta$-function.}
$\partial_\mu A^{0}_{\mu}$ can and will mix into $\omega$:\footnote{On the lattice
  with Wilson fermions, in principle $\omega$ can also mix with $a^{-1}P^0$.
  However, such power divergent terms are removed if we define
  $\omega$ after a gradient flow time $t$ that we keep fixed in physical units
  as the continuum limit is approached~\cite{Luscher:2010iy}.
  Likewise, $\langle Q^2\rangle$ will contain a contact term
  $\langle \omega(0)\omega(0)\rangle$, which diverges in the continuum
  limit. This divergence is in fact required to
  reproduce the correct topological susceptibility. However,
  this term could
  in principle also mix with lower dimensional operators. The latter
  possibility is eliminated too, by virtue of the gradient flow.

  Also note
  that $\omega=\partial_{\mu}K_{\mu}$ will not mix into
  lower dimensional operators and neither does the gauge non-invariant
  Chern-Simons current $K_{\mu}$ interfere with the
  renormalization of $A^{0}_{\mu}$.}
\begin{align}
  \widehat{\omega}=Z_{\omega}\omega+Z_{\omega A}\partial_{\mu}A^{0}_{\mu},
  \label{eq:rentop}
\end{align}
up to gradient flow time dependent $\mathcal{O}(a)$ corrections.
We remark that the anomalous dimensions of $A^{0}_{\mu}$ and of
$\omega$ differ from each other in lattice regularization
as well as in naive dimensional regularization.
The singlet AWI holds exactly when defining the topological
charge density using overlap fermions~\cite{Giusti:2001xh}, without any
factor $Z_{\omega}$ and the term containing $Z_{\omega A}$ cancels when
computing the topological susceptibility $\tau$, defined in eq.~\eqref{eq:topol},
with periodic boundary conditions. Since the topological susceptibility 
obtained from employing the overlap definition and the
field theoretical definition after cooling (which is equivalent
to the gradient flow) appear
to agree in the continuum limit~\cite{DelDebbio:2003rn},
it is likely that actually $Z_{\omega}=1$.

Note that the running of $Z_{\omega A}$ with
the scale is the same as that of $Z^s_A$, which
is consistent with eq.~\eqref{eq:massl}.\footnote{%
  In~\cite{Larin:1993tq} somewhat different conventions are used that
  correspond to $4\pi^2a_sG\widetilde{G}=F_{\mu\nu}^aF^a_{\mu\nu}$,
  where $a_s=\alpha_s/\pi=g_s^2/(4\pi^2)$.
  Therefore, in that case the $\gamma$-function for $G\widetilde{G}$ reads
  $-\beta(a_s)/a_s$, while our $Z_{\omega}$ does not carry any anomalous
  dimension. Likewise, in that article the anomalous dimension of the
  off-diagonal element is proportional to $\gamma_A^s/a_s$, while here
  $\gamma_{\omega A}=\gamma_A^s$.}
An alternative scheme of renormalizing the
singlet axialvector current is discussed in sec.~\ref{sec:singletrenorm}.
In that case, both $Z_A^{s}$ and $Z_{\omega A}$ have no anomalous dimension.
We remark that the $Z_{\omega A}\partial_{\mu}A^0_{\mu}$ term will not affect
the topological susceptibility
\begin{equation}
  \label{eq:topol}
  \hat{\tau}=\sum_x\langle \widehat{\omega}(0)\widehat{\omega}(x)\rangle
  =\frac{1}{V}\sum_{x,y}\langle \widehat{\omega}(x)\widehat{\omega}(y)\rangle
  =\frac{\langle \widehat{Q}^2\rangle}{V}
\end{equation}
since this term does not contribute to the volume sum,
due to translational invariance.\footnote{This also holds approximately for open
  boundary conditions in time, provided $L_t-2b$ is much larger than the relevant
  correlation lengths:
  \(
  \sum_{x_0=-L_t/2+b}^{L_t/2-b}\sum_{\vec{x}}\partial_{\mu}A^{0}_{\mu}(x)=
  \frac{1}{2a}\sum_{\vec{x}}\left[A^0_0(L_t/2-b)-
  A^0_0(-L_t/2+b)\right]\rightarrow 0\) for \(L_t\rightarrow\infty\).}

\begin{figure}
  \includegraphics[width=\linewidth]{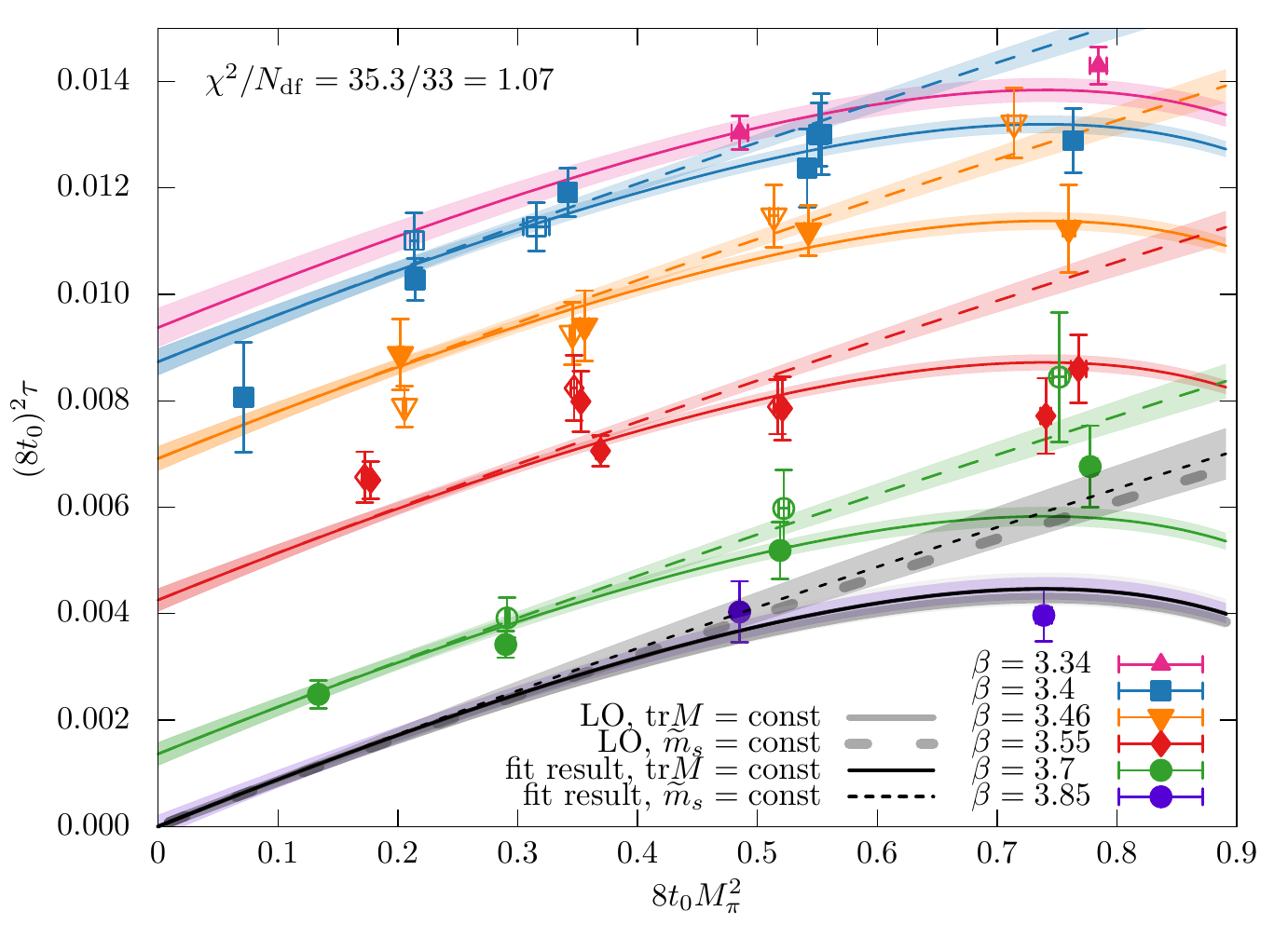}
  \caption{\label{fig:chi}Topological susceptibility for many
    of the CLS ensembles described in~\cite{spectrum}. Filled symbols mark
    ensembles that are simulated with a
    constant sum of quark masses (solid lines), open symbols correspond to
    ensembles with the strange quark mass fixed to approximately the physical
    value (dashed lines).
    Lines and shaded regions are the result of a fit to eq.~\eqref{eq:chifit}.
    The continuum limit result (black lines) is very close to both the fit
    result at
    $\beta=3.85$ as well as the leading order
    expectation (grey lines), when using $\sqrt{8 t_0} F = 0.1866$,
    see~eq.~\eqref{eq:lecsmassprior},
    and setting $Z_\omega = 1$.}
\end{figure}
Within the numerical computations, we use the field-theoretical definition of
$\omega$ extracted after evolving the gauge fields to the gradient flow time
$\sqrt{8 t} = \sqrt{8 t_0^{*}}\approx 0.413\,\mathrm{fm}$.
For ensembles with open boundary conditions we keep
the same distance $b \gtrsim 1.9\,\mathrm{fm}$ to the boundaries when
computing eq.~\eqref{eq:topol} as we did in
the computation of the fermion loops, cf. sec.~\ref{sec:stochmeas}.

As a first sanity check, we plot the topological susceptibility in
fig.~\ref{fig:chi}, where we include most of the CLS ensembles analysed
in~\cite{spectrum}, which adds additional points at finer and coarser lattice spacings.
We find large cut-off effects with our definition of the
susceptibility, shifting points considerably away from the $N_f=3$ continuum
expectation~\cite{DiVecchia:1980yfw,Leutwyler:1992yt},
\begin{equation}
    \hat{\tau} = \frac{F^2}{2} \left(\frac{1}{2 M_K^2 - M_\pi^2} + \frac{2}{M_\pi^2} \right)^{-1}.
\end{equation}
Indeed, large cut-off effects have been reported in unquenched
simulations previously~\cite{Bazavov:2010xr,Chowdhury:2011yj,Cichy:2013rra,Bruno:2014ova,Bonati:2015vqz,Borsanyi:2016ksw,Alexandrou:2017bzk}.
To confirm $Z_\omega = 1$ numerically, we attempt a simple fit to
\begin{equation}
  \label{eq:chifit}
  (8 t_0)^2 \tau =
    \frac{(8 t_0)^2 F^2}{2 Z_\omega^2} \left(\frac{1}{2 M_K^2 - M_\pi^2} + \frac{2}{M_\pi^2} \right)^{-1} + l^{(2)}_\tau \frac{a^2}{t_0^*} + l_\tau^{(3)} \frac{a^3}{(t_0^*)^{3/2}} + l_\tau^{(4)} \frac{a^4}{(t_0^*)^{2}}.
\end{equation}
From this four parameter fit with $\chi^2 / N_{\rm df} \approx 35.3/33 \approx
1.07$, we obtain in the continuum limit
\begin{equation}
  \frac{\sqrt{8 t_0^{\chi}}F}{Z_\omega} = 0.190(13).
\end{equation}
When assuming $Z_{\omega} = 1$, this value agrees with our previous
result $\sqrt{8 t_{0}^{\chi}}F=0.1866(48)$
(see eq.~\eqref{eq:lecsmassprior}
of sec.~\ref{sec:lecresults}).
The coefficients of the terms parameterizing the lattice spacing dependence are
\begin{equation}
  l_\tau^{(2)} = -0.072(10), \quad
  l_\tau^{(3)} = 0.355(34) \quad \text{and}\quad
  l_\tau^{(4)} = -0.324(30),
\end{equation}
resulting in the non-monotonous behaviour observed in fig.~\ref{fig:chi}.
The alternating sign
also explains how the susceptibilities at our finest lattice spacing
$a\approx 0.039\,\mathrm{fm}$ ($\beta=3.85$) can agree with the
continuum limit expectation. We also tried to add mass-dependent
terms to our parametrization of lattice artefacts, however, the
resulting coefficients turned out to be small and the quality
of the fit did not improve. Equation~\eqref{eq:chifit} with
four parameters turned out to be the minimal ansatz that resulted
in a valid description of all our 37 data points. Interestingly,
the leading order continuum limit expectation for the
dependence of $\hat{\tau}$ on the pion and kaon masses already
gives a very adequate description of the data.

\subsection{Direct determination of the gluonic matrix elements}
\begin{figure}
  \includegraphics[width=0.49\linewidth]{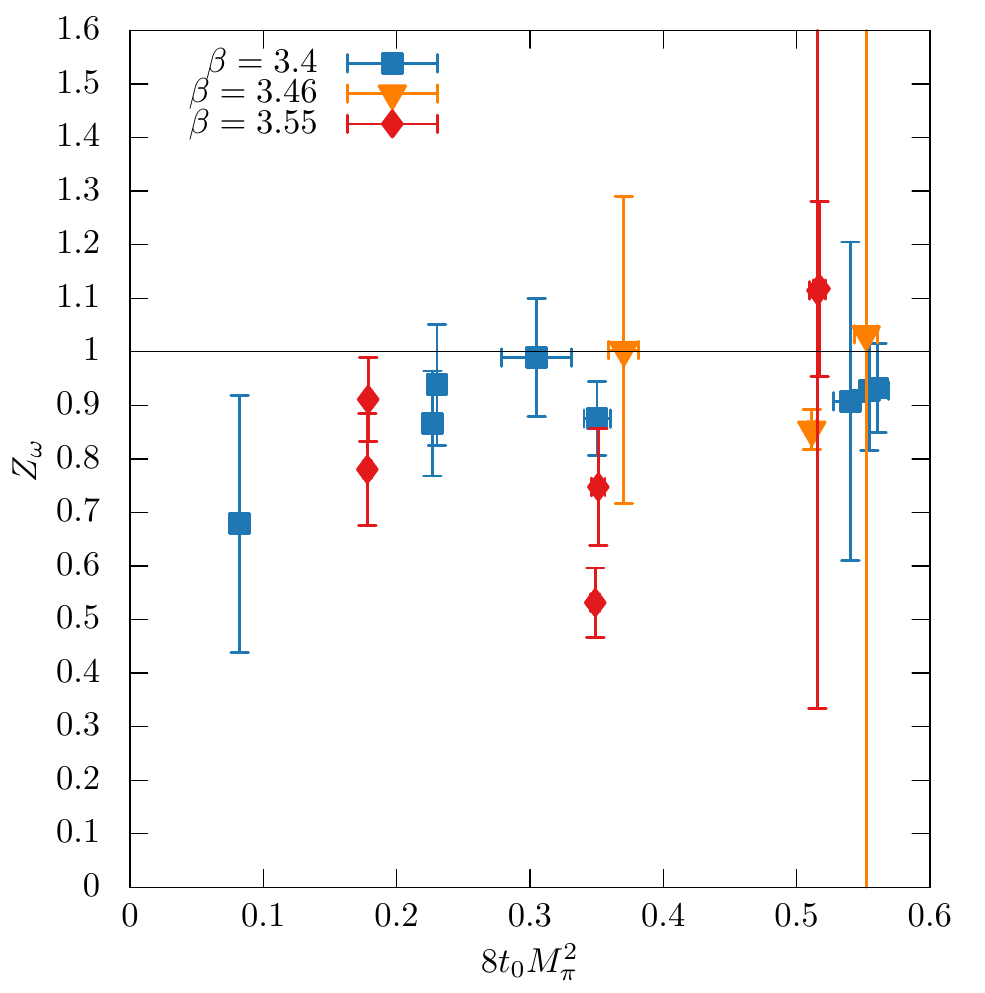}
  \includegraphics[width=0.49\linewidth]{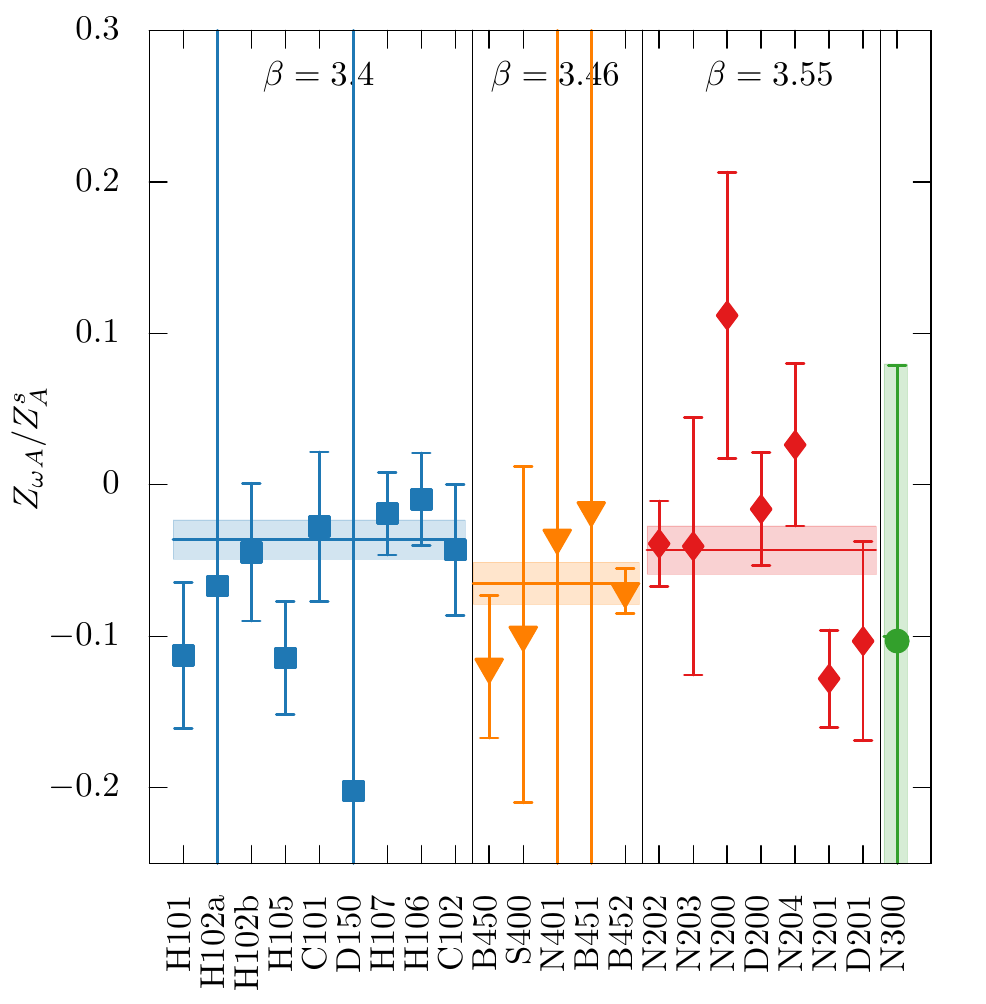}
  \caption{\label{fig:zomegaa} (Left) $Z_\omega$ from solving
    eq.~\eqref{eq:gluonmegluonic}. The $m_s=m_{\ell}$ points are not shown since
    in these cases $F_\eta^0 = 0$ and the equation system is singular.
    (Right) Values of $Z_{\omega
      A} / Z_A^s$ from eq.~\eqref{eq:zomegaA}, assuming $Z_{\omega}=1$.
  }
\end{figure}

We wish to check if the fermionic results that were
obtained in sec.~\ref{sec:gluonmefermionic} from
employing the singlet AWI are consistent with
a direct determination of the gluonic matrix elements. 
The renormalized matrix elements are given as
\begin{equation}
  a_{\mathcal{M}}(\mu)=2\,Z_{\omega}\langle\Omega|\omega|\mathcal{M}\rangle+
  2\frac{Z_{\omega A}}{Z_A^s}M_{\mathcal{M}}^2F_{\mathcal{M}}^0(\mu),\label{eq:gluonmegluonic}
\end{equation}
see eq.~\eqref{eq:rentop}. In the previous section, we have found $Z_{\omega} = 1$ from a fit
to the topological susceptibility. As an additional cross check, we also
simultaneously solve the above equation
for $\mathcal{M} = \eta$ and $\mathcal{M}=\eta^{\prime}$ to obtain
$Z_{\omega A}$ and $Z_{\omega}$. We plot the
resulting values for $Z_\omega$ in the left panel of
fig.~\ref{fig:zomegaa}. Qualitatively these are in agreement with
$Z_{\omega} = 1$ and we suspect that the two outliers are due to
lattice artefacts.

\begin{figure}
  \includegraphics[width=0.49\linewidth]{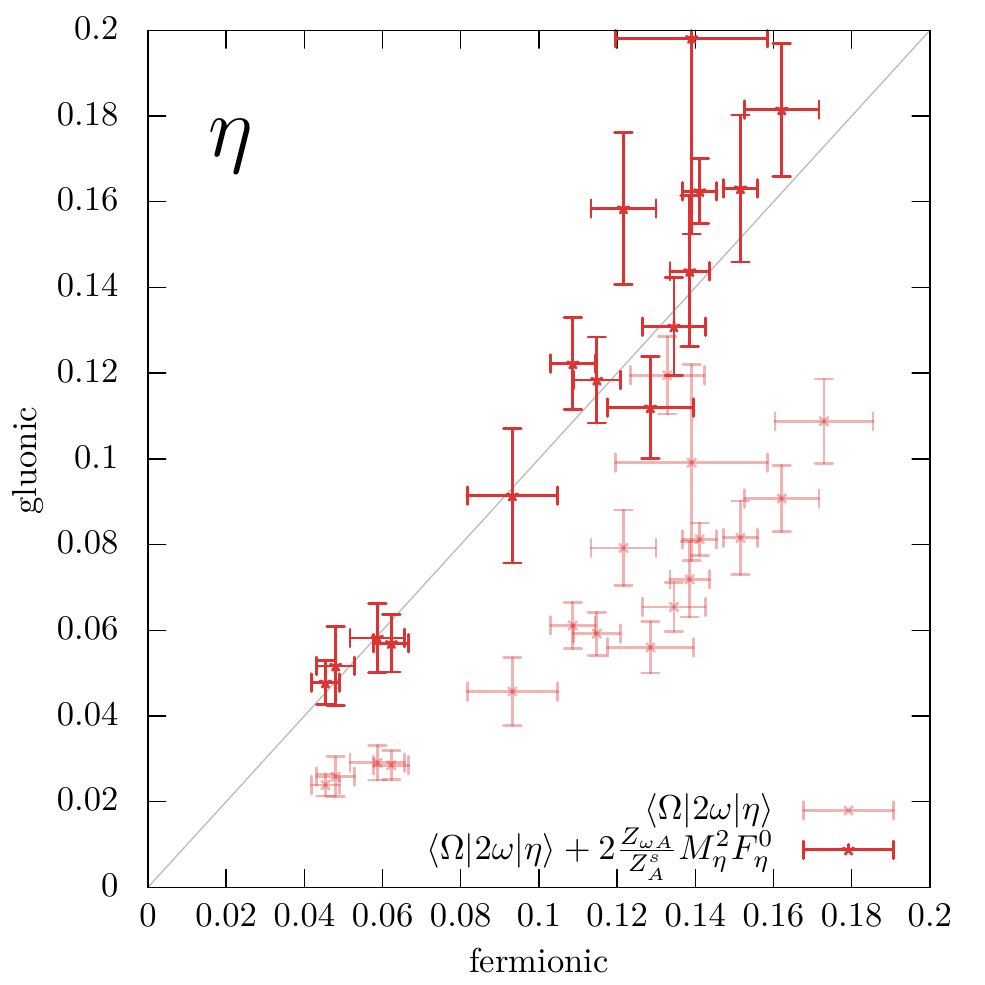}
  \includegraphics[width=0.49\linewidth]{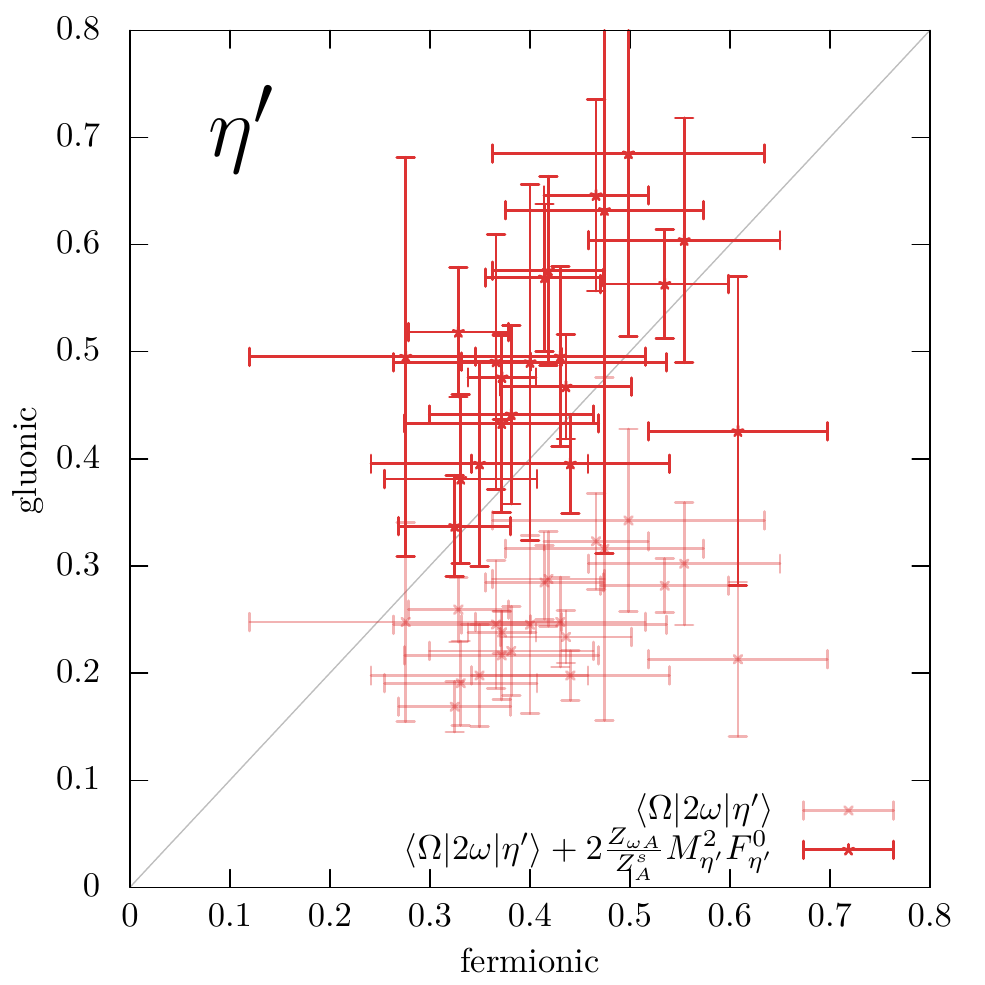}
  \caption{Scatter plot of the fermionic (eq.~\eqref{eq:gluonmefermionicpartimp},
    horizontally) and gluonic (eq.~\eqref{eq:gluonmegluonic}, vertically)
    determinations of the gluonic matrix elements $a_{\eta^{(\prime)}}$. The
    mixing with the derivative of the axialvector current is non-negligible:
    the unrenormalized lattice matrix elements (pale red points) do not
    agree with the
  fermionic definition.\label{fig:gluonmecmp}}
\end{figure}

Based on the evidence presented above, we assume $Z_{\omega}=1$, however,
$Z_{\omega A}$ is not known and
therefore comparing the direct determination~\eqref{eq:gluonmegluonic}
of the anomaly terms with the corresponding predictions from the
singlet AWI eq.~\eqref{eq:gluonmefermionic} cannot
be entirely independent. Fortunately, the
ratio $Z_{\omega A}/Z_A^s$ only depends
on the inverse lattice coupling, $\beta$, but not
on the pion and kaon masses. Moreover, the renormalization is
independent of the meson.
Rearranging eq.~\eqref{eq:gluonmegluonic}, we can isolate the renormalization
scale independent ratio
\begin{align}
  \frac{Z_{\omega A}}{Z_A^s}=\frac{a_{\mathcal{M}}-2\,Z_{\omega}\langle\Omega|\omega|\mathcal{M}\rangle}{2M_{\mathcal{M}}^2F_{\mathcal{M}}^0}\label{eq:zomegaA}.
\end{align}
Since $F_\eta^0$ in the denominator is
close to zero and has large relative errors, we only use
the $\eta^\prime$ matrix
elements. We plot this ratio for $Z_{\omega}=1$
in the right panel of fig.~\ref{fig:zomegaa}.
Indeed, the data for each $\beta$-value are compatible with a constant.
Taking a weighted average over all points
at each of our four lattice spacing, we obtain
\begin{align}
  \left( Z_{\omega_A}/Z_A^s \right)(\beta=3.4)  & =  -0.036(13), & \left( Z_{\omega_A}/Z_A^s \right)(\beta=3.46) & = -0.065(14),\nonumber\\
  \left( Z_{\omega_A}/Z_A^s \right)(\beta=3.55) & = -0.043(16),  & \left( Z_{\omega_A}/Z_A^s \right)(\beta=3.7) & =  -0.10(18).
\end{align}
Using these values (and $Z_{\omega} = 1$), we evaluate
eq.~\eqref{eq:gluonmegluonic} with
the anomalous matrix elements computed
at the gradient flow time $t \approx t_{0}^*$ on the individual ensembles.
We compare our results on every ensemble with the fermionic
determination of sec.~\ref{sec:gluonmefermionic} in the
scatter plot fig.~\ref{fig:gluonmecmp}.
Our gluonic results agree qualitatively with the fermionic
determination.
The mixing with the axialvector current is non-negligible,
i.e.~$Z_{\omega A} \neq 0$.
Had we ignored this mixing, the gluonic determinations would have undershot the
fermionic ones by roughly 30\,\%  both for the $\eta$ and the $\eta^{\prime}$.
We stress that agreement can only be expected in the continuum limit since both
definitions are subject to different discretization effects.
We have observed considerable
lattice spacing effects both for the topological susceptibility
in sec.~\ref{sec:ren0awi} and the singlet pseudoscalar matrix elements
in sec.~\ref{sec:gluonmefermionic}. The qualitative agreement suggests
that some of the discretization effects may be similar for both
definitions.

\section{Summary and comparison to other results}
\label{sec:summary}
In this section we summarize our results and compare them to other
determinations from lattice or phenomenological studies.  The meson
masses, decay constants, large-$N_c$ U($3$) ChPT LECs (and their
relation to their SU($3$) equivalents) and pseudoscalar gluonic and
fermionic matrix elements are presented. In addition, we study
the implications of our findings on the photoproduction transition
form factors of the $\eta$ and $\eta^\prime$ mesons.

The results are converted into physical units, using $(8t_0^{\rm
  ph})^{-1/2}=475(6)\,\mathrm{MeV}$~\cite{Bruno:2016plf} and
$(8t_0^{\chi})^{-1/2}=470(7)$, see eqs.~\eqref{eq:t0ph}
and~\eqref{eq:t0ch}.  For some of our results the uncertainty of this
scale significantly contributes to the total error. Since improved
determinations may become available in the future, we quote this
uncertainty separately to the statistical and other systematic errors.

\subsection{The $\eta$ and $\eta^\prime$ meson masses\label{sec:masssummary}}
Our final results for the masses of the $\eta$ and $\eta^{\prime}$ mesons are  (see sec.~\ref{sec:massdec} and eq.~\eqref{eq:massfitresults})
\begin{align}
  M_\eta &=
  554.7\left(\substack{4.0\\6.6}\right)_\mathrm{stat} \left(\substack{2.4\\2.7}\right)_{\mathrm{syst}}(7.0)_{t_0} ,\,\mathrm{MeV}
  \quad\text{and}\\
  M_{\eta^\prime} &=
    929.9\left(\substack{12.9\\6.0}\right)_\mathrm{stat} \left(\substack{22.9\\3.3}\right)_{\mathrm{syst}}(11.7)_{t_0}
  \,\mathrm{MeV},
\end{align}
where we added the systematic errors associated with the continuum and physical quark
mass point extrapolations in quadrature.
We find reasonably good agreement when
comparing these results of $N_{f} = 2+1$ QCD with the known experimental masses,
\begin{equation}
  \text{PDG\,\cite{PDG}}:\qquad
  M^{\mathrm{ph}}_\eta = 547.862(17)\,\mathrm{MeV}
  \quad\text{and}\quad
  M^{\mathrm{ph}}_{\eta^\prime} = 957.78(6)\,\mathrm{MeV}\label{eq:physetaetaprimemasses}.
\end{equation}
The masses are 0.7 standard errors above and one standard error below
the experimental values for the $\eta$ and $\eta^{\prime}$,
respectively.
For $M_{\eta}$, the combined relative error is 1.7\,\% with the statistical
and scale setting uncertainties forming the biggest contributions.
Our value for $M_{\eta^{\prime}}$ has a total uncertainty of 2.3\,\%, where
in this case the statistical error and the uncertainty
from the quark mass extrapolation dominate.
In both cases, lattice spacing effects are less significant.
This reflects the fact that we are not able to resolve any such effects
in the masses, see sec.~\ref{sec:fitselection}. For
$M_{\eta^{\prime}}$ this is not so surprising considering the
relatively large statistical error obtained on the individual
ensembles. We remark that the precision of the final results was
achieved by utilizing NLO large-$N_{c}$ ChPT to simultaneously fit the
two masses and four decay constants~(summarized in the next
subsection) determined on twenty-one ensembles lying along two trajectories in
the quark mass plane and comprising four lattice spacings.

\begin{figure}
  \resizebox{\linewidth}{!}{\input{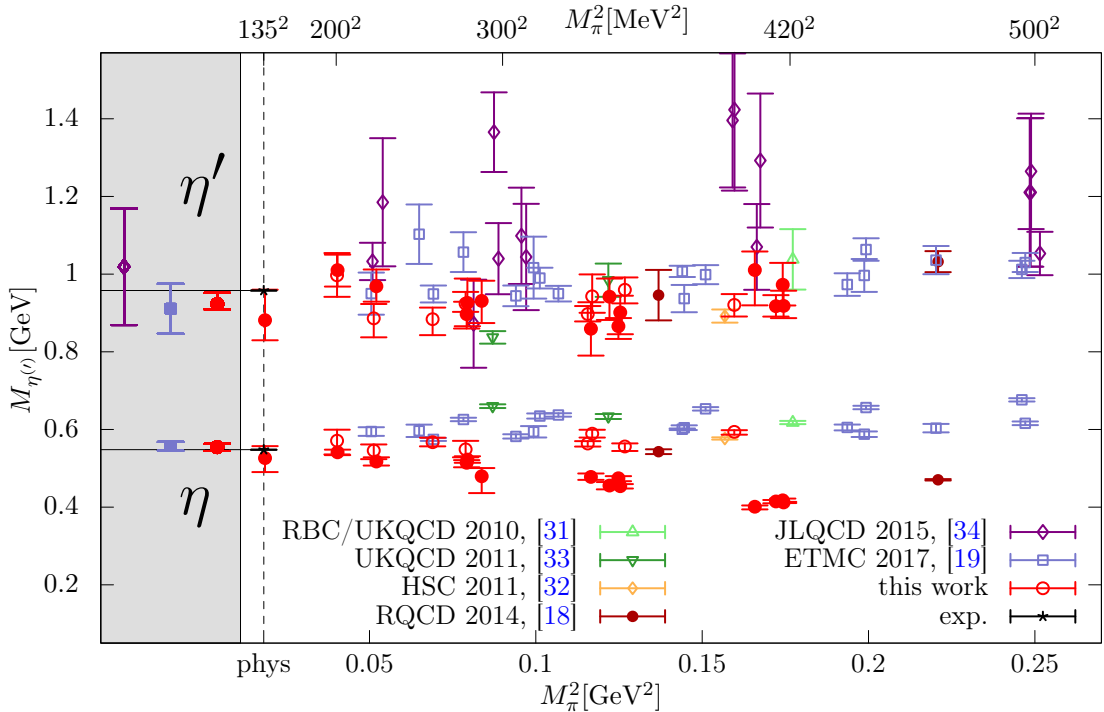}}
  \caption{Recent $N_f=2+1(+1)$ lattice results for the masses of the $\eta$
    and $\eta^\prime$ mesons. Most points have been simulated at approximately
    physical strange quark masses (open symbols), whereas in this work we also include an additional
    trajectory along which the average of the quark masses is kept constant (filled
    symbols). The three sets of points in the shaded regions left of the physical point (dashed line)
    correspond to the continuum and chirally extrapolated results of JLQCD~\cite{Fukaya:2015ara} (who do not give an estimate of $M_{\eta}$), ETMC~\cite{Ottnad:2017bjt} and this work. \label{fig:masssummary}}
\end{figure}
Our results at unphysical quark masses as well as in the physical
limit are displayed in fig.~\ref{fig:masssummary}, together with
$N_f=2+1(+1)$ results of other groups that we are aware of.  The $\eta$
mass is sensitive to the masses of the light and strange quarks and
the data points clearly fall along two lines which converge at the
physical point, reflecting the two sets of ensembles employed: for one
set the physical strange quark mass is kept approximately constant
while for the other the flavour average of the strange and light quark
masses is held fixed.  The singlet contribution to the mass of the
$\eta^{\prime}$ is significant and no clear quark mass dependence is
observed.

Overall, the results for the $\eta^{\prime}$ are consistent across
different collaborations and actions~(also at larger quark masses),
whereas for the $\eta$ some scatter is visible. The latter may be due
to mistuning of the strange quark mass and/or lattice spacing
effects. In particular, a previous exploratory study of our
group~\cite{Bali:2014pva} is affected by mistuning.  The
ETMC~\cite{Ottnad:2017bjt} and JLQCD~\cite{Fukaya:2015ara}
collaborations employ pion masses reaching down to approximately
$220\,\mathrm{MeV}$. In this work, we obtain results close to the physical point
for the first time. While the errors are relatively large for our
$M_\pi\approx 126\,\mathrm{MeV}$ ensemble, the results are in good agreement
with the quark mass extrapolation.

To our knowledge the only other studies which attempt a physical limit
extrapolation are those of ETMC~\cite{Ottnad:2012fv,Michael:2013gka,Ottnad:2017bjt}
and JLQCD~\cite{Fukaya:2015ara}.  The latter $N_f=2+1$ work utilizes
gluonic correlation functions to determine the $\eta^{\prime}$ mass. A
simple linear extrapolation is performed which is justified in view of
the large statistical errors. ETMC~\cite{Ottnad:2017bjt} employ the
twisted mass fermion formulation and simulate $N_{f}=2+1+1$ QCD\@.
The physical point is approached keeping the strange quark mass
approximately equal to its physical value, although some mistuning is
visible in the results for $M_\eta$.  This is compensated for
by including terms proportional to $m_{s}$ in the quark mass extrapolation,
in addition to terms proportional to $m_{\ell}$ and $a^{2}$.  This
leading order ansatz yields an effective parametrization of the data,
however, the $\eta$ and $\eta^{\prime}$ masses are assumed to be independent of each
other and are fitted separately, ignoring potential correlations in
the data.  The final errors for $M_{\eta}$ and $M_{\eta^{\prime}}$ at
the physical point are larger than ours, in particular for the latter,
although the uncertainties on the individual ensembles are similar and
at the percent and few-percent level, for the $\eta$ and
$\eta^{\prime}$, respectively.  We achieve smaller final errors by
simultaneously fitting the quark mass and lattice spacing dependence
of six observables (two masses and four decay constants), which
have been determined on ensembles following two trajectories to the
physical point. This, together with including ensembles with small
quark masses, enables the quark mass extrapolation to be tightly
constrained.  The results from ETMC at the physical point are in
agreement with our estimates and the experimental values within the
quoted errors.

\subsection{Decay constants\label{sec:decsummary}}
We carry out two sets of fits to extract the four decay constants, one
where we simultaneously fit to our lattice results for the masses and
decay constants from which the values of the masses at the physical
point (presented in the previous subsection) are taken and another
set where we constrain the masses to reproduce the physical values by
adding prior terms to the $\chi^2$ function. The latter fits enable
the LECs to be better constrained, see sec.~\ref{sec:lecresults}.  The
two sets of results, detailed in tabs.~\ref{tab:alldecresults}
and~\ref{tab:alldecresultspriors} in app~\ref{app:decayresults}, are
consistent within errors.  In the following, we will only discuss the
second set of results.  Since the singlet axialvector current has an
anomalous dimension in the $\overline{\mathrm{MS}}$ scheme, some of
the results depend on the QCD scale, see
sec.~\ref{sec:singletrenorm}. Although the fits were performed setting
$\mu=\infty$ in $N_f=3$ QCD, in this section we will mostly quote results at
$\mu=2\,\mathrm{GeV}$. This simplifies a comparison to
literature values, as discussed below. The results obtained for a
range of scales are listed in tabs.~\ref{tab:alldecresults}
and~\ref{tab:alldecresultspriors}.

\paragraph{Summary of the results.}
The decay constants, converted to the angle representation of the
octet/singlet basis, read at $\mu = 2\,\mathrm{GeV}$ in $N_f=3$ QCD
\begin{align}
  F^8 &=  \phantom{-}115.0\left(\substack{1.1\\1.2}\right)_\mathrm{stat} \left(\substack{1.6\\2.4}\right)_{\mathrm{syst}}(1.5)_{t_0}\,\mathrm{MeV},\\
  \theta_8 &=   -25.8\left(\substack{1.2\\2.1}\right)_{\mathrm{stat}}\left(\substack{2.2\\0.3}\right)_{\mathrm{syst}}^\circ,\\
  F^{0}(\mu = 2\,\mathrm{GeV})   &=  \phantom{-}100.1\left(\substack{7\\1.9}\right)_\mathrm{stat} \left(\substack{2.0\\2.7}\right)_{\mathrm{syst}}(1.3)_{t_0}\,\mathrm{MeV}, \\
  \theta_0 &=  -8.1\left(\substack{1.0\\1.1}\right)_{\mathrm{stat}}\left(\substack{1.5\\1.5}\right)_{\mathrm{syst}}^\circ, \end{align}
where we added the systematic errors arising from the continuum and
chiral extrapolation in quadrature.  This representation has the
advantage that only $F^0$ depends on the scale,
however, often the flavour basis in the angle representation
is employed in the literature. We find 
\begin{align}
      F^\ell(\mu = 2\,\mathrm{GeV}) &=  88.28\left(\substack{1.20\\2.02}\right)_\mathrm{stat} \left(\substack{3.00\\1.74}\right)_{\mathrm{syst}}(1.12)_{t_0}\,\mathrm{MeV}, \label{eq:resultfl}\\
      \phi_\ell(\mu = 2\,\mathrm{GeV}) &=  36.2\left(\substack{1.1\\2.0}\right)_{\mathrm{stat}}\left(\substack{1.3\\0.4}\right)_{\mathrm{syst}}^\circ, \\
      F^s(\mu = 2\,\mathrm{GeV}) &=  124.3\left(\substack{1.7\\1.6}\right)_\mathrm{stat} \left(\substack{2.7\\4.3}\right)_{\mathrm{syst}}(1.6)_{t_0}\,\mathrm{MeV}, \\
      \phi_s(\mu = 2\,\mathrm{GeV}) &=  37.9\left(\substack{1.0\\1.3}\right)_{\mathrm{stat}}\left(\substack{1.4\\0.8}\right)_{\mathrm{syst}}^\circ\label{eq:resultphis},
\end{align}
where all quantities depend non-trivially on the scale.
The popularity of the flavour representation is due to the similarity of the two angles which suggests that the four (independent) decay constants can be described by only three parameters, setting
$\phi_{\ell} = \phi_{s}=\phi$.
This approximation is made in the Feldmann-Kroll-Stech (FKS) scheme~\cite{Feldmann:1998vh,Feldmann:1998sh,Feldmann:1999uf} and to NLO in large-$N_c$ ChPT it is equivalent to neglecting OZI-rule violating terms, specifically those involving $\Lambda_{1}$. At NLO the latter parameter is related to the angles and decay
constants via~\cite{Leutwyler:1997yr,Feldmann:1998vh}
\begin{equation}
\frac{\sqrt{2}}{3} F_{\pi}^{2} \Lambda_{1}=  F^{\ell} F^{s} \sin(\phi_{\ell} - \phi_{s}).\label{eq:anglediffnlo}
\end{equation}
Thus, if $\Lambda_{1}$ is set to zero, then within this approximation $\phi_{\ell} = \phi_{s}$.

\begin{figure}
  \includegraphics[width=0.49\textwidth]{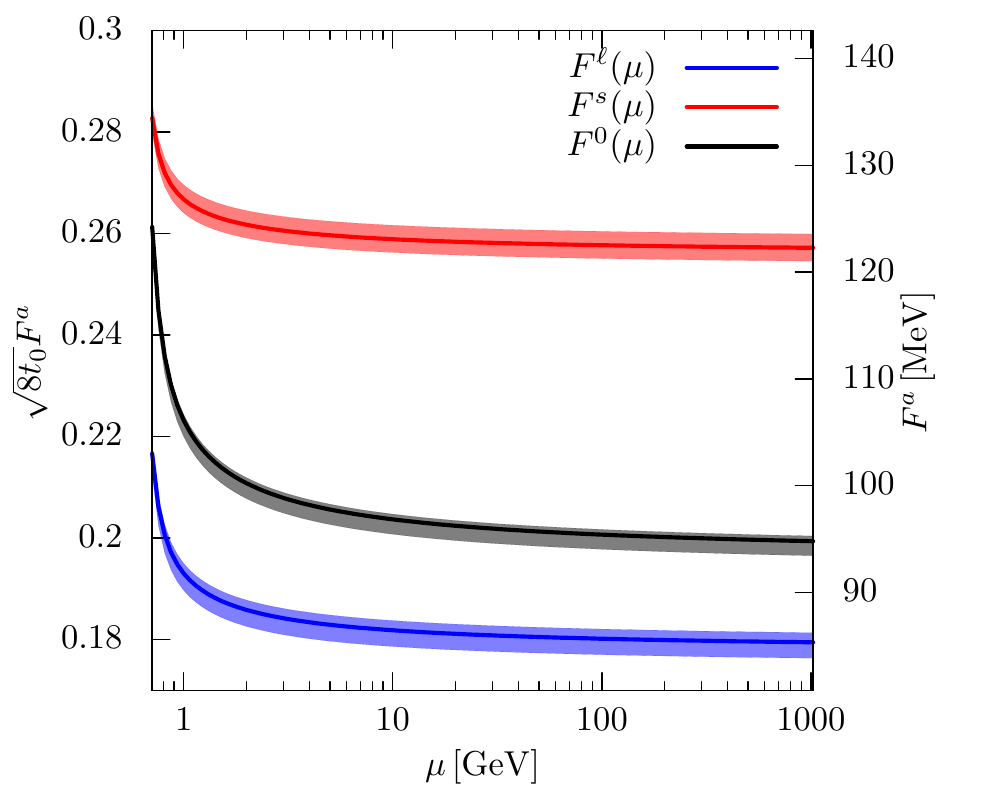}
  \includegraphics[width=0.49\textwidth]{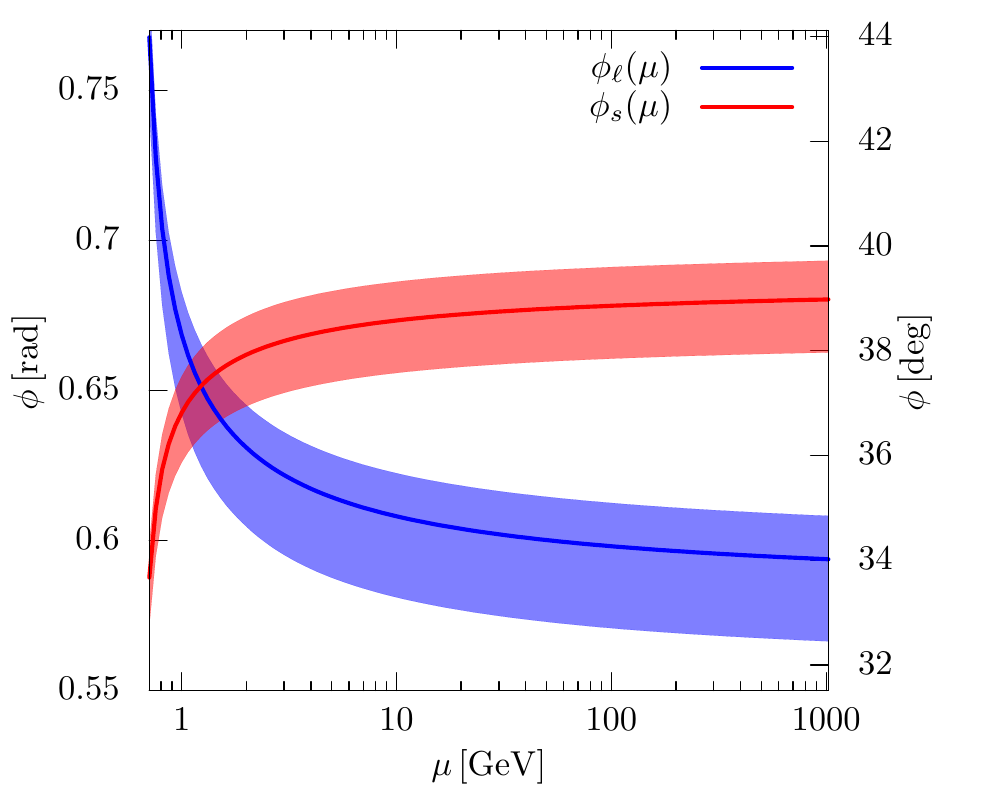}
  \caption{Scale dependence of the decay constants and their mixing
    angles in the flavour basis.
    The approximation $\phi_\ell \approx \phi_s \approx \phi$
    is only valid for $\mu$ inbetween about $1\,\mathrm{GeV}$ and $2\,\mathrm{GeV}$. In the same region the decay constants
    vary considerably with the scale. The (asymmetric) errors indicated by the coloured bands
    are statistical only.\label{fig:decsAtScale}}
\end{figure}

\begin{figure}
  \centerline{\includegraphics[width=0.6\linewidth]{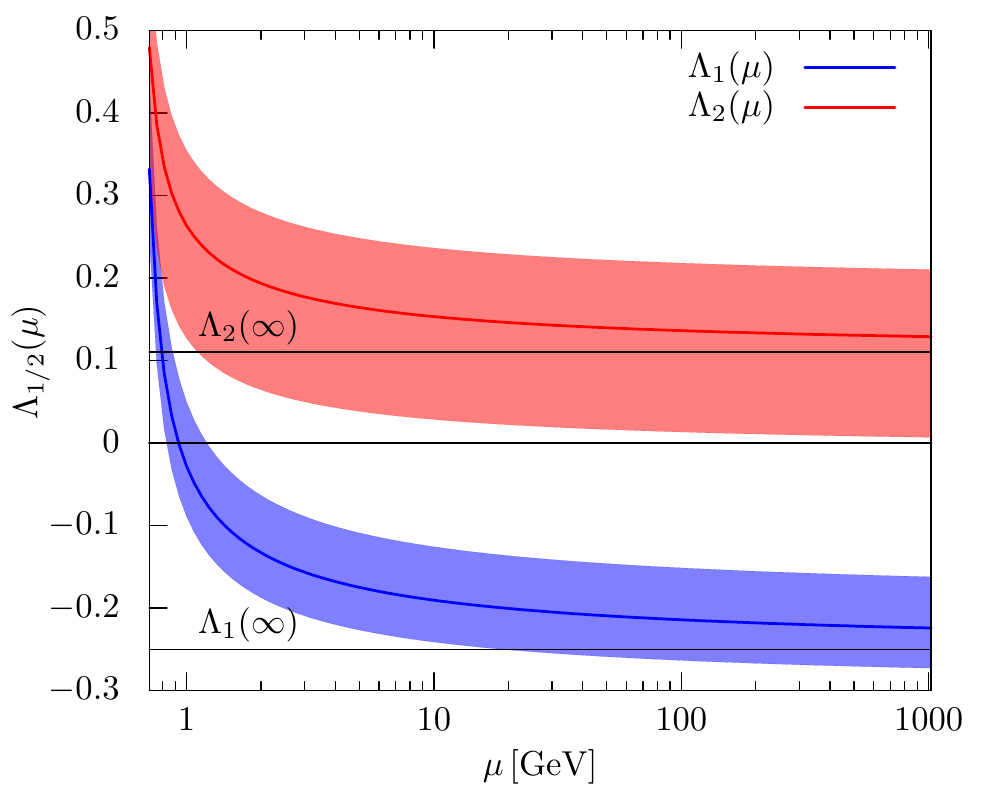}}
  \caption{Scale dependence of the large-$N_{c}$ ChPT LECs $\Lambda_{1}$ and $\Lambda_{2}$.
    \label{fig:lambdasAtScale}}
\end{figure}

\paragraph{Dependence on the QCD scale.}
In effect, the assumption $\Lambda_1=0$ renders the singlet decay constant independent of the scale since~\cite{Leutwyler:1997yr,Escribano:2015yup}
\begin{equation}
  \mu \frac{\rm d}{\mathrm{d}\mu} \frac{F_{0}(\mu)}{\sqrt{1 + \Lambda_{1}(\mu)}} = 0.
  \label{eq:lambda1scale}
\end{equation}
The results in eqs.~\eqref{eq:resultfl}--\eqref{eq:resultphis} show
that at $\mu = 2\,\mathrm{GeV}$ the angles almost agree within
errors. However, our estimate for $\Lambda_{1} = -0.25(5)$ (see
sec.~\ref{sec:lecresults}) determined at $\mu = \infty$, suggests that
this approximation cannot hold at high scales.  We display the scale
dependent decay constants and angles as a function of $\mu$ in
fig.~\ref{fig:decsAtScale}. The two angles are significantly different
at large scales where the combination $2 (\phi_{s} -
\phi_{\ell}) / (\phi_{s} + \phi_{\ell})$ approaches 16\,\%. However,
this difference decreases towards lower $\mu$ and in the range 
$0.9\,\mathrm{GeV}\lesssim \mu \lesssim 2\,\mathrm{GeV}$
then $\phi_{\ell}\approx\phi_s$. This is due to
$\Lambda_{1}(\mu)$ crossing zero around $1\,\mathrm{GeV}$ as shown in
fig.~\ref{fig:lambdasAtScale}, where we display both OZI violating
LECs~\cite{Kaiser:1998ds},
\begin{equation}
  \Lambda_{1}(\mu) = \left(\frac{Z_{A}^{s}(\mu)}{Z_{A}^{s\prime}}\right)^{2}(1 + \Lambda_{1}(\mu = \infty)) - 1,\quad
  \Lambda_{2}(\mu) = \frac{Z_{A}^{s}(\mu)}{Z_{A}^{s\prime}}(1 + \Lambda_{2}(\mu = \infty)) - 1.
\end{equation}
The LEC $\Lambda_{2}$, which mostly impacts on the masses, becomes small at
high scales but should not be neglected at $\mu <
4\,\mathrm{GeV}$. This provides an explanation for the observation
of some ChPT studies that $\Lambda_{2}$ plays a more important role
than $\Lambda_{1}$ in terms of reproducing the physical $\eta$ and
$\eta^{\prime}$ masses~\cite{Guo:2011pa,Guo:2015xva}.

The scale dependence of some observables complicates direct
comparisons to phenomenology in many studies employing ChPT, where the
relevant QCD scale depends on the processes
that are considered to fix the LECs and on the order of ChPT.
Typically, the LECs are determined
using experimental input from, e.g., the $\eta$ and
$\eta^{\prime}$ masses and the widths of radiative decays. This
implies a low QCD renormalization scale. $\Lambda_{1}$ varies rapidly
in this region which means that if this LEC is determined using
physical processes dominated by different physical scales, the
predictions for, e.g., $F^0$ will be affected.  Clearly, results
obtained using the FKS scheme should be compared at the scale where
$\Lambda_{1}$ vanishes.  One possibility to mitigate this problem is
to compare results in the octet/singlet basis~(where only $F^0$
depends on $\mu$) and form the scale independent combination
$F^0/\sqrt{1 + \Lambda_1}$~\cite{Leutwyler:1997yr}. Our result for the
latter reads
\begin{equation}
F^{0}/\sqrt{1 + \Lambda_{1}} = 107.3\left(\substack{1.5\\7}\right)_\mathrm{stat} \left(\substack{1.3\\1.3}\right)_{\mathrm{syst}}(1.4)_{t_0}\,\mathrm{MeV}.
\end{equation}

\begin{table*}
  \footnotesize
  \centering
  \pgfplotstabletypeset[columns={ref,Fqerr,Fqstr,Fserr,Fsstr},
  column type=,
  begin table={\begin{tabularx}{\textwidth}{l p{\summarygraphplotwidth} X p{\summarygraphplotwidth} X}},
                 end table={\end{tabularx}},
               every head row/.style = {before row=\toprule, after row=\midrule},
               every last row/.style = {after row=[3ex]\bottomrule},
               every row no 6/.style = {before row=\hline},
               every row no 13/.style = {before row=\hline},
               every row no 14/.style = {before row=\hline},
               every row no 16/.style = {before row=\hline},
               skip rows between index = {16}{17},
               columns/ref/.style = {string type, column name = ref},
               columns/Fqerr/.style = {%
                 column name = {},
                 assign cell content/.code = {%
                   \ifnum\pgfplotstablerow=0
                   \pgfkeyssetvalue{/pgfplots/table/@cell content}
                   {\multirow{\numberofflavourcmprows}{\summarygraphplotwidth}{\flcmpplot}}%
                   \else
                   \pgfkeyssetvalue{/pgfplots/table/@cell content}{}%
                   \fi
                 }
               },
               columns/Fqstr/.style = {string type, column name = $F^\ell / \mathrm{MeV}$},
               columns/Fserr/.style = {%
                 column name = {},
                 assign cell content/.code = {%
                   \ifnum\pgfplotstablerow=0
                   \pgfkeyssetvalue{/pgfplots/table/@cell content}
                   {\multirow{\numberofflavourcmprows}{\summarygraphplotwidth}{\fscmpplot}}%
                   \else
                   \pgfkeyssetvalue{/pgfplots/table/@cell content}{}%
                   \fi
                 }
               },
               columns/Fsstr/.style = {string type, column name = $F^s / \mathrm{MeV}$},
               ]{\deccmpdata}
  \pgfplotstabletypeset[columns={ref,philerr,philstr,phiserr,phisstr},
  column type=,
  begin table={\begin{tabularx}{\textwidth}{l p{\summarygraphplotwidth} X p{\summarygraphplotwidth} X}},
                 end table={\end{tabularx}},
               every head row/.style = {after row=\midrule},
               every last row/.style = {after row=[3ex]\bottomrule},
               every row no 6/.style = {before row=\hline},
               every row no 13/.style = {before row=\hline},
               every row no 14/.style = {before row=\hline},
               every row no 16/.style = {before row=\hline},
               skip rows between index = {16}{17},
               columns/ref/.style = {string type, column name = ref},
               columns/philerr/.style = {%
                 column name = {},
                 assign cell content/.code = {%
                   \ifnum\pgfplotstablerow=0
                   \pgfkeyssetvalue{/pgfplots/table/@cell content}
                   {\multirow{\numberofflavourcmprows}{\summarygraphplotwidth}{\philcmpplot}}
                   \else
                   \pgfkeyssetvalue{/pgfplots/table/@cell content}{}%
                   \fi
                 }
               },
               columns/philstr/.style = {string type, column name = $\phi_\ell$},
               columns/phiserr/.style = {%
                 column name = {},
                 assign cell content/.code = {%
                   \ifnum\pgfplotstablerow=0
                   \pgfkeyssetvalue{/pgfplots/table/@cell content}
                   {\multirow{\numberofflavourcmprows}{\summarygraphplotwidth}{\phiscmpplot}}%
                   \else
                   \pgfkeyssetvalue{/pgfplots/table/@cell content}{}%
                   \fi
                 }
               },
               columns/phisstr/.style = {string type, column name = $\phi_s$},
]{\deccmpdata}
  \caption{Comparison of recent phenomenological and lattice results for the
(scale dependent) decay constants in the angle representation for the light/strange flavour basis. The
results from this work are presented at three different scales.
\label{tab:cmpflavour}}
\end{table*}
\begin{table*}
  \setlength\tabcolsep{0pt}
  \footnotesize
  \centering
  \pgfplotstabletypeset[columns={ref,F8err,F8str,F0err,F0str},
  column type=,
  begin table={\begin{tabularx}{\textwidth}{l p{\summarygraphplotwidth} X p{\summarygraphplotwidth} X}},
                 end table={\end{tabularx}},
               every head row/.style = {before row=\toprule, after row=\midrule},
               every last row/.style = {after row=[3ex]\bottomrule},
               every row no 6/.style = {before row=\hline},
               every row no 13/.style = {before row=\hline},
               every row no 14/.style = {before row=\hline},
               every row no 16/.style = {before row=\hline},
               every row no 17/.style = {before row=\hline},
               columns/ref/.style = {string type, column name = ref},
               columns/F8err/.style = {%
                 column name = {},
                 assign cell content/.code = {%
                   \ifnum\pgfplotstablerow=0
                   \pgfkeyssetvalue{/pgfplots/table/@cell content}
                   {\multirow{\numberofcmprows}{\summarygraphplotwidth}{\foctcmpplot}}%
                   \else
                   \pgfkeyssetvalue{/pgfplots/table/@cell content}{}%
                   \fi
                 }
               },
               columns/F8str/.style = {string type, column name = $F^8 / \mathrm{MeV}$},
               columns/F0err/.style = {%
                 column name = {},
                 assign cell content/.code = {%
                   \ifnum\pgfplotstablerow=0
                   \pgfkeyssetvalue{/pgfplots/table/@cell content}
                   {\multirow{\numberofcmprows}{\summarygraphplotwidth}{\fsinglcmpplot}}%
                   \else
                   \pgfkeyssetvalue{/pgfplots/table/@cell content}{}%
                   \fi
                 }
               },
               columns/F0str/.style = {string type, column name = $F^0 / \mathrm{MeV}$},
               ]{\deccmpdata}

  \pgfplotstabletypeset[columns={ref,theta8err,theta8str,theta0err,theta0str},
  column type=,
  begin table={\begin{tabularx}{\textwidth}{l p{\summarygraphplotwidth} X p{\summarygraphplotwidth} X}},
                 end table={\end{tabularx}},
               every head row/.style = {after row=\midrule},
               every last row/.style = {after row=[3ex]\bottomrule},
               every row no 6/.style = {before row=\hline},
               every row no 13/.style = {before row=\hline},
               every row no 14/.style = {before row=\hline},
               every row no 16/.style = {before row=\hline},
               every row no 17/.style = {before row=\hline},
               columns/ref/.style = {string type, column name = ref},
               columns/theta8err/.style = {%
                 column name = {},
                 assign cell content/.code = {%
                   \ifnum\pgfplotstablerow=0
                   \pgfkeyssetvalue{/pgfplots/table/@cell content}
                   {\multirow{\numberofcmprows}{\summarygraphplotwidth}{\phioctcmpplot}}%
                   \else
                   \pgfkeyssetvalue{/pgfplots/table/@cell content}{}%
                   \fi
                 }
               },
               columns/theta8str/.style = {string type, column name = $\theta_8$},
               columns/theta0err/.style = {%
                 column name = {},
                 assign cell content/.code = {%
                   \ifnum\pgfplotstablerow=0
                   \pgfkeyssetvalue{/pgfplots/table/@cell content}
                   {\multirow{\numberofcmprows}{\summarygraphplotwidth}{\phisinglcmpplot}}%
                   \else
                   \pgfkeyssetvalue{/pgfplots/table/@cell content}{}%
                   \fi
                 }
               },
               columns/theta0str/.style = {string type, column name = $\theta_0$},
]{\deccmpdata}\\[-3pt]
\caption{Comparison of recent phenomenological and lattice results for the
  decay constants in the angle representation for the octet/singlet
  basis, where only $F^{0}$ depends on the scale. We use $F_{\pi^{+}} = 92.1\,\mathrm{MeV}$~\cite{PDG} to convert decay constants given as a multiple of $F_{\pi}$
  and
  eq.~\eqref{eq:leutwyler} refers to the NLO result from literature
  pion and kaon decay constants. The results from this work are presented
  at three different scales.
  \label{tab:cmpoctsinglet}}
\end{table*}
\begin{table*}
\setlength\tabcolsep{0pt}
\pgfplotstabletypeset[columns={ref,Lamerr,Lamstr,F0Lamerr,F0Lamstr},
  column type=,
  begin table={\begin{tabularx}{\textwidth}{l p{\summarygraphplotwidth} X p{\summarygraphplotwidth} X}},
                 end table={\end{tabularx}},
               every head row/.style = {before row=\toprule, after row=\midrule},
               every last row/.style = {after row=[3ex]\bottomrule},
               every row no 6/.style = {before row=\hline},
               every row no 13/.style = {before row=\hline},
               every row no 14/.style = {before row=\hline},
               every row no 16/.style = {before row=\hline},
               every row no 17/.style = {before row=\hline},
               columns/ref/.style = {string type, column name = ref},
               columns/Lamerr/.style = {%
                 column name = {},
                 assign cell content/.code = {%
                   \ifnum\pgfplotstablerow=0
                   \pgfkeyssetvalue{/pgfplots/table/@cell content}
                   {\multirow{\numberofcmprows}{\summarygraphplotwidth}{\lamcmpplot}}%
                   \else
                   \pgfkeyssetvalue{/pgfplots/table/@cell content}{}%
                   \fi
                 }
               },
               columns/Lamstr/.style = {string type, column name = $\Lambda_1$},
               columns/F0Lamerr/.style={%
                 column name = {},
                 assign cell content/.code = {%
                   \ifnum\pgfplotstablerow=0
                   \pgfkeyssetvalue{/pgfplots/table/@cell content}
                   {\multirow{\numberofcmprows}{\summarygraphplotwidth}{\fsingllamcmpplot}}%
                   \else
                   \pgfkeyssetvalue{/pgfplots/table/@cell content}{}%
                   \fi
                 }
               },
               columns/F0Lamstr/.style = {string type, column name = \hspace{-1.2cm}$F^{0}/\sqrt{1+\Lambda_{1}} / \mathrm{MeV}$},
]{\deccmpdata}
\caption{Comparison of determinations of the (scale dependent)
  large-$N_c$ ChPT LEC $\Lambda_1$ and the scale
  independent combination of $\Lambda_1$ and $F^0$, where
  eq.~\eqref{eq:leutwyler} refers to the NLO result from literature
  pion and kaon decay constants. The values
  indicated in Italics have been computed by us from $F_0$ and
  $\Lambda_1$ with naive error propagation. The
  results from this work are presented at three different scales.
  \label{tab:cmpLambda1}}
\end{table*}

\paragraph{Comparison with phenomenological results.}
A comparison with a variety of results for the decay constants in the
light/strange flavour and octet/singlet bases is shown
in tabs.~\ref{tab:cmpflavour}
and~\ref{tab:cmpoctsinglet}, respectively.  Most of the results rely
on large-$N_{c}$ ChPT using experimental input to fix the LECs.  One
of the first such computations was undertaken at NLO by
Leutwyler~\cite{Leutwyler:1997yr}, using predominantly pseudoscalar
meson masses and non-singlet decay constants to fix the LECs.  Only
scale independent combinations are quoted and a result for the singlet
decay constant is not given.  Feldmann~\cite{Feldmann:1999uf} then
employed the FKS scheme discussed above to give values also for the
scale dependent decay constants and the single flavour mixing angle~(in
this scheme).  This approximate scheme is also used on the lattice by
ETMC~\cite{Michael:2013gka,Ottnad:2017bjt} to relate the pseudoscalar
matrix elements to the (axial) decay constants, which will be discussed
further below.  In the first NNLO large-$N_{c}$ ChPT calculation, Guo
et al.~\cite{Guo:2015xva} take lattice input for $M_{\eta},
M_{\eta^{\prime}}$ and the non-singlet pseudoscalar masses and decay
constants at unphysical quark masses from the literature. This
allows them to constrain the LECs to NLO, but further assumptions are
needed for the many NNLO coefficients.  Subsequently, Gu et
al.~\cite{Gu:2018swy} extended the analysis by also utilizing the
decay constant results from ETMC~\cite{Ottnad:2017bjt}. However,
additional constraints on the parameters still seem to be necessary in
order to obtain stable NNLO results.  Bickert et
al.~\cite{Bickert:2016fgy} also perform an NNLO analysis, in this case
combining LECs obtained from the literature and derived from
experimental input for the masses and non-singlet decay
constants. Again only QCD scale independent combinations are given.

There exist a number of other studies, some of which are based on models, for
instance, Benayoun~\cite{Benayoun:1999fv,Benayoun:1999au} employs
vector meson dominance, while others involve more
phenomenologically driven extractions, for example, Escribano et
al.~\cite{Escribano:2005qq,Escribano:2013kba,Escribano:2015nra,Escribano:2015yup},
use experimental data on, e.g., the transition form factors
$\gamma\gamma^{*}\to\eta$ and $\gamma\gamma^{*}\to\eta^{\prime}$.  However, a
connection to NLO large-$N_{c}$ ChPT is made and allows to predict
some of the LECs. Chen et al.~\cite{Chen:2014yta} couple
large-$N_c$ ChPT at NLO to vector resonances
and extract the LECs, including these additional couplings, by simultaneously
analysing in this framework radiative decay form factors of light vector
mesons and charmonia into pseudoscalar final states.
Finally, in their calculation Ding et al.~\cite{Ding:2018xwy} employ
coupled gap and Bethe-Salpeter equations.

We also include the values $F^8=115.2(1.3)\,\text{MeV}$ and
$F^0/\sqrt{1+\Lambda_1}=104.3(1.1)\,\text{MeV}$
in tabs.~\ref{tab:cmpoctsinglet} and~\ref{tab:cmpLambda1}
(labelled as ``eq.~\eqref{eq:leutwyler}'').
These are obtained from the identities
\begin{align}
(F^8)^2=\frac{4F_K^2-F_\pi^2}{3}\quad\text{and}\quad
(F^0)^2=\frac{2F_K^2+F_{\pi}^2}{3}\left(1+\Lambda_1\right),
\label{eq:leutwyler}
\end{align}
which hold at NLO in large-$N_c$ ChPT~\cite{Leutwyler:1997yr},
using the values
$F_{K^{+}}/F_{\pi^{+}} = 1.193(2)$~\cite{Aoki:2019cca} and
$F_{\pi^{+}}=92.1(8)\,\text{MeV}$~\cite{Aoki:2019cca,PDG} as input, neglecting electromagnetic and isospin breaking effects.
The perfect agreement with our lattice QCD determination of $F^8$
and the agreement on the $1.3\,\sigma$ level for $F^0/\sqrt{1+\Lambda_1}$
indicates that NLO large-$N_c$ ChPT is a good approximation
for these quantities, at least near the physical quark mass point.

Overall, we find reasonable agreement between our results and phenomenological
NLO large-$N_{c}$ ChPT determinations. However, in the latter case
the errors are often not easily quantifiable.
For quantities that depend on the scale, the comparison should be made
with our values determined at low scales (for which $\Lambda_{1}$ vanishes).
Note that the mixing angles and fundamental decay constants $F^q$
in the flavour basis all depend on the QCD scale,
unless the FKS approximation is used. The other
determinations, which rely on more complicated experimental analyses
(incorporating processes at various scales), differ more significantly, in
particular, for the scale independent octet decay constant and
the octet mixing angle.
The theoretical effort involved in working out the higher orders not
withstanding, in general, it seems that the data available~(lattice or
experimental) are not sufficient to constrain the many parameters of
NNLO large-$N_{c}$ ChPT. Hence, these studies tend to have larger
errors while still relying on assumptions, such as setting individual
parameters to zero~\cite{Guo:2015xva,Bickert:2016fgy}. The lattice can
help, for example, by determining the (QCD renormalization scale
dependent) OZI violating parameters $\Lambda_{1}$ and $\Lambda_{2}$.

\paragraph{Relation to other lattice results.}
Matrix elements  of the $\eta$ and $\eta^{\prime}$ have previously
been computed on the lattice in the context of an exploratory study of $D_{s}\to
\eta,\eta^{\prime}$ semileptonic decays~\cite{Bali:2014pva}. However,
only two ensembles were utilized, with pion masses far away from the
physical point. ETMC calculated the decay constants utilizing an
indirect approach in~\cite{Michael:2013gka}. This study was updated
in~\cite{Ottnad:2017bjt} to include a continuum extrapolation,
employing seventeen $N_{f}=2+1+1$ gauge ensembles at three different lattice
spacings. Their results for the masses were discussed in
sec.~\ref{sec:masssummary}.  Due to the level of noise in the
axialvector channels, they utilize the FKS scheme to access the decay
constants via the pseudoscalar matrix elements of the $\eta$ and
$\eta^{\prime}$ states~\cite{Feldmann:1999uf}. Before summarizing
their results, we will discuss the assumptions they make.

The FKS approximation neglects all OZI violating terms. 
This amounts to setting $\Lambda_1=0$.
Rotating eq.~\eqref{eq:leutwyler} into the flavour basis results in
\begin{equation}
  (F^{\ell})^{2} = F_{\pi}^{2}+\frac{2}{3} \Lambda_{1}(2F_K^2+F_{\pi}^2)
    \quad\text{and}\quad
    (F^{s})^{2} = 2 F_{K}^{2} - F_{\pi}^{2}- \frac{1}{3}\Lambda_{1}(2F_K^2+F_{\pi}^2).
  \label{eq:decsfromfpik}
\end{equation}
Setting $\Lambda_1=0$ and plugging the experimental ratio
$F_{K^{+}}/F_{\pi^{+}} = 1.193(2)$~\cite{Aoki:2019cca} into
eq.~\eqref{eq:decsfromfpik} gives
\begin{equation}
F^{\ell}/F_{\pi}=1\quad\text{and}\quad  F^{s} / F_{K} = \sqrt{2 - \frac{F_{\pi^{+}}^{2}}{F_{K^{+}}^{2}}} = 1.139(2)\label{eq:fsoverfk}.
\end{equation}
In the FKS model the flavour basis AWIs~\eqref{eq:awiflavour}
factorize into anomalous and
non-anomalous contributions and one obtains the
relations~\cite{Feldmann:1998sh,Feldmann:1999uf}
\begin{equation}
  2 m_{\ell} H^{\ell} = M_{\pi}^{2} F^{\ell}
  \qquad\text{and}\qquad
  2 m_{s} H^{s} = (2 M_{K}^{2} - M_{\pi}^{2}) F^{s},
  \label{eq:decsfromps}
\end{equation}
where $H^{q} = \sqrt{(H_{\eta}^{q})^{2} +
  (H^{q}_{\eta^{\prime}})^{2}}$ for $q = \ell, s$ and the
$H_{\eta^{(\prime)}}^{q}$ are defined in eq.~\eqref{eq:psdecs}.
Therefore, in this approximation, in the flavour basis
the mass mixing angle $\phi_{\rm SU(3)}$, the mixing angles
$\phi^{\ell}$ and $\phi^s$ and the mixing angles for pseudoscalar
matrix elements are all equal:
$\phi_{\rm SU(3)}=\phi = \phi_{s} = \phi_{\ell} = \phi_{\rm PS}$,
where~\cite{Feldmann:2002kz}
\begin{equation}
  \phi_{\mathrm{SU(3)}} = \arcsin\sqrt{\frac{\left(M_{\eta^{\prime}}^{2} - (2 M_{K}^{2} - M_{\pi}^{2})\right)(M_{\eta}^{2}-M_{\pi})}{(M_{\eta^{\prime}}^{2}-M_{\eta}^{2})(2 M_{K}^{2} - 2 M_{\pi}^{2})}}= 42.4^{\circ}.
  \label{eq:phisu3}
\end{equation}

ETMC~\cite{Ottnad:2017bjt}
compute the mixing angle from the averaged ratios of their pseudoscalar
matrix elements:
\begin{equation}
  \tan\phi_{\rm PS} = \sqrt{\tan\phi_{\rm PS}^{\ell}\tan\phi_{\rm PS}^{s}}\quad\text{with}\quad
  \tan\phi_{\rm PS}^{\ell} = \frac{H_{\eta^{\prime}}^{\ell}}{H_{{\eta}}^{\ell}}\quad\text{and}\quad
  \tan\phi_{\rm PS}^{s} = - \frac{H_{\eta}^{s}}{H_{\eta^{\prime}}^{s}}.\label{eq:phips}
\end{equation}
The decay constants are obtained, using eq.~\eqref{eq:decsfromps}.
Subsequently, the ratios $F^{\ell}/F_{\pi}$, $F^{s}/F_{K}$ and the angle
$\phi$ are extrapolated,
using a fit that is linear in the quark masses
and quadratic in the lattice spacing. At the physical point ETMC obtain
\begin{equation}
  \text{ETMC~\cite{Ottnad:2017bjt}:}\qquad
  F^{\ell} / F_{\pi} = 0.960(59),\quad F^{s}/ F_{K} = 1.143(24)\quad\text{and}\quad \phi = 38.8(3.3)^{\circ}\label{eq:ETMCdecresult},
\end{equation}
which within errors agrees with the values computed within the FKS
model in eqs.~\eqref{eq:fsoverfk} and~\eqref{eq:phisu3} from the experimental
kaon and pion decay constants and the experimental meson
masses, respectively.
The ETMC results also agree well
with our results, that are
obtained directly from the axialvector matrix elements,
at the scale $\mu=1\,\mathrm{GeV}$~(see tab.~\ref{tab:cmpflavour}),
where we find $\Lambda_1\approx 0$, whereas
at higher scales the two sets of results differ somewhat.

\begin{figure}
  \includegraphics[width=0.49\linewidth]{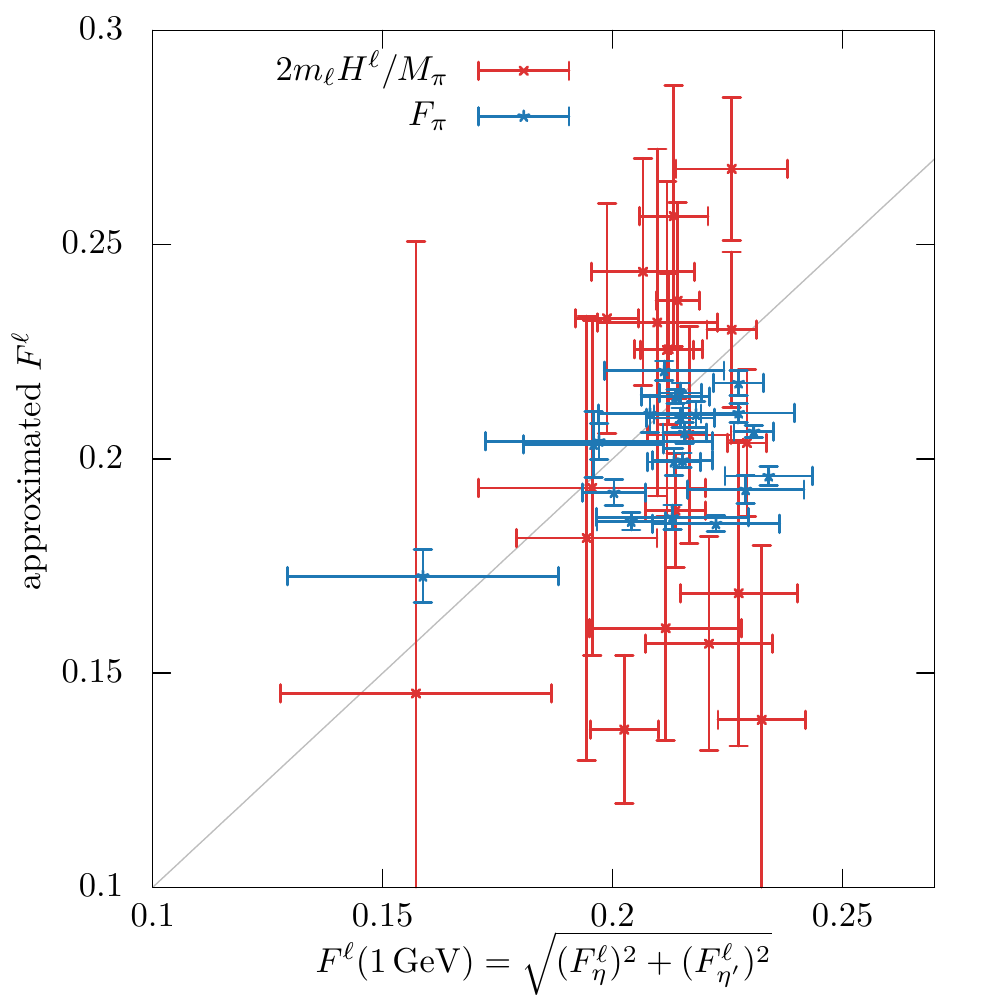}
  \includegraphics[width=0.49\linewidth]{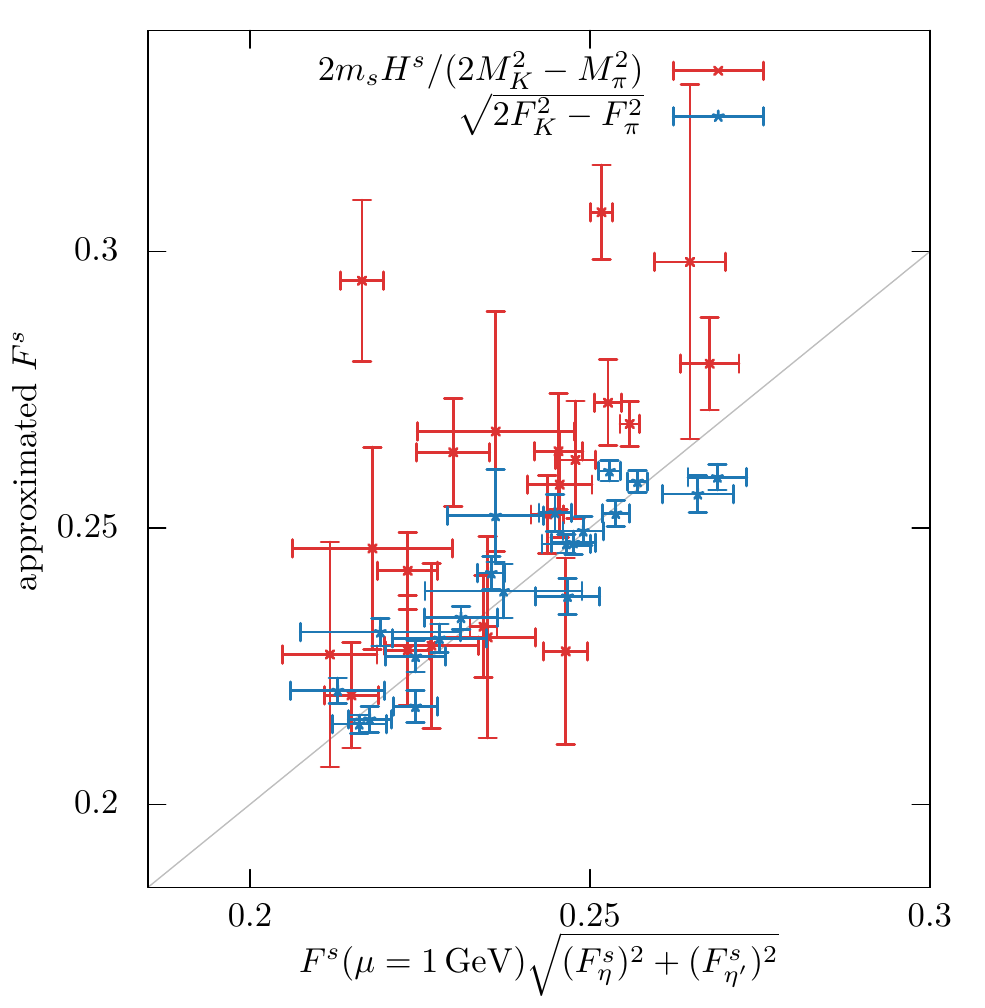}\\
  \includegraphics[width=0.49\linewidth]{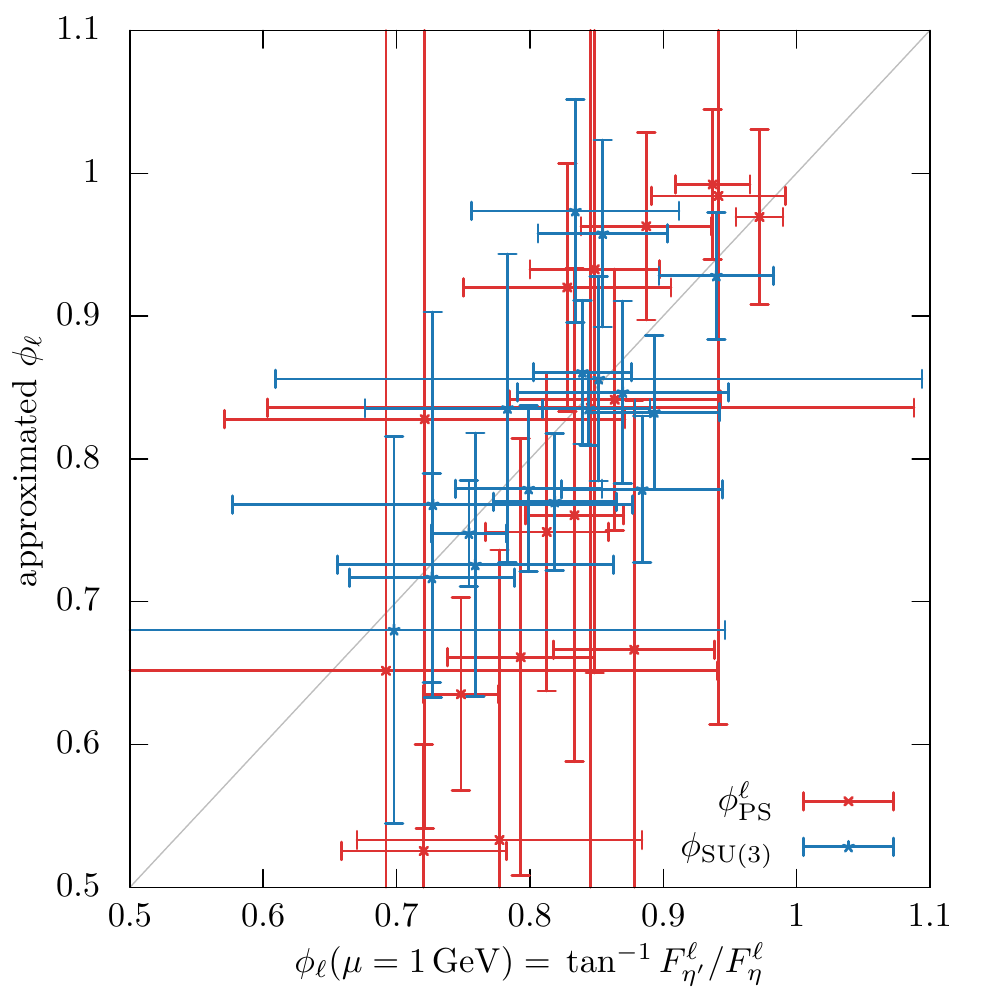}
  \includegraphics[width=0.49\linewidth]{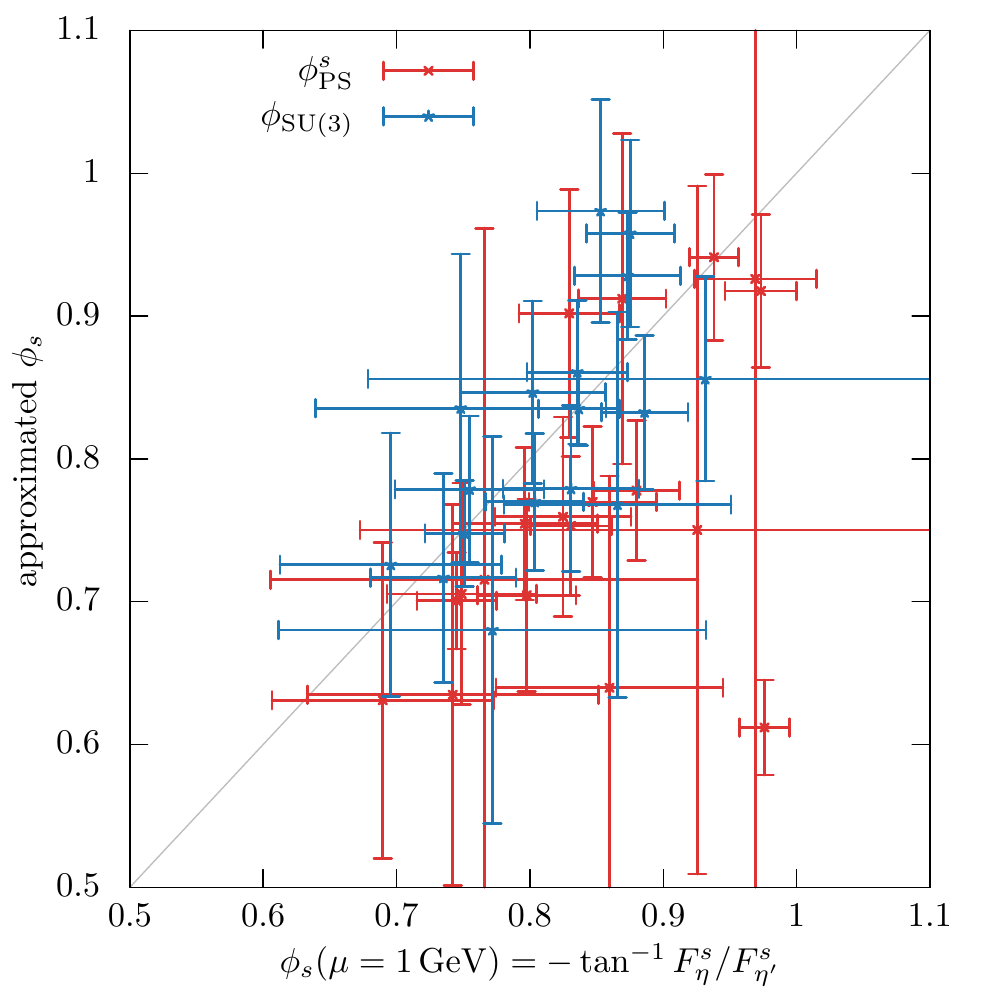}
  \caption{\label{fig:avvsfksdecs} Light and strange decay constants
    (top) and angles (bottom) determined on each ensemble at the QCD scale
    $\mu=1\,\mathrm{GeV}$. The central
    values and errors in the $x$-direction indicate the results obtained
    directly from the axialvector matrix elements, while
    the position and error in the $y$-direction indicates the values
    constructed from the pseudoscalar matrix elements (red,
    eqs.~\eqref{eq:decsfromps} and~\eqref{eq:phips}). For the decay constants, the expectations
    derived from combinations of the pion and kaon masses
    and decay constants (blue, eq.~\eqref{eq:decsfromfpik}) are also displayed,
    while for the angles $\phi_{\mathrm{SU(3)}}$~(eq.~\eqref{eq:phisu3}) is also shown.  The blue
    points have been shifted slightly to the right for better
    visibility.  }
\end{figure}

\paragraph{Test of the FKS approximation, away from the physical point.}
The results in tabs.~\ref{tab:cmpflavour}, \ref{tab:cmpoctsinglet}
and~\ref{tab:cmpLambda1} (see also eqs.~\eqref{eq:fsoverfk}
and~\eqref{eq:phisu3}) show that our ab-initio values determined at
$\mu=1\,\mathrm{GeV}$ agree well with those derived by employing the
FKS scheme. We can go further and directly check the
relations~\eqref{eq:decsfromps}, \eqref{eq:phips} and~\eqref{eq:decsfromfpik},
also away from the physical point. Figure~\ref{fig:avvsfksdecs}
displays the two decay constants and two angles in the flavour basis
determined from the pseudoscalar matrix elements against the direct
results for a range of ensembles at $\mu=1\,\mathrm{GeV}$.  The values for $F^\ell$ and $F^s$
obtained from the pion and kaon decay constants are also shown. Modulo
the large errors for some ensembles, there is reasonable agreement
between the direct results for the angles and the FKS expectation, with
$\phi_{\mathrm{SU(3)}}\approx \phi_{\rm PS}$. Qualitative agreement is also found
for the decay constants, however, some scatter in the results is
visible, which may be due to discretization effects and/or
the limitations of the FKS approximation. This is less significant for
$F^s$ and it is striking how well this quantity is reproduced by the
combination $\sqrt{2F_K^2-F_\pi^2}$. However, at higher scales, where
$\Lambda_1$ can no longer be neglected, scale dependent quantities
cannot be reliably extracted with any precision using the FKS method,
as indicated in the comparison tables.  In
particular, for the singlet decay constant we observe the difference
between the results at high and low scales $2
(F^{0}(1\,\mathrm{GeV}) - F^{0}(\infty))/(F^{0}(1\,\mathrm{GeV}) +
F^{0}(\infty)) = 12.9\,\%$. We remark that previously it was unclear
at what scale the FKS approximation holds and this led to an
additional unquantifiable uncertainty in phenomenological
analyses.

Our direct QCD results can be used as input to theory
calculations and we consider one important example in
sec.~\ref{sec:tff}, namely the light-cone sum rule computation of
the $\gamma \gamma^{*} \to \eta^{(\prime)}$ transition form factors.

\subsection{Large-$N_c$ low energy constants\label{sec:leccomparison}}
As part of our analysis we are able to extract the
large-$N_{c}$ ChPT LECs up to NLO.  The singlet mass
in the chiral limit $M_{0}$ and the two OZI-rule violating parameters
$\Lambda_{1}$ and $\Lambda_{2}$ are all $\mathcal{O}(1/N_{c})$ in the
power counting. Besides these large-$N_{c}$ specific LECs, $L_{5}$,
$L_{8}$ and the decay constant in the chiral limit, $F$, also
appear. These are present in ordinary SU($3$) ChPT\@, although, their
values can differ. In particular, $L_{5}$ and $L_{8}$ depend on the
ChPT renormalization
scale in the SU($3$) theory, however, such scale dependence only
arises at NNLO in large-$N_{c}$ ChPT.

The $\mathcal{O}(1/N_{c})$ LECs extracted from our NLO fits at $\mu = \infty$
for $N_f=3$ read:
\begin{align}
  M_{0} &=  761\left(\substack{13\\21}\right)_\mathrm{stat} \left(\substack{18\\11}\right)_{\mathrm{syst}}(11)_{t_0}\,\mathrm{MeV},\\
  \Lambda_{1} &=  -0.25\left(\substack{1\\4}\right)_\mathrm{stat} \left(\substack{6\\2}\right)_{\rm syst},\\
  \Lambda_{2} &=  0.11\left(\substack{5\\5}\right)_\mathrm{stat} \left(\substack{7\\10}\right)_{\rm syst},
\end{align}
where the ChPT and lattice spacing errors have been combined into a single systematic uncertainty.
The dependence of these quantities on the QCD renormalization scale is
discussed in sec.~\ref{sec:decsummary}.  To aid comparison with
literature values we consider the scale independent
combinations~\cite{Leutwyler:1997yr}:
\begin{equation}
  M_{0} / \sqrt{1+\Lambda_{1}} = 877\left(\substack{12\\10}\right)_\mathrm{stat} \left(\substack{21\\8}\right)_{\mathrm{syst}}(13)_{t_0}\,\mathrm{MeV}\quad\text{and}\quad
  \tilde{\Lambda}= \Lambda_{1} - 2 \Lambda_{2} = -0.46(19).
\end{equation}
Previous determinations of these quantities include:
\begin{align}
  &\text{Leutwyler~\cite{Leutwyler:1997yr}:} & M_{0}/\sqrt{1 + \Lambda_{1}} &\approx 899\,\mathrm{MeV} & \text{and}\quad & \tilde{\Lambda} =  -0.31,\\
  &\text{Benayoun et al.~\cite{Benayoun:1999au}:} & & & & \tilde{\Lambda} =  -0.42(6),\\
  &\text{Guo et al.~\cite{Guo:2015xva}:} & M_{0}/\sqrt{1+\Lambda_{1}} &=  804(80)\,\mathrm{MeV} & \text{and}\quad & \tilde{\Lambda} =  -0.37(17),\\
  &\text{Bickert et al.~\cite{Bickert:2016fgy}:} & M_{0}/\sqrt{1+\Lambda_{1}} &=  950(7)\,\mathrm{MeV} & \text{and}\quad & \tilde{\Lambda} = -0.34(5),
\end{align}
where except for~\cite{Bickert:2016fgy}, we have constructed these scale independent quantities from the individual results quoted in the publications.
Our central value for $M_0/\sqrt{1+\Lambda_{1}}$ is larger than the result
of~\cite{Guo:2015xva}, which utilizes lattice data, however,
considering the large uncertainty quoted in this reference,
there is no significant
disagreement. The determination from~\cite{Bickert:2016fgy} lies
roughly two standard deviations higher, where the LECs in this study are
determined from experimental input which includes the singlet and
non-singlet meson masses and non-singlet decay constants.
We also find some disagreement with~\cite{Bickert:2016fgy} when
comparing predictions for the decay constants and angles,
cf.\ tab.~\ref{tab:cmpoctsinglet}.  Interestingly, our value for the
combination of OZI-violating LECs $\tilde{\Lambda}$ is in good
agreement with the above determinations.

For the decay constant in the chiral limit, we obtain 
\begin{equation}
  F = 87.71\left(\substack{1.44\\1.57}\right)_\mathrm{stat} \left(\substack{2.69\\81}\right)_{\mathrm{syst}}(1.31)_{t_0}\,\mathrm{MeV}.\label{eq:8771}
\end{equation}
This result agrees within errors with the NLO values presented
in~\cite{Guo:2015xva} and~\cite{Bickert:2016fgy}:
\begin{equation}
  \text{Guo et al.~\cite{Guo:2015xva}:}\quad F = 92.1(6)\,\mathrm{MeV},\quad
  \text{Bickert et al.~\cite{Bickert:2016fgy}:}\quad F = 90.73(11)\,\mathrm{MeV}.
\end{equation}
However, the corresponding
NNLO analyses give somewhat lower values of $F$,
\begin{equation}
  \text{Guo et al.~\cite{Guo:2015xva}:}\quad F = 80.8(6.3)\,\mathrm{MeV},\quad
  \text{Bickert et al.~\cite{Bickert:2016fgy}:}\quad F = 79.46(6.59)\,\mathrm{MeV},
\end{equation}
which within errors still agree with our result~\eqref{eq:8771}.
In the simulations of~\cite{Hernandez:2019qed}
with $N_f=4$ sea quarks, the number of colours $N_c\in\{2, 3, 4, 5, 6\}$ is
varied. The pion decay constant and its mass
are then fitted to the NNLO large-$N_c$ U($4$) ChPT prediction.
From the expected dependence on $N_f/N_c$ and $1/N_c^2$ (neglecting
$N_f^2/N_c^2$ terms), the even lower value
\begin{equation}
  \text{Hernandez et al.~\cite{Hernandez:2019qed}: }\quad F(N_{f} = 3, N_{c} = 3) = 68(7)\,\mathrm{MeV}
\end{equation}
is inferred for $N_f=N_c=3$ at the lattice spacing $a\approx 0.075\,\mathrm{fm}$.

The additional terms appearing at NNLO comprise chiral logs and expressions
which include the LECs $L_{4}$, $L_{6}$, $L_{7}$, $L_{18}$ and
$L_{25}$. In particular, in~\cite{Bijnens:2014lea},
it is argued that $L_{4}$ is anti-correlated with the decay constant in the chiral limit
as seen, e.g., in fits to experimental data in~\cite{Bijnens:2011tb}. Thus, neglecting NNLO contributions including $L_{4}$-terms may lead to larger values of $F$.
However, also our fits in app.~\ref{sec:loopfit}, including only the
NNLO loop contributions (see the discussion of
sec.~\ref{sec:fitselection} as well as below),
gives $F= 79.0(3.8)\,\mathrm{MeV}$.
In conclusion, both these effects may
account for the reduction of the value of $F$ within NNLO analyses, in
comparison to results from NLO parametrizations.

For the LECs $L_5$ and $L_8$, we find
\begin{equation}
  L_5 = 1.66\left(11\right)_\mathrm{stat} \left(20\right)_{\rm syst}\cdot 10^{-3}\quad\text{and}\quad
  L_8 = 1.08\left(9\right)_\mathrm{stat} \left(9\right)_{\rm syst}\cdot 10^{-3}.
  \label{eq:nlolec}
\end{equation}
These values agree reasonably
well with those obtained from other NLO large-$N_{c}$ ChPT studies, e.g.,
\begin{align}
  &\text{Leutwyler~\cite{Leutwyler:1997yr}:} & L_{5} &= 2.2\cdot 10^{-3} & \text{and}\quad & L_{8} = 1.0\cdot 10^{-3},\\
  &\text{Guo et al.~\cite{Guo:2015xva}:} & L_{5} &= 1.47(29)\cdot 10^{-3} & \text{and}\quad & L_{8} = 1.08(6)\cdot 10^{-3},\\
  &\text{Bickert et al.~\cite{Bickert:2016fgy}:} & L_{5} &= 1.86(6)\cdot 10^{-3} & \text{and}\quad & L_{8} = 0.78(5)\cdot 10^{-3}.
\end{align}
A comparison can also be made with the LECs obtained
using SU($3$) ChPT.  The LECs in the SU($3$) and large-$N_c$ theories are related
via~\cite{Kaiser:2000gs,HerreraSiklody:1996pm,Bickert:2016fgy}
\begin{align}
  L_{5}(\mu_{\rm EFT}) &= L_{5}^{\rm SU(3)}(\mu_{\rm SU(3)}) + \frac{3}{8}\frac{1}{16\pi^{2}}\ln\left(\frac{\mu_{\rm SU(3)}}{\mu_{\rm EFT}}\right)\\
  L_{8}(\mu_{\rm EFT}) &= L_{8}^{\rm SU(3)}(\mu_{\rm SU(3)}) + \frac{5}{48}\frac{1}{16\pi^{2}}\ln\left(\frac{\mu_{\rm SU(3)}}{\mu_{\rm EFT}}\right)+
                           \frac{1}{12}\frac{1}{16\pi^{2}}\ln\left(\frac{\mu_{\rm match}}{\mu_{\rm EFT}}\right),
\end{align}
where $\mu_{{\rm SU(3)}}$ is the SU($3$) ChPT scale, $\mu_{\rm EFT}$ is the scale of large-$N_{c}$ ChPT~(which is ill-defined at NLO) and $\mu_{\rm match}$ is the scale at which the two theories are matched. 
We set $\mu_{\rm EFT} = \mu_{\rm SU(3)} = \mu_{\rm match} = 0.770\,\mathrm{GeV}$ such that
$L_{5,8}(\mu_{\rm EFT}) = L_{5,8}^{\rm SU(3)}(\mu_{\rm SU(3)})$. 
A direct comparison can then be made with the SU($3$) values obtained in~\cite{Bazavov:2010hj}
from a $N_{f}=2+1$ lattice study of the pion and kaon masses and decay constants.
Here, we quote the values presented in the FLAG review~\cite{Aoki:2019cca} for $\mu_{\rm SU(3)} = 0.770\,\mathrm{GeV}$  
\begin{equation}
  \text{MILC~\cite{Bazavov:2010hj}:}\quad L_{5}^{\rm SU(3)} = 0.98(38)\cdot 10^{-3}\quad \text{and}\quad L_{8}^{\rm SU(3)} = 0.42(27)\cdot 10^{-3}.
\end{equation}
The agreement with our (scale independent) results improves for $\mu_{\rm EFT} < 0.770\,\mathrm{GeV}$ and $\mu_{\rm match} > 0.770\,\mathrm{GeV}$.

Overall, our results for the large-$N_{c}$ ChPT LECs are reasonably
consistent with literature values.  A direct comparison of NLO and
NNLO results is difficult due to the scale dependence which arises
at NNLO\@. Results from our fits including the NNLO loop contributions
can be found in
app.~\ref{sec:loopfit}.
The inferior $\chi^{2}/N_{\rm df} = 2.56$ indicates that
this parametrization does not
describe our data well and additional NNLO terms are required for
a consistent description of the data. Note that this analysis
gives values for the
LECs $L_5$ and $L_8$ (see eq.~\eqref{eq:nnlolec})
that are slightly smaller and larger than our
NLO values quoted in eq.~\eqref{eq:nlolec}, respectively.
As also observed in our analysis, it appears
difficult to reliably pin down the many NNLO LECs, and
usually priors or assumptions are needed to
carry out such fits, giving rise to additional uncertainties, see for
example the scatter of NNLO results
in~\cite{Guo:2015xva,Bickert:2016fgy}.

\subsection{Pseudoscalar gluonic and fermionic matrix elements\label{sec:gluonmesummary}}
\begin{table*}
  \begin{tabularx}{\textwidth}{l|X|X}
    \toprule
    & $a_{\eta} $ & $a_{\eta^{\prime}}$ \\
    \midrule
 $\mu=1\,\mathrm{GeV}$ & $0.01720\left(\substack{40\\69}\right)_{\rm stat}\left(48\right)_{\rm syst}\left(67\right)_{t_{0}}\,\mathrm{GeV}^3$ & $0.0424\left(\substack{19\\17}\right)_{\rm stat}\left(80\right)_{\rm syst}\left(19\right)_{t_{0}}\,\mathrm{GeV}^3$\\
 $\mu=2\,\mathrm{GeV}$ & $0.01700\left(\substack{40\\69}\right)_{\rm stat}\left(48\right)_{\rm syst}\left(66\right)_{t_{0}}\,\mathrm{GeV}^3$ & $0.0381\left(\substack{18\\17}\right)_{\rm stat}\left(80\right)_{\rm syst}\left(17\right)_{t_{0}}\,\mathrm{GeV}^3$\\
 $\mu=10\,\mathrm{GeV}$ & $0.01688\left(\substack{40\\69}\right)_{\rm stat}\left(48\right)_{\rm syst}\left(66\right)_{t_{0}}\,\mathrm{GeV}^3$ & $0.0356\left(\substack{18\\17}\right)_{\rm stat}\left(80\right)_{\rm syst}\left(17\right)_{t_{0}}\,\mathrm{GeV}^3$\\
    $\mu=\infty$        & $0.01676\left(\substack{40\\67}\right)_{\rm stat}\left(48\right)_{\rm syst}\left(65\right)_{t_{0}}\,\mathrm{GeV}^3$ & $0.0330\left(\substack{18\\17}\right)_{\rm stat}\left(80\right)_{\rm syst}\left(16\right)_{t_{0}}\,\mathrm{GeV}^3$\\
    \toprule
    & $\theta_{y} $ & $a^{2}_{\eta^{\prime}}/a^{2}_{\eta}$ \\
    \midrule
 $\mu=1\,\mathrm{GeV}$ &  $-22.1\left(\substack{3\\5}\right)_{\rm stat}\left(2.8\right)_{\rm syst}^{\circ}$      & $6.09\left(\substack{27\\53}\right)_{\rm stat}\left(2.05\right)_{\rm syst}$\\
 $\mu=2\,\mathrm{GeV}$ &  $-24.0\left(\substack{4\\1.0}\right)_{\rm stat}\left(3.2\right)_{\rm syst}^{\circ}$    & $5.03\left(\substack{19\\45}\right)_{\rm stat}\left(1.94\right)_{\rm syst}$\\
 $\mu=10\,\mathrm{GeV}$ & $-25.3\left(\substack{4\\1.1}\right)_{\rm stat}\left(3.6\right)_{\rm syst}^{\circ}$    & $4.46\left(\substack{16\\41}\right)_{\rm stat}\left(1.86\right)_{\rm syst}$\\
    $\mu=\infty$           & $-26.9\left(\substack{4\\1.2}\right)_{\rm stat}\left(4.1\right)_{\rm syst}^{\circ}$ & $3.88\left(\substack{14\\38}\right)_{\rm stat}\left(1.78\right)_{\rm syst}$\\
    \bottomrule
  \end{tabularx}
  \caption{Gluonic matrix elements of the $\eta$ and $\eta^{\prime}$  and combinations thereof at various scales.\label{tab:gluonmeAtScale}}
\end{table*}

We determined the anomaly
matrix elements $a_{\eta}$ and $a_{\eta^{\prime}}$ in
sec.~\ref{sec:gluonmefermionic} from a fit to combinations of
axialvector and pseudoscalar matrix elements,
eq.~\eqref{eq:gluonmefermionic}.  The fit is performed for data at the
QCD renormalization scale $\mu = \infty$ and we carry out the
conversion to lower scales, using the fact that the combinations
$m_{f}H^{f}_{n}$ are scale independent.  We first determine these
combinations by plugging our physical point results on the masses,
decay constants and the gluonic matrix elements into the AWIs in the
flavour basis, eq.~\eqref{eq:awiflavour}.  Following this, we
reconstruct $a_{\eta^{(\prime)}}$ at different scales using the
known running of the singlet axialvector current.  With $N_f=3$
active quark flavours, at
$2\,\mathrm{GeV}$ we obtain:
\begin{align}
  a_\eta(\mu = 2\,\mathrm{GeV}) &= 0.01700\left(\substack{40\\69}\right)_{\rm stat}\left(48\right)_{\rm syst}\left(66\right)_{t_{0}}\,\mathrm{GeV}^3,\\
  a_{\eta^\prime}(\mu = 2\,\mathrm{GeV}) &= 0.0381\left(\substack{18\\17}\right)_{\rm stat}\left(80\right)_{\rm syst}\left(17\right)_{t_{0}}\,\mathrm{GeV}^3.
\end{align}
The systematic error is computed as the difference between our results
from a direct NLO fit to the $a_{\eta^{(\prime)}}$ data (see
eq.~\eqref{eq:gluonmefitresults}), that included lattice correction
terms, and the continuum NLO large-$N_{c}$ ChPT prediction (see
eq.~\eqref{eq:gluonmechptresults}), based on the set of LECs that we
obtained from our simultaneous fits to the masses and decay constants.
We list our results at various scales in tab.~\ref{tab:gluonmeAtScale}
and compare to literature values in tab.~\ref{tab:cmpgluonme}, where
the scale is not specified. These analyses are based on, e.g., QCD sum
rule calculations~\cite{Novikov:1979uy,Singh:2013oya}, large-$N_{c}$
ChPT~\cite{Feldmann:1999uf,Beneke:2002jn} and related state
mixing models that include a pseudoscalar
glueball~\cite{Cheng:2008ss,Qin:2017qes}.
We find agreement with the references that give error estimates,
with the exception of~\cite{Beneke:2002jn}.

Combining our physical point results on $a_{\eta^{(\prime)}}$,
the $\eta^{(\prime)}$ masses and their decay constants with eq.~\eqref{eq:awiflavour}
gives the following predictions
\begin{align}
m_{\ell} H^{\ell}_{\eta} &= 0.0021\left(\substack{3\\2}\right)_{\rm stat}(13)_{\rm syst}(0)_{t_{0}}\,\mathrm{GeV}^{3},&\quad
m_{s} H^{s}_{\eta} &= - 0.0173\left(\substack{3\\2}\right)_{\rm stat}(17)_{\rm syst}(7)_{t_{0}}\,\mathrm{GeV}^{3}, \nonumber\\
m_{\ell} H^{\ell}_{\eta^{\prime}} &= 0.0045\left(\substack{10\\8}\right)_{\rm stat}(40)_{\rm syst}(0)_{t_{0}}\,\mathrm{GeV}^{3},&\quad
m_{s} H^{s}_{\eta^{\prime}} &= \phantom{-}0.0309\left(\substack{15\\5}\right)_{\rm stat}(50)_{\rm syst}(10)_{t_{0}}\,\mathrm{GeV}^{3}
\end{align}
for the pseudoscalar fermionic matrix elements,
where again the systematic error is the difference with respect
to the NLO ChPT predictions
eqs.~\eqref{eq:mpslighteta}--\eqref{eq:mpsstrangeetaprime}, obtained using
our set of LECs. Since the values of the above combinations are smaller
in the light quark sector than for strange quarks, and the
absolute error on $a_{\eta^{(\prime)}}$ is the major contribution to
their uncertainty, the relative precision that we can achieve is limited for
the light quark combinations. Note that this is a statement about the
physical mass continuum limit; on individual ensembles also the
light quark matrix elements can be quite precise.
While there is some tension for the combination $m_{s}H^{s}_{\eta}$,
most of our results agree with the estimate in the FKS
approximation, where the pseudoscalar matrix element is taken
in the SU($2$) isospin limit,
\begin{align}
  \text{Feldmann~\cite{Feldmann:1998sh}:}\quad m_{\ell}H^{\ell}_{\eta} &=  0.0010\,\mathrm{GeV}^{3}, &
                                           m_{s} H^{s}_{\eta} &=  -0.026\,\mathrm{GeV}^{3},\nonumber\\
                                        m_{\ell}H^{\ell}_{\eta^{\prime}} &=  0.0008\,\mathrm{GeV}^{3}, &
                                       m_{s}H^{s}_{\eta^{\prime}} &=  \phantom{-}0.032\,\mathrm{GeV}^{3},
\end{align}
and the very similar numbers of a QCD sum rule calculation,
\begin{align}
  \text{Singh~\cite{Singh:2013oya}:}\quad m_{\ell}H^{\ell}_{\eta} &= 0.00105(14)\,\mathrm{GeV}^{3}, &
                                           m_{s} H^{s}_{\eta} &= -0.0284(55)\,\mathrm{GeV}^{3},\nonumber\\
                                        m_{\ell}H^{\ell}_{\eta^{\prime}} &= 0.000782(250)\,\mathrm{GeV}^{3}, &
                                       m_{s}H^{s}_{\eta^{\prime}} &= \phantom{-}0.0379(71)\,\mathrm{GeV}^{3}.
\end{align}
We again emphasize that the above combinations are renormalization
group invariants.

\begin{table*}
\setlength\tabcolsep{0pt}
\pgfplotstabletypeset[columns={ref,aetaerr,aetastr,aetaprimeerr,aetaprimestr},
  column type=,
  begin table={\begin{tabularx}{\textwidth}{l p{\summarygraphplotwidth} X p{\summarygraphplotwidth} X}},
                 end table={\end{tabularx}},
               every head row/.style = {before row=\toprule, after row=\midrule},
               every last row/.style = {after row=[3ex]\bottomrule},
               every row no 7/.style = {before row=\hline},
               columns/ref/.style = {string type, column name = ref},
               columns/aetaerr/.style = {%
                 column name = {},
                 assign cell content/.code = {%
                   \ifnum\pgfplotstablerow=0
                   \pgfkeyssetvalue{/pgfplots/table/@cell content}
                   {\multirow{\numberofgluoncmprows}{\summarygraphplotwidth}{\aetacmpplot}}%
                   \else
                   \pgfkeyssetvalue{/pgfplots/table/@cell content}{}%
                   \fi
                 }
               },
               columns/aetastr/.style = {string type, column name = $a_{\eta} / \mathrm{GeV}^{3}$},
               columns/aetaprimeerr/.style={%
                 column name = {},
                 assign cell content/.code = {%
                   \ifnum\pgfplotstablerow=0
                   \pgfkeyssetvalue{/pgfplots/table/@cell content}
                   {\multirow{\numberofgluoncmprows}{\summarygraphplotwidth}{\aetaprimecmpplot}}%
                   \else
                   \pgfkeyssetvalue{/pgfplots/table/@cell content}{}%
                   \fi
                 }
               },
               columns/aetaprimestr/.style = {string type, column name = $a_{\eta^{\prime}} / \mathrm{GeV}^{3}$},
]{\gluoncmpdata}
\caption{Literature values for the anomaly matrix elements in comparison with our results at various scales. Note that the error bars of~\cite{Cheng:2008ss} are cut off at both ends.\label{tab:cmpgluonme}}
\end{table*}

It is particularly interesting to inspect
the ratio of the gluonic matrix elements that can be used to define
a mixing angle in the gluonic sector~\cite{Feldmann:1999uf},
\begin{equation}
  \theta_{y}(\mu = 2\,\mathrm{GeV}) = - \arctan\left(\frac{a_{\eta}(2\,\mathrm{GeV})}{a_{\eta^{\prime}}(2\,\mathrm{GeV})}\right) = -24.0\left(\substack{4\\1.0}\right)_{\rm stat}\left(3.2\right)_{\rm syst}^{\circ}.
\end{equation}
The squared ratio $(a_{\eta^\prime} / a_\eta)^2 = \left(\cot \theta_{y}\right)^{2}$ is closely related to the ratio of decay widths of $J/\psi \to \eta^{(\prime)}\gamma$ when assuming that the anomaly dominates~\cite{Novikov:1979uy,Goldberg:1979qv},
\begin{equation}
  R(J/\psi) = \frac{\Gamma[J/\psi \to \eta^{\prime}\gamma]}{\Gamma[J/\psi \to \eta \gamma]} = \frac{a_{\eta^{\prime}}^{2}}{a_{\eta}^{2}} \left(\frac{k_{\eta^{\prime}}}{k_{\eta}}\right)^{3},
\end{equation}
where $k_{\mathcal{M}} =\frac12 \frac{M_{J/\psi}^{2} - M_{\mathcal{M}}^{2}}{2\,M_{J/\psi}}$ is the momentum of the meson $\mathcal{M}$ in the rest frame of the
$J/\psi$. Using the experimental masses of~\cite{PDG},
we obtain $(k_{\eta^{\prime}}/k_{\eta})^{3} \approx 0.8137$.
Our result for the ratio $a_{\eta^{\prime}}^{2} / a_{\eta}^{2}$, listed in~tab.~\ref{tab:gluonmeAtScale}, gives at $\mu=2\,\mathrm{GeV}$
\begin{equation}
  R(J/\psi, \mu = 2\,\mathrm{GeV}) = 5.03\left(\substack{19\\45}\right)_{\rm stat}\left(1.94\right)_{\rm syst}.
\end{equation}
Note that $a_{\eta^{\prime}}$ depends strongly on the scale,
see.~tab.~\ref{tab:gluonmeAtScale}, and the most relevant
scale for this decay is probably below $2\,\mathrm{GeV}$, which would 
somewhat increase the prediction.  The most recent PDG averages for the partial
widths $\Gamma(J/\psi\to\eta\gamma)/\Gamma_{\rm total} =
1.108(27)\cdot 10^{-3}$ and $\Gamma(J/\psi\to\eta^{\prime}\gamma) /
\Gamma_{\rm total} = 5.25(7)\cdot 10^{-3}$ result in
\begin{equation}
  \text{PDG~\cite{PDG}:} \quad R(J/\psi) = 4.74(13),
\end{equation}
which is very close to our value with $N_f=3$ at
$\mu = 2\,\mathrm{GeV}$.  Clearly,
a more precise comparison should take $\mathcal{O}(\alpha_s)$
corrections into account.

\subsection{Transition form factors $F_{\gamma\gamma^*\to\eta}(Q^2)$ and $F_{\gamma\gamma^*\to\eta^\prime}(Q^2)$\label{sec:tff}}
Photoproduction is the simplest hard process involving $\eta^{(\prime)}$
mesons. The corresponding transition form factors
$F_{\gamma\gamma^*\to\eta}(Q^2)$ and
$F_{\gamma\gamma^*\to\eta^\prime}(Q^2)$ at large photon virtualities
$Q^2 = |q^2|$
have been studied, e.g., in~\cite{Kroll:2002nt,Agaev:2014wna,Ding:2018xwy}
and assessed phenomenologically for instance
in~\cite{Escribano:2015nra,Escribano:2015yup}.
The special role of the transition form factors as ``gold plated''
observables for the study of meson light-cone distribution amplitudes (LCDAs)
is widely recognized. The corresponding theory is similar to that for
the more easily accessible pion transition form factor
$F_{\gamma\gamma^*\to\pi^0}(Q^2)$ but the non-perturbative input encoded in the
LCDAs is more complicated.
The two outstanding issues are, first, whether the $\eta$ and $\eta^{\prime}$ LCDAs follow
the same mixing pattern as the decay constants at a low scale, and,
second, whether $\eta^{\prime}$ contains a significant
two-gluon component, see, e.g.,~\cite{Kroll:2002nt,Agaev:2014wna} for a discussion.

\begin{figure}
  \includegraphics[width=0.49\linewidth]{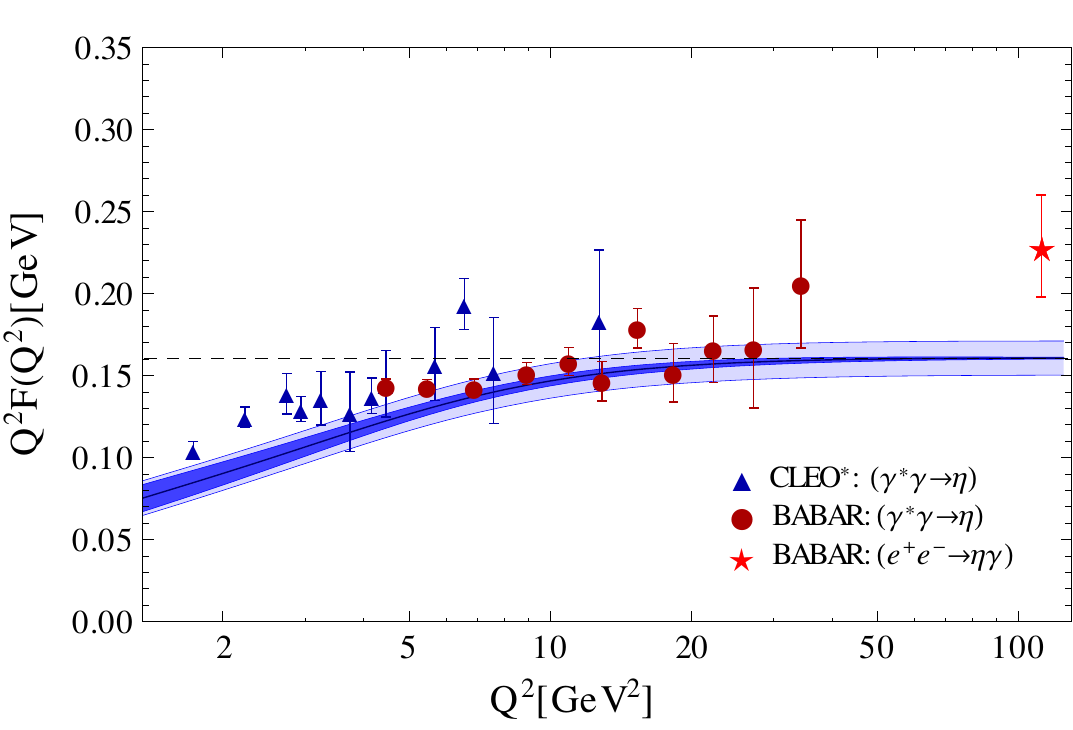}
  \includegraphics[width=0.49\linewidth]{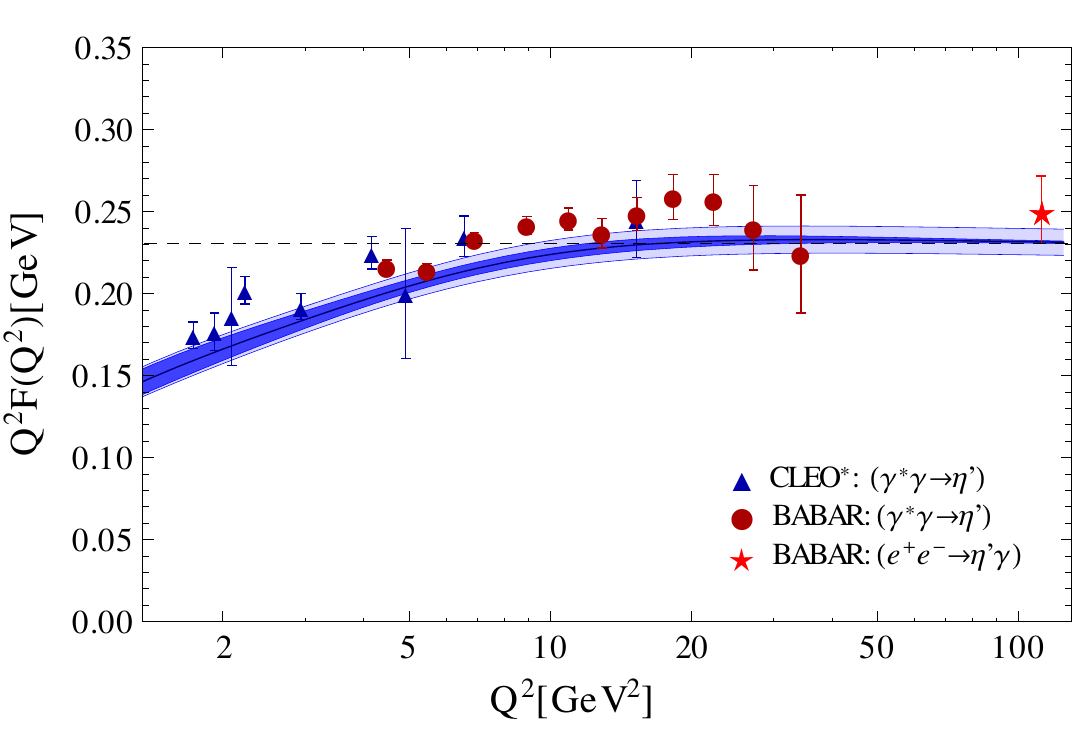}
  \caption{\label{fig:transitionff}Predictions for the transition form factor
    $F_{\gamma\gamma^*\to\eta}(Q^2)$ (left) and $F_{\gamma\gamma^*\to\eta^\prime}(Q^2)$
    (right) following the approach in~\cite{Agaev:2014wna}, using the decay
    constants determined in our work as input parameters.}
\end{figure}

In what follows we briefly discuss the impact of our results on
predictions of these transition form factors.
One has to keep in mind that such predictions are
affected by higher-twist and end-point (``Feynman'') contributions that
are formally suppressed by a power of the photon virtuality $Q^2$. However,
at the virtualities covered by present day experiments these corrections
are still significant.
In fig.~\ref{fig:transitionff} we show the QCD prediction for the form
factors, complemented by taking into account power-suppressed contributions,
using dispersion techniques and quark-hadron duality (light-cone sum rules,
LCSRs), see~\cite{Agaev:2014wna} for details.
The results are compared 
to the experimental data for the space-like form factors by the
CLEO~\cite{Gronberg:1997fj} and BaBar~\cite{BABAR:2011ad} 
collaborations, and we also include BaBar's
time-like data points at $q^2 = 112\,\text{GeV}^2$~\cite{Aubert:2006cy}
as stars.

The difference with respect to the original calculation
is that in fig.~\ref{fig:transitionff} 
the lattice values of the decay constants determined in this work
are used as an input, whereas in~\cite{Agaev:2014wna}
the decay constants were taken from~\cite{Feldmann:1998sh}
under the (ad hoc) assumption that they correspond to the 
scale $1\,\mathrm{GeV}$. Using lattice results removes this scale uncertainty.
In view of the experimentally available range of $Q^{2}$, employing
$N_{f} = 4$ appears reasonable.
Therefore, we run our $N_f=3$ values for $F^0_{\eta^{(\prime)}}$
(see tab.~\ref{tab:alldecresultspriors})
down to $\mu_{0}=1.51\,\mathrm{GeV}$~\cite{deFlorian:2016spz}
(see sec.~\ref{sec:singletrenorm}), where we match to the
$N_f=4$ theory. This value is then taken as an input
for the LCSR calculation.

The LCSR technique involves a certain model dependence in the calculation of the
power-suppressed contributions to the form factors. This is indicated by
the dark blue shaded regions in fig.~\ref{fig:transitionff} 
and can be regarded as an (at present) irreducible uncertainty of
such calculations. The total uncertainty including that of the lattice
values for the decay constants is shown in light blue (added in quadrature).
Starting around $Q^2\sim 10\,\mathrm{GeV}^2$ this
uncertainty dominates over the model dependence. 

The calculation is carried out assuming that the shapes of the LCDAs
of the $\eta$ and the
$\eta^{\prime}$ at low scales are the same as that of the pion.
In our calculation, following~\cite{Agaev:2014wna}, the corresponding
parameters are chosen from the fit to the pion transition  
form factor in the same approach~\cite{Agaev:2012tm}. Moreover, the
two-gluon LCDA at the low scale is set to zero.
Under these assumptions the only additional non-perturbative input
at the leading-twist level are the decay constants that
we computed here. The comparison of fig.~\ref{fig:transitionff}
between the predictions for
$F_{\gamma\gamma^*\to\eta}(Q^2)$ and $F_{\gamma\gamma^*\to\eta^\prime}(Q^2)$
and experimental
data shows that the above approximation
appears to works relatively well, 
although there is some tension with the available data.
In the asymptotic limit $Q^2\to \infty$ the dependence on the shape of the
LCDAs is removed and the decay constants provide the only necessary input:
\begin{equation}
  \lim_{Q^2\to\infty} Q^2 F_{\gamma\gamma^*\to\eta^{(\prime)}} =
  \frac{2}{\sqrt{3}}\left( F_{\eta^{(\prime)}}^8 + 2 \sqrt{2} F_{\eta^{(\prime)}}^0(\mu=\infty,N_f=4)\right).\label{eq:ffasy}
\end{equation}
We obtain for $N_f=4$:
\begin{align}
  \lim_{Q^2\to\infty} Q^2 F_{\gamma\gamma^*\to\eta}(Q^2) &= 160.5(10.0) \,\mathrm{MeV},
\nonumber\\
  \lim_{Q^2\to\infty} Q^2 F_{\gamma\gamma^*\to\eta^\prime}(Q^2) &= 230.5(10.1) \,\mathrm{MeV}.
\end{align}
These asymptotic values are shown as dashed lines in
fig.~\ref{fig:transitionff}. Regarding the latter form factor,
it is particularly important to take the scale dependence of the
singlet decay constant into account.
This explains the relatively large values
obtained for $Q^2 F_{\gamma\gamma^*\to\eta^\prime}(Q^2)$ when
neglecting the scale evolution, see, e.g.,
the predictions in~\cite{Ottnad:2017bjt}. Also the matching
to the $N_f=4$ theory somewhat reduces the value.
As already emphasized in~\cite{Agaev:2014wna}, the effect due
to the scale dependence is enhanced for
the $\eta'$ form factor because in this case the
two terms in eq.~\eqref{eq:ffasy} have opposite signs.

The current experimental accuracy is not yet sufficient to draw definite
conclusions.
In the future, due to an increase of the statistics by a large factor and
improved particle identification, the Belle~II collaboration will be
able to measure the pseudoscalar meson transition form
factors with much higher precision~\cite{Kou:2018nap}.
A disagreement with QCD calculations using lattice input for the decay
constants would either indicate qualitative differences between the
shapes of the LCDAs for different pseudoscalar mesons or
the presence of a large two-gluon contribution. Both
would have important consequences for other hard processes involving
$\eta$ and $\eta^\prime$ mesons, e.g., in weak $B$-decays.

\section{Conclusions}
\label{sec:conclude}
In this study we determined the $\eta$ and $\eta^{\prime}$ masses,
their singlet and octet decay constants and gluonic
matrix elements without model assumptions in $N_{f} = 2+1$
QCD as well as the LECs of large-$N_c$ ChPT at NLO.
This was achieved by analysing several gauge ensembles,
employing non-perturbatively improved Sheikholeslami-Wilson fermions.
The twenty-one large volume CLS ensembles that were used
are distributed across four lattice
spacings $0.050\,\mathrm{fm}\lesssim a \lesssim 0.086\,\mathrm{fm}$,
along two distinct quark mass trajectories that both lead down to and
include the physical point, which enables a controlled
continuum extrapolation of the quark mass dependence of all
the observables.

The main results on the masses and decay constants
are shown in fig.~\ref{fig:nlodecfit} and summarized
in secs.~\ref{sec:masssummary}--\ref{sec:decsummary}.
For the masses we agree with experiment within about one
standard deviation and achieve a precision that
has considerably improved with respect to previous lattice studies,
while fully controlling all systematic errors.
Adding all errors in quadrature, we
obtain in the continuum limit
\begin{align}
  M_\eta = 554.7(9.2)\,\text{MeV}\quad\text{and}\quad
  M_{\eta^\prime} = 930(21)\,\text{MeV}.
\end{align}

Our results for the decay constants are the first to be directly
determined from the axialvector matrix elements and they are at a
similar level of accuracy in terms of the quoted errors as existing results
from the literature that rely on model assumptions and experimental
data. The reasonable agreement found with many of these values
confirms some of the approximations made and sheds light on their
range of validity. In the octet/singlet mixing scheme, defined in
eq.~\eqref{eq:octsingletanglerep}, the $\eta$ and $\eta^\prime$ decay
constants can be parameterized as follows:%
\begin{align}
 F^8 &= 115.0(2.8)\,\text{MeV},\quad & \theta_{8} &= -25.8(2.3)^{\circ}, \nonumber\\
 F^{0}(\mu=2\,\mathrm{GeV}) &= 100.1(3.0)\,\text{MeV},\quad & \theta_0 &= -8.1(1.8)^{\circ},\label{eq:Feta-FetaP}
\end{align}
where $F_\pi\approx 92$\,MeV and
the value of the scale dependent singlet decay constant is given
in the $\overline{\mathrm{MS}}$ scheme for $N_f=3$.
The corresponding results in different parametrizations and at various
renormalization scales are given in tab.~\ref{tab:alldecresultspriors}
of app.~\ref{app:decayresults}. Computing for the first time
a value for $F^0$ at a definite QCD scale, enabled us to
improve on the prediction~\cite{Agaev:2014wna} of the transition form
factors $\gamma\gamma^*\rightarrow \eta^{(\prime)}$,
which is presented in sec.~\ref{sec:tff}.

The continuum extrapolation and physical mass point interpolation
benefit from the large parameter space explored. We carry out
simultaneous NLO large-$N_{c}$ ChPT fits to the
two masses and four decay constants, while including all possible
correlations among these. A number of different fits are performed in
order to quantify the systematic errors and the parametrization
employed yields a consistent set of low energy constants (LECs), see
sec.~\ref{sec:leccomparison}. By taking the renormalization group
running of the singlet axialvector current into account, we can for
the first time determine the OZI-rule violating LECs $\Lambda_{1}$ and
$\Lambda_{2}$ at well-defined scales and find that $\Lambda_1$ is
small only around $\mu=1\,\mathrm{GeV}$, while $\Lambda_2$ cannot be
neglected at any scale, see fig.~\ref{fig:lambdasAtScale}.  The NLO
large-$N_{c}$ ChPT LECs read:
\begin{align}
  M_{0}(\mu = 2\,\mathrm{GeV}) &= \phantom{-}818(27)\,\mathrm{MeV},\quad &
  F &= 87.7(2.8)\,\mathrm{MeV},\nonumber\\
  \Lambda_{1}(\mu = 2\,\mathrm{GeV}) &= -0.13(5),\quad &
  L_5 &= 1.66(23)\cdot 10^{-3},\nonumber\\
  \Lambda_{2}(\mu = 2\,\mathrm{GeV}) &= \phantom{-}0.19(10),\quad &
  L_8 &= 1.08(13)\cdot 10^{-3}.
\end{align}
Note that $M_{0}$, $\Lambda_{1}$ and $\Lambda_{2}$ depend
on the QCD scale. Meson loops do not contribute
at NLO and, therefore, the above LECs are
independent of the ChPT renormalization scale.

Using the axialvector and pseudoscalar matrix elements of the $\eta$
and $\eta^{\prime}$ mesons, we were able to test the octet
and singlet AWIs and to determine the gluonic
matrix elements $a_{\eta^{(\prime)}}=\langle \Omega | 2 \omega |
\eta^{(\prime)}\rangle$. The results are discussed in
sec.~\ref{sec:gluonmesummary}. We found consistency with the prediction
from large-$N_{c}$ ChPT (derived in app.~\ref{sec:nloglue}), see
fig.~\ref{fig:gluonme}, and we successfully checked our computation against
the same matrix elements determined directly from the gluonic
correlation functions, after carrying out the appropriate renormalization. As a
by-product we also confirmed for the first time in $N_f=2+1$ QCD that
the topological susceptibility, while significantly affected by
lattice corrections, is well described by the leading order ChPT
expectation with only one LEC, $F$, see fig.~\ref{fig:chi}.  Our first
ab-initio calculation of the anomaly matrix elements gives at
the physical point in the continuum limit, in the
$\overline{\mathrm{MS}}$ scheme for $N_f=3$, 
\begin{align}
  a_\eta(\mu = 2\,\mathrm{GeV}) = 0.0170(10)\,\mathrm{GeV}^3\quad\text{and}\quad
  a_{\eta^\prime}(\mu = 2\,\mathrm{GeV}) = 0.0381(84)\,\mathrm{GeV}^3,
\end{align}
where the mixing angle
\begin{align}
  \theta_{y}(\mu = 2\,\mathrm{GeV}) = -\arctan\left(\frac{a_{\eta}}{a_{\eta^{\prime}}}\right) = -24.0(3.3)^{\circ}
\end{align}
at this scale is close to $\theta_8$ as expected in the FKS
state mixing model~\cite{Feldmann:1998sh,Feldmann:2002kz}.
While $\theta_8$ is scale independent,
the value of $-\theta_y$ increases with
the renormalization scale, see tab.~\ref{tab:gluonmeAtScale}.
Using the above result, we find excellent agreement with the ratio
of the experimental
decay rates for $J/\psi\rightarrow\eta^\prime\gamma$
and $J/\psi\rightarrow\eta\gamma$.

In general, we find NLO large-$N_c$ U($3$) ChPT to describe our data reasonably
well, however, there is some tension regarding the LECs between
the gluonic and fermionic matrix elements, in particular regarding
$\Lambda_1$. In view of this, a NNLO
description may be desirable, also with respect to a matching to
SU($3$) ChPT, where meson loops already contribute at NLO\@. Simply adding
the meson loop contributions that enter at NNLO in large-$N_c$ ChPT to
the NLO parametrization gives a less satisfactory description of the
data. Therefore, ideally, one would carry out a full NNLO
analysis. Constraining the additional LECs in this case will require
data on additional ensembles, in particular along the $m_s=m_{\ell}$
line in the quark mass plane, and a simultaneous analysis of the
masses and decay constants of the whole nonet of light mesons.

Many phenomenological descriptions of experimental data give numbers
for the matrix elements that agree with or are close to
those of our QCD calculation. For the first time, we presented
results at an unambiguous QCD renormalization scale with a
reliable quantification of the systematic uncertainties.
The values of the decay constants and anomaly matrix elements
calculated here, therefore, constitute very valuable input to theory predictions
that are related to upcoming experiments, e.g., at
Belle~II~\cite{Kou:2018nap}. 
\acknowledgments
The authors thank T.~Feldmann, K.~Ottnad and S.~Scherer for discussions
and clarifications regarding their articles.
We thank F. Joswig and I. Kanamori for alerting us of misprints in an earlier
draft.
J.~S. thanks W.~S\"oldner for discussions. This work was supported by
the Deutsche Forschungsgemeinschaft through the collaborative research
centre SFB/TRR-55 and the Research Unit FOR 2926 “Next Generation pQCD
for Hadron Structure: Preparing for the EIC” and the European Union’s
Horizon 2020 research and innovation programme under the Marie
Sk{\l}odowska-Curie grant agreement no.~813942 (ITN EuroPLEx)
and grant agreement no.~824093 (STRONG-2020). We thank
our colleagues in CLS [\url{http://wiki-zeuthen.desy.de/CLS/CLS}] for
the joint effort in the generation of the gauge field ensembles. The
authors gratefully acknowledge the Gauss Centre for Supercomputing
(GCS) for providing computing time through the John von Neumann
Institute for Computing (NIC) on the the super-computers
JUQUEEN~\cite{juqueen}, JUWELS~\cite{juwels} and in particular
JURECA-Booster~\cite{jureca}
at Jülich Supercomputing Centre (JSC). GCS is the alliance of
the three national supercomputing centres HLRS (Universität
Stuttgart), JSC (Forschungszentrum Jülich), and LRZ (Bayerische
Akademie der Wissenschaften), funded by the German Federal Ministry of
Education and Research (BMBF) and the German State Ministries for
Research of Baden-Württemberg (MWK), Bayern (StMWFK) and
Nordrhein-Westfalen (MIWF). Simulations were also performed on the
Regensburg iDataCool and Athene2 clusters, and the SFB/TRR 55
QPACE~2~\cite{Arts:2015jia} and QPACE~3 machines. We use the multigrid
solver of~\cite{Heybrock:2014iga,Heybrock:2015kpy,Richtmann:2016kcq,Georg:2017diz}
for the inversion of the Dirac operator.

\appendix
\section{NNLO loop corrections: parametrization and fit results\label{sec:loopfit}}
Unlike in SU($3$) ChPT,
in large-$N_c$ U($3$) ChPT meson loops only enter at NNLO
in the power counting
because formally these contributions
are of  $\mathcal{O}(\delta^2)$. Therefore,
the expressions in sec.~\ref{sec:nlochpt} do not contain
chiral logarithms or a dependence on the EFT renormalization
scale $\mu_{\rm EFT}$.
We define the loop functions
\begin{equation}
  A_0(M^2) = - M^2 \log\left( \frac{M^2}{\mu_{\rm EFT}^2} \right).
\end{equation}
The octet and singlet decay constants of
eqs.~\eqref{eq:nlodecf08}--\eqref{eq:nlodecf10}
receive the additional contributions~\cite{Guo:2015xva,Bickert:2016fgy}
\begin{align}
  {F^8_{\eta}}^{\mathrm{NLO}+\mathrm{loops}}&= {F^8_{\eta}}^{\mathrm{NLO}}+
  \frac{3}{32 \pi^2 F}\cos(\theta)A_0(M_K^2),\\
  {F^8_{\eta^\prime}}^{\mathrm{NLO}+\mathrm{loops}}&= {F^8_{\eta^\prime}}^{\mathrm{NLO}}+
  \frac{3}{32 \pi^2 F}\sin(\theta)A_0(M_K^2),\\
  {F^0_{\eta}}^{\mathrm{NLO}+\mathrm{loops}}&= {F^0_{\eta}}^{\mathrm{NLO}}-
  \frac{1}{32 \pi^2 F}\sin(\theta)A_0(M_{\pi}^2),\\
  {F^0_{\eta^\prime}}^{\mathrm{NLO}+\mathrm{loops}}&= {F^0_{\eta^\prime}}^{\mathrm{NLO}}+
  \frac{1}{32 \pi^2 F}\cos(\theta)A_0(M_{\pi}^2).
\end{align}
Moreover, the quark mass dependence of the
mass mixing angle $\theta$, defined in eq.~\eqref{eq:massangle},
changes as the entries of the square mass matrix
eq.~\eqref{eq:chptmassmatrix} also
receive additional contributions.
Specifically, we have to add to
eqs.~\eqref{eq:nlomu8}--\eqref{eq:nlomu80}~\cite{Bickert:2016fgy}:\footnote{Note that there are misprints within the normalizations of eqs.~(C11)--(C13)
  in~\cite{Bickert:2016fgy}.}
\begin{align}
  (\mu_{8}^{\mathrm{NLO}+\mathrm{loops}})^2 &= (\mu_8^{\mathrm{NLO}})^2 + \frac{1}{48 \pi^2 F^2}\Bigg[\left( \frac{3}{2} \overline{M}^2 - \frac{1}{2}\delta M^2 \right) A_0(M_\pi^2)\nonumber\\
                                              & - \left( 4 \overline{M}^2 + \frac{2}{3} \delta M^2 \right) A_0(M_K^2) \nonumber\\
                                              & + \left( \frac{5}{4} \overline{M}^2 + \frac{7}{12}\delta M^2 \right)\left( A_0(\tilde{M}_\eta^2) + A_0(\tilde{M}_{\eta^\prime}^2) \right)\nonumber\\
                                              & + \frac{2 \sqrt{2} \sin(2\theta^{\mathrm{LO}}) + \cos(2 \theta^{\mathrm{LO}})}{4}\left( \overline{M}^2 + \delta M^2 \right)\left( A_0(\tilde{M}_\eta^2) - A_0(\tilde{M}_{\eta^\prime}^2) \right)\Bigg],\\
  (\mu_{0}^{\mathrm{NLO}+\mathrm{loops}})^2 &= (\mu_0^{\mathrm{NLO}})^2 + \frac{1}{48 \pi^2 F^2} \Bigg[\left( 3 \overline{M}^2 - \delta M^2 \right) A_0(M_\pi^2) \nonumber\\
                                              &+ \left( 4 \overline{M}^2 + \frac{2}{3} \delta M^2 \right) A_0(M_K^2) \nonumber\\
                                              &+ \left( \overline{M}^2 + \frac{1}{6} \delta M^2\right)\left( A_0(\tilde{M}_\eta^2) + A_0(\tilde{M}_{\eta^\prime}^2) \right)\nonumber\\
                                              & + \frac{2\sqrt{2}\sin(2\theta^{\mathrm{LO}})+\cos(2\theta^{\mathrm{LO}})}{6}\delta M^2 \left( A_0(\tilde{M}_\eta^2) - A_0(\tilde{M}_{\eta^\prime}^2) \right)\Bigg],\displaybreak\\
  (\mu_{80}^{\mathrm{NLO}+\mathrm{loops}})^2 = & (\mu_{80}^{\mathrm{NLO}})^2 + \frac{ \sqrt{2}}{48 \pi^2 F^2} \Bigg[\left( \frac{3}{2} \overline{M}^2 - \frac{1}{2} \delta M^2 \right) A_0(M_\pi^2) \nonumber\\
                                              & - \left( \overline{M}^2 + \frac{1}{6} \delta M^2 \right) A_0(M_K^2) \nonumber\\
                                              & - \left( \frac{1}{4} \overline{M}^2 + \frac{5}{12}\delta M^2 \right)\left( A_0(\tilde{M}_\eta^2) + A_0(\tilde{M}_{\eta^\prime}^2) \right)\nonumber\\
                                              & - \frac{2 \sqrt{2} \sin(2\theta^{\mathrm{LO}}) + \cos(2 \theta^{\mathrm{LO}})}{4}\left( \overline{M}^2 + \frac{1}{3}\delta M^2 \right)\left( A_0(\tilde{M}_\eta^2) - A_0(\tilde{M}_{\eta^\prime}^2) \right)\Bigg].
\end{align}
$\theta^{\mathrm{LO}}$ corresponds to the mass mixing angle
eq.~\eqref{eq:massangle}, evaluated at LO, eqs.~\eqref{eq:lomasses}--\eqref{eq:lomasses3}.
$\tilde{M}_{\eta}$ and
$\tilde{M}_{\eta^\prime}$ denote the $\eta$ and $\eta^\prime$ masses,
computed at LO via
eqs.~\eqref{eq:mminus}--\eqref{eq:etaetaprmasses}
and~\eqref{eq:lomasses}--\eqref{eq:lomasses3}.

Carrying out the analysis of our masses and decay constants,
including the NNLO loops,
we obtain for the LECs at $\mu_{\rm EFT} = 0.770\,\mathrm{GeV}$ in the
$N_f=3$ $\overline{\mathrm{MS}}$ scheme at $\mu=\infty$:
\begin{align}
  L_5 & = 1.97\left(\substack{16\\11}\right)_\mathrm{stat} \left(\substack{0\\19}\right)_a\left(\substack{0\\23}\right)_\chi\times 10^{-3},\nonumber\\
L_8 &=  0.848\left(\substack{126\\109}\right)_\mathrm{stat} \left(\substack{0\\124}\right)_a\left(\substack{0\\113}\right)_\chi\times 10^{-3},\nonumber\\
  M_0 &= 1.78\left(\substack{3\\5}\right)_\mathrm{stat} \left(\substack{2\\2}\right)_a\left(\substack{0\\0}\right)_\chi\left(\substack{1\\1}\right)_{\mathrm{renorm}}\,(8 t_0^{\chi})^{-1/2} \nonumber\\
 &=  837\left(\substack{13\\23}\right)_\mathrm{stat} \left(\substack{11\\11}\right)_{\mathrm{syst}}(12)_{t_0}\,\mathrm{MeV},\nonumber\\
  F & = 0.1680\left(\substack{38\\66}\right)_\mathrm{stat} \left(\substack{69\\0}\right)_a\left(\substack{39\\0}\right)_\chi\left(\substack{4\\1}\right)_{\mathrm{renorm}}\,(8 t_0^{\chi})^{-1/2} \nonumber\\
    &= 78.97\left(\substack{1.78\\3.10}\right)_\mathrm{stat} \left(\substack{3.71\\17}\right)_{\mathrm{syst}}(1.18)_{t_0}\,\mathrm{MeV},\nonumber\\
\Lambda_1 &= -0.10\left(\substack{2\\4}\right)_\mathrm{stat} \left(\substack{5\\4}\right)_a\left(\substack{0\\2}\right)_\chi\left(\substack{4\\2}\right)_{\mathrm{renorm}},\nonumber\\
\tilde{\Lambda}&= -1.0\left(\substack{2\\3}\right)_\mathrm{stat} \left(\substack{4\\0}\right)_a\left(\substack{2\\0}\right)_\chi\left(\substack{0\\1}\right)_{\mathrm{renorm}},\nonumber\\
  \Lambda_2 &= 0.45\left(\substack{16\\12}\right)_\mathrm{stat} \left(\substack{1\\19}\right)_a\left(\substack{0\\12}\right)_\chi\left(\substack{6\\3}\right)_{\mathrm{renorm}} \label{eq:nnlolec}
\end{align}
with $\chi^2/N_{\rm df} \approx 312 / 122 \approx 2.56$. Note that only $M_0$, $\Lambda_1$ and $\Lambda_2$
depend on the QCD scale $\mu$.

\section{NLO expressions for pseudoscalar and gluonic matrix elements\label{sec:nlogluonme}}\label{sec:nloglue}
We start from the octet and singlet AWIs eqs.~\eqref{eq:8awi0} and~\eqref{eq:0awi}. We apply these to states $|n\rangle\in\{|\eta\rangle,|\eta^\prime\rangle\}$
(see eq.~\eqref{eq:8awi} for the octet case) and replace
$\langle\Omega | \partial_{\mu}A_{\mu}^{a} | n\rangle = M_n^2 F^{a}_n$, where
$a\in\{8,0\}$. This gives
\begin{align}\label{eq:fmwi8}
  F_n^8M_n^2&=\frac{2}{\sqrt{3}}m_{\ell}H_n^{\ell}-2\sqrt{\frac{2}{3}}m_sH_n^s,\\
  F_n^0M_n^2&=2\sqrt{\frac{2}{3}}m_{\ell}H_n^{\ell}+\frac{2}{\sqrt{3}}m_sH_n^s
  +\sqrt{\frac{2}{3}}a_n,\label{eq:fmwi0}
\end{align}
where $H_n^q=\langle\Omega| P^q|n\rangle$ are the pseudoscalar matrix elements
and $H_n^{\ell}=(H_n^u+H_n^d)/\sqrt{2}$.
The anomaly terms $a_n=2\langle\Omega|\omega|n\rangle$ are the matrix elements
of the topological charge density $\omega$. The left hand sides of the above
equations are functions of $\overline{M}\vphantom{M}^2$
and $\delta M^2$ and can be parameterized in terms of the six LECs $F$, $M_0^2$,
$L_5$, $L_8$, $\Lambda_1$ and $\Lambda_2$.

In terms of the large-$N_c$ ChPT power counting, one finds
$\{L_5, L_8\}\sim\delta^{-1}$,
$F \sim \delta^{-1/2}$, $\{\sin\theta, \cos\theta, B_0\}\sim \delta^0$
and $\{m_q, M^2, M_0^2, 
\Lambda_1, \Lambda_2\}\sim\delta^1$.
This counting is consistent with the LO
GMOR relation $M_\pi^2=2B_0 m_{\ell}$, where
$\langle S\rangle =\langle\bar{q}q\rangle
\sim N_c=\mathcal{O}(\delta^{-1})$ and $B_0=-\langle\bar{q}q\rangle/F^2
=\mathcal{O}(\delta^0)$. Using the AWIs, $P\sim \delta^{-1}$ implies
that $A_{\mu}\sim\delta^{-1/2}$ and $\omega\sim\delta^0$. The latter is
consistent with the topological susceptibility $\tau\sim\delta^0$ 
as one would expect from the Witten-Veneziano
relation $M_0^2=6\tau_0/F^2$, where $M_0^2\sim\delta^1$ and
$F^2\sim\delta^{-1}$.
Finally, the parametrization of the axial matrix elements
$\langle \Omega|A^a_{\mu}|n\rangle= ip_\mu F^a_n\sim\delta^0$
means that $|n\rangle\sim \delta^{1/2}$. Note that for
the meson masses and the matrix elements the above counting
applies to the LO expressions and there will
be higher order corrections in $\delta$.

Using these counting rules,
eqs.~\eqref{eq:fmwi8} and~\eqref{eq:fmwi0} start at
$\mathcal{O}(\delta^{1/2})$ and should be expanded up to
$\mathcal{O}(\delta^{3/2})$ to obtain predictions at NLO.
We carried out the matching, first at LO and then at NLO.
The NLO results are presented below.
The decay constants can be expressed
in terms of the LECs and meson masses at $\mathcal{O}(\delta^{-1/2})$
and $\mathcal{O}(\delta^{1/2})$
(LO and NLO) via eqs.~\eqref{eq:nlodecf08}--\eqref{eq:nlodecf10} while
$M^2_{\eta}$ and $M^2_{\eta^\prime}$ can be parameterized in terms of the squared mass
matrix (with elements eqs.~\eqref{eq:nlomu8}--\eqref{eq:nlomu80}) via the
rotation~\eqref{eq:ort}--\eqref{eq:ort2} with the angle
defined in eq.~\eqref{eq:massangle} at $\mathcal{O}(\delta^1)$
and $\mathcal{O}(\delta^2)$. We truncate the product
at $\mathcal{O}(\delta^{3/2})$. Accordingly, we replace the quark
masses on the right hand sides by combinations of
$\overline{M}^2$ and $\delta M^2$ via the NLO large-$N_c$ GMOR relations
\begin{align}
  2 m_{\ell} B_{0 } &=  M_{\pi}^{2} \left(1 - 8 \frac{2 L_{8} - L_{5}}{F^{2}}M_{\pi}^{2}\right),\\
  \left(m_{\ell} + m_{s}\right)B_{0 } &=  M_{K}^{2} \left(1 - 8 \frac{2 L_{8} - L_{5}}{F^{2}}M_{K}^{2}\right).
\end{align}
Then both sides are polynomials in $\delta M^2$ and
$\overline{M}\vphantom{M}^2$
of degree one and two at LO and at NLO, respectively.
The pion masses also enter through $\sin\theta$
and $\cos\theta$.
Since we carry out the matching in terms of powers of $\delta$,
we keep the sine and the cosine (that are of $\mathcal{O}(\delta^0)$)
in the coefficient functions.
Note that at LO $\theta$ only depends on $\delta M^2$
(as well as on the LEC $M_0^2$).

Equations~\eqref{eq:fmwi8}--\eqref{eq:fmwi0} amount to
four identities ($n\in\{\eta,\eta^\prime\}$ and
$a\in\{8,0\}$) but we have six unknown functions on the right hand
sides (four $H_n^{q}$ and two $a_n$). Nevertheless, we are able to
determine these unambiguously since the relations
should hold for any combination of $\delta M^2$
and $\overline{M}\vphantom{M}^2>3\delta M^2$. It is instructive
first to inspect the special case $\delta M=0$, where $\sin\theta=0$.
Then the substitution of the meson masses
simplifies:
$M^2_{\eta}=\mu_8^2=\overline{M}\vphantom{M}^2+\delta M^2/3+\ldots$,
$M_{\eta^\prime}^2=\mu_0^2=\overline{M}^2+M_0^2+\ldots$.
Moreover, in this limit there exist only two non-trivial relations
(for $F^8_{\eta}$ and $F^0_{\eta^\prime}$) since
$H_{\eta}^s+\sqrt{2}H_{\eta}^{\ell}=0$,
$H_{\eta^\prime}^{\ell}-\sqrt{2}H_{\eta}^s=0$ and $a_{\eta}=0$.
These three equalities are also obvious from the
respective Wick contractions.
In the vicinity of this limit, to leading order, these
combinations must be proportional to $\sin\theta$ or
to $\delta M^2$, where $\sin\theta\propto\delta M^2$
for small $\theta$.
One can easily see that in the limit $\delta M=0$, to leading
order also $H_{\eta}^{s}=-(\sqrt{2}/3)B_0F=-\sqrt{2}H_{\eta^\prime}^s$
holds and therefore, $a_{\eta^\prime}=\sqrt{2/3}\,F M_{\eta^\prime}^2=
\sqrt{2/3}\,F (M_0^2+\overline{M}^2)$.
Starting from these identifications and sorting all terms accordingly,
where in the end we substitute back the GMOR relations and eliminate
$B_0$, gives
\begin{align}
  \label{eq:mpslighteta}
  m_{\ell} H_{\eta}^{\ell} & = \cos\theta F \Bigg\{
                             \begin{aligned}[t]
                               & \left(\overline{M}\vphantom{M}^{2} - \frac{\delta M^{2}}{3}\right)\frac{\sqrt{3}}{12} \left[\left(1 - 2 \Lambda_{1} + 4 \Lambda_{2}\right)\left(1 -\ \frac{1}{\cos\left(2 \theta\right)}\right)+ 2\right]\nonumber\\
& + M_{0}^{2}\left(1 - \frac{3 \overline{M}\vphantom{M}^{2}}{\delta M^{2}}\right) \frac{\sqrt{3}}{12}\left(1 - \Lambda_{1}\right)\left(1 - \frac{1}{\cos\left(2 \theta\right)}\right)\nonumber\\
  & + \frac{\overline{M}\vphantom{M}^{2} \delta M^{2}}{3}
    \begin{aligned}[t]
      & \left[\frac{5 \sqrt{3}}{3}\left(2 \frac{2 L_{8} - L_{5}}{F^{2}} + \frac{L_{5}}{F^{2}}\right)\left(1 + \frac{1}{5 \cos\left(2 \theta \right)}\right)\right.\nonumber\\
      &\left. + \sqrt{6} \frac{L_{5}}{F^{2}} \tan\theta \left(1 - \frac{1}{3\cos\left(2 \theta\right)}\right)\right]
    \end{aligned}\nonumber\\
    & + \overline{M}\vphantom{M}^{2}M_{0}^{2}
    \begin{aligned}[t]
      & \left[\frac{2L_{8} - L_{5}}{F^{2}}\left(\frac{5 \sqrt{3}}{3} + 2 \sqrt{3}\frac{\overline{M}\vphantom{M}^{2}}{\delta M^{2}}\right)\right. \nonumber\\
      &\left. - \left( \frac{\sqrt{6}}{3} \tan\theta + \sqrt{3}\right) \frac{L_{5}}{F^{2}}\right]\left(1 - \frac{1}{\cos\left(2 \theta\right)}\right)
    \end{aligned}\nonumber\\
  & - \frac{\delta M^{2} M_{0}^{2}}{3} \left[\frac{7\sqrt{3}}{3}\frac{2 L_{8} - L_{5}}{F^{2}} - \left(\sqrt{3} + \frac{\sqrt{6}}{3}\tan\theta\right)\frac{L_{5}}{F^{2}}\right]\left(1 - \frac{1}{\cos\left(2 \theta\right)}\right)\nonumber\\
  & - 2 \sqrt{3}\,\overline{M}\vphantom{M}^{4}\frac{2 L_{8} - L_{5}}{F^{2}} \left(1 - \frac{1}{3\cos\left(2 \theta\right)}\right)\nonumber\\
  & - \delta M^{4}
  \begin{aligned}[t]
    & \left[\frac{4\sqrt{3}}{27}\frac{2 L_{8} - L_{5}}{F^{2}}\left(1 + \frac{1}{\cos\left(2\theta\right)}\right) + \frac{\sqrt{3}}{27}\frac{L_{5}}{F^{2}} \left(5 + \frac{1}{\cos\left(2\theta\right)}\right)\right.\nonumber\\
    &\left.\left. + \frac{\sqrt{6}}{9}\tan\theta \frac{L_{5}}{F^{2}}\left(1 - \frac{1}{3\cos\left(2\theta\right)}\right)\right]\right\},\\
  \end{aligned}
  \end{aligned}\\
\end{align}
\begin{align}
  m_{s} H_{\eta}^{s} = & \cos \theta F \Bigg\{
                         \begin{aligned}[t]
                           & - \frac{\sqrt{6}}{8}\left(\overline{M}\vphantom{M}^{2} + \frac{2}{3}\delta M^{2}\right)\left[\left(1 - \frac{\Lambda_{1}}{3} + \frac{2 \Lambda_{2}}{3}\right)\left(1 - \frac{1}{\cos\left( 2\theta \right)}\right) + \frac{4}{3 \cos\left(2\theta\right)}\right] \nonumber\\
  & - \frac{\sqrt{6}}{12} M_{0}^{2} \left(1 - \Lambda_{1}\right)\left(1 + \frac{3 \overline{M}\vphantom{M}^{2}}{2 \delta M^{2}}\right)\left(1 - \frac{1}{\cos\left(2\theta\right)}\right)\nonumber\\
  & + \frac{\overline{M}\vphantom{M}^{2} \delta M^{2}}{3}
  \begin{aligned}[t]
    & \left[ \frac{5\sqrt{6}}{3}\frac{2 L_{8} - L_{5}}{F^{2}}\left(1 + \frac{1}{5\cos\left(\theta\right)}\right)\right. \nonumber\\
    & - \frac{L_{5}}{F^{2}}
      \begin{aligned}[t]
        & \left(\frac{2\sqrt{6}}{3}\left(1 + \frac{1}{2\cos\left(2 \theta\right)}\right)\right.\\
        & +\sqrt{3} \tan\theta \left(1 + \frac{1}{3\cos\left(2\theta\right)}\right)\Bigg)\Bigg] +
        \end{aligned}\\
      \end{aligned}
  \end{aligned} \nonumber\displaybreak\\
\phantom{m_{s} H_{\eta}^{s} =} & \phantom{\cos \theta F \Bigg\{}
  \begin{aligned}
  & + \overline{M}\vphantom{M}^{2}M_{0}^{2}\left[\left(\frac{5\sqrt{6}}{6} + \sqrt{6}\frac{\overline{M}\vphantom{M}^{2}}{\delta M^{2}}\right)\frac{2L_{8} - L_{5}}{F^{2}} - \frac{\sqrt{3}}{3}\frac{L_{5}}{F^{2}} \tan\theta \right]\left(1-\frac{1}{\cos\left(2\theta\right)}\right)\nonumber\\
  & - \delta M^{2}M_{0}^{2}  \left(\frac{7\sqrt{6}}{18}\frac{2L_{8} - L_{5}}{F^{2}} + \frac{2\sqrt{3}}{9}\frac{L_{5}}{F^{2}}\tan\theta \right)\left(1 - \frac{1}{\cos\left(2\theta\right)}\right)\nonumber\\
  & + \sqrt{6} \overline{M}\vphantom{M}^{4} \frac{2 L_{8} - L_{5}}{F^{2}} \left(1 + \frac{1}{3\cos\left(2\theta\right)}\right)\nonumber\\
  & - \frac{\delta M^{4}}{9}
    \begin{aligned}[t]
      & \left[\frac{\sqrt{6}}{6}\frac{2 L_{8} - L_{5}}{F^{2}}\left(31 + \frac{13}{\cos\left(2 \theta\right)}\right) + \frac{2 \sqrt{6}}{3} \frac{L_{5}}{F^{2}}\left(2 + \frac{1}{\cos\left(2 \theta\right)}\right)\right.\\
      & + 2 \sqrt{3} \tan\theta \frac{L_{5}}{F^{2}}\left(1 + \frac{1}{3\cos\left(2 \theta\right)}\right)\Bigg]\Bigg\},\nonumber\\
  \end{aligned}
  \end{aligned}
      \label{eq:mpsstrangeeta}\\
\end{align}
\begin{align}
  m_{\ell} H_{\eta^{\prime}}^{\ell} & = \sin\theta F \Bigg\{
                                      \begin{aligned}[t]
                                        & \left(\overline{M}\vphantom{M}^{2} - \frac{\delta M^{2}}{3}\right)\frac{\sqrt{3}}{12} \left[\left(1 - 2 \Lambda_{1} + 4 \Lambda_{2}\right)\left(1 + \frac{1}{\cos\left(2 \theta\right)}\right)+ 2\right]\nonumber\\
  & + M_{0}^{2}\left(1 - \frac{3 \overline{M}\vphantom{M}^{2}}{\delta M^{2}}\right) \frac{\sqrt{3}}{12}\left(1 - \Lambda_{1}\right)\left(1 + \frac{1}{\cos\left(2 \theta\right)}\right)\nonumber\\
  & +\frac{\overline{M}\vphantom{M}^{2} \delta M^{2}}{3}
  \begin{aligned}[t]
    &\left[\frac{5 \sqrt{3}}{3}\left(2 \frac{2 L_{8} - L_{5}}{F^{2}} + \frac{L_{5}}{F^{2}}\right)\left(1 - \frac{1}{5 \cos\left(2 \theta \right)}\right)\right.\\
    & - \sqrt{6} \frac{L_{5}}{F^{2}} \cot\theta \left(1 + \frac{1}{3\cos\left(2 \theta\right)}\right)\Bigg]\nonumber\\
  \end{aligned}\\
  & + \overline{M}\vphantom{M}^{2} M_{0}^{2}
  \begin{aligned}[t]
    & \left[\frac{2L_{8} - L_{5}}{F^{2}}\left(\frac{5 \sqrt{3}}{3} + 2 \sqrt{3} \frac{\overline{M}\vphantom{M}^{2}}{\delta M^{2}}\right)\right. \\
      &\left. + \left( \frac{\sqrt{6}}{3} \cot\theta - \sqrt{3}\right) \frac{L_{5}}{F^{2}}\right]\left(1 + \frac{1}{\cos\left(2 \theta\right)}\right)\\
    \end{aligned}\\
  & - \frac{\delta M^{2}M_{0}^{2}}{3} \left[\frac{7\sqrt{3}}{3}\frac{2 L_{8} - L_{5}}{F^{2}} + \left(-\sqrt{3} + \frac{\sqrt{6}}{3}\cot\theta\right)\frac{L_{5}}{F^{2}}\right]\left(1 + \frac{1}{\cos\left(2 \theta\right)}\right)\nonumber\\
  & - 2 \sqrt{3} \overline{M}\vphantom{M}^{4}\frac{2 L_{8} - L_{5}}{F^{2}} \left(1 + \frac{1}{3\cos\left(2 \theta\right)}\right)\nonumber\\
  & - \frac{\sqrt{3}}{27} \delta M^{4}
  \begin{aligned}[t]
    & \left[ 4 \frac{2 L_{8} - L_{5}}{F^{2}}\left(1 - \frac{1}{\cos\left(2 \theta\right)}\right) + \left(5 - \frac{1}{\cos\left(2 \theta\right)}\right) \frac{L_{5}}{F^{2}}\right.\\
    & \left. - 3 \sqrt{2} \cot\theta \frac{L_{5}}{F^{2}}\left(1 + \frac{1}{3\cos\left( 2 \theta \right)}\right) \right]\Bigg\},
  \end{aligned}
  \end{aligned} \label{eq:mpslightetaprime}\\
\end{align}
\begin{align}
  m_{s}H_{\eta^{\prime}}^{s} & = \sin\theta F \Bigg\{
                                 \begin{aligned}[t]
                                   & - \frac{\sqrt{6}}{8}\left(\overline{M}\vphantom{M}^{2} + \frac{2}{3}\delta M^{2}\right)\left[\left(1 - \frac{\Lambda_{1}}{3} + \frac{2 \Lambda_{2}}{3}\right)\left(1 + \frac{1}{\cos\left( 2\theta \right)}\right) - \frac{4}{3 \cos\left(2\theta\right)}\right] \nonumber\\
  & - \frac{\sqrt{6}}{12} M_{0}^{2} \left(1 - \Lambda_{1}\right)\left(1 + \frac{3 \overline{M}\vphantom{M}^{2}}{2 \delta M^{2}}\right)\left(1 + \frac{1}{\cos\left(2\theta\right)}\right)\nonumber\\
  & + \frac{\overline{M}\vphantom{M}^{2} \delta M^{2}}{3}
    \begin{aligned}[t]
      & \left[ \frac{5\sqrt{6}}{3}\frac{2 L_{8} - L_{5}}{F^{2}}\left(1 - \frac{1}{5\cos\left(\theta\right)}\right)\right. \nonumber\\
      & - \left.\frac{L_{5}}{F^{2}}\left(\frac{2\sqrt{6}}{3}\left(1 - \frac{1}{2\cos\left(2 \theta\right)}\right) - \sqrt{3} \cot\theta \left(1 - \frac{1}{3\cos\left(2\theta\right)}\right)\right)\right]\nonumber\\
      \end{aligned}\\
  & + \overline{M}\vphantom{M}^{2}M_{0}^{2}\left[\left(\frac{5\sqrt{6}}{6} + \sqrt{6}\frac{\overline{M}\vphantom{M}^{2}}{\delta M^{2}}\right)\frac{2L_{8} - L_{5}}{F^{2}} + \frac{\sqrt{3}}{3}\frac{L_{5}}{F^{2}} \cot\theta \right]\left(1+\frac{1}{\cos\left(2\theta\right)}\right)\nonumber\\
  & - \delta M^{2} M_{0}^{2} \left(\frac{7\sqrt{6}}{18}\frac{2L_{8} - L_{5}}{F^{2}} - \frac{2\sqrt{3}}{9}\frac{L_{5}}{F^{2}}\cot\theta \right)\left(1 + \frac{1}{\cos\left(2\theta\right)}\right)\nonumber\\
  & + \sqrt{6}\,\overline{M}\vphantom{M}^{4} \frac{2 L_{8} - L_{5}}{F^{2}} \left(1 - \frac{1}{3\cos\left(2\theta\right)}\right)\nonumber\\
  & - \frac{\delta M^{4}}{9}
  \begin{aligned}[t]
    & \left[\frac{\sqrt{6}}{6}\frac{2 L_{8} - L_{5}}{F^{2}}\left(31 - \frac{13}{\cos\left(2 \theta\right)}\right) + \frac{4 \sqrt{6}}{6} \frac{L_{5}}{F^{2}}\left(2 - \frac{1}{\cos\left(2 \theta\right)}\right)\right.\\
    & \left.- 2 \sqrt{3} \frac{L_{5}}{F^{2}}\cot\theta\left(1 - \frac{1}{3 \cos\left(2 \theta\right)}\right)\right]\Bigg\}.\\
    \end{aligned}
  \end{aligned}
      \label{eq:mpsstrangeetaprime}\\
\end{align}
The NLO
expressions contain $M^4$ terms because the left hand sides
are already proportional to quark masses. One remark is in order:
the $\eta^\prime$ matrix elements all start with $\sin\theta$. This
does not mean that they vanish in the limit $\delta M=0$ (where
$\sin\theta=0$). For small $\delta M^2$ one can expand
$\sin\theta=-\sqrt{2}\delta M^2/(3M_0^2)+\ldots$, which cancels against
a term $\propto \overline{M}\vphantom{M}^2 M_0^2/\delta M^2$,
resulting in the limiting case discussed above.

Finally, the gluonic matrix elements can be obtained via the singlet
AWI:
\begin{align}
  a_{\eta} & = \cos\theta F \Bigg\{
               \begin{aligned}[t]
                 & \frac{\sqrt{3}}{6} \overline{M}\vphantom{M}^{2}\left(-1 + \Lambda_{1}- 2 \Lambda_{2}+3\frac{M_{0}^{2}}{\delta M^{2}}\left(1 - \Lambda_{1}\right)\right)\left(1 - \frac{1}{\cos\left(2\theta\right)}\right)\nonumber\\
  & + \frac{\sqrt{3}}{9}\delta M^{2}\left[\left(-\Lambda_{1} + 2 \Lambda_{2}\right)\left(1 - \frac{1}{\cos\left(2\theta\right)}\right) + 2\right]\nonumber\\
  & - \frac{4\sqrt{3}}{9} \overline{M}\vphantom{M}^{2}\delta M^{2}\left[\frac{2 L_{8} - L_{5}}{F^{2}}\left(5 + \frac{1}{\cos\left(2\theta\right)}\right) + 3 \frac{L_{5}}{F^{2}}\right]\nonumber\\
  & - \frac{2\sqrt{3}}{3} \overline{M}\vphantom{M}^{2} M_{0}^{2}\left( 5 \frac{2 L_{8} - L_{5}}{F^{2}} - 2 \frac{L_{5}}{F^{2}}\right)\left(1 - \frac{1}{\cos\left(2\theta\right)}\right)\nonumber\\
  & + \frac{2\sqrt{3}}{9} \delta M^{2} M_{0}^{2} \left(7 \frac{2 L_{8} - L_{5}}{F^{2}} - 4 \frac{L_{5}}{F^{2}}\right)\left(1 - \frac{1}{\cos\left(2\theta\right)}\right)\nonumber\\
  & + \frac{4\sqrt{3}}{3} \overline{M}\vphantom{M}^{4}\left(1 - 3\frac{M_{0}^{2}}{\delta M^{2}}\right)\frac{2 L_{8}-L_{5}}{F^{2}}\left(1 - \frac{1}{\cos\left(2\theta\right)}\right)\nonumber\\
  & + \frac{\sqrt{3}}{27} \delta M^{4}\left[\frac{2L_{8}- L_{5}}{F^{2}}\left(26 + \frac{14}{\cos\left(2\theta\right)}\right)+ 8 \frac{L_{5}}{F^{2}}\right]\Bigg\}\nonumber\\
  \end{aligned}\\
           & + \sin\theta F \Bigg\{
             \begin{aligned}[t]
               & - \frac{\sqrt{6}}{3} \overline{M}\vphantom{M}^{2}\left[1 + \frac{\Lambda_{1}}{2 \cos\left(2 \theta\right)} + \Lambda_{2}\left(1 - \frac{1}{\cos\left(2 \theta\right)}\right)\right]\nonumber\\
  & - \frac{\sqrt{6}}{18}\delta M^{2} \left(1 + \frac{\Lambda_{1}}{2}\right)\left(1 + \frac{1}{\cos\left(2 \theta\right)}\right)\nonumber\\
  & - \frac{\sqrt{6}}{6} M_{0}^{2} \left(1 - \frac{\Lambda_{1}}{2}\right)\left(1 - \frac{1}{\cos\left(2\theta\right)}\right)\nonumber\\
  & - \frac{4 \sqrt{6}}{9}\frac{L_{5}}{F^{2}} \overline{M}\vphantom{M}^{2}\delta M^{2}
  + \frac{4\sqrt{6}}{3} \overline{M}\vphantom{M}^{2} M_{0}^{2} \frac{L_{5}}{F^{2}} \left(1 - \frac{1}{\cos\left(2\theta\right)}\right) - \frac{4\sqrt{6}}{3} \frac{L_{5}}{F^{2}} \overline{M}\vphantom{M}^{4}\nonumber\\
  & - \frac{2\sqrt{6}}{3}\delta M^{4}\left[\left(1 + \frac{1}{3\cos\left(2\theta\right)}\right)\frac{2L_{8} - L_{5}}{F^{2}} - \frac{4}{9} \frac{L_{5}}{F^{2}}\right]\Bigg\},\\
  \end{aligned}\label{eq:aetanlo}\\
\end{align}
\begin{align}
  a_{\eta^{\prime}} & = \cos\theta F \Bigg\{
                      \begin{aligned}[t]
                        & \frac{\sqrt{6}}{3}\overline{M}\vphantom{M}^{2}\left[1 - \frac{\Lambda_{1}}{2 \cos\left(2 \theta\right)} + \Lambda_{2}\left(1 + \frac{1}{\cos\left(2\theta\right)}\right)\right]\nonumber\\
  & + \frac{\sqrt{6}}{18} \delta M^{2} \left(1 + \frac{\Lambda_{1}}{2}\right)\left(1 - \frac{1}{\cos\left(2\theta\right)}\right)\nonumber\\
  & + \frac{\sqrt{6}}{6}M_{0}^{2}\left(1 - \frac{\Lambda_{1}}{2}\right)\left(1 + \frac{1}{\cos\left(2\theta\right)}\right)\nonumber\\
  & + \frac{4\sqrt{6}}{9}\overline{M}\vphantom{M}^{2}\delta M^{2}\frac{L_{5}}{F^{2}} - \frac{4 \sqrt{6}}{3}\overline{M}\vphantom{M}^{2}M_{0}^{2}\frac{L_{5}}{F^{2}}\left(1 + \frac{1}{\cos\left(2\theta\right)}\right)\nonumber\\
  & + \frac{4 \sqrt{6}}{3} \overline{M}\vphantom{M}^{4} \frac{L_{5}}{F^{2}}\nonumber\\
  & + \frac{2\sqrt{6}}{3}\delta M^{4}\left[ \frac{2 L_{8}- L_{5}}{F^{2}} \left(1 - \frac{1}{3 \cos\left(2\theta\right)}\right)- \frac{4}{9} \frac{L_{5}}{F^{2}}\right]
    \Bigg\}\nonumber\\
    \end{aligned}\\
                    & + \sin\theta F \Bigg\{
                      \begin{aligned}[t]
  & - \frac{\sqrt{3}}{6}\overline{M}\vphantom{M}^{2}\left(1 - \Lambda_{1} + 2 \Lambda_{2}-3\frac{M_{0^{2}}}{\delta M^{2}}\left(1-\Lambda_{1}\right)\right)\left(1 + \frac{1}{\cos\left(2\theta\right)}\right)\nonumber\\
  & + \frac{2\sqrt{3}}{9}\delta M^{2} \left[1+\left(-\frac{\Lambda_{1}}{2}+\Lambda_{2}\right)\left(1 + \frac{1}{\cos\left(2\theta\right)}\right)\right]\nonumber\\
  & - \frac{4\sqrt{3}}{9}\overline{M}\vphantom{M}^{2}\delta M^{2}\left[\frac{2 L_{8} - L_{5}}{F^{2}}\left(5 - \frac{1}{\cos\left(2\theta\right)}\right)+ 3 \frac{L_{5}}{F^{2}}\right]\nonumber\\
  & - \frac{2 \sqrt{3}}{3} \overline{M}\vphantom{M}^{2}M_{0}^{2} \left(5 \frac{2 L_{8}- L_{5}}{F^{2}} - 2 \frac{L_{5}}{F^{2}}\right)\left(1 + \frac{1}{\cos\left(2\theta\right)}\right)\nonumber\\
  & + \frac{2 \sqrt{3}}{9}\delta M^{2} M_{0}^{2} \left(7 \frac{2 L_{8} - L_{5}}{F^{2}}- 4 \frac{L_{5}}{F^{2}}\right)\left(1 + \frac{1}{\cos\left(2\theta\right)}\right)\nonumber\\
  & + \frac{4 \sqrt{3}}{3} \overline{M}\vphantom{M}^{4} \frac{2 L_{8}-L_{5}}{F^{2}}\left(1 - 3 \frac{M_{0}^{2}}{\delta M^{2}}\right)\left(1 + \frac{1}{\cos\left(2\theta\right)}\right)\nonumber\\
  & + \frac{2\sqrt{3}}{27}\delta M^{4}\left[\frac{2 L_{8} - L_{5}}{F^{2}}\left(13 - \frac{7}{\cos\left(2 \theta\right)}\right)+ 4 \frac{L_{5}}{F^{2}}\right]
  \Bigg\}.
  \end{aligned}\label{eq:aetaprimenlo}\\
\end{align}
Note that $a_{\eta}$ vanishes for $\theta=0$, as it should.

The LO results can easily be obtained, setting $L_5=L_8=\Lambda_1=
\Lambda_2=0$. These only depend on the LECs $F$ and $M_0$. The
mass mixing angle $\theta$ at LO is
given in eqs.~\eqref{eq:massangle}
and~\eqref{eq:lomasses}--\eqref{eq:lomasses3} as a function of the
$\delta M^2$ and $M_0^2$. Above, we use the NLO expression
for $\theta$, eqs.~\eqref{eq:massangle}
and~\eqref{eq:nlomu8}--\eqref{eq:nlomu80}.

\section{The leading order fit\label{sec:lofitres}}\label{sec:lo}
\begin{figure}
  \includegraphics[width=\linewidth]{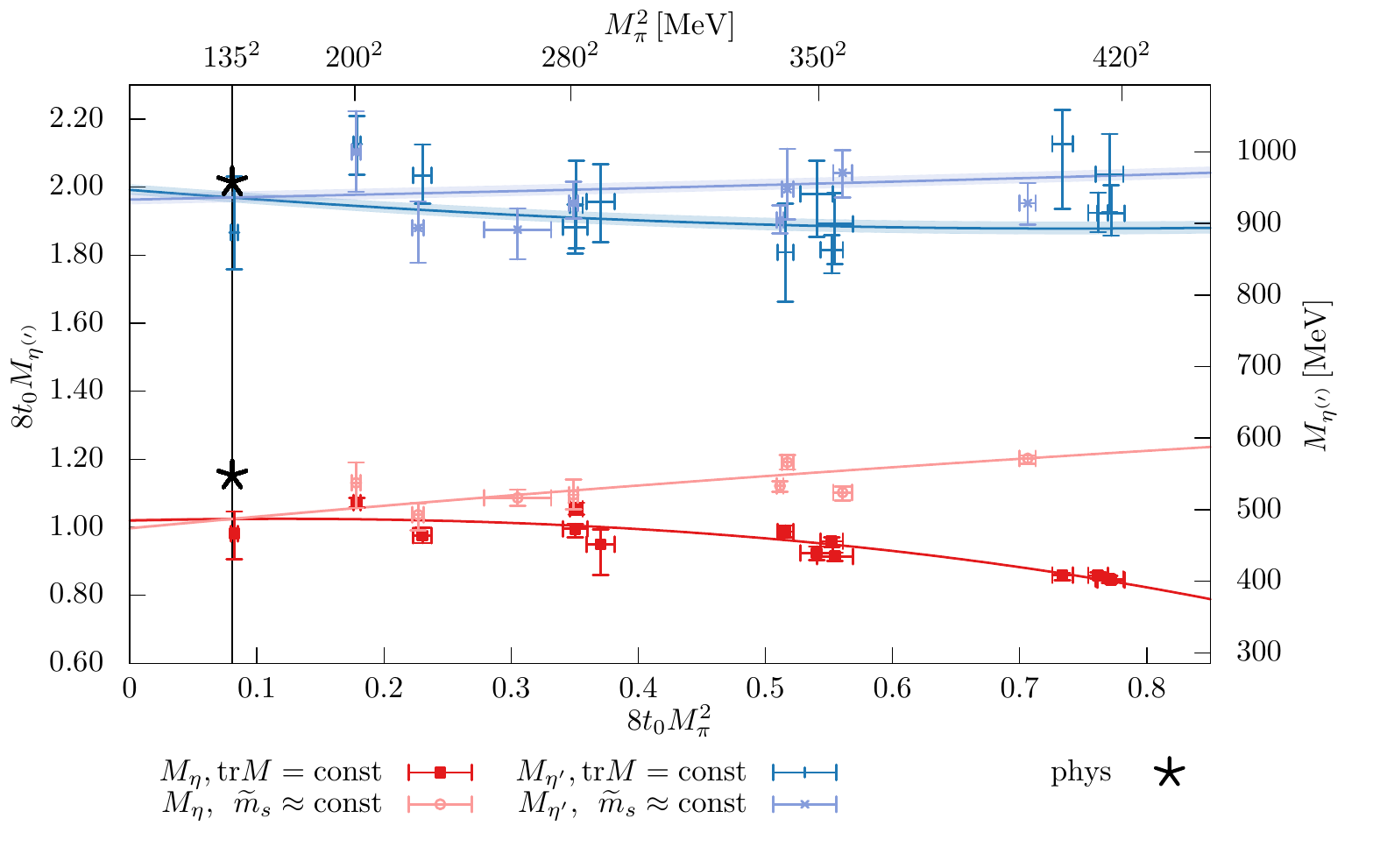}
  \caption{\label{fig:lomassfit}LO parametrization of the masses of
    the $\eta$ and the
    $\eta^\prime$ mesons for our two trajectories in the
    quark mass plane. The data are corrected for lattice spacing
  effects according to the fit.}
\end{figure}
We show in fig.~\ref{fig:lomassfit}
the analogue of fig.~\ref{fig:nlodecfit} for our fit to the
LO parametrization
eqs.~\eqref{eq:mminus}--\eqref{eq:etaetaprmasses}
and~\eqref{eq:lomasses}--\eqref{eq:lomasses3}.
No simultaneous
fit of the masses and the decay constants can be carried out since
$F_{\eta}^8\neq F_{\eta}^0$ and $F_{\eta}^0\neq -F_{\eta^\prime}^8$, which
is why only the masses are included. The continuum parametrization shown
depend on a single parameter,
the LEC $M_0\approx 785\,\mathrm{MeV}$, as detailed
in sec.~\ref{sec:massdec}. Although the raw lattice data fall
onto continuous curves (see the upper panel of fig.~\ref{fig:nlodecfit}),
lattice correction terms $\propto a^2\delta M^2$ had to be added for each
particle to obtain $\chi^2/N_{\rm df} \approx 91/41$.
In fig.~\ref{fig:lomassfit} the shifted data are shown, along with
the continuum limit curves that depend only on the parameter $M_0$.
Since in our NLO fits no lattice spacing dependent terms had to be
added for the masses, we suspect that in the LO case the $a^2\delta M^2$ terms
mostly compensate for a shortcoming of the continuum parametrization.

\section{Continuum limit fit parameters}\label{sec:contlimitparams}
\begin{table*}
  \centering
\pgfplotstabletypeset[
  column type=,
  begin table={\begin{tabularx}{0.8\textwidth}{l X | X X X X}},
  end table={\end{tabularx}},
  columns={fitid,chi2perdof,dA,dtildeA,deltacA,fA},
  every head row/.style = {before row=\toprule, after row=\midrule},
  every last row/.style = {after row=[1ex]\bottomrule},
  font=\small,
  every column/.style = {string type},
  columns/fitid/.style = {column name=id},
  columns/chi2perdof/.style = {string type, column name=$\chi^2/N_{\mathrm{df}}$},
  columns/dA/.style = {string type, column name=$d_A^l$},
  columns/dtildeA/.style = {string type, column name=$\tilde{d}^l_A$},
  columns/deltacA/.style = {string type, column name=$\delta c_A^l$},
  columns/fA/.style = {string type, column name=$f_A^l$},
  ]{\nloextrap}
\caption{Fit results for the unknown $\mathcal{O}(a)$ improvement coefficients,
  see eqs.~\eqref{eq:linearoctetaeffects},
  \eqref{eq:linearsingletaeffects} and~\eqref{eq:imprcoefsinglparam}.
  The fit ids are defined in tab.~\ref{tab:fitids}.\label{tab:linearfx}}
\end{table*}
\begin{table*}
  \centering
\pgfplotstabletypeset[
  column type=,
  begin table={\begin{tabularx}{\textwidth}{l X | X X X X X X}},
  end table={\end{tabularx}},
  columns={fitid,chi2perdof,Af080,Af081,Af082,Af180,Af181,Af182},
  every head row/.style = {before row=\toprule, after row=\midrule},
  every last row/.style = {after row=[1ex]\bottomrule},
  font=\small,
  every column/.style = {string type},
  columns/fitid/.style = {column name=id},
  columns/chi2perdof/.style = {string type, column name=$\chi^2/N_{\mathrm{df}}$},
  columns/Af080/.style = {string type, column name=$l_{F^8_{\eta}}$},
  columns/Af081/.style = {string type, column name=$m_{F^8_{\eta}}$},
  columns/Af082/.style = {string type, column name=$n_{F^8_{\eta}}$},
  columns/Af180/.style = {string type, column name=$l_{F^8_{\eta^\prime}}$},
  columns/Af181/.style = {string type, column name=$m_{F^8_{\eta^\prime}}$},
  columns/Af182/.style = {string type, column name=$n_{F^8_{\eta^\prime}}$},
  ]{\nloextrap}
\caption{Fit parameters, accompanying quadratic lattice effects
  for the octet decay constants, see eq.~\eqref{eq:quadraticafx}.
  The fit ids are defined in tab.~\ref{tab:fitids}.\label{tab:f8contlimitfits}}
\end{table*}
\begin{table*}
  \centering
\pgfplotstabletypeset[
  column type=,
  begin table={\begin{tabularx}{\textwidth}{l X | X X X X X X}},
  end table={\end{tabularx}},
  columns={fitid,chi2perdof,Af000,Af001,Af002,Af100,Af101,Af102},
  every head row/.style = {before row=\toprule, after row=\midrule},
  every last row/.style = {after row=[1ex]\bottomrule},
  font=\small,
  every column/.style = {string type},
  columns/fitid/.style = {column name=id},
  columns/chi2perdof/.style = {string type, column name=$\chi^2/N_{\mathrm{df}}$},
  columns/Af000/.style = {string type, column name=$l_{F^0_{\eta}}$},
  columns/Af001/.style = {string type, column name=$m_{F^0_{\eta}}$},
  columns/Af002/.style = {string type, column name=$n_{F^0_{\eta}}$},
  columns/Af100/.style = {string type, column name=$l_{F^0_{\eta^\prime}}$},
  columns/Af101/.style = {string type, column name=$m_{F^0_{\eta^\prime}}$},
  columns/Af102/.style = {string type, column name=$n_{F^0_{\eta^\prime}}$},
  ]{\nloextrap}
  \caption{Fit parameters, accompanying quadratic lattice effects
  for the singlet decay constants, see eq.~\eqref{eq:quadraticafx}.
  The fit ids are defined in tab.~\ref{tab:fitids}.
  \label{tab:f0contlimitfits}}
\end{table*}
The parametrizations of lattice artefacts within our simultaneous
fits to the masses and decay constants are defined in
sec.~\ref{sec:contlimit}. In sec.~\ref{sec:fitselection}
we explain how 17 different parametrizations were selected.
These are enumerated and defined in tab.~\ref{tab:fitids}.
The six continuum limit fit parameters (LECs) for each of these
fits are given in tab.~\ref{tab:contlecs}. Here, in tab.~\ref{tab:linearfx}
we list the results for the unknown $\mathcal{O}(a)$
improvement coefficients within eqs.~\eqref{eq:linearoctetaeffects}
and~\eqref{eq:linearsingletaeffects}.
Their parametrizations are given in
eq.~\eqref{eq:imprcoefsinglparam}.
In tabs.~\ref{tab:f8contlimitfits} and~\ref{tab:f0contlimitfits}
we list the $\mathcal{O}(a^2)$ coefficients,
defined in eq.~\eqref{eq:quadraticafx} for both octet and
both singlet decay constants, respectively.
\clearpage

\section{Decay constants in various representations}
\label{app:decayresults}
We list the four decay constants,
in units of $(8t_0^{\rm ph})^{-1/2}$ and in MeV.
In tab.~\ref{tab:alldecresults} we collect the results of
our simultaneous fits to the masses and decay constants,
including the statistical and systematic errors, while in
tab.~\ref{tab:alldecresultspriors} the corresponding results are
shown, using the experimental masses of the $\eta$ and the
$\eta^\prime$ mesons as an additional input (priors).

In each table
we list the decay constants in both the octet/singlet and
the light/strange flavour bases. The conversion is given
in eq.~\eqref{eq:convertdecay}. In addition, we give the
parameters of the corresponding two-angle representations
eqs.~\eqref{eq:octsingletanglerep} and~\eqref{eq:flavouranglerep}.
All these results are given at four distinct
renormalization scales: $\mu=1\,\mathrm{GeV}$,
$\mu=2\,\mathrm{GeV}$, $\mu=10\,\mathrm{GeV}$ and
$\mu=\infty$, where all the values refer to the
$\overline{\mathrm{MS}}$ scheme for $N_f=3$ active flavours.
Only the octet decay constants $F^8_{\eta}$, $F^8_{\eta^\prime}$ and $F^8$
as well as the angles $\theta_8$ and $\theta_0$
are scale independent. We remark that in the latter case the scale
dependence cancels since
$\tan(\theta_0) = -F_\eta^0/F_{\eta^\prime}^0$.

\setlength\LTcapwidth{\textwidth} 
\begin{scriptsize}
\begin{center}
\begin{longtable}{ll|l|l}
  \caption{Decay constants in various representations and
    at several renormalization scales.\label{tab:alldecresults}}
  \endfirsthead
  \multicolumn{4}{l}{\textbf{Table~\ref{tab:alldecresults} (continued):} Decay constants at various scales.}\\
  \midrule
  \endhead
  \midrule\multicolumn{4}{r}{{Continued on next page}}
  \endfoot

  \bottomrule
    \endlastfoot
  \toprule
      \multicolumn{4}{l}{\textbf{octet/singlet basis}, state representation}\\\hline\rowcolor{gray!20}
      $F_\eta^8$ & & $  0.2219\left(\substack{18\\37}\right)_\mathrm{stat} \left(\substack{17\\24}\right)_a\left(\substack{10\\2}\right)_\chi (8 t_0^{\rm ph})^{-1/2}$ & $105.4\left(\substack{9\\1.8}\right)_\mathrm{stat} \left(\substack{9\\1.1}\right)_{\mathrm{syst}}(1.3)_{t_0}\,\mathrm{MeV}$ \\\hline\rowcolor{gray!20}
      $F_{\eta^\prime}^8$ & & $-0.0939\left(\substack{28\\100}\right)_\mathrm{stat} \left(\substack{84\\0}\right)_a\left(\substack{58\\82}\right)_\chi\,(8 t_0^{\rm ph})^{-1/2}$ & $-44.6\left(\substack{1.3\\4.8}\right)_\mathrm{stat} \left(\substack{4.9\\3.9}\right)_{\mathrm{syst}}(6)_{t_0}\,\mathrm{MeV}$ \\\hline\rowcolor{gray!20}
      $F_\eta^0\quad \mu =$& $\infty$ & $0.0224\left(\substack{53\\30}\right)_\mathrm{stat} \left(\substack{28\\0}\right)_a\left(\substack{5\\21}\right)_\chi\left(\substack{20\\8}\right)_{\mathrm{renorm}}\,(8 t_0^{\rm ph})^{-1/2}$ & $10.6\left(\substack{2.5\\1.4}\right)_\mathrm{stat} \left(\substack{1.4\\1.4}\right)_{\mathrm{syst}}(1)_{t_0}\,\mathrm{MeV}$ \\
                 & $1\,\mathrm{GeV}$ & $0.0255\left(\substack{60\\35}\right)_\mathrm{stat} \left(\substack{32\\0}\right)_a\left(\substack{5\\24}\right)_\chi\left(\substack{23\\9}\right)_{\mathrm{renorm}}\,(8 t_0^{\rm ph})^{-1/2}$ & $12.1\left(\substack{2.9\\1.6}\right)_\mathrm{stat} \left(\substack{1.6\\1.6}\right)_{\mathrm{syst}}(2)_{t_0}\,\mathrm{MeV}$ \\
                 & $2\,\mathrm{GeV}$ & $0.0241\left(\substack{57\\33}\right)_\mathrm{stat} \left(\substack{30\\0}\right)_a\left(\substack{5\\23}\right)_\chi\left(\substack{22\\9}\right)_{\mathrm{renorm}}\,(8 t_0^{\rm ph})^{-1/2}$ & $11.4\left(\substack{2.7\\1.5}\right)_\mathrm{stat} \left(\substack{1.5\\1.5}\right)_{\mathrm{syst}}(1)_{t_0}\,\mathrm{MeV}$ \\
                 & $10\,\mathrm{GeV}$ & $0.0233\left(\substack{55\\32}\right)_\mathrm{stat} \left(\substack{29\\0}\right)_a\left(\substack{5\\22}\right)_\chi\left(\substack{21\\8}\right)_{\mathrm{renorm}}\,(8 t_0^{\rm ph})^{-1/2}$ & $11.1\left(\substack{2.6\\1.5}\right)_\mathrm{stat} \left(\substack{1.5\\1.5}\right)_{\mathrm{syst}}(1)_{t_0}\,\mathrm{MeV}$ \\\hline\rowcolor{gray!20}
      $F_{\eta^\prime}^0\quad \mu =$ & $ \infty$ & $0.1974\left(\substack{14\\48}\right)_\mathrm{stat} \left(\substack{0\\31}\right)_a\left(\substack{4\\27}\right)_\chi\left(\substack{52\\26}\right)_{\mathrm{renorm}}\,(8 t_{0}^{\rm ph})^{-1/2}$ & $93.77\left(\substack{67\\2.29}\right)_\mathrm{stat} \left(\substack{1.24\\3.18}\right)_{\mathrm{syst}}(1.18)_{t_0}\,\mathrm{MeV}$ \\
                 & $1\,\mathrm{GeV}$ & $0.2247\left(\substack{16\\55}\right)_\mathrm{stat} \left(\substack{0\\36}\right)_a\left(\substack{5\\31}\right)_\chi\left(\substack{60\\29}\right)_{\mathrm{renorm}}\,(8 t_0^{\rm ph})^{-1/2}$ & $106.7\left(\substack{8\\2.6}\right)_\mathrm{stat} \left(\substack{1.4\\3.6}\right)_{\mathrm{syst}}(1.3)_{t_0}\,\mathrm{MeV}$ \\
                 & $2\,\mathrm{GeV}$ & $0.2122\left(\substack{15\\52}\right)_\mathrm{stat} \left(\substack{0\\34}\right)_a\left(\substack{5\\29}\right)_\chi\left(\substack{56\\28}\right)_{\mathrm{renorm}}\,(8 t_0^{\rm ph})^{-1/2}$ & $100.8\left(\substack{7\\2.5}\right)_\mathrm{stat} \left(\substack{1.3\\3.4}\right)_{\mathrm{syst}}(1.3)_{t_0}\,\mathrm{MeV}$ \\
                 & $10\,\mathrm{GeV}$ & $0.2051\left(\substack{15\\50}\right)_\mathrm{stat} \left(\substack{0\\33}\right)_a\left(\substack{5\\28}\right)_\chi\left(\substack{54\\27}\right)_{\mathrm{renorm}}\,(8 t_0^{\rm ph})^{-1/2}$ & $97.41\left(\substack{69\\2.38}\right)_\mathrm{stat} \left(\substack{1.29\\3.30}\right)_{\mathrm{syst}}(1.23)_{t_0}\,\mathrm{MeV}$ \\\midrule
      \multicolumn{4}{l}{\textbf{octet/singlet basis}, angle representation}\\\hline\rowcolor{gray!20}
      $F^8$ & & $0.2410\left(\substack{23\\16}\right)_\mathrm{stat} \left(\substack{11\\50}\right)_a\left(\substack{38\\12}\right)_\chi\,(8 t_0^{\rm ph})^{-1/2}$ & $114.5\left(\substack{1.1\\8}\right)_\mathrm{stat} \left(\substack{1.9\\2.5}\right)_{\mathrm{syst}}(1.4)_{t_0}\,\mathrm{MeV}$ \\\hline\rowcolor{gray!20}
      $\theta_8$ &  & $-0.400\left(\substack{9\\45}\right)_\mathrm{stat} \left(\substack{30\\0}\right)_a\left(\substack{24\\30}\right)_\chi$ & $-22.9\left(\substack{5\\2.6}\right)_{{\mathrm{stat}}}\left(\substack{2.2\\1.7}\right)_{{\mathrm{syst}}}^\circ$ \\\hline\rowcolor{gray!20}
      $F^0\quad \mu =$ & $ \infty$ & $0.1987\left(\substack{12\\42}\right)_\mathrm{stat} \left(\substack{0\\31}\right)_a\left(\substack{2\\29}\right)_\chi\left(\substack{54\\27}\right)_{\mathrm{renorm}}\,(8 t_0^{\rm ph})^{-1/2}$ & $94.37\left(\substack{57\\2.01}\right)_\mathrm{stat} \left(\substack{1.26\\3.26}\right)_{\mathrm{syst}}(1.19)_{t_0}\,\mathrm{MeV}$ \\
                 & $1\,\mathrm{GeV}$ & $0.2262\left(\substack{14\\48}\right)_\mathrm{stat} \left(\substack{0\\35}\right)_a\left(\substack{2\\33}\right)_\chi\left(\substack{62\\30}\right)_{\mathrm{renorm}}\,(8 t_0^{\rm ph})^{-1/2}$ & $107.4\left(\substack{6\\2.3}\right)_\mathrm{stat} \left(\substack{1.4\\3.7}\right)_{\mathrm{syst}}(1.4)_{t_0}\,\mathrm{MeV}$ \\
                 & $2\,\mathrm{GeV}$ & $0.2136\left(\substack{13\\46}\right)_\mathrm{stat} \left(\substack{0\\33}\right)_a\left(\substack{2\\31}\right)_\chi\left(\substack{58\\29}\right)_{\mathrm{renorm}}\,(8 t_0^{\rm ph})^{-1/2}$ & $101.5\left(\substack{6\\2.2}\right)_\mathrm{stat} \left(\substack{1.4\\3.5}\right)_{\mathrm{syst}}(1.3)_{t_0}\,\mathrm{MeV}$ \\
                 & $10\,\mathrm{GeV}$ & $0.2064\left(\substack{12\\44}\right)_\mathrm{stat} \left(\substack{0\\32}\right)_a\left(\substack{2\\30}\right)_\chi\left(\substack{56\\28}\right)_{\mathrm{renorm}}\,(8 t_0^{\rm ph})^{-1/2}$ & $98.04\left(\substack{59\\2.09}\right)_\mathrm{stat} \left(\substack{1.31\\3.39}\right)_{\mathrm{syst}}(1.24)_{t_0}\,\mathrm{MeV}$ \\\hline\rowcolor{gray!20}
      $\theta_0$ & & $-0.113\left(\substack{15\\29}\right)_\mathrm{stat} \left(\substack{0\\15}\right)_a\left(\substack{11\\3}\right)_\chi$ & $-6.5\left(\substack{9\\1.7}\right)_{{\mathrm{stat}}}\left(\substack{6\\9}\right)_{{\mathrm{syst}}}^\circ$ \\
      \pagebreak
      \multicolumn{4}{l}{\textbf{light/strange basis}, state representation}\\\hline\rowcolor{gray!20}
      $F_\eta^\ell\quad \mu =$ & $ \infty$ & $0.1464\left(\substack{27\\21}\right)_\mathrm{stat} \left(\substack{23\\2}\right)_a\left(\substack{3\\14}\right)_\chi\left(\substack{17\\7}\right)_{\mathrm{renorm}}\,(8 t_0^{\rm ph})^{-1/2}$ & $69.56\left(\substack{1.28\\1.01}\right)_\mathrm{stat} \left(\substack{1.13\\1.05}\right)_{\mathrm{syst}}(88)_{t_0}\,\mathrm{MeV}$ \\\
                 & $1\,\mathrm{GeV}$ & $0.1490\left(\substack{34\\23}\right)_\mathrm{stat} \left(\substack{25\\2}\right)_a\left(\substack{3\\17}\right)_\chi\left(\substack{19\\8}\right)_{\mathrm{renorm}}\,(8 t_0^{\rm ph})^{-1/2}$ & $70.76\left(\substack{1.60\\1.10}\right)_\mathrm{stat} \left(\substack{1.24\\1.21}\right)_{\mathrm{syst}}(89)_{t_0}\,\mathrm{MeV}$ \\
                 & $2\,\mathrm{GeV}$ & $0.1478\left(\substack{31\\23}\right)_\mathrm{stat} \left(\substack{24\\2}\right)_a\left(\substack{3\\16}\right)_\chi\left(\substack{18\\7}\right)_{\mathrm{renorm}}\,(8 t_0^{\rm ph})^{-1/2}$ & $70.21\left(\substack{1.46\\1.10}\right)_\mathrm{stat} \left(\substack{1.19\\1.13}\right)_{\mathrm{syst}}(89)_{t_0}\,\mathrm{MeV}$ \\
                 & $10\,\mathrm{GeV}$ & $0.1471\left(\substack{29\\22}\right)_\mathrm{stat} \left(\substack{23\\2}\right)_a\left(\substack{3\\15}\right)_\chi\left(\substack{17\\7}\right)_{\mathrm{renorm}}\,(8 t_0^{\rm ph})^{-1/2}$ & $69.89\left(\substack{1.37\\1.05}\right)_\mathrm{stat} \left(\substack{1.16\\1.09}\right)_{\mathrm{syst}}(88)_{t_0}\,\mathrm{MeV}$ \\\hline\rowcolor{gray!20}
      $F_{\eta^\prime}^\ell\quad \mu =$ & $ \infty$  & $0.1070\left(\substack{24\\88}\right)_\mathrm{stat} \left(\substack{26\\3}\right)_a\left(\substack{11\\44}\right)_\chi\left(\substack{37\\20}\right)_{\mathrm{renorm}}\,(8 t_0^{\rm ph})^{-1/2}$ & $50.82\left(\substack{1.16\\4.18}\right)_\mathrm{stat} \left(\substack{1.65\\2.75}\right)_{\mathrm{syst}}(64)_{t_0}\,\mathrm{MeV}$ \\
                 & $1\,\mathrm{GeV}$ & $0.1293\left(\substack{24\\92}\right)_\mathrm{stat} \left(\substack{23\\5}\right)_a\left(\substack{8\\43}\right)_\chi\left(\substack{43\\22}\right)_{\mathrm{renorm}}\,(8 t_0^{\rm ph})^{-1/2}$ & $61.41\left(\substack{1.15\\4.36}\right)_\mathrm{stat} \left(\substack{1.59\\2.93}\right)_{\mathrm{syst}}(78)_{t_0}\,\mathrm{MeV}$ \\
                 & $2\,\mathrm{GeV}$ & $0.1191\left(\substack{24\\89}\right)_\mathrm{stat} \left(\substack{25\\4}\right)_a\left(\substack{10\\44}\right)_\chi\left(\substack{41\\21}\right)_{\mathrm{renorm}}\,(8 t_0^{\rm ph})^{-1/2}$ & $56.57\left(\substack{1.16\\4.23}\right)_\mathrm{stat} \left(\substack{1.62\\2.84}\right)_{\mathrm{syst}}(71)_{t_0}\,\mathrm{MeV}$ \\
                 & $10\,\mathrm{GeV}$ & $0.1132\left(\substack{24\\88}\right)_\mathrm{stat} \left(\substack{26\\4}\right)_a\left(\substack{10\\44}\right)_\chi\left(\substack{39\\20}\right)_{\mathrm{renorm}}\,(8 t_0^{\rm ph})^{-1/2}$ & $53.79\left(\substack{1.16\\4.16}\right)_\mathrm{stat} \left(\substack{1.64\\2.80}\right)_{\mathrm{syst}}(68)_{t_0}\,\mathrm{MeV}$ \\\hline\rowcolor{gray!20}
      $F_{\eta}^s\quad \mu =$ & $ \infty$ & $\,(-0.1683\left(\substack{59\\30}\right)_\mathrm{stat} \left(\substack{36\\5}\right)_a\left(\substack{4\\19}\right)_\chi\left(\substack{12\\4}\right)_{\mathrm{renorm}}8 t_0^{\rm ph})^{-1/2}$ & $-79.93\left(\substack{2.78\\1.43}\right)_\mathrm{stat} \left(\substack{1.72\\1.09}\right)_{\mathrm{syst}}(1.01)_{t_0}\,\mathrm{MeV}$ \\
                 & $1\,\mathrm{GeV}$ & $-0.1665\left(\substack{63\\32}\right)_\mathrm{stat} \left(\substack{38\\4}\right)_a\left(\substack{5\\20}\right)_\chi\left(\substack{13\\5}\right)_{\mathrm{renorm}}\,(8 t_0^{\rm ph})^{-1/2}$ & $-79.08\left(\substack{3.01\\1.50}\right)_\mathrm{stat} \left(\substack{1.83\\1.17}\right)_{\mathrm{syst}}(100)_{t_0}\,\mathrm{MeV}$ \\
                 & $2\,\mathrm{GeV}$ & $-0.1673\left(\substack{61\\31}\right)_\mathrm{stat} \left(\substack{37\\5}\right)_a\left(\substack{5\\20}\right)_\chi\left(\substack{13\\5}\right)_{\mathrm{renorm}}\,(8 t_0^{\rm ph})^{-1/2}$ & $-79.46\left(\substack{2.91\\1.47}\right)_\mathrm{stat} \left(\substack{1.78\\1.13}\right)_{\mathrm{syst}}(1.00)_{t_0}\,\mathrm{MeV}$ \\
                 & $10\,\mathrm{GeV}$ & $-0.1678\left(\substack{60\\30}\right)_\mathrm{stat} \left(\substack{36\\5}\right)_a\left(\substack{4\\19}\right)_\chi\left(\substack{12\\4}\right)_{\mathrm{renorm}}\,(8 t_0^{\rm ph})^{-1/2}$ & $-79.69\left(\substack{2.85\\1.45}\right)_\mathrm{stat} \left(\substack{1.75\\1.11}\right)_{\mathrm{syst}}(1.01)_{t_0}\,\mathrm{MeV}$ \\\hline\rowcolor{gray!20}
      $F_{\eta^\prime}^s\quad \mu =$ & $ \infty$ & $0.1906\left(\substack{71\\31}\right)_\mathrm{stat} \left(\substack{0\\75}\right)_a\left(\substack{70\\63}\right)_\chi\left(\substack{38\\17}\right)_{\mathrm{renorm}}\,(8 t_0^{\rm ph})^{-1/2}$ & $90.55\left(\substack{3.37\\1.48}\right)_\mathrm{stat} \left(\substack{3.41\\5.00}\right)_{\mathrm{syst}}(1.14)_{t_0}\,\mathrm{MeV}$ \\
                 & $1\,\mathrm{GeV}$ & $0.2064\left(\substack{68\\32}\right)_\mathrm{stat} \left(\substack{0\\77}\right)_a\left(\substack{70\\65}\right)_\chi\left(\substack{42\\19}\right)_{\mathrm{renorm}}\,(8 t_0^{\rm ph})^{-1/2}$ & $98.04\left(\substack{3.21\\1.53}\right)_\mathrm{stat} \left(\substack{3.45\\5.20}\right)_{\mathrm{syst}}(1.24)_{t_0}\,\mathrm{MeV}$ \\
                 & $2\,\mathrm{GeV}$ & $0.1992\left(\substack{70\\33}\right)_\mathrm{stat} \left(\substack{0\\76}\right)_a\left(\substack{70\\64}\right)_\chi\left(\substack{40\\18}\right)_{\mathrm{renorm}}\,(8 t_0^{\rm ph})^{-1/2}$ & $94.62\left(\substack{3.30\\1.58}\right)_\mathrm{stat} \left(\substack{3.43\\5.11}\right)_{\mathrm{syst}}(1.20)_{t_0}\,\mathrm{MeV}$ \\
                 & $10\,\mathrm{GeV}$ & $0.1951\left(\substack{70\\33}\right)_\mathrm{stat} \left(\substack{0\\76}\right)_a\left(\substack{70\\64}\right)_\chi\left(\substack{39\\18}\right)_{\mathrm{renorm}}\,(8 t_0^{\rm ph})^{-1/2}$ & $92.65\left(\substack{3.34\\1.55}\right)_\mathrm{stat} \left(\substack{3.42\\5.05}\right)_{\mathrm{syst}}(1.17)_{t_0}\,\mathrm{MeV}$ \\\midrule
      \multicolumn{4}{l}{\textbf{light/strange basis}, angle representation}\\\hline\rowcolor{gray!20}
      $F^\ell\quad \mu =$ & $ \infty$ & $0.1814\left(\substack{12\\49}\right)_\mathrm{stat} \left(\substack{25\\1}\right)_a\left(\substack{0\\37}\right)_\chi\left(\substack{35\\17}\right)_{\mathrm{renorm}}\,(8 t_0^{\rm ph})^{-1/2}$ & $86.14\left(\substack{59\\2.32}\right)_\mathrm{stat} \left(\substack{1.45\\2.45}\right)_{\mathrm{syst}}(1.09)_{t_0}\,\mathrm{MeV}$ \\
                 & $1\,\mathrm{GeV}$ & $0.1972\left(\substack{12\\56}\right)_\mathrm{stat} \left(\substack{27\\0}\right)_a\left(\substack{0\\41}\right)_\chi\left(\substack{43\\21}\right)_{\mathrm{renorm}}\,(8 t_0^{\rm ph})^{-1/2}$ & $93.69\left(\substack{59\\2.67}\right)_\mathrm{stat} \left(\substack{1.62\\2.81}\right)_{\mathrm{syst}}(1.18)_{t_0}\,\mathrm{MeV}$\\
                 & $2\,\mathrm{GeV}$ & $0.1898\left(\substack{13\\53}\right)_\mathrm{stat} \left(\substack{27\\1}\right)_a\left(\substack{0\\39}\right)_\chi\left(\substack{39\\19}\right)_{\mathrm{renorm}}\,(8 t_0^{\rm ph})^{-1/2}$ & $90.16\left(\substack{62\\2.52}\right)_\mathrm{stat} \left(\substack{1.55\\2.65}\right)_{\mathrm{syst}}(1.14)_{t_0}\,\mathrm{MeV}$ \\
                 & $10\,\mathrm{GeV}$ & $0.1857\left(\substack{13\\51}\right)_\mathrm{stat} \left(\substack{26\\1}\right)_a\left(\substack{0\\38}\right)_\chi\left(\substack{37\\18}\right)_{\mathrm{renorm}}\,(8 t_0^{\rm ph})^{-1/2}$ & $88.20\left(\substack{61\\2.42}\right)_\mathrm{stat} \left(\substack{1.50\\2.55}\right)_{\mathrm{syst}}(1.11)_{t_0}\,\mathrm{MeV}$ \\\hline\rowcolor{gray!20}
      $\phi_\ell$ & $\mu = \infty$ & $0.631\left(\substack{12\\44}\right)_\mathrm{stat} \left(\substack{8\\6}\right)_a\left(\substack{8\\15}\right)_\chi\left(\substack{11\\6}\right)_{\mathrm{renorm}}$ & $36.2\left(\substack{7\\2.5}\right)_{{\mathrm{stat}}}\left(\substack{9\\1.0}\right)_{{\mathrm{syst}}}^{\circ}$ \\
                 & $1\,\mathrm{GeV}$ & $0.715\left(\substack{12\\42}\right)_\mathrm{stat} \left(\substack{5\\6}\right)_a\left(\substack{7\\11}\right)_\chi\left(\substack{11\\6}\right)_{\mathrm{renorm}}$ & $41.0\left(\substack{7\\2.4}\right)_{{\mathrm{stat}}}\left(\substack{8\\8}\right)_{{\mathrm{syst}}}^\circ$ \\
                 & $2\,\mathrm{GeV}$ & $0.678\left(\substack{12\\43}\right)_\mathrm{stat} \left(\substack{6\\6}\right)_a\left(\substack{7\\13}\right)_\chi\left(\substack{11\\6}\right)_{\mathrm{renorm}}$ & $38.8\left(\substack{7\\2.5}\right)_{{\mathrm{stat}}}\left(\substack{8\\9}\right)_{{\mathrm{syst}}}^\circ$ \\
                 & $10\,\mathrm{GeV}$ & $0.656\left(\substack{12\\44}\right)_\mathrm{stat} \left(\substack{7\\6}\right)_a\left(\substack{7\\14}\right)_\chi\left(\substack{11\\6}\right)_{\mathrm{renorm}}$ & $37.6\left(\substack{7\\2.5}\right)_{{\mathrm{stat}}}\left(\substack{8\\9}\right)_{{\mathrm{syst}}}^\circ$ \\\hline\rowcolor{gray!20}
      $F^s\quad \mu =$ & $ \infty$ & $0.2543\left(\substack{40\\28}\right)_\mathrm{stat} \left(\substack{0\\81}\right)_a\left(\substack{64\\34}\right)_\chi\left(\substack{20\\10}\right)_{\mathrm{renorm}}\,(8 t_0^{\rm ph})^{-1/2}$ & $120.8\left(\substack{1.9\\1.3}\right)_\mathrm{stat} \left(\substack{3.1\\4.3}\right)_{\mathrm{syst}}(1.5)_{t_0}\,\mathrm{MeV}$ \\
                 & $1\,\mathrm{GeV}$ & $0.2652\left(\substack{40\\30}\right)_\mathrm{stat} \left(\substack{0\\85}\right)_a\left(\substack{66\\38}\right)_\chi\left(\substack{24\\12}\right)_{\mathrm{renorm}}\,(8 t_0^{\rm ph})^{-1/2}$ & $126.0\left(\substack{1.9\\1.4}\right)_\mathrm{stat} \left(\substack{3.2\\4.6}\right)_{\mathrm{syst}}(1.6)_{t_0}\,\mathrm{MeV}$ \\
                 & $2\,\mathrm{GeV}$ & $0.2601\left(\substack{40\\29}\right)_\mathrm{stat} \left(\substack{0\\83}\right)_a\left(\substack{65\\36}\right)_\chi\left(\substack{22\\11}\right)_{\mathrm{renorm}}\,(8 t_0^{\rm ph})^{-1/2}$ & $123.6\left(\substack{1.9\\1.4}\right)_\mathrm{stat} \left(\substack{3.1\\4.4}\right)_{\mathrm{syst}}(1.6)_{t_0}\,\mathrm{MeV}$ \\
                 & $10\,\mathrm{GeV}$ & $0.2573\left(\substack{40\\29}\right)_\mathrm{stat} \left(\substack{0\\82}\right)_a\left(\substack{64\\35}\right)_\chi\left(\substack{21\\11}\right)_{\mathrm{renorm}}\,(8 t_0^{\rm ph})^{-1/2}$ & $122.2\left(\substack{1.9\\1.4}\right)_\mathrm{stat} \left(\substack{3.1\\4.4}\right)_{\mathrm{syst}}(1.5)_{t_0}\,\mathrm{MeV}$ \\\hline\rowcolor{gray!20}
      $\phi_s\quad \mu =$ & $ \infty$       & $0.723\left(\substack{13\\30}\right)_\mathrm{stat} \left(\substack{20\\3}\right)_a\left(\substack{22\\13}\right)_\chi\left(\substack{6\\13}\right)_{\mathrm{renorm}}$ & $41.4\left(\substack{7\\1.7}\right)_{{\mathrm{stat}}}\left(\substack{1.7\\1.1}\right)_{{\mathrm{syst}}}^\circ$ \\
                 & $1\,\mathrm{GeV}$  & $0.679\left(\substack{12\\29}\right)_\mathrm{stat} \left(\substack{18\\4}\right)_a\left(\substack{22\\11}\right)_\chi\left(\substack{6\\14}\right)_{\mathrm{renorm}}$ & $38.9\left(\substack{7\\1.7}\right)_{{\mathrm{stat}}}\left(\substack{1.7\\1.0}\right)_{{\mathrm{syst}}}^\circ$ \\
                 & $2\,\mathrm{GeV}$  & $0.699\left(\substack{12\\29}\right)_\mathrm{stat} \left(\substack{19\\4}\right)_a\left(\substack{22\\12}\right)_\chi\left(\substack{6\\14}\right)_{\mathrm{renorm}}$ & $40.0\left(\substack{7\\1.7}\right)_{{\mathrm{stat}}}\left(\substack{1.7\\1.1}\right)_{{\mathrm{syst}}}^\circ$ \\
                 & $10\,\mathrm{GeV}$ & $0.710\left(\substack{12\\30}\right)_\mathrm{stat} \left(\substack{19\\3}\right)_a\left(\substack{22\\12}\right)_\chi\left(\substack{6\\14}\right)_{\mathrm{renorm}}$ & $40.7\left(\substack{7\\1.7}\right)_{{\mathrm{stat}}}\left(\substack{1.7\\1.1}\right)_{{\mathrm{syst}}}^\circ$ \\
\end{longtable}
\pagebreak
\begin{longtable}{ll|l|l}
  \caption{Decay constants, using the experimental $\eta$ and
    $\eta^\prime$ masses as additional input (priors, see
    sec.~\ref{sec:lecresults}) in various representations and
    at several renormalization scales.\label{tab:alldecresultspriors}}
  \endfirsthead
  \multicolumn{4}{l}{\textbf{Table~\ref{tab:alldecresultspriors} (continued):} Decay constants at various scales.}\\
  \midrule
  \endhead
  \midrule\multicolumn{4}{r}{{Continued on next page}}
  \endfoot
  \bottomrule
  \endlastfoot
  \toprule
  \multicolumn{4}{l}{\textbf{octet/singlet basis}, state representation}\\\hline\rowcolor{gray!20}
      $F_\eta^8$ & & $0.2180\left(\substack{25\\32}\right)_\mathrm{stat} \left(\substack{15\\22}\right)_a\left(\substack{33\\0}\right)_\chi\,(8 t_0^{\rm ph})^{-1/2}$ & $103.5\left(\substack{1.2\\1.5}\right)_\mathrm{stat} \left(\substack{1.7\\1.1}\right)_{\mathrm{syst}}(1.3)_{t_0}\,\mathrm{MeV}$ \\\hline\rowcolor{gray!20}
      $F_{\eta^\prime}^8$ & & $-0.105\left(\substack{5\\9}\right)_\mathrm{stat} \left(\substack{7\\0}\right)_a\left(\substack{7\\0}\right)_\chi\,(8 t_0^{\rm ph})^{-1/2}$ & $-50.0\left(\substack{2.2\\4.2}\right)_\mathrm{stat} \left(\substack{4.8\\3}\right)_{\mathrm{syst}}(6)_{t_0}\,\mathrm{MeV}$ \\\hline\rowcolor{gray!20}
      $F_\eta^0\quad \mu =$ & $ \infty$ & $0.0276\left(\substack{34\\36}\right)_\mathrm{stat} \left(\substack{52\\0}\right)_a\left(\substack{0\\50}\right)_\chi\left(\substack{26\\11}\right)_{\mathrm{renorm}}\,(8 t_0^{\rm ph})^{-1/2}$ & $13.1\left(\substack{1.6\\1.7}\right)_\mathrm{stat} \left(\substack{2.5\\2.7}\right)_{\mathrm{syst}}(2)_{t_0}\,\mathrm{MeV}$ \\
                 & $1\,\mathrm{GeV}$ & $0.0314\left(\substack{39\\41}\right)_\mathrm{stat} \left(\substack{60\\0}\right)_a\left(\substack{0\\56}\right)_\chi\left(\substack{30\\12}\right)_{\mathrm{renorm}}\,(8 t_0^{\rm ph})^{-1/2}$ & $14.9\left(\substack{1.9\\2.0}\right)_\mathrm{stat} \left(\substack{2.9\\3.0}\right)_{\mathrm{syst}}(2)_{t_0}\,\mathrm{MeV}$ \\
                & $2\,\mathrm{GeV}$ & $0.0297\left(\substack{37\\39}\right)_\mathrm{stat} \left(\substack{56\\0}\right)_a\left(\substack{0\\53}\right)_\chi\left(\substack{28\\11}\right)_{\mathrm{renorm}}\,(8 t_0^{\rm ph})^{-1/2}$ & $14.1\left(\substack{1.7\\1.8}\right)_\mathrm{stat} \left(\substack{2.7\\2.9}\right)_{\mathrm{syst}}(2)_{t_0}\,\mathrm{MeV}$ \\
                & $10\,\mathrm{GeV}$ & $0.0287\left(\substack{36\\38}\right)_\mathrm{stat} \left(\substack{54\\0}\right)_a\left(\substack{0\\51}\right)_\chi\left(\substack{27\\11}\right)_{\mathrm{renorm}}\,(8 t_0^{\rm ph})^{-1/2}$ & $13.6\left(\substack{1.7\\1.8}\right)_\mathrm{stat} \left(\substack{2.6\\2.8}\right)_{\mathrm{syst}}(2)_{t_0}\,\mathrm{MeV}$ \\\hline\rowcolor{gray!20}
      $F_{\eta^\prime}^0\quad \mu =$ & $ \infty$ & $0.1941\left(\substack{15\\39}\right)_\mathrm{stat} \left(\substack{4\\10}\right)_a\left(\substack{34\\4}\right)_\chi\left(\substack{44\\23}\right)_{\mathrm{renorm}}\,(8 t_0^{\rm ph})^{-1/2}$ & $92.21\left(\substack{69\\1.85}\right)_\mathrm{stat} \left(\substack{1.96\\2.14}\right)_{\mathrm{syst}}(1.16)_{t_0}\,\mathrm{MeV}$ \\
                & $1\,\mathrm{GeV}$ & $0.2210\left(\substack{17\\44}\right)_\mathrm{stat} \left(\substack{5\\34}\right)_a\left(\substack{39\\4}\right)_\chi\left(\substack{50\\26}\right)_{\mathrm{renorm}}\,(8 t_0^{\rm ph})^{-1/2}$ & $105.0\left(\substack{8\\2.1}\right)_\mathrm{stat} \left(\substack{2.2\\2.9}\right)_{\mathrm{syst}}(1.3)_{t_0}\,\mathrm{MeV}$ \\
                & $2\,\mathrm{GeV}$ & $0.2087\left(\substack{16\\42}\right)_\mathrm{stat} \left(\substack{4\\32}\right)_a\left(\substack{37\\4}\right)_\chi\left(\substack{47\\24}\right)_{\mathrm{renorm}}\,(8 t_0^{\rm ph})^{-1/2}$ & $99.14\left(\substack{74\\1.99}\right)_\mathrm{stat} \left(\substack{2.11\\2.70}\right)_{\mathrm{syst}}(1.25)_{t_0}\,\mathrm{MeV}$ \\
                & $10\,\mathrm{GeV}$ & $0.2017\left(\substack{15\\40}\right)_\mathrm{stat} \left(\substack{4\\31}\right)_a\left(\substack{36\\4}\right)_\chi\left(\substack{45\\24}\right)_{\mathrm{renorm}}\,(8 t_0^{\rm ph})^{-1/2}$ & $95.79\left(\substack{72\\1.92}\right)_\mathrm{stat} \left(\substack{2.04\\2.61}\right)_{\mathrm{syst}}(1.21)_{t_0}\,\mathrm{MeV}$ \\\midrule
      \multicolumn{4}{l}{\textbf{octet/singlet basis}, angle representation}\\\hline\rowcolor{gray!20}
      $F^8$ & & $0.2421\left(\substack{22\\26}\right)_\mathrm{stat} \left(\substack{8\\50}\right)_a\left(\substack{32\\12}\right)_\chi\,(8 t_0^{\rm ph})^{-1/2}$ & $115.0\left(\substack{1.1\\1.2}\right)_\mathrm{stat} \left(\substack{1.6\\2.4}\right)_{\mathrm{syst}}(1.5)_{t_0}\,\mathrm{MeV}$ \\\hline\rowcolor{gray!20}
      $\theta_8$ &  & $-0.450\left(\substack{21\\36}\right)_\mathrm{stat} \left(\substack{24\\0}\right)_a\left(\substack{29\\0}\right)_\chi\left(\substack{1\\5}\right)_{\mathrm{renorm}}$ & $-25.8\left(\substack{1.2\\2.1}\right)_{\mathrm{stat}}\left(\substack{2.2\\0.3}\right)_{\mathrm{syst}} ^\circ$ \\\hline\rowcolor{gray!20}
      $F^0\quad \mu =$ & $ \infty$ & $0.1961\left(\substack{13\\37}\right)_\mathrm{stat} \left(\substack{12\\6}\right)_a\left(\substack{28\\7}\right)_\chi\left(\substack{47\\24}\right)_{\mathrm{renorm}}\,(8 t_0^{\rm ph})^{-1/2}$ & $93.14\left(\substack{62\\1.75}\right)_\mathrm{stat} \left(\substack{1.83\\2.27}\right)_{\mathrm{syst}}(1.18)_{t_0}\,\mathrm{MeV}$ \\
                 & $1\,\mathrm{GeV}$ & $0.2232\left(\substack{15\\42}\right)_\mathrm{stat} \left(\substack{13\\26}\right)_a\left(\substack{32\\8}\right)_\chi\left(\substack{53\\27}\right)_{\mathrm{renorm}}\,(8 t_0^{\rm ph})^{-1/2}$ & $106.0\left(\substack{7\\2.0}\right)_\mathrm{stat} \left(\substack{2.1\\2.9}\right)_{\mathrm{syst}}(1.3)_{t_0}\,\mathrm{MeV}$ \\
                 & $2\,\mathrm{GeV}$ & $0.2108\left(\substack{14\\40}\right)_\mathrm{stat} \left(\substack{12\\25}\right)_a\left(\substack{30\\8}\right)_\chi\left(\substack{50\\26}\right)_{\mathrm{renorm}}\,(8 t_0^{\rm ph})^{-1/2}$ & $100.1\left(\substack{7\\1.9}\right)_\mathrm{stat} \left(\substack{2.0\\2.7}\right)_{\mathrm{syst}}(1.3)_{t_0}\,\mathrm{MeV}$ \\
                 & $10\,\mathrm{GeV}$ & $0.2037\left(\substack{14\\38}\right)_\mathrm{stat} \left(\substack{12\\24}\right)_a\left(\substack{29\\7}\right)_\chi\left(\substack{49\\25}\right)_{\mathrm{renorm}}\,(8 t_0^{\rm ph})^{-1/2}$ & $96.76\left(\substack{65\\1.82}\right)_\mathrm{stat} \left(\substack{1.90\\2.60}\right)_{\mathrm{syst}}(1.22)_{t_0}\,\mathrm{MeV}$ \\\hline\rowcolor{gray!20}
      $\theta_0$ & & $-0.141\left(\substack{18\\20}\right)_\mathrm{stat} \left(\substack{0\\27}\right)_a\left(\substack{27\\0}\right)_\chi$ & $-8.1\left(\substack{1.0\\1.1}\right)_{\mathrm{stat}}\left(\substack{1.5\\1.5}\right)_{\mathrm{syst}}^\circ$ \\\midrule
      \multicolumn{4}{l}{\textbf{light/strange basis}, state representation}\\\hline\rowcolor{gray!20}
      $F_\eta^\ell\quad \mu =$ & $ \infty$ & $0.1484\left(\substack{20\\21}\right)_\mathrm{stat} \left(\substack{31\\0}\right)_a\left(\substack{0\\21}\right)_\chi\left(\substack{21\\9}\right)_{\mathrm{renorm}}\,(8 t_0^{\rm ph})^{-1/2}$ & $70.48\left(\substack{93\\99}\right)_\mathrm{stat} \left(\substack{1.53\\1.42}\right)_{\mathrm{syst}}(89)_{t_0}\,\mathrm{MeV}$ \\\
                 & $1\,\mathrm{GeV}$ & $0.1515\left(\substack{23\\26}\right)_\mathrm{stat} \left(\substack{37\\0}\right)_a\left(\substack{0\\27}\right)_\chi\left(\substack{24\\10}\right)_{\mathrm{renorm}}\,(8 t_0^{\rm ph})^{-1/2}$ & $71.96\left(\substack{1.11\\1.22}\right)_\mathrm{stat} \left(\substack{1.81\\1.71}\right)_{\mathrm{syst}}(91)_{t_0}\,\mathrm{MeV}$ \\
                 & $2\,\mathrm{GeV}$ & $0.1501\left(\substack{22\\23}\right)_\mathrm{stat} \left(\substack{34\\0}\right)_a\left(\substack{0\\24}\right)_\chi\left(\substack{22\\9}\right)_{\mathrm{renorm}}\,(8 t_0^{\rm ph})^{-1/2}$ & $71.29\left(\substack{1.03\\1.10}\right)_\mathrm{stat} \left(\substack{1.68\\1.58}\right)_{\mathrm{syst}}(90)_{t_0}\,\mathrm{MeV}$ \\
                 & $10\,\mathrm{GeV}$ & $0.1493\left(\substack{21\\22}\right)_\mathrm{stat} \left(\substack{33\\0}\right)_a\left(\substack{0\\23}\right)_\chi\left(\substack{22\\9}\right)_{\mathrm{renorm}}\,(8 t_0^{\rm ph})^{-1/2}$ & $70.90\left(\substack{98\\1.04}\right)_\mathrm{stat} \left(\substack{1.61\\1.50}\right)_{\mathrm{syst}}(90)_{t_0}\,\mathrm{MeV}$ \\\hline\rowcolor{gray!20}
      $F_{\eta^\prime}^\ell\quad \mu =$ & $ \infty$  & $0.09773\left(\substack{362\\686}\right)_\mathrm{stat} \left(\substack{441\\12}\right)_a\left(\substack{364\\0}\right)_\chi\left(\substack{285\\164}\right)_{\mathrm{renorm}}\,(8 t_0^{\rm ph})^{-1/2}$ & $46.42\left(\substack{1.72\\3.26}\right)_\mathrm{stat} \left(\substack{2.83\\1.35}\right)_{\mathrm{syst}}(59)_{t_0}\,\mathrm{MeV}$ \\
                 & $1\,\mathrm{GeV}$ & $0.1197\left(\substack{40\\71}\right)_\mathrm{stat} \left(\substack{44\\2}\right)_a\left(\substack{36\\0}\right)_\chi\left(\substack{33\\19}\right)_{\mathrm{renorm}}\,(8 t_0^{\rm ph})^{-1/2}$ & $56.84\left(\substack{1.88\\3.38}\right)_\mathrm{stat} \left(\substack{2.86\\1.59}\right)_{\mathrm{syst}}(72)_{t_0}\,\mathrm{MeV}$ \\
                 & $2\,\mathrm{GeV}$ & $0.1096\left(\substack{40\\70}\right)_\mathrm{stat} \left(\substack{44\\1}\right)_a\left(\substack{36\\0}\right)_\chi\left(\substack{31\\18}\right)_{\mathrm{renorm}}\,(8 t_0^{\rm ph})^{-1/2}$ & $52.08\left(\substack{1.90\\3.32}\right)_\mathrm{stat} \left(\substack{2.84\\1.48}\right)_{\mathrm{syst}}(66)_{t_0}\,\mathrm{MeV}$ \\
                 & $10\,\mathrm{GeV}$ & $0.1039\left(\substack{38\\69}\right)_\mathrm{stat} \left(\substack{44\\1}\right)_a\left(\substack{36\\0}\right)_\chi\left(\substack{30\\17}\right)_{\mathrm{renorm}}\,(8 t_0^{\rm ph})^{-1/2}$ & $49.35\left(\substack{1.82\\3.29}\right)_\mathrm{stat} \left(\substack{2.83\\1.42}\right)_{\mathrm{syst}}(62)_{t_0}\,\mathrm{MeV}$ \\\hline\rowcolor{gray!20}
      $F_{\eta}^s\quad \mu =$ & $ \infty$ & $-0.1620\left(\substack{44\\40}\right)_\mathrm{stat} \left(\substack{48\\7}\right)_a\left(\substack{0\\56}\right)_\chi\left(\substack{16\\6}\right)_{\mathrm{renorm}}\,(8 t_0^{\rm ph})^{-1/2}$ & $-76.97\left(\substack{2.07\\1.91}\right)_\mathrm{stat} \left(\substack{2.29\\2.76}\right)_{\mathrm{syst}}(97)_{t_0}\,\mathrm{MeV}$ \\
                 & $1\,\mathrm{GeV}$ & $-0.1598\left(\substack{46\\42}\right)_\mathrm{stat} \left(\substack{52\\6}\right)_a\left(\substack{0\\60}\right)_\chi\left(\substack{18\\7}\right)_{\mathrm{renorm}}\,(8 t_0^{\rm ph})^{-1/2}$ & $-75.93\left(\substack{2.20\\2.00}\right)_\mathrm{stat} \left(\substack{2.49\\2.97}\right)_{\mathrm{syst}}(96)_{t_0}\,\mathrm{MeV}$ \\
                 & $2\,\mathrm{GeV}$ & $-0.1609\left(\substack{45\\41}\right)_\mathrm{stat} \left(\substack{50\\6}\right)_a\left(\substack{0\\58}\right)_\chi\left(\substack{17\\7}\right)_{\mathrm{renorm}}\,(8 t_0^{\rm ph})^{-1/2}$ & $-76.41\left(\substack{2.14\\1.96}\right)_\mathrm{stat} \left(\substack{2.40\\2.87}\right)_{\mathrm{syst}}(97)_{t_0}\,\mathrm{MeV}$ \\
                 & $10\,\mathrm{GeV}$ & $-0.1614\left(\substack{44\\41}\right)_\mathrm{stat} \left(\substack{49\\7}\right)_a\left(\substack{0\\57}\right)_\chi\left(\substack{16\\7}\right)_{\mathrm{renorm}}\,(8 t_0^{\rm ph})^{-1/2}$ & $-76.68\left(\substack{2.11\\1.93}\right)_\mathrm{stat} \left(\substack{2.35\\2.82}\right)_{\mathrm{syst}}(97)_{t_0}\,\mathrm{MeV}$ \\\hline\rowcolor{gray!20}
      $F_{\eta^\prime}^s\quad \mu =$ & $ \infty$ & $0.1980\left(\substack{59\\44}\right)_\mathrm{stat} \left(\substack{0\\74}\right)_a\left(\substack{24\\58}\right)_\chi\left(\substack{35\\16}\right)_{\mathrm{renorm}}\,(8 t_0^{\rm ph})^{-1/2}$ & $94.06\left(\substack{2.80\\2.08}\right)_\mathrm{stat} \left(\substack{1.37\\4.77}\right)_{\mathrm{syst}}(1.19)_{t_0}\,\mathrm{MeV}$ \\
                 & $1\,\mathrm{GeV}$ & $0.2135\left(\substack{58\\47}\right)_\mathrm{stat} \left(\substack{0\\76}\right)_a\left(\substack{27\\58}\right)_\chi\left(\substack{39\\18}\right)_{\mathrm{renorm}}\,(8 t_0^{\rm ph})^{-1/2}$ & $101.4\left(\substack{2.8\\2.2}\right)_\mathrm{stat} \left(\substack{1.5\\4.9}\right)_{\mathrm{syst}}(1.3)_{t_0}\,\mathrm{MeV}$ \\
                 & $2\,\mathrm{GeV}$ & $0.2064\left(\substack{58\\45}\right)_\mathrm{stat} \left(\substack{0\\75}\right)_a\left(\substack{25\\58}\right)_\chi\left(\substack{37\\17}\right)_{\mathrm{renorm}}\,(8 t_0^{\rm ph})^{-1/2}$ & $98.06\left(\substack{2.78\\2.15}\right)_\mathrm{stat} \left(\substack{1.45\\4.85}\right)_{\mathrm{syst}}(1.24)_{t_0}\,\mathrm{MeV}$ \\
                 & $10\,\mathrm{GeV}$ & $0.2024\left(\substack{59\\45}\right)_\mathrm{stat} \left(\substack{0\\74}\right)_a\left(\substack{25\\58}\right)_\chi\left(\substack{36\\17}\right)_{\mathrm{renorm}}\,(8 t_0^{\rm ph})^{-1/2}$ & $96.13\left(\substack{2.79\\2.12}\right)_\mathrm{stat} \left(\substack{1.41\\4.81}\right)_{\mathrm{syst}}(1.21)_{t_0}\,\mathrm{MeV}$ \\
      \pagebreak
      \multicolumn{4}{l}{\textbf{light/strange basis}, angle representation}\\\hline\rowcolor{gray!20}
      $F^\ell\quad \mu =$ & $ \infty$ & $0.1777\left(\substack{25\\40}\right)_\mathrm{stat} \left(\substack{55\\0}\right)_a\left(\substack{12\\4}\right)_\chi\left(\substack{33\\16}\right)_{\mathrm{renorm}}\,(8 t_0^{\rm ph})^{-1/2}$ &
                              $84.40\left(\substack{1.17\\1.91}\right)_\mathrm{stat} \left(\substack{2.78\\1.58}\right)_{\mathrm{syst}}(1.07)_{t_0}\,\mathrm{MeV}$ \\
                 & $1\,\mathrm{GeV}$ & $0.1931\left(\substack{26\\44}\right)_\mathrm{stat} \left(\substack{63\\0}\right)_a\left(\substack{13\\3}\right)_\chi\left(\substack{39\\19}\right)_{\mathrm{renorm}}\,(8 t_0^{\rm ph})^{-1/2}$ & $91.70\left(\substack{1.22\\2.08}\right)_\mathrm{stat} \left(\substack{3.17\\1.87}\right)_{\mathrm{syst}}(1.16)_{t_0}\,\mathrm{MeV}$\\
                 & $2\,\mathrm{GeV}$ & $0.1859\left(\substack{25\\43}\right)_\mathrm{stat} \left(\substack{59\\0}\right)_a\left(\substack{13\\3}\right)_\chi\left(\substack{36\\18}\right)_{\mathrm{renorm}}\,(8 t_0^{\rm ph})^{-1/2}$ & $88.28\left(\substack{1.20\\2.02}\right)_\mathrm{stat} \left(\substack{3.00\\1.74}\right)_{\mathrm{syst}}(1.12)_{t_0}\,\mathrm{MeV}$ \\
                 & $10\,\mathrm{GeV}$ & $0.1819\left(\substack{25\\42}\right)_\mathrm{stat} \left(\substack{57\\0}\right)_a\left(\substack{13\\3}\right)_\chi\left(\substack{35\\17}\right)_{\mathrm{renorm}}\,(8 t_0^{\rm ph})^{-1/2}$ & $86.38\left(\substack{1.19\\1.99}\right)_\mathrm{stat} \left(\substack{2.90\\1.66}\right)_{\mathrm{syst}}(1.09)_{t_0}\,\mathrm{MeV}$ \\\hline\rowcolor{gray!20}
      $\phi_\ell\quad \mu =$ & $ \infty$ & $0.582\left(\substack{19\\36}\right)_\mathrm{stat} \left(\substack{11\\4}\right)_a\left(\substack{20\\0}\right)_\chi\left(\substack{7\\5}\right)_{\mathrm{renorm}} $ & $33.3\left(\substack{1.1\\2.1}\right)_{\mathrm{stat}}\left(\substack{1.3\\0.4}\right)_{\mathrm{syst}}^\circ$ \\
                 & $1\,\mathrm{GeV}$ & $0.669\left(\substack{19\\34}\right)_\mathrm{stat} \left(\substack{6\\5}\right)_a\left(\substack{21\\0}\right)_\chi\left(\substack{6\\5}\right)_{\mathrm{renorm}}$ & $38.3\left(\substack{1.1\\1.9}\right)_{\mathrm{stat}}\left(\substack{1.3\\0.4}\right)_{\mathrm{syst}}^\circ$ \\
                 & $2\,\mathrm{GeV}$ & $0.631\left(\substack{19\\35}\right)_\mathrm{stat} \left(\substack{8\\4}\right)_a\left(\substack{20\\0}\right)_\chi\left(\substack{7\\5}\right)_{\mathrm{renorm}}$ & $36.2\left(\substack{1.1\\2.0}\right)_{\mathrm{stat}}\left(\substack{1.3\\0.4}\right)_{\mathrm{syst}}^\circ$ \\
                 & $10\,\mathrm{GeV}$ & $0.608\left(\substack{19\\35}\right)_\mathrm{stat} \left(\substack{9\\4}\right)_a\left(\substack{19\\0}\right)_\chi\left(\substack{7\\5}\right)_{\mathrm{renorm}}$ & $34.8\left(\substack{1.1\\2.0}\right)_{\mathrm{stat}}\left(\substack{1.3\\0.4}\right)_{\mathrm{syst}}^\circ$ \\\hline\rowcolor{gray!20}
      $F^s\quad \mu =$ & $ \infty$ & $0.2559\left(\substack{35\\34}\right)_\mathrm{stat} \left(\substack{0\\83}\right)_a\left(\substack{54\\25}\right)_\chi\left(\substack{17\\9}\right)_{\mathrm{renorm}}\,(8 t_0^{\rm ph})^{-1/2}$ & $121.5\left(\substack{1.7\\1.6}\right)_\mathrm{stat} \left(\substack{2.6\\4.2}\right)_{\mathrm{syst}}(1.5)_{t_0}\,\mathrm{MeV}$ \\
                 & $1\,\mathrm{GeV}$ & $0.2667\left(\substack{37\\35}\right)_\mathrm{stat} \left(\substack{0\\86}\right)_a\left(\substack{57\\27}\right)_\chi\left(\substack{20\\10}\right)_{\mathrm{renorm}}\,(8 t_0^{\rm ph})^{-1/2}$ & $126.7\left(\substack{1.8\\1.7}\right)_\mathrm{stat} \left(\substack{2.8\\4.4}\right)_{\mathrm{syst}}(1.6)_{t_0}\,\mathrm{MeV}$ \\
                 & $2\,\mathrm{GeV}$ & $0.2617\left(\substack{37\\34}\right)_\mathrm{stat} \left(\substack{0\\85}\right)_a\left(\substack{56\\26}\right)_\chi\left(\substack{19\\9}\right)_{\mathrm{renorm}}\,(8 t_0^{\rm ph})^{-1/2}$ & $124.3\left(\substack{1.7\\1.6}\right)_\mathrm{stat} \left(\substack{2.7\\4.3}\right)_{\mathrm{syst}}(1.6)_{t_0}\,\mathrm{MeV}$ \\
                 & $10\,\mathrm{GeV}$ & $0.2589\left(\substack{36\\34}\right)_\mathrm{stat} \left(\substack{0\\84}\right)_a\left(\substack{55\\26}\right)_\chi\left(\substack{18\\9}\right)_{\mathrm{renorm}}\,(8 t_0^{\rm ph})^{-1/2}$ & $123.0\left(\substack{1.7\\1.6}\right)_\mathrm{stat} \left(\substack{2.6\\4.3}\right)_{\mathrm{syst}}(1.6)_{t_0}\,\mathrm{MeV}$ \\\hline\rowcolor{gray!20}
      $\phi_s\quad \mu =$ & $ \infty$ & $0.686\left(\substack{17\\23}\right)_\mathrm{stat} \left(\substack{6\\4}\right)_a\left(\substack{24\\0}\right)_\chi\left(\substack{6\\14}\right)_{\mathrm{renorm}}$ & $39.3\left(\substack{1.0\\1.3}\right)_{\mathrm{stat}}\left(\substack{1.5\\0.8}\right)_{\mathrm{syst}}^\circ$ \\
                 & $1\,\mathrm{GeV}$  & $0.643\left(\substack{16\\22}\right)_\mathrm{stat} \left(\substack{6\\4}\right)_a\left(\substack{23\\0}\right)_\chi\left(\substack{6\\14}\right)_{\mathrm{renorm}}$ & $36.8\left(\substack{0.9\\1.3}\right)_{\mathrm{stat}}\left(\substack{1.4\\0.8}\right)_{\mathrm{syst}}^\circ$ \\
                 & $2\,\mathrm{GeV}$  & $0.662\left(\substack{17\\23}\right)_\mathrm{stat} \left(\substack{6\\4}\right)_a\left(\substack{23\\0}\right)_\chi\left(\substack{6\\14}\right)_{\mathrm{renorm}}$ & $37.9\left(\substack{1.0\\1.3}\right)_{\mathrm{stat}}\left(\substack{1.4\\0.8}\right)_{\mathrm{syst}}^\circ$ \\
                 & $10\,\mathrm{GeV}$ & $0.673\left(\substack{17\\23}\right)_\mathrm{stat} \left(\substack{6\\4}\right)_a\left(\substack{23\\0}\right)_\chi\left(\substack{6\\14}\right)_{\mathrm{renorm}}$ & $38.6\left(\substack{1.0\\1.3}\right)_{\mathrm{stat}}\left(\substack{1.4\\0.8}\right)_{\mathrm{syst}}^\circ$ \\
\end{longtable}
\end{center}
\end{scriptsize}

\bibliographystyle{jhep}
\bibliography{bibliography.bib}
\end{document}